# Maximizing Science in the Era of LSST: A Community-Based Study of Needed US OIR Capabilities

*A report on the Kavli Futures Symposium organized by NOAO and LSST*

Joan Najita (NOAO) and Beth Willman (LSST)
Douglas P. Finkbeiner (Harvard University)
Ryan J. Foley (University of California, Santa Cruz)
Suzanne Hawley (University of Washington)
Jeffrey Newman (University of Pittsburgh)
Gregory Rudnick (University of Kansas)
Joshua D. Simon (Carnegie Observatories)
David Trilling (Northern Arizona University)
Rachel Street (Las Cumbres Observatory Global Telescope Network)
Adam Bolton (NOAO)
Ruth Angus (University of Oxford)
Eric F. Bell (University of Michigan)
Derek Buzasi (Florida Gulf Coast University)
David Ciardi (IPAC, Caltech)
James R. A. Davenport (Western Washington University)
Will Dawson ((Lawrence Livermore National Laboratory)
Mark Dickinson (NOAO)
Alex Drlica-Wagner (Fermilab)
Jay Elias (NOAO)
Dawn Erb (University of Wisconsin-Milwaukee)
Lori Feaga (University of Maryland)
Wen-fai Fong (University of Arizona)
Eric Gawiser (The State University of New Jersey, Rutgers)
Mark Giampapa (National Solar Observatory)
Puragra Guhathakurta (University of California, Santa Cruz)
Jennifer L. Hoffman (University of Denver)
Henry Hsieh (Planetary Science Institute)
Elise Jennings (Fermilab)
Kathryn V. Johnston (Columbia University)
Vinay Kashyap (Harvard-Smithsonian CfA)
Ting S. Li (Texas A&M University)
Eric Linder (Lawrence Berkeley National Laboratory)
Rachel Mandelbaum (Carnegie Mellon University)
Phil Marshall (SLAC National Accelerator Laboratory)
Thomas Matheson (National Optical Astronomy Observatory)
Søren Meibom (Harvard-Smithsonian CfA)
Bryan W. Miller (Gemini Observatory)

*John O'Meara (Saint Michael's College)*
*Vishnu Reddy (University of Arizona)*
*Steve Ridgway (NOAO)*
*Constance M. Rockosi (University of California, Santa Cruz)*
*David J. Sand (Texas Tech University)*
*Chad Schafer (Carnegie Mellon University)*
*Sam Schmidt (UC Davis)*
*Branimir Sesar (Max Planck Institute for Astronomy)*
*Scott S. Sheppard (Carnegie Institute for Science/Department of Terrestrial Magnetism)*
*Cristina A. Thomas (Planetary Science Institute)*
*Erik J. Tollerud (Space Telescope Science Institute)*
*Jon Trump (Penn State, Hubble Fellow)*
*Anja von der Linden (SUNY)*
*Benjamin Weiner (Steward Observatory)*

# Executive Summary

The Large Synoptic Survey Telescope (LSST) will be a discovery machine for the astronomy and physics communities, revealing astrophysical phenomena from the Solar System to the outer reaches of the observable Universe. While many discoveries will be made using LSST data alone, maximizing the science from LSST will require ground-based optical-infrared (OIR) supporting capabilities, e.g., observing time on telescopes, instrumentation, computing resources, and other infrastructure. In the 2015 OIR System Report (*Optimizing the U.S. Optical and Infrared System in the Era of LSST*, Elmegreen et al. 2015), led by the National Research Council's Committee on Astronomy and Astrophysics and sponsored by the NSF, initial steps were taken toward identifying the required supporting capabilities that US astronomers will need to take full advantage of LSST discoveries.

In August 2015, NSF-AST asked NOAO and LSST to expand on the recommendations from the OIR System Report by carrying out a detailed quantitative study, organized around six to eight representative science cases, that (1) quantifies and prioritizes the resources needed to accomplish the science cases and (2) highlights ways that existing and planned resources could be positioned to accomplish the science goals.

In response to this charge, NOAO and LSST convened in February 2016 study groups that were drawn from the broad US community and focused around broad astrophysical themes. The groups selected six LSST-enabled science cases that connect closely with scientific priorities from the 2010 decadal surveys (*New Worlds, New Horizons* and *Vision and Voyages for Planetary Sciences in the Decade 2013–2022*) and carried out detailed studies of the OIR resources needed to accomplish these. The study culminated in a workshop, supported by The Kavli Foundation, that was held 2–4 May 2016 near Tucson at Biosphere 2. This document presents the detailed findings of each study group and the collective study recommendations that result from these studies. The results overlap closely with and expand on those of the OIR System Report.

The study recommendations listed below are grouped into 3 tiers by the development status of the resource. The recommendations relate to the capabilities that were found to have particularly high priority and high demand from multiple communities. Taking full advantage of LSST data also entails OIR system infrastructure developments as well as computing and analysis resources. We investigated these topics less fully than the telescope- and instrument-related resources, and further work is needed to specify these needs. We make several recommendations along these lines in a fourth tier below.

## Critical resources in need of a prompt development path

**Develop or obtain access to a highly multiplexed, wide-field optical multi-object spectroscopic capability on an 8m-class telescope, preferably in the Southern Hemisphere.** This high priority, high-demand capability is not currently available to the broad US community. Given the long lead time to develop any new capability, there is an urgent need to investigate possible development pathways now, so that the needed capabilities can be available in the LSST

era. Possibilities include implementing a new wide-field, massively multiplexed optical spectrograph on a Southern Hemisphere 6-8m telescope, e.g., as in the Southern Spectroscopic Survey Instrument, a project recommended for consideration by the DOE's Cosmic Visions panel (arxiv.org/abs/1604.07626 and arxiv.org/abs/1604.07821); open access to the PFS instrument on the Subaru telescope in order to propose and execute new large surveys; and alternatively, joining an international effort to implement a wide-field spectroscopic survey telescope (e.g., the Maunakea Spectroscopic Explorer at CFHT or a future ESO wide-field spectroscopic facility) if the facility will deliver data well before the end of the LSST survey.

## Critical resources that have a development path

**Deploy a broad wavelength coverage, moderate-resolution (R = 2000 or larger) OIR spectrograph on Gemini South.** The Gen 4#3 instrument is an ideal opportunity. It is critical that development plans for these capabilities proceed in a timely way so that the capabilities are available when LSST operations begin. A basic, workhorse instrument, deployed early in the LSST mission, is greatly preferred to a multi-mode instrument that arrives later in the mission. A wavelength range of at least 0.36–2.5 microns would provide the highest scientific impact.

**Ensure the development and early deployment of an alert broker, scalable to LSST.** Public broker(s), and supporting community data and filtering resources, are essential to select priority targets for follow-up. The development of an alert broker that can process the LSST alert stream has challenges beyond the field of astronomy alone. The key questions can be best addressed by computer scientists working with astronomers on this multi-disciplinary problem, and support is needed to enable effective collaboration across the relevant fields.

## Critical resources that exist today

**Support into the LSST era high-priority capabilities that are currently available.** Wide-field optical imaging (e.g., DECam on the Blanco 4m at CTIO) is one valuable, but relatively uncommon, capability, as is AO-fed diffraction limited imaging (e.g., NIFS on the 8m Gemini telescope). Other important capabilities are standard on many facilities. Those called out in this report include

- *single-object, multi-color imaging on < 5m facilities*
- *single-object R = 100–5000 spectroscopy on 3–5m facilities*

Support costs for these capabilities include those associated with routine operations as well as timely repair and refurbishment.

## Infrastructure resources and processes in need of timely development

**Support OIR system infrastructure developments that enable efficient follow-up programs.** Two of LSST's strengths are the large statistical samples it will produce and LSST's ability to provide rapid alerts for a wide variety of time domain phenomena. An efficient OIR system can capitalize on these strengths by (i) developing target and observation management software and increasing the availability of (ii) follow-up telescopes accessible in queue-scheduled modes, as well as (iii) data reduction pipelines that provide rapid access to data products. Following up large samples will be time and cost prohibitive if on-site observing is required and/or large programs and triage observations are not part of the time allocation infrastructure. To develop and prioritize community needs along these lines, we recommend a study aimed at developing a follow-up system for real-time, large-volume, time domain observations. As part of this study, discussions with the operators of observing facilities (e.g., through targeted workshops) are important in developing workable, cost-efficient procedures.

**Study and prioritize needs for computing, software, and data resources.** LSST is the most data-intensive project in the history of optical astronomy. To maximize the science from LSST, support is needed for (i) the development and deployment of data analysis and exploration tools that work at the scale of LSST; (ii) training for scientists at all career stages in LSST-related analysis techniques and computing technologies; (iii) cross-disciplinary workshops that facilitate the cross-pollination of ideas and tools between astronomy and other fields. We recommend a follow-on systematic study to prioritize community needs for computing, software, and data resources. The study should account for the capabilities that will be delivered by the LSST project and other efforts, the demands of forefront LSST-enabled research, and the opportunities presented by new technology.

**Continue community planning and development.** It is critical to continue the community-wide planning process, begun here, to motivate and review the development of the ground-based OIR System capabilities that will be needed to maximize LSST science. The current study focused primarily on instrumentation. Further work is needed to define the needs for observing infrastructure and computing, as described above. Regular review of progress (and lack thereof) in all of these areas is important to ensure the development of an OIR System that does maximize LSST science. Studies like these form the basis for a development roadmap and take a step in the direction envisioned by the Elmegreen committee that "a system organizing committee, chosen to represent all segments of the community ... would produce the prioritized plan. NSF would then solicit, review, and select proposals to meet those capabilities, within available funding."

# Table of Contents

# Chapter 1: Introduction and Background

## The Large Synoptic Survey Telescope—A discovery machine for the 2020s

The Large Synoptic Survey Telescope (LSST) is an 8.4m (6.7 m effective) telescope under construction on Cerro Pachón in Chile. Starting in 2022, LSST will conduct a 10-year survey over more than 18,000 deg$^2$ of the southern sky in six broadband optical filters (*ugrizy*). This wide field will be observed nearly 1000 times over the six filters, yielding a static census of ~ 40 billion objects as well as a dynamic time domain census over an unprecedented range of timescales and flux limits. This survey was designed to provide a dataset to uniquely enable the four LSST science objectives (measuring dark energy, cataloging the Solar System, exploring the transient sky, mapping the Milky Way) to be met and to engage the public in the exploration of the dynamic Universe.

LSST's primary deliverables to the scientific community will be
- *time domain event alerts, streamed within 60 seconds of the camera shutter closing*
- *a catalog of Solar System objects and their orbits*
- *annual catalog data releases that process all data taken up to the point of initiation*
- *a science user interface for accessing and interacting with the data*
- *infrastructure for community-generated data products*
- *documentation for all of the above*

LSST's Data Products Definition document provides details of the data deliverables for the scientific community. The LSST Overview paper and LSST's Science Requirements Document connect LSST's scientific goals with its technical requirements, and the LSST Science Book details numerous LSST science use cases.

## Community resources needed to maximize LSST science

LSST's dataset will be a discovery machine for the entire U.S. astronomical community, revealing astrophysical phenomena from the Solar System to the outer reaches of the visible Universe. Although much of this science can be accomplished using LSST data alone, optical and infrared photometric and spectroscopic follow-up of new discoveries on 4–30m-class telescopes will enormously extend LSST's potential. Community tools to filter the millions of time domain alerts ("event brokers") expected each night, and a national infrastructure to enable the timely follow-up of transient objects will also be needed to maximize the U.S. community's participation in LSST-era time domain science.

A report on *Optimizing the U.S. Ground-Based Optical and Infrared Astronomy System in the Era of LSST* was recently commissioned by the National Science Foundation) and the National Research Council of the National Academies (referred to hereafter as the OIR System Report). This report, also referred to as "the Elmegreen report," made several specific recommendations for the OIR capabilities needed to enable science in the LSST era, including wide-field, multiplexed spectroscopy on medium- to large-aperture telescopes; a high-throughput, moderate-resolution spectrograph on Gemini South; and U.S. involvement in one or more GSMT projects.

This report also made several recommendations about the infrastructure necessary to maximize the US community's participation in time domain science in the LSST era and about the importance of engaging the community in a planning process to crystallize the critical needs (Recommendations 2, 3, and 4a–d; see Appendix A).

## A study concept endorsed by NSF/AST and supported by The Kavli Foundation

To take the next steps, NOAO and LSST developed the concept for a follow-on study in concert with The Kavli Foundation. The study was proposed to the National Science Foundation's Astronomy division (NSF/AST), which endorsed the study concept. In August 2015, NSF/AST issued a letter asking NOAO and LSST to jointly conduct a study and workshop to build upon the recommendations of the OIR System Report (Appendix B).

In response, NOAO and LSST undertook a quantitative study that sharply focused on the ground-based OIR capabilities needed to accomplish six representative LSST-enabled science programs. The science programs were drawn from the scientific priorities outlined in *New Worlds, New Horizons, Vision and Voyages for Planetary Sciences in the Decade 2013–2022* and from community input (see Chapter 2 for details of the study process and the culminating workshop). In addition to a thorough investigation of six LSST-enabled science programs, we convened two additional groups charged with studying the time domain and computing infrastructure needed to maximize LSST science. This study culminated in a workshop, funded by The Kavli Foundation, at Biosphere 2 in Arizona, 2–4 May 2016. A subset of all study participants (40 of 53) attended the workshop.

This study's primary goal is to develop a report (this document), primarily organized by representative science program, that quantifies the resources needed for each program (including resources such as telescope apertures, wavelength ranges, instrument capabilities, number of nights, software, and computing resources). As a starting point for this study, we relied on the groundwork laid by the OIR System Report and the 2013 *NOAO Spectroscopy in the Era of LSST* report. In this document, we also highlight ways that existing and planned resources could be positioned to accomplish these science goals, and identify high-priority future investments for OIR infrastructure (see Chapter 9). For example, such implementation may include suggestions for efficient cross-field implementation (e.g., specific multiple programs that could efficiently be conducted simultaneously on massively multiplexed spectrographs), partnerships among facilities, and/or data sharing (i.e., archival data access).

We aim for this report to be informational to federal and private funding sources and public and private observatories in both the U.S. and international communities to (i) guide funding priorities and (ii) facilitate cross-facility and cross-science field collaborations.

## Acknowledgements

We thank Chris Martin and Miyoung Chun from The Kavli Foundation for the consideration and support for this study and workshop. We thank Jim Ulvestad, Nigel Sharp, and Ed Ajhar (NST/AST) for soliciting this study and for their attendance at the workshop. We thank Daniel Calabrese (LSST) for handling the logistics of the Biosphere 2 workshop, Iain Goodenow and Joseph Cockrum (LSST) for advising on IT support issues for the Biosphere 2 workshop, and Sandra Ortiz and Libby Petrick (LSST) for providing administrative support for the study and workshop. Jane Price (NOAO) provided administrative support for the workshop, and Sharon Hunt (NOAO) assisted with formatting and editing this report. Adam Bolton (NOAO), Cesar Briceño (CTIO), David Ciardi (IPAC/LSST), Mark Dickinson (NOAO), Jay Elias (SOAR), Bryan Miller (Gemini), Steve Ridgway (NOAO), and Rachel Street (LCOGT) provided summaries and supporting information on the existing and planned telescope/instrumentation, time domain, and computational infrastructure that the US community may be able to access. Thank you to Kathy Flanagan (STScI), Marcia Rieke (UA), Sidney Wolff (NOAO/LSST), and Dennis Zaritsky (UA) for supporting the workshop with your attendance and to Buell Januzzi (UA) for providing an overview of the OIR System Study and its key recommendations. We thank Biosphere 2 and its employees for their accommodating hospitality during the workshop and Gallery of Food for ensuring we were well fed.

Michael Strauss (Princeton) and Sidney Wolff (NOAO) provided an internal review of this report.

# Chapter 2: Study Structure and Process

To identify and assess quantitatively the resources needed to accomplish LSST-enabled science, we adopted a science-driven, community-based approach. The study was led by a Study Organizing Committee (SOC), which was responsible for fostering community participation in the study, organizing and leading primarily science-based study groups, and summarizing their findings and recommendations in this report.

## Study Organizing Committee

The SOC was drawn from the broad community with an eye toward diversity in terms of scientific expertise, gender, institution type and geographic location, and extent of previous engagement with LSST. Recruited by co-chairs Joan Najita (NOAO) and Beth Willman (LSST), SOC members were Douglas Finkbeiner (Harvard University), Ryan Foley (University of Illinois), Suzanne Hawley (University of Washington), Jeff Newman (University of Pittsburgh), Gregory Rudnick (University of Kansas), Josh Simon (Carnegie Observatories), and David Trilling (Northern Arizona University).

## Community Input

Following the assembly of the SOC in late fall 2015, the US community was invited to express interest in participating in a study group and/or to provide input on the supporting capabilities needed for the LSST-enabled science they each hope to accomplish. The study and the opportunity to provide input were advertised broadly through the
- *NOAO e-newsletter* Currents *(December 2015, January 2016)*
- *AAS News Digest (10 December 2015)*
- *NOAO Town Hall, LSST Town Hall, and LSST booth at the January 2016 meeting of the American Astronomical Society*

In addition, the SOC reached out to the LSST science collaborations, and to their scientific, institutional, and collaboration colleagues (e.g., Apache Point Observatory and SDSS communities) for their input. Given the short timescale for the study, the deadline for input and expressions of interest was 15 January 2016, with input received from over 100 individuals.

## Study Groups

Based on the community input received, six study groups were formed around broad astrophysical topics (Solar System, stars, the Milky Way and dwarf galaxies, explosive transients, galaxy formation and evolution, cosmology), each led by one or more members of the SOC. In recruiting the study participants, close attention was paid to diversity considerations identical to those described above.

From February through April 2016, the study groups selected compelling science questions drawn from the scientific priorities outlined in *New Worlds, New Horizons* and enabled by LSST. They then identified the resource needs for their topic and developed illustrative science cases that they worked out in quantitative detail to estimate the type and quantity of the capability required. Each group had the freedom to adjust the scope of their study, with the result that some studies were broad in scope and others more narrowly focused and detailed. Because LSST enables a broader range of science cases than those considered here, the full set of capabilities needed to maximize LSST science is broader than those described in this report.

The study group topics and members were
- Characterizing Primitive Small Bodies of the Solar System: David Trilling (study lead; Northern Arizona University), Lori Feaga (University of Maryland), Henry Hsieh (Planetary Science Institute), Vishnu Reddy (Planetary Science Institute), Scott Sheppard (Carnegie Institute of Washington/Department of Terrestrial Magnetism), Christina Thomas (Planetary Science Institute)
- Stellar Rotation and Magnetic Activity in the Field and Open Clusters: Suzanne Hawley (study lead; University of Washington), Ruth Angus (University of Oxford), Derek Buzasi (Florida Gulf Coast University), James Davenport (Western Washington University), Mark Giampapa (National Solar Observatory), Vinay Kashyap (Harvard-Smithsonian CfA), Soren Meibom (Harvard-Smithsonian CfA)
- Mapping Galaxies to Dark Matter Halos: Joshua D. Simon (study lead, Carnegie Observatories) and Douglas Finkbeiner (study lead; Harvard University), Eric F. Bell (University of Michigan), Alex Drlica-Wagner (Fermilab), Puragra Guhathakurta (University of California, Santa Cruz), Kathryn V. Johnston (Columbia University), Ting S. Li (Texas A&M University), Bryan W. Miller (Gemini Observatory), Constance M. Rockosi (University of California, Santa Cruz), Branimir Sesar (Max Planck Institute for Astronomy), and Erik J. Tollerud (Space Telescope Science Institute)
- Explosive Transients: Ryan J. Foley (study lead; University of California, Santa Cruz), Wen-fai Fong (University of Arizona), Jennifer Hoffman (University of Denver), Thomas Matheson (NOAO), David J. Sand (Texas Tech University), Rachel Street (Las Cumbres Observatory Global Telescope Network)
- Co-evolution of Baryons, Black Holes and Cosmic Structure: Gregory Rudnick (study lead; University of Kansas), Mark Dickinson (NOAO), Dawn Erb (University of Wisconsin–Milwaukee), John O'Meara (Saint Michael's College), Jon Trump (Penn State, Hubble Fellow), Benjamin Weiner (Steward Observatory), Adam Bolton (NOAO)
- Cosmology: Jeffrey Newman (study lead; University of Pittsburgh), Adam Bolton (NOAO), Will Dawson (Lawrence Livermore National Laboratory), Mark Dickinson (NOAO), Eric Gawiser (The State University of New Jersey, Rutgers), Elise Jennings (Fermilab), Eric Linder (Lawrence Berkeley National Laboratory), Rachel Mandelbaum (Carnegie Mellon University), Phil Marshall (SLAC National Accelerator Laboratory), Chad Schafer (Carnegie Mellon University), Sam Schmidt, Anja von der Linden (SLAC National Accelerator Laboratory), Ben Weiner (Steward Observatory)

To supplement the work of the science-focused study groups, two additional study groups were convened to review the OIR observing resources (telescopes and instruments) that are likely to be available in the era of LSST and to address common infrastructure issues:
- *Review of OIR Resources: Jay Elias (SOAR), Cesar Briceño (NOAO), Mark Dickinson (NOAO), Bryan Miller (Gemini Observatory), Stephen Ridgway (NOAO), Rachel Street (Las Cumbres Observatory Global Telescope Network)*
- *Time Domain Follow-up and Evolution of Observing Paradigms: Rachel Street (study*

*lead; Las Cumbres Observatory Global Telescope Network), Steve Ridgway (NOAO), David Ciardi (California Institute of Technology), Adam Bolton (NOAO), Chad Schafer, Jay Elias (SOAR), Tom Matheson (NOAO), Erik Tollerud (Space Telescope Science Institute), Bryan Miller (Gemini Observatory)*

## Workshop

Following the initial study period, the study participants gathered at a workshop to synthesize the results of the individual study groups and to develop a prioritization of needed capabilities for the study overall. Held at Biosphere 2 near Tucson, AZ, 2–4 May 2016, the workshop also provided an opportunity to discuss ways that existing and planned resources could be positioned to accomplish the study groups' science goals and to identify high-priority future investments for ground-based OIR infrastructure. In addition to the study groups, participants included observers from the NSF (Nigel Sharp, Ed Ajhar), science discussion facilitators (Marcia Rieke, Dennis Zaritsky), and additional advisors (Kathy Flanagan, Sidney Wolff).

The workshop format included detailed presentations by the study groups as well as breakout discussions on individual resources that were identified as needs by multiple study groups. The breakout groups explored the extent to which their needs could be met by a common capability. They refined the specifications for the capability, the demand for it (e.g., the amount of observing time on an instrument that would be needed to meet their science objectives), and possible paths to achieve it.

Breakout topics included both instruments (e.g., high-resolution spectroscopy, IFUs, optical multi-object spectroscopy, broad-wavelength high-throughput spectroscopy, optical imaging, near-infrared imaging), and other infrastructure (e.g., a time domain follow-up system, computing resources and statistical methods, observing modes, and reduction pipelines).

The preliminary work of the study groups was critical to the success of the workshop. Participants arrived with a detailed understanding of the flowdown from science to capabilities in their own areas, which allowed for quick progress in the breakout discussions. The workshop results revealed remarkable synergy, both among the resources needs of different science areas, and with the recommendations of earlier studies and reports.

## Report and Flowdown to Recommendations

Based on their discussions and input from the workshop, the study groups each contributed a chapter to this report that describes their science goals and the resulting quantitative flowdown to prioritized resource requirements. These chapters describe the capabilities needed to study the Milky Way and local dwarf galaxies (Chapter 3), explore the variable Universe, both explosive phenomena (Chapter 4) and variable stars (Chapter 5), investigate small bodies in the Solar System (Chapter 6), examine the evolution of galaxies (Chapter 7), and enable fundamental cosmological measurements (Chapter 8). Two additional chapters discuss the infrastructure needed to develop a time domain follow-up system (Chapter 9) and computing infrastructure needs to maximize LSST science (Chapter 10). The findings and recommendations that summarize the overall outcome of the study are described in Chapter 11.

Each science chapter (Chapter 3 to Chapter 8) concludes with tables summarizing the resource needs of the science cases described, both the capabilities and observing time required. The observing time requirements for the entire program are totaled at the bottom of the latter table, assuming 10 hours of observing time per night and 365 clear nights in an observing year, except where indicated. Of course, poor weather, instrument downtime, observatory maintenance, calibration observations, and other factors increase the calendar nights/years needed to complete each program. The reader can apply an appropriate scaling factor to obtain the true number of calendar years needed for any project; standard assumptions yield a factor of 1.5.

Based on our synthesis of the primary resource needs identified by the study groups (Chapter 11), we identified high-priority, high-demand capabilities that would significantly enrich and diversify the science reach of LSST. While our recommendations focus on these capabilities, the summary tables at the end of each science chapter provide a more complete list of the capabilities identified in this study.

# Chapter 3: Mapping Galaxies to Dark Matter Halos

*Probing Galaxy Formation and the Nature of Dark Matter and Gravity in the Local Group*

*Joshua D. Simon (Carnegie Observatories), Douglas P. Finkbeiner (Harvard University), Eric F. Bell (University of Michigan), Alex Drlica-Wagner (Fermilab), Puragra Guhathakurta (University of California, Santa Cruz), Kathryn V. Johnston (Columbia University), Ting S. Li (Texas A&M University), Bryan W. Miller (Gemini Observatory), Constance M. Rockosi (University of California, Santa Cruz), Branimir Sesar (Max Planck Institute for Astronomy), and Erik J. Tollerud (Space Telescope Science Institute)*

### Executive Summary

LSST's combination of depth, photometric accuracy, cadence, and time baseline will provide an unprecedented dataset for studying stars in the local universe. In uncrowded regions, LSST will detect individual red giants out to distances beyond 5 Mpc, while RR Lyrae variables will be visible throughout most of the Local Group. These observations will provide unique opportunities for constraining the nature of dark matter, the structure of the Milky Way's dark matter halo, and the fossil record of galaxy formation. Specifically, LSST will detect a complete sample of Milky Way satellite galaxies in the southern sky down to the expected limit of the galaxy luminosity function, map the tidal debris from disrupted dwarf galaxies and globular clusters out to distances of ~ 100 kpc, and provide an unprecedented view of stellar populations beyond 100 kpc. The dark matter content of LSST dwarf galaxies can be determined by measuring the radial velocities of their stars. These observations will place a lower limit on the mass of the dark matter particle and an upper limit on the dark matter self-interaction cross-section per unit mass and will be sensitive to modified gravity theories. Spectroscopy of stars in tidal streams can help determine the mass function of dark subhalos, which must be present if the LCDM model is correct. Six-dimensional phase space measurements can be obtained for RR Lyrae stars, stellar streams, and dwarf galaxies out to the virial radius of the Milky Way, which will determine the total mass and three-dimensional shape of the Galaxy's dark matter halo with an accuracy of better than 10%. Finally, multiplexed spectroscopic surveys of Milky Way halo stars out to 300 kpc at both medium and high spectral resolution will provide up to 14 (6 dynamical and 8 chemical) phase space dimensions per star, enabling the reconstruction of the entire accretion history of the Galaxy. The required capabilities for achieving these science goals include (1) multi-object optical spectroscopy at R ~ 5000 to R > 20,000 on an 8m or larger telescope, (2) wide-field optical imaging with medium-band filters on a 4m-class telescope, and (3) narrow-field near-IR imaging on an 8m or larger telescope. The planned medium-resolution spectroscopy needs a minimum spectral resolution of R = 5000 to provide 1 km/s velocity accuracy, a wavelength range of 8450–8700 Å, a field-of-view of ~ 1 degree, and a minimum target spacing of 10" or less, with overall multiplexing of > 100. High-resolution spectroscopy requires the same basic parameters, but with resolution R > 20,000 and wavelength coverage of

~ 4000–6000 Å. The medium- and high-resolution spectroscopic surveys can be executed in ~ 10,000 hours on an 8m telescope or ~ 1000 hours on a 25m telescope. A medium-band imaging survey of the LSST footprint with a DECam-like instrument would require ~ 5300 hours, and high-S/N near-IR photometry of RR Lyrae stars would take at least 300 hours with an efficient imager on an 8m telescope. Multi-object spectroscopy is critical for all Local Group science, wide-field medium-band imaging is very important for efficient spectroscopic target selection, and near-IR imaging is important for determining the distances to RR Lyrae stars throughout the Milky Way's halo.

## Science Goals

One of the primary missions of the LSST project is to produce a three-dimensional map of the Milky Way. This immense dataset will provide key tools for addressing some of the most pressing questions in galaxy formation and cosmology. In particular, LSST discoveries will pave the way toward improving our understanding of the nature of dark matter, unraveling the formation of our Galaxy through the fossil record of galaxy assembly, and testing theories of modified gravity. It has been clear for nearly two decades that the most serious challenges faced by the LCDM model, most notably the missing satellite problem and the TBTF problem, occur on the smallest scales accessible to astronomical observations (e.g., Flores & Primack 1994; Moore 1994; Moore et al. 1999; Klypin et al. 1999; Boylan-Kolchin et al. 2011). By dramatically expanding the census of nearby dwarf galaxies, LSST represents a crucial next step in using these issues, as well as dark matter indirect detection experiments, to constrain the properties of dark matter. With the imminent release of the first Gaia measurements, the field of galactic archaeology is also poised for a fundamental transformation. LSST will unveil vastly larger volumes by probing many magnitudes fainter than Gaia, opening the possibility of completely reconstructing the assembly history of the Milky Way. In this chapter, we describe the progress in these science areas that will be made possible by LSST and summarize the additional observational capabilities that will be necessary to fully realize the potential science gain from LSST.

*Missing Satellite Problem* — The number of low-mass dark matter subhalos orbiting a Milky Way–size galaxy is a sensitive function of the power spectrum on small scales. The existence of abundant dark substructure is perhaps the most important testable prediction of the LCDM model. LCDM N-body simulations predict that the Milky Way should have ~ 1000 satellites more massive than $10^7 M_\odot$ (e.g., Diemand et al. 2007), but observational searches have only identified 50 dwarfs (or dwarf candidates) within 400 kpc. A complete census of Milky Way satellite galaxies immediately places a lower limit on the dark matter particle mass, potentially ruling out substantial portions of parameter space for dark matter candidates with less small-scale power (e.g., Lovell et al. 2014). The subhalo mass function can be probed more directly by using cold stellar streams as accelerometers; the gravity of a dark matter subhalo passing near a stream will perturb the stream stars in a detectable way regardless of whether the halo formed stars itself (Ibata et al. 2002; Johnston, Spergel, & Haydn 2002). Deep photometry and kinematics of stars in multiple streams thus provide a measurement of the presence and nature of a population of low-mass dark matter halos in the Milky Way (Yoon, Johnston & Hogg 2011; Carlberg 2012; Erkal & Belokurov 2015; Bovy et al. 2016). By probing fainter magnitudes than current surveys, LSST will greatly increase the number of known thin streams, as it will reveal structures at larger distances and with lower surface brightnesses. Subhalos within tens of kpc of the Galactic center may be destroyed by the Galactic disk (d'Onghia et al. 2010), so the current population of known thin streams within 30 kpc of the Sun does not yet provide the critical test of LCDM that those revealed by LSST will enable.

*Too Big to Fail Problem (TBTF)* — A related, but distinct, problem facing LCDM on small scales is that the stellar kinematics of the classical ($M_V < -8.5$) dwarf spheroidals (dSphs) suggest that they do <u>*not*</u> reside in the most massive subhalos predicted by simulations (Boylan-Kolchin et al. 2011, 2012). Those subhalos are expected to have circular velocities above 20 km/s, in contrast to the observed velocity dispersions below 11 km/s for the brightest dSphs. Halos with such large masses should be too big to have failed to form stars, and yet we have not identified any dwarf galaxies consistent with residing in such halos. Since dwarfs in this luminosity range should already have been detected near the Milky Way, the next key question is whether TBTF also afflicts field dwarf galaxies. Beyond the Milky Way's virial radius, such objects should have formed free from any environmental influences from nearby massive galaxies that could have affected their star formation or dynamics. LSST will be sensitive to classical dSphs out to distances of ~ 5 Mpc via counts of resolved stars (cf. Crnojevic et al. 2016), and objects located beyond the Milky Way virial radius but within the Local Group (400 kpc < d < 1 Mpc) will be excellent spectroscopic targets. The persistence of TBTF in dwarfs beyond the influence of the Milky Way would represent a significant piece of evidence favoring modifications of dark matter halo masses or density profiles relative to LCDM predictions. Such modifications could be explained by dark matter particle properties such as a non-negligible self-interaction cross-section (additional tests of dark matter self-interaction are discussed in Chapter 8).

*Indirect Detection of Dark Matter* — Current and near-future facilities are sensitive to the gamma-rays that could be produced from the annihilation of standard thermal relic dark matter particles. Excitingly, a possible signal of this annihilation radiation has been detected toward the Galactic center (e.g., Hooper & Goodenough 2011; Abazajian & Kaplinghat 2012; Daylan et al. 2016). However, alternative models such as a large and spatially extended population of faint millisecond pulsars can also explain the observed emission (e.g., Abazajian 2011; Petrovic et al. 2015; Brandt & Kocsis 2015). Observations of dwarf galaxies, where the millisecond pulsar population should be minimal and little foreground gamma-ray contamination is present, provide a clean test of dark matter interpretations of the Galactic center excess (e.g., Ackermann et al. 2015; Geringer-Sameth et al. 2015). The dwarf galaxies with the highest expected gamma-ray fluxes from dark matter annihilation are those at the smallest distances and with the largest dark matter densities (the most compact, lowest-luminosity systems; Simon et al. 2011). The discovery of new ultra-faint dwarfs nearby (D < 50 kpc) and more massive dwarfs out to larger distances (D < 100 kpc) promises to increase the sensitivity of indirect searches for dark matter annihilation (Charles et al. 2016). Accurate determinations of the J-factor (the integral of the dark matter density squared along the line of sight through the dwarf galaxy; Bergström et al. 1998) are essential to properly weight each dwarf in stacking analyses and to choose the best objects for targeted observations (e.g., with the upcoming Cherenkov Telescope Array). While the majority of recent efforts have focused specifically on gamma-ray searches for WIMP dark matter, the proximity and astrophysical inactivity of dwarf galaxies make them an important population for other dark matter search techniques as well (e.g., Colafrancesco et al. 2015; Ruchayskiy et al. 2016; Jeltema & Profumo 2016). As additional instruments come online, dwarf galaxies will continue to provide some of the most compelling targets for the indirect detection of particle dark matter.

*Probing Modified Gravity* — Observations of dwarf galaxies beyond the gravitational influence of the Milky Way provide the opportunity to test a variety of proposed models for modifications to the behavior of gravity. Specifically, Local Group systems can play an important role in distinguishing between dark energy and other modifications to general relativity (GR) that could manifest as an accelerating expansion on cosmological scales. Any modification to gravity must be indistinguishable from GR on Solar System scales, a problem conventionally solved by imposing a screening mechanism that suppresses the influence of additional forces in regions of high density / small scales. In contrast, it is possible that small halos, the outer parts of halos, or regions where the mass distribution differs significantly from the local value could be unscreened and would experience non-GR forces. The simplest and best-studied theoretical modifications to GR are scalar-tensor theories with non-linear screening (e.g., reviews by Silvestri & Trodden 2009, Jain & Khoury 2010). Several of these models have been tested with isolated dwarf galaxies on cosmological scales (Jain & VanderPlas 2011; Vikram et al. 2013). However, the length scales probed by these studies are currently limited by the sizes of the dwarf spiral galaxies studied ($V_{circ} \sim 50$–$100$ km/s). The expansion of the population of lower-mass Local Group dwarf galaxies provided by LSST can probe length scales an order of magnitude smaller. Many models of modified gravity predict both spatial and velocity offsets in the statistical distribution of main sequence and giant branch stars, since higher-density main sequence stars will be self-screened even within unscreened dwarf galaxies. Similarly, MOND and MOND-like models predict different velocity dispersions for dwarfs that are far from any massive galaxies compared to those that are within a strong external potential. Radial velocity measurements of stars in the Local Group galaxies that will be discovered by LSST will therefore enable novel tests of cosmic acceleration through modified gravity.

*Measuring the Three-Dimensional Structure of a Dark Matter Halo* — Both the missing satellite problem and the TBTF problem depend critically on the mass of the Milky Way, which embarrassingly is currently uncertain at the factor of two level (e.g., Boylan-Kolchin et al. 2013; Kafle et al. 2014; Eadie et al. 2015; Peñarrubia et al. 2016; Huang et al. 2016). If the mass of the Milky Way is at the low end of the range of recent determinations, these problems would be significantly alleviated, if not solved entirely. Moreover, it will likely never be possible to determine the three-dimensional shape- and mass-*profile* of the dark matter halos of any external galaxies. A realistic expectation in these cases is to accurately measure projected shapes, as well as the mass enclosed within smaller radii (e.g., the Einstein radius for galaxy-galaxy lenses). Our privileged position within the Milky Way's halo, however, will make precision measurements of the shape, orientation, and total mass feasible for our Galaxy in the LSST era. As the *only dark matter halo* for which we can access this information from the inner galaxy to the virial radius, the Milky Way can provide key constraints on structure formation models and the interplay between baryons and dark matter that shapes the visible structures of galaxies. Useful probes of the Milky Way's gravitational potential at large radii include individual stars, dwarf galaxies, globular clusters, and tidal streams or other debris structures. While individual stars and satellites can provide crucial constraints (e.g., Deason et al. 2012), stars tidally stripped from dwarf galaxies and globular clusters are uniquely powerful tools for constraining the Galactic potential (e.g., Price-Whelan et al. 2014; Sanders 2014;

Sanderson 2016). Unlike stars and dwarfs, each debris structure simultaneously probes a range of radii and angles; as an extreme example, the Sagittarius stream wraps entirely around the Milky Way and covers distances from 10–100 kpc. LSST will provide identifications for all of these structures, bound and unbound, out to 300 kpc. Simulations suggest that we may expect to find ~ 1000 RR Lyrae beyond 100 kpc clumped into several structures (Bullock & Johnston 2005), each of which can be used as a potential probe (see Figure 3.2). LSST alone will produce useful proper motions for stars brighter than ~ 20th magnitude (giants within ~ 100 kpc), but radial velocities and accurate distances will require additional observations.

*Recovering the Accretion History of the Milky Way* — The improved understanding of dark matter and gravity that LSST and LSST follow-up will provide can be matched by an analogous leap forward in our knowledge of the formation of the Milky Way. The entire assembly history of the accreted baryonic component of the Milky Way is encoded in its stellar halo, where current associations of stars apparent locally in space and/or motion or globally in orbits and/or abundances can be attributed to the common (orbit and enrichment) histories they shared in accreting dwarf galaxies (e.g., Bullock & Johnston 2005; Cooper et al. 2010). Debris structures (e.g., streams and shells) that are still detectable as overdensities in space have (by definition) not had time to fully phase-mix and are expected to correspond to fairly recent accretion events from the last 5–8 Gyr (Sharma et al. 2011). Follow-up measurements of kinematics and chemistry of large samples (tens of thousands) of halo stars in the smooth inner halo (within tens of kpc) will allow constraints on the accretion history to extend back further to the main epoch of galaxy formation. Even stars that are fully mixed in phase space retain a memory of their origins in their chemical abundances and can be *chemically tagged* (e.g., similar to the stronger experiment discussed for clusters in the disk by Bland-Hawthorn & Freeman 2004) to the type of object in which they were born.  Indeed, there are clear and systematic distinctions between abundance patterns observed for different types of satellites of our Galaxy (Venn et al. 2004), and mock studies of models of accreted stellar halos have demonstrated in principle how such differences might allow the reconstruction of the luminosity function of accreting dwarfs both currently and in the past  (Lee et al. 2015).  LSST will increase the volume surveyed in the Milky Way halo by several orders of magnitude, mapping stellar tracers to the virial radius of the Milky Way and beyond.  It is therefore nearly guaranteed to discover many new stellar substructures, probing to very low luminosities and surface brightnesses with MSTO stars within ~ 100 kpc (see Figure 3.1 for what can currently be done with SDSS) and mapping higher mass accretion events through RR Lyrae and giant stars at larger distances (see Figure 3.2).

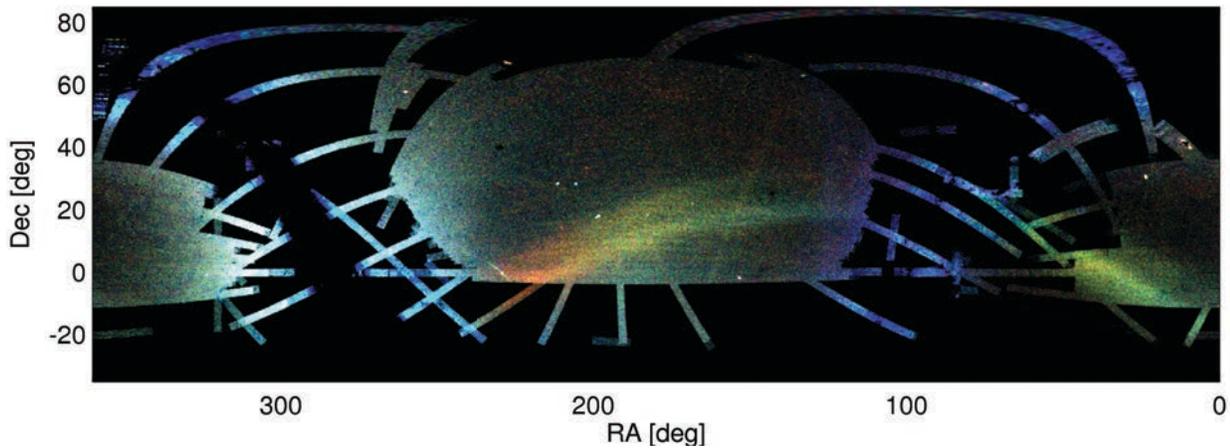

*Figure 3.1. The Sloan Digital Sky Survey (SDSS) "Field of Streams" (Belokurov et al. 2006), updated to include data through SDSS DR8 (Aihara et al. 2011). This map illustrates the surface density of stars near the main sequence turnoff detected by SDSS, with brighter areas indicating higher surface densities. The data are color coded according to distance, where blue corresponds to distances of ~ 15 kpc, green to distances of ~ 20 kpc, and red to distances of ~ 27 kpc. Prominent features visible in the SDSS data include the Sagittarius tidal stream, the Palomar 5 stream, the Orphan Stream, the Monoceros ring, and a number of globular clusters and ultra-faint dwarf galaxies. LSST will provide similar maps of the southern sky, covering a factor of ~ 2 larger area and ~ 10 in volume (in first-year data alone) and probing structures with surface brightnesses as much as 100 times lower. (Figure courtesy of Vasily Belokurov)*

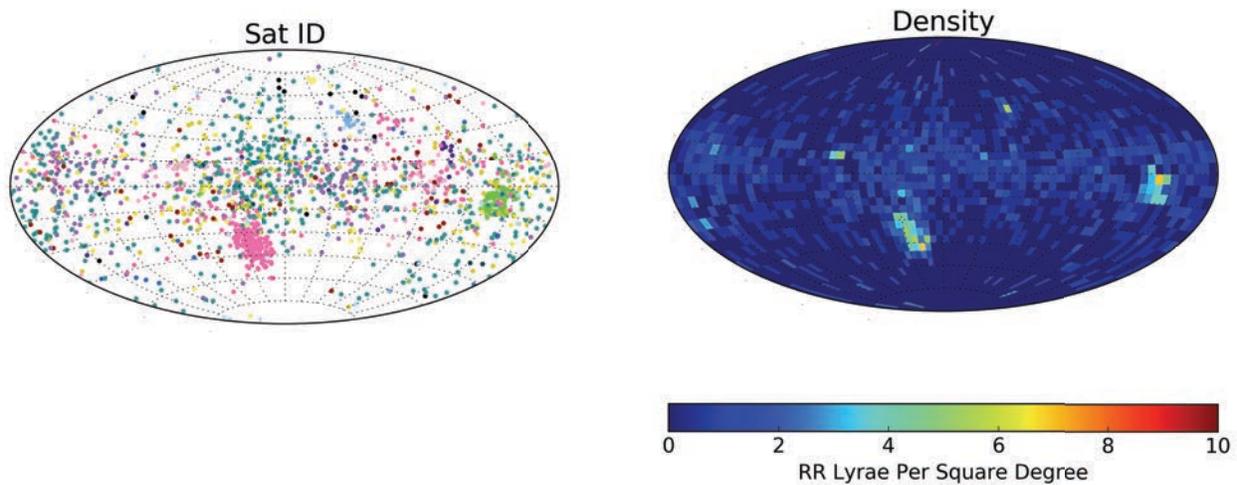

*Figure 3.2. Simulated all-sky view of RR Lyrae at distances of more than 100 kpc from the Sun generated from one stellar halo model from Bullock & Johnston (2005).  The halo model was constructed entirely from accreted dwarf galaxies, including only dwarfs with stellar masses greater than $10^5 M_\odot$. The left-hand panel is color coded by the dwarf galaxy in which the stars were formed and the right-hand panel is color coded by sky surface density. Such models suggest that ~ 1000 halo RR Lyrae will be found by LSST in the distance range 100–300 kpc; they will be non-uniformly distributed on the sky with typical densities much less than 1 per square degree; many will be clumped in debris structures corresponding to just a handful of accretion events; and debris structures at these distances are more typically cloud-like in morphology rather than stream-like.  (Image credit: Amy Secunda and Robyn Sanderson)*

*Connecting the Local Volume to the high-redshift universe* — Follow-up observations to provide abundance patterns for dwarf galaxies as well as for spatial, kinematic, or chemical substructures can be used to recover star formation and chemical enrichment histories for surviving satellites and recent (debris apparent as substructures in space) and ancient (debris only apparent in chemistry) accretion events.  Analyses of cosmological N-body simulations demonstrate that these small objects are unobservable in situ in the high-redshift universe even with upcoming facilities such as JWST.  Furthermore, the number of such objects in the Local Group provides a sample that is representative of the broader population over larger volumes (Okrochkov & Tumlinson 2010; Boylan-Kolchin et al. 2016). The smallest satellite galaxies of the Milky Way correspond to the low end of the galaxy luminosity function at high redshift and are believed to be responsible for re-ionization and metal-enrichment in the early universe.  Using local data to take a full account of old stellar populations in both surviving and disrupted dwarfs is an essential approach for constraining their effect at early times (e.g., Weisz, Johnson, & Conroy 2014). Additionally, the progenitors of the different populations (surviving satellites, recent accretion events and long-dead dwarfs) occupy different volumes in N-body simulations around the main Milky Way progenitor at high redshift; any systematic differences between stellar populations in these structures today therefore probe the spatial scales of variation in metallicity in the early universe (Corlies et al. 2013). LSST provides the essential starting point for constructing a more complete inventory of halo structures with which to examine the relationships between these systems.

## Technical Description

All of the science goals described above will require multi-object medium-resolution spectroscopy of stars out to distances of ~ 1 Mpc. In most cases, the primary aim of spectroscopy is to measure radial velocities at the ~ 1 km/s level, although obtaining simultaneous chemical abundance information is interesting for all targets and essential for the outer halo spectroscopic survey. Ensuring the accuracy of velocity measurements sets a lower limit on the spectral resolution of R = 5000 (e.g., Simon & Geha 2007; Koposov et al. 2011). Because of the presence of multiple strong lines spaced closely in wavelength, the most efficient spectral region for measuring stellar velocities is around the Ca triplet lines (8450–8700 Å). For a slit spectrograph (as opposed to a fiber-fed instrument) or for different wavelength ranges (e.g., the Mg triplet lines from 5160–5190 Å), higher spectral resolution would likely be necessary to reach 1 km/s velocity uncertainties (e.g., Simon et al. 2015; Simon et al. in prep.). Since even luminous red giants are faint at the relevant distances, spectroscopy must be obtained using telescopes with a diameter of at least 8m. We note that all of the calculations below assume seeing-limited observations. A ground-layer adaptive optics capability that can operate over wide fields would substantially increase the spectroscopic efficiency for such faint targets.

To place the tightest constraints on the missing satellite problem, the dark matter content of every dwarf galaxy candidate identified by LSST within the virial radius of the Milky Way must be measured spectroscopically. Typically, this determination is made by measuring the velocity dispersion and/or metallicity dispersion of stars in a system and demonstrating that dark matter is necessary for the object to be gravitationally bound or for it to have retained supernova ejecta from earlier generations of stars (Willman & Strader 2012). Using projections for the total satellite population of the Milky Way from Tollerud et al. (2008), Hargis et al. (2014), and Garrison-Kimmel et al. (2014), we expect that spectroscopy will be needed for at least ~ 200 dwarf galaxies (Figure 3.3), with a minimum sample of 10 stars per dwarf needed to constrain the velocity and metallicity dispersions. Because the velocity dispersions of these systems can be as small as ~ 2 km/s (Kirby et al. 2013), the individual velocity measurements of each star need to be at this level or better to guarantee that the dispersion is measurable. For a dwarf comparable to the faintest currently known galaxy, Segue 1, obtaining a sample of 10 stars requires observing to a limiting magnitude of g = 20.5 + ($\mu$ − 16.8), where $\mu$ is the distance modulus (Simon et al. 2011). At d = 100 kpc, this limit corresponds to g = 23.7 (see Figure 3.4).

For a more typical ultra-faint dwarf with a stellar mass of $10^4$ $M_\odot$ ($M_V$ ~ −5), the tenth brightest star has a magnitude of g = 21.5 + ($\mu$ − 21.0), corresponding to g = 20.5 at d = 100 kpc and g =

22.9 at d = 300 kpc (scaling from Simon & Geha 2007). Spectroscopic confirmation of the full population of Milky Way satellites in the LSST era will also provide good measurements of the radial velocity of each satellite, which can then be used to improve constraints on the total mass, density profile, and shape of the Milky Way's dark matter halo.

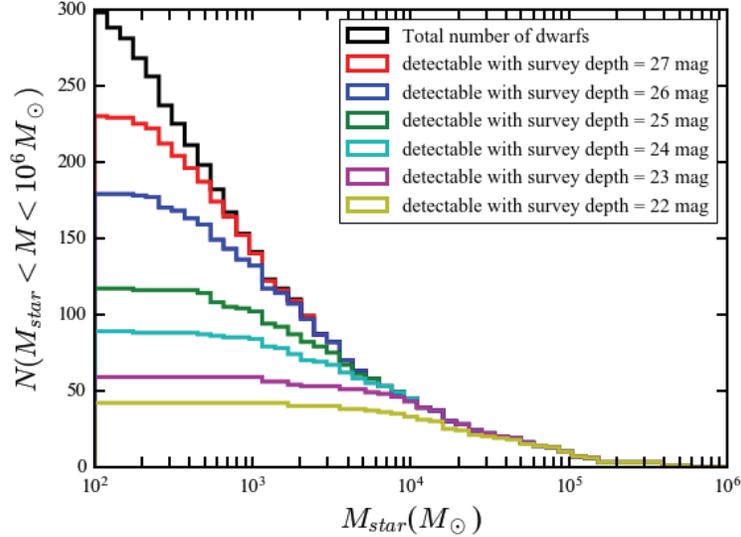

Figure 3.3. *Predicted cumulative stellar mass function of Milky Way satellite galaxies that can be detected for various survey depths in the LSST footprint. The three lowest curves correspond to surveys similar to SDSS, Pan-STARRS, and the Dark Energy Survey [DES], respectively, while stacked LSST data (red curve) will reveal nearly the entire population of nearby dwarfs. The total number of dwarfs within the virial radius of the Milky Way shown by the black curve is based on the ELVIS suite of N-body simulations (Garrison-Kimmel et al. 2014).*

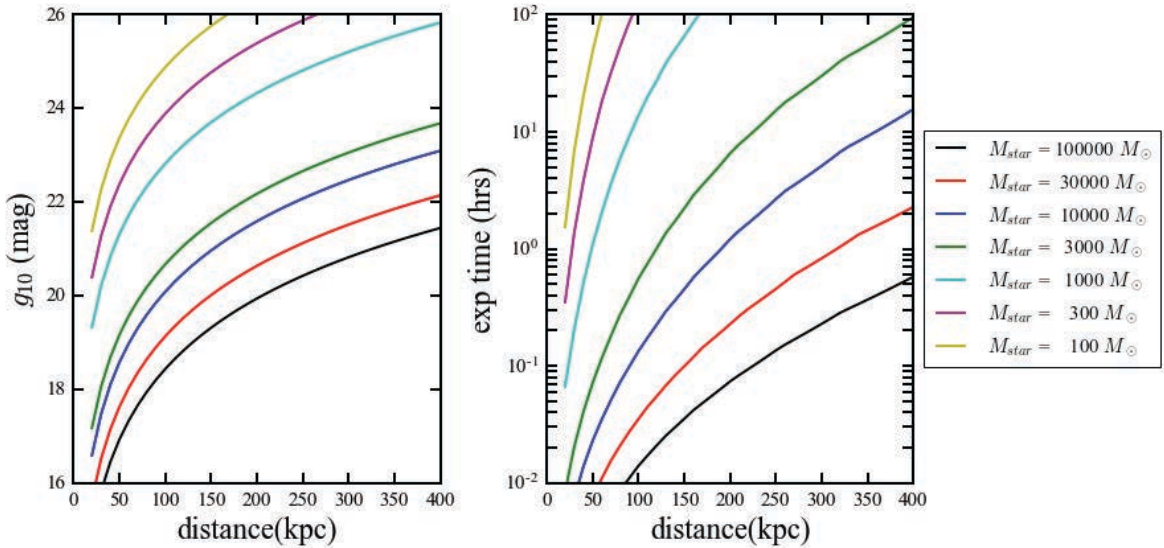

Figure 3.4. *(left) Estimated g-band magnitude limit in order to provide a sample of 10 stars in a dwarf galaxy as a function of heliocentric distance. The curves of different colors show results for different stellar masses. (right) The exposure time needed to reach S/N ~ 10 on the tenth brightest star in a galaxy with a spectrograph similar to Keck/DEIMOS.*

Other science goals will demand more intensive spectroscopy of specific subsets of LSST dwarf galaxies. Dwarfs at distances of up to ~ 100 kpc are relevant targets for indirect detection of dark matter, and those within ~ 50 kpc are particularly valuable. Velocity measurements of at least 30 stars in each of these dwarfs will constrain the J-factor to ~ 0.25 dex (Wang et al. 2016, in prep.), allowing appropriate targeting decisions and signal weighting. A total population of ~ 60 dwarf galaxies within this volume are expected. For the TBTF problem and modified gravity experiments, on the other hand, larger spectroscopic samples will be needed in distant dwarfs beyond the virial radius of the Milky Way. By measuring velocities for ~ 20 stars per dwarf with individual velocities accurate to ~ 2 km/s or better for a sample of ~ 20 dwarfs, we can determine velocity dispersions, and hence the mass function, for isolated low-mass dark matter halos. The limiting magnitude for a spectroscopic sample of this size is g = 22.4 + ($\mu$ −23). At 400 kpc, the observations can be made with existing instruments on 8m-class telescopes; by 1 Mpc, larger telescope apertures will be required. Because many dwarf galaxies are relatively compact (half-light radii of ~1 arcmin), the fiber or slit density may be a limiting factor on the efficiency of spectroscopic observations. The ability to target stars with separations of ~ 10 arcsec is a requirement for dwarf galaxy spectroscopy, and observing stars 5 arcsec apart is desirable.

Spectroscopy of stars in tidal streams generally requires similar spectroscopic capabilities as observations of dwarf galaxies. For the purpose of constraining the presence and mass function of dark matter subhalos it is critical to be able to measure velocities with an accuracy of ~ 1 km/s, because the streams are cold and the perturbations from the subhalos are small (Erkal & Belokurov 2015). Kinematic surveys of streams could also be used to separate genuine members from foreground contamination and allow more precise number density maps to search for spatial inhomogeneities. Perhaps the most interesting regime for this experiment is at distances of ~ 50 kpc where subhalos are less likely to be destroyed by the disk, yet streams are still expected to be extended enough to encounter them. Ignoring the effect of the disk, Pal 5 (orbiting at ~ 25 kpc, with a current length of ~ 20 degrees suggesting an age of ~ 6 Gyr) might be expected to have 20 (~ 70) encounters with $10^6 – 10^7$ M$_\odot$ ($10^5 – 10^6$ M$_\odot$) subhalos leading to ~ few (~ 1) degree-scale disturbances during its lifetime (Yoon, Johnston, & Hogg 2011). An analogous stream (same age, mass of progenitor and mass loss rate) at 50 kpc would likely have an ~ 8 times lower encounter rate as it would have a shorter length, a lower density of subhalos to encounter, and a longer orbital period. Moreover, the spatial scales of the disturbances would be lowered by a factor of 2 to ~ 1 (sub-) degree scales for $10^6 – 10^7$ M$_\odot$ ($10^5 – 10^6$ M$_\odot$) subhalos. The requirement to survey stars to map the stream in density and velocity suggests a field-of-view of order ~ 1 deg$^2$ to enable efficient observations of multiple stream stars at a time. In most streams, reaching a non-negligible number of spectroscopic targets dictates that turnoff stars will be the primary targets. At d = 100 kpc, MSTO stars have magnitudes of g ~ 24, so that spectroscopy requires an 8m telescope. Based on the expectation of order unity detectable subhalo encounters per stream if the LCDM subhalo abundance is correct, observations of at least ~ 10 streams will be needed in order to draw confident conclusions about the true subhalo mass function.

Realizing the full potential of LSST for determining the dark halo structure of the Milky Way requires combining multiple tracers of the gravitational potential, as measured by both radial velocities and proper motions. These include point-like objects, such as individual RR Lyrae stars, globular clusters, and dwarf galaxies, as well as spatially extended stellar streams. Single objects, however, and even streams, can be highly biased (Bonaca et al. 2014), with errors on the total mass of the Galaxy of up to 50%. Indeed, analyses of known streams often produce mutually inconsistent results (e.g., Law et al. 2009; Koposov et al. 2010; Pearson et al. 2015; Küpper et al. 2015). LSST will provide the first significant samples of stars at distances beyond 100 kpc, will greatly expand the population of known dwarfs beyond 200 kpc, and will discover the first streams beyond ~ 50 kpc. Follow-up observations of the full set of dwarf galaxies in the outer halo (as described above) as well as several hundred RR Lyrae stars will increase the number of objects constraining the total mass and outer density profile of the Galaxy by ~ 2 orders of magnitude. LSST will identify all RR Lyrae stars within the survey footprint out to well beyond the virial radius and provide 5–10% distances for each star. Improving the distance accuracy to 1–2% requires deep photometry (S/N = 50) in the H-band or at longer wavelengths. For RR Lyrae within 100 kpc these measurements are possible from the ground (e.g., with VLT/HAWK-I), although not with any instruments in the US system, but for more distant stars, it is only possible to reach the necessary depth from space in a plausible amount of observing time. Radial velocity measurements for RR Lyrae stars can be obtained either with a single-object medium-resolution spectrograph (since the surface density of targets is low; Figure 3.2) on an 8m telescope or with a multi-object spectrograph as part of a larger program (see below). New radial velocity measurements for stars in ~ 10 streams (which are also needed for the missing satellite problem experiment) combined with LSST photometry and proper motions will determine the flattening of the Milky Way's dark halo.

The comprehensive archaeological survey of the stellar halo we envision for unraveling the formation of the Galaxy has two components. Stars in the outer halo (d > 25 kpc) are faint, but can be observed at medium spectral resolution (R ~ 5000) to measure radial velocities, metallicities, and abundances of several other elements (at minimum, [α/Fe] and [C/Fe]). With these data, one can produce maps like that shown in Figure 3.2 reaching much lower-mass structures that do not contain significant numbers of RR Lyrae. A survey of ~ $10^6$ stars could reveal the remnants of as many as ~ 100 progenitors of the halo (e.g., Ting et al. 2015), with accretion times measured via their kinematics. LSST photometry, proper motions, and wide-field DDO51 medium-band imaging reaching S/N = 20 at g = 23 to separate foreground main sequence stars from halo giants would be sufficient to select a clean target sample for spectroscopy. The second component of the survey would focus on the smooth inner halo population. Rare, bright (g < 18) halo stars in the solar neighborhood can be observed at high spectral resolution (R > 20,000) to produce accurate chemical abundances for a wide array of elements. These data will add up to eight additional phase space dimensions (Ting et al. 2012) to the 6D position and kinematic measurements available from Gaia and LSST. With a sample of ~10,000 stars, Lee et al. (2015) show that the luminosity function of destroyed dwarf galaxies can be recovered down to stellar masses of ~ $10^4$ M$_\odot$. To reach the edge of the stellar halo for luminous red giant stars, the medium-resolution survey would need to observe to g ~ 23, requiring several hour exposures on an 8m telescope. The surface density of halo stars to that limit is ~ 125 deg$^{-2}$, demanding a wide-field instrument and a survey covering ~ 8000 deg$^2$ to build up the desired sample size. Because g < 18 halo stars have a surface density of ~ 20 deg$^{-2}$, a survey of many such stars is only feasible as part of a larger survey. An appealing approach would be for a wide-field fiber array to feed both medium-resolution and high-resolution spectrographs simultaneously, so that observations could be done in both modes at the same time.

## Needed Capabilities and Estimate of Demand

The primary and highest-priority capability required for achieving the main LSST science goals related to dark matter, the Milky Way, and the Local Group is a **wide-field, multi-object, medium- to high-resolution spectrograph on a large (8+m) telescope**. Medium spectral resolution (R = 5000) is necessary for all of the science described in this chapter, while high resolution (R > 20,000) is essential for studying the formation of the Milky Way through chemical tagging. To estimate the observing time necessary for dwarf galaxy spectroscopy, we use the dark matter subhalo population from one of the ELVIS Milky Way analogs (Garrison-Kimmel et al. 2014) to create a realization of Milky Way satellites. We create a mock color-magnitude diagram for each satellite with an assumed theoretical isochrone and initial mass function. We scale exposure times to reach the desired depth for each object (based on the number of member stars for which velocities are needed) with Keck/DEIMOS observations from Simon & Geha (2007) and Simon et al. (2011) and Magellan/IMACS observations from Simon et al. (in prep.). With a moderate field-of-view (> 10–20 arcmin) spectrograph targeting the Ca triplet lines at R = 5000, and with a minimum fiber or slit spacing of 5–10 arcsec, the total time needed is dominated by confirmation of the faintest dwarfs (up to 100 hours per target for M < $10^3$ M$_\odot$ galaxies at d < 100 kpc and up to 10 hours per target for M < $10^4$ M$_\odot$ galaxies at 100 kpc < d < 300 kpc). We estimate that radial velocity and metallicity measurements for stars in dwarf galaxies will require ~ 3200 hours on an 8m telescope, and the same observations for stellar streams will take an additional ~ 800 hours. These observations could be obtained ~ 10 times faster with a 25m telescope assuming equivalent multiplexing. Scaling from DESI projections, at g = 23 the lowest useful signal-to-noise ratio of ~ 3 per resolution element requires 2.5 hours of integration time with an 8m telescope. A wide-field spectroscopic survey of ~ 1 million halo stars at medium resolution will therefore take 2.5 million fiber-hours. With a target surface density of 125 stars deg$^{-2}$, this translates to 20,000 hours of 8m telescope time per square degree of the spectrograph field-of-view, that is, 6700 hours for an instrument with a 3 deg$^2$ field-of-view, or 2500 hours for an instrument with an 8 deg$^2$ field-of-view. The companion high-resolution survey requires 30,000 fiber-hours for 3-hour integrations on 10,000 stars. Given the similar integration times, these surveys could easily operate simultaneously if fibers can be used to feed medium- and high-resolution spectrographs at the same time. Depending on the total number of fibers, it may also be possible to carry out dwarf galaxy/stellar stream observations concurrently with the halo spectroscopy. The Prime Focus Spectrograph (PFS) on Subaru would satisfy some of these needs, but given its location in the Northern Hemisphere it would be unable to observe approximately half of the dwarfs, streams, and halo structures discovered by LSST, compromising science goals related to the census of dwarf galaxies and the mass and shape of the Milky Way's dark matter halo. PFS also does not allow high-resolution spectroscopy to determine detailed chemical abundance patterns.

Characterizing the stellar halo and the dark matter halo of the Galaxy also depends on several other capabilities. In addition to LSST photometry and proper motions, efficiently selecting halo stars for spectroscopy requires **medium-band DDO51 imaging over a substantial fraction of the LSST footprint**. A wide-field optical imaging capability on a 4m-class telescope is the second priority for the science described in this chapter. With an imager such as  on the Blanco telescope, these observations would require the production of a new DECam filter and exposure times of 40 min per 3 deg$^2$ pointing (based on the DECam exposure time calculator) to reach S/N

= 20 for accurate dwarf/giant separation down to g = 23. Covering the full LSST footprint would require ~ 6000 pointings, amounting to ~ 5300 hours of observing time with weather losses and overheads included. Maximizing the potential of distant RR Lyrae stars to provide the most accurate possible 6D phase space information demands radial velocity measurements and improved distances for each star. RRL velocities can be obtained either with a **single-object medium-resolution spectrograph on an 8m telescope** or as part of a broader medium survey of the halo. In the former case, ~ 1200 hours would be required. The most accurate distances for RR Lyrae variables necessitate observations at H-band or longer wavelengths. High signal-to-noise ratio photometry of RR Lyrae within 100 kpc can be obtained with ~ 300 hours of time on a **sensitive near-infrared imager on an 8m telescope** such as HAWK-I. These two capabilities rank third for LSST science goals related to the Milky Way and dark matter. For more distant stars, the needed photometric accuracy can only be obtained from space, perhaps with a mission such as Euclid or WFIRST.

For dark matter, Milky Way, and Local Group science, the timing of the availability of these capabilities relative to the LSST project is not critical, except inasmuch as it is always preferable to begin observations sooner rather than later.

## Summary Tables

### Table 3.1. Needed Capabilities

| | Infrastructure | < 3m | 3–5m | 8m | 25m |
|---|---|---|---|---|---|
| **Dark Matter** | | | | **Medium-resolution (R ~ 5000) and high-resolution (R > 20,000) multi-object spectroscopy**<br><br>*Narrow-field near-IR imaging; single-object medium-resolution spectroscopy* | **Medium- and high-resolution multi-object spectroscopy** |
| **Milky Way Halo Formation** | | | Wide-field medium-band optical imaging | **Medium-resolution (R ~ 5000) and high-resolution (R > 20,000) multi-object spectroscopy** | **Medium-resolution and high-resolution multi-object spectroscopy** |

Entries in boldface type indicate that the capability is **Priority 1 (critical)**.
Roman type indicates Priority 2 (very important).
Italic type indicates *Priority 3 (important)*.

*Table 3.2. Resource Demand*

| | Infrastructure | < 3m | 3–5m | 8m | 25m |
|---|---|---|---|---|---|
| **Dark Matter** | | | | **Multi-object spectroscopy: 4000 hrs**<br><br>*Near-IR imaging: 300 hrs*<br><br>*Single-object spectroscopy: 1200 hrs* | **Multi-object spectroscopy: 400 hrs** |
| **Milky Way Halo Formation** | | | Medium-band imaging: 5300 hrs | **Multi-object spectroscopy: 6700 hrs** | **Multi-object spectroscopy: 670 hrs** |
| **Total On Sky Time** | | | ~ 1.5 years | ~ 3.3 years | ~ 0.3 year |

Entries in boldface type indicate that the capability is **Priority 1 (critical).**
Roman type indicates Priority 2 (very important).
Italic type indicates *Priority 3 (important)*.

# Chapter 4: Characterizing the Transient Sky

*Ryan J. Foley (University of California, Santa Cruz), Wen-fai Fong (University of Arizona), Jennifer L. Hoffman (University of Denver), Thomas Matheson (National Optical Astronomy Observatory), David J. Sand (Texas Tech University), Rachel Street (Las Cumbres Observatory Global Telescope Network)*

### Executive Summary

The LSST is an inherently transient survey which, through its discovery of millions of supernovae (SNe) and other exotic transients, will greatly expand our knowledge of dark energy, explosion physics, and the changing sky. Concurrently, a new era of physics is beginning with the first gravitational-wave (GW) sources discovered by Advanced LIGO (aLIGO), with transient electromagnetic counterparts being key to maximizing the GW science. By working at the edge of these scientific frontiers, LSST will soon open a new era in transient astronomy.

One of LSST's primary goals is to characterize the entire transient sky, improving our understanding of which stars explode and how. Most of the explosive transients detected over the past century are either Type Ia supernovae, which have white dwarf progenitors, or core-collapse supernovae, which have massive-star progenitors. However, in recent years we have begun to discover many types of "exotic" transients that do not easily fall into these traditional classes. Simply discovering such transients is insufficient for achieving broader scientific goals; follow-up observations are critical for understanding the diversity of explosions. Below, we suggest four primary science priorities focusing on building our understanding of the transient zoo: characterizing the transient sky, thoroughly investigating the causes of the diversity of Type Ia supernovae, opening the new window of very early-time observations for all transients, and discovering the first electromagnetic counterparts of gravitational wave sources.

We find that LSST will be transformative for transient science, but follow-up observations will be critical to achieve its potential. Since we will only be able to follow a small subset of the transients discovered by LSST, quick prioritization and dissemination by a transient broker system will be necessary. Immediate target-of-opportunity (ToO) spectroscopy of high-priority targets with medium-resolution, 0.32–2 µm spectrographs on 8–30m telescopes will provide fast classification for further triage. Triggering LSST itself in a ToO mode may result in the first detection of gravitational wave counterparts. The entire suite of available telescopes (from 1 through 30 meters) will then be coordinated, hopefully through a newly developed software platform, to maximize the transient science results from LSST.

## Science Case: Characterizing the Transient Sky

For millennia, people have been fascinated by *stellae novae*, new stars that occasionally appear and fade away on human timescales. We now know that these historical events are linked to the deaths of stars. Over the last century, we have discovered thousands of supernovae and identified many classes of explosive transients that illuminate the broad diversity of ways that stars can die. LSST will discover orders of magnitude more transients than ever before, revealing even rarer classes and allowing for the precise physical characterization of all types of explosive transients.

Over the last decade, we have begun to discover many explosive transients that do not fall into the traditional supernova classification system (Ia, II, Ib/c, etc.). These new discoveries hint at several new classes of "exotic" or "peculiar" explosive transients that were previously lurking in the shadows, including Type Ia SNe, Ca-rich SNe, fallback SNe, kilonovae, luminous red novae, luminous SNe IIn, pair-instability SNe, SN impostors, SNe Iax, SN 2006bt-like SNe, and tidal disruption flares (e.g., Berger et al. 2009; Bildsten et al. 2007; Foley et al. 2010; Gal-Yam et al. 2009; Gezari et al. 2012; Li et al. 2003; Perets et al. 2010; Quimby et al. 2011; Smith et al. 2007). These newly identified classes have very diverse observational and physical properties: some are 1/100 as bright as typical SNe with durations of a few days instead of weeks, while others are 100 times brighter than the average SN and last for years (see Table 4.1 for a non-exhaustive sample). Some of these classes only have a handful of known members, but with LSST, we expect all these classes to accumulate hundreds or thousands of members in less than a decade. In addition to allowing more robust characterization of these recently discovered phenomena, LSST will most likely reveal entirely new classes of transients as it probes regions of parameter space that are not well investigated by current telescopes or modes of observation.

*Table 4.1.*

| Transient Class | Approximate Timescale | Peak Brightness (mag) | Probe of? |
|---|---|---|---|
| SNe .Ia | ≈ 1 week | $M_V \sim -16$ | WD-WD physics |
| SNe Iax | ≈ 15 days | $-14 > M_V > -19$ | Deflagration flames |
| Ca-rich transients | ≈ 10 days | $-15 \gtrsim M_V \gtrsim -16$ | Unknown stellar death |
| Kilonovae | ≈ 1 day | $M_I \approx -15?$ | GW physics |
| Off-Axis/Dirty GRBs | ≈ 2 days | Varies | Relativistic explosion physics |
| Fallback SNe | ≈ 1 week | $M_V \approx -15$ | BH formation |
| SLSNe | ≈ 1 – 5 months | $M_V < -21$ | Most massive stars, pair instability, magnetars |

Such exotic transients are not just an idle curiosity. These events are the result of the most extreme physical conditions imaginable. They arise from the most massive stars, or the weakest explosions, or truly unexplored phenomena. Explosive transients explore new endpoints of stellar evolution and the boundaries of possible physical conditions of stellar systems. They also probe interesting corners of physics such as *r*-process element creation and the nature of the first stars in the Universe. Finally, these unusual events illuminate the processes that drive more typical stellar explosions. Exploring such extreme cases in greater detail can help us contextualize common but important phenomena such as neutron stars, black holes, heavy element abundances, and dark energy.

## Technical Description

LSST will detect $\sim 10^6$ SNe in the deep drilling fields during its 10-year survey, which is two orders of magnitude more than the total number of SNe yet discovered. As a result, LSST will produce the most detailed survey of the transient sky to date not only revealing new classes of objects but also significantly increasing the number of SNe in known classes. Taking full advantage of these discoveries will require a comprehensive and diverse observing program. Our goals for these newly detected objects will be to determine their compositions, morphologies, expansion velocities, and intrinsic luminosities. We will also investigate the nature of their progenitor environments and characterize the ways and timescales on which all these properties evolve over time. As part of this effort, we will undoubtedly learn much more about the more common SN classes as well, which will provide a foundation for understanding the rarer objects.

Because we still have scant information about many exotic transients, their early identification is particularly important. Rapidly determining which transients belong to unusual classes will not only maximize the efficiency of follow-up observing efforts but also lead to groundbreaking scientific results. The earliest epochs of such transients probe their most extreme (hottest, smallest, most compact) properties and provide the most direct connections to their progenitor systems and the physical mechanisms driving the explosions.

## Needed Capabilities and Estimate of Demand

This program has two distinct phases. The first is fast classification of transient events suspected to be exotic, while the second consists of additional follow-up observations required to fully characterize the objects determined to be of interest. Exotic transients represent $\sim 1\%$ of a magnitude-limited survey (e.g., Li et al. 2011), and thus we expect LSST to discover $\sim 10^4$ such objects over the 10-year survey. Considering the number of different classes and the potential to discover new classes, we propose to study 10% of LSST-discovered exotic transients ($\sim 1000$ objects) in greater detail.

Previous experience suggests an efficiency of at most 33% when attempting to select exotic transients based on only the data available at discovery. To achieve the goals of this program, then, we must obtain spectra of roughly 3000 transients simply to select our final sample. With $\sim 30$-minute exposures (assuming fainter objects are sent to larger telescopes), this will require a total of 1500 hours or 150 nights of telescope time. The bona fide exotic transients will require an additional 3 hours of spectroscopy each, or 3000 hours for the full sample. Thus the spectroscopy we propose will total roughly 4500 hours, or 450 nights. We also recommend obtaining spectropolarimetric time series data for $\sim 5\%$ of the 1000 intensely followed exotic transients, which will require $\sim 300$ hours, or 30 nights. Given the predicted brightness distribution of the sample, we expect 8m-class telescopes to do the bulk ($\sim 60\%$) of the observing with 2m-, 4m-, and 30m-class telescopes contributing at rates of $\sim 10\%$, 20%, and 10%, respectively. This results in roughly 5, 10, 30, and 5 nights per year on 2m-, 4m-, 8m-, and 30m-class telescopes, respectively.

The optical photometry we propose may be available directly from the LSST survey itself, but that depends strongly on the exact cadence of observations. If the optical photometry is not available from LSST (or not in the proper cadence or filter set), then essentially all the early photometry can be done on 1–4m-class telescopes. In addition, we estimate that about half the infrared photometry can be done with 2–8m facilities. With 15 minutes per SN for an optical photometric observation and 30 minutes for an IR observation, the photometric study will require a total of 750 hours, or 75 nights.

Our specific recommendations are detailed at the end of this section.

## Science Case: Type Ia Supernova Demographics

Supernovae of Type Ia have been well studied in certain regimes, mainly because of their utility as distance indicators over cosmological scales (e.g., Betoule et al. 2014; Perlmutter et al. 1999; Rest et al. 2014; Riess et al. 1998; Suzuki et al. 2012). They are generally believed to be thermonuclear disruptions of white dwarf stars, but many of the basic facts about progenitors are still unknown (Maoz et al. 2014). Progenitors may include double degenerate (DD) systems with two white dwarfs that coalesce and explode (Iben & Tutukov 1984; Webbink 1984) and single degenerate (SD) systems in which a white dwarf explodes as a result of accretion from a companion (e.g., Whelan and Iben 1973). Evidence exists for both of these possibilities, but other as-yet-unexplored mechanisms may also be feasible. A detailed and systematic study of the demographics of SNe Ia would reveal the underlying physics that drive these explosions.

Although SNe Ia can provide precise distances, dark energy constraints are not currently limited by statistics. Even $10^3$ SNe Ia are more than necessary to reach the current systematic floor (Betoule et al. 2014; Scolnic et al. 2014). While LSST will certainly reduce some systematic uncertainties such as those related to calibration (e.g., Burke et al. 2014), those related to astrophysics will require more knowledge of the SNe Ia themselves. This broad and detailed demographic study of SNe Ia can address the nature of the astrophysical systematic uncertainties in distances that afflict dark energy constraints.

## Technical Description

A key observable describing the known diversity of SNe Ia is the decline rate of the brightness of the SN (e.g., Phillips 1993). This light-curve shape can be parameterized in many ways, such as by stretch (that changes the width of the lightcurve, Goldhaber et al. 2001) or by $\Delta m_{15}$ (the change in magnitude over the 15 days past maximum brightness in a specific filter, often $B$, see, e.g., Jha et al. 2007). For the purposes of this discussion, we will use $\Delta m_{15}$ as the diversity parameter. In general, a large value of $\Delta m_{15}$ indicates a rapid fading associated with fainter, redder SNe Ia, while a small value of $\Delta m_{15}$ indicates a brighter, bluer SN Ia. Figure 4.1 illustrates the different light-curve shapes. Even with a similar light-curve shape in one band, diversity can appear in other bands (Figure 4.1). Studying the physical mechanisms that produce this distribution is a direct approach to understanding the ultimate utility of SNe Ia as dark energy probes and the physics that drives the explosions.

A sample of SNe Ia distributed over the known $\Delta m_{15}$ range with $\sim 10$ bins would enable such a study. Within each $\Delta m_{15}$ bin, the SNe Ia should be distributed across other significant observables, such as host galaxy type (e.g., Childress et al. 2013; Wolf et al. 2016) and expansion velocity of the SN itself (e.g., Foley et al. 2011; Silverman et al. 2015). With $\sim 4$ bins for host galaxy type and $\sim 3$ bins for velocity, and desiring 10 SNe per combination of parameters, such a study requires 1200 SNe Ia. In addition, many unusual SNe Ia do not necessarily fall into the typical $\Delta m_{15}$ distribution. These include faint objects like SNe Iax (Foley et al. 2013) and bright "super-Chandra" events (e.g., Howell et al. 2006; Scalzo et al. 2010). If we distribute these unusual objects over hosts with 10 examples of each combination, we add another $\sim 300$ SNe Ia for a total of 1500. Distribution over redshift would be extremely useful to test evolution, but that would require a similar, complementary study that would use more resources, as objects at higher redshift will be fainter.

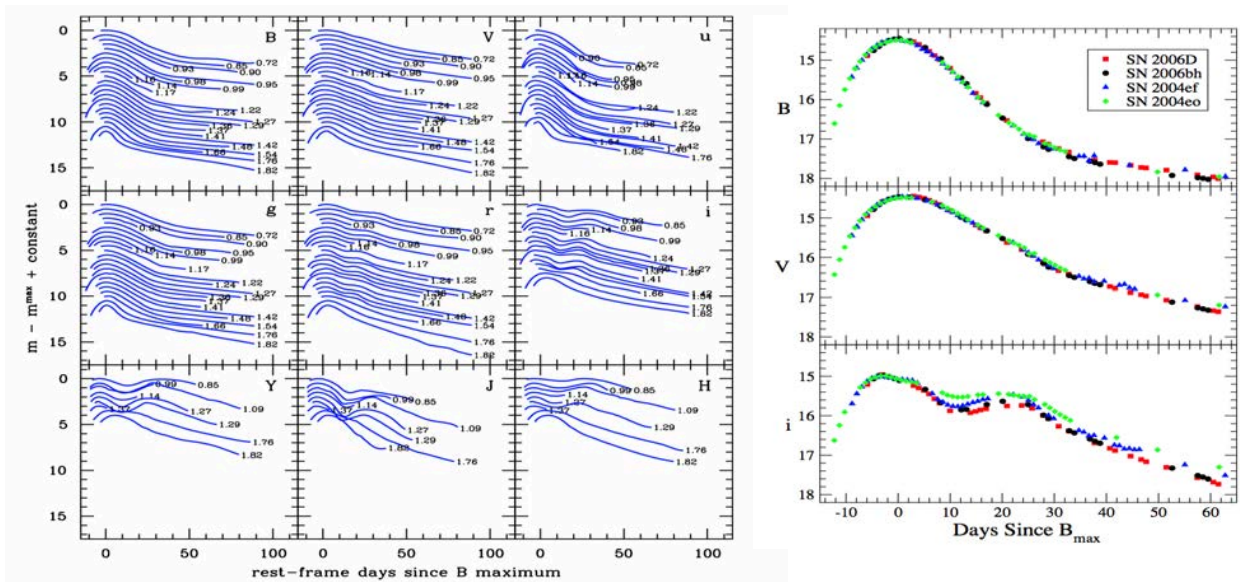

Figure 4.1. Left panel: Lightcurves of SNe Ia in many optical/infrared bands, each labeled by its $\Delta m_{15}$ parameter in the B band. There is a great range of lightcurve shapes. Right panel: Lightcurves of four SNe Ia in the B, V, and i bands. The $\Delta m_{15}$ values for the four SNe are essentially identical in the B band, but the i-band lightcurves demonstrate considerable dispersion, indicating the range of diversity that must be adequately sampled. (Figures from Folatelli et al. 2010)

We would need well-sampled lightcurves for each of the SNe in both the optical and infrared bands. This would necessitate dense coverage over the rise and fall of the SN (an observation every 2–3 days) and less frequent coverage as the SN fades along the radioactive tail (every $\sim 7$ days). We estimate $\sim 20$ lightcurve points in all. In addition, late-time photometry ($\sim 3$ epochs), both optical (*ugriz*, although splitting the *g* band into *BV* is particularly useful for both the physics of SNe Ia and constraining dust properties) and infrared (*JHK*), is useful to constrain explosion energetics. We would require relatively dense spectroscopic coverage over the rise and fall of the lightcurve (every 3–4 days), as well as a few at later times, for a total of 13 epochs. Some optical nebular-phase spectroscopy would also be informative ($\sim 3$ epochs). For a small subset of the SNe Ia, a few spectropolarimetric observations per event will help characterize the nature of the explosions (e.g., Wang & Wheeler 2008; Maund et al. 2010).

## Needed Capabilities and Estimate of Demand

If we adopt the magnitude distribution of SNe Ia from Pan-STARRS (where the majority of spectroscopically identified SNe Ia were in the range $18 < r < 21$ mag; Rest et al. 2014), we can estimate the time necessary to obtain the observations described above. The optical photometry may be available directly from the LSST survey itself, but that depends strongly on the exact cadence of observations. (Note that LSST will not be capable of providing $BV$ photometry.) It may be possible to observe most of these SNe in the deep drilling fields. If the optical photometry is not available from LSST (or not in the proper cadence or filter set), then essentially all the early photometry can be done on 1–2m-class telescopes. In addition, approximately half the infrared photometry can be done with 1–2m facilities. Assuming 15 minutes per optical photometric observation and 30 minutes per infrared for each SN, the photometry requires 1500 nights, or 150 nights per year. On 3–6m-class telescopes, we can obtain all the late-time optical photometry (30 minutes per SN), the second half of the early infrared photometry (30 minutes per SN), half the late-time infrared photometry (60 minutes per SN), and two-thirds of the early optical/infrared spectroscopy (30 minutes per SN) for a total of 1850 nights, or 185 nights per year. We will need 8–10m-class facilities for half the late-time infrared photometry (60 minutes per SN), one-third of the early optical/infrared spectroscopy (30 minutes per SN), and all the late-time spectroscopy (60 minutes per SN). This is a total of 730 nights, or 73 nights per year (1–2 nights each month on both telescopes at Gemini and Keck). Note that for the 2016A semester, the two Gemini telescopes are already scheduled with 18 nights or 1.5 nights per month of transient and variable science. The addition of spectropolarimetric observations of $\sim 100$ of the brighter SNe Ia with $\sim 3$ epochs each would add 60 nights of 8–10m telescope time, although some may be faint enough to require a 20m-class facility.

It may be that data for some subset of this SNe Ia demographic study will be available from time domain surveys and follow-up programs that precede LSST. It is unlikely that the full range of properties and the quality of the data will exist prior to LSST, but the follow-up time estimated above is an upper limit. Before including other data sources, though, detailed consideration must be applied to calibration issues across disparate observing programs. Cross-calibration is also a concern for this study, but careful attention to such issues would be integral to its design.

Our specific recommendations are detailed at the end of this section.

## Science Case: Early Evolution of Supernovae

The key questions for our third area of focus are **What are the progenitors of the various supernova types?** and **How do they explode?** Mapping which stellar deaths lead to the different SN types and characterizing the physics of the explosions themselves are fundamental quests of astrophysics, linking stellar evolution with SN research, and ultimately to galaxy evolution. One of the best ways to gain insight into a SN's progenitor and explosion physics is to study it in the first hours to days after the explosion when the outer layers of the progenitor still leave a spectroscopic imprint, the explosion is the dominant heat source rather than radioactive decay, and any circumstellar material has not been overtaken by the SN ejecta.

Discovering the youngest SNe and studying them in detail is an observational challenge requiring fast reaction and pre-planning. However, when luck strikes and a SN is caught early with prompt follow-up, the results are exceptional. For instance, the "once-a-generation" event SN 2011fe exploded in the nearby galaxy M101 and was first observed roughly 11 hours after explosion (Nugent et al. 2011). These early data led to constraints on the progenitor's size and composition—it had to be $\lesssim 0.02\ R_\odot$ in size (Bloom et al. 2012) and be composed of carbon and oxygen—proof that the progenitor was a carbon-oxygen white dwarf. Early radio and X-ray non-detections of the SN ruled out most scenarios in which the white dwarf was accreting material from a non-degenerate companion (Chomiuk et al. 2012; Horesh et al. 2012; Margutti et al. 2012), while the shape of the early optical lightcurve also put constraints on any companions to the exploding white dwarf (Bloom et al. 2012).

*Table 4.2.*

| SN Feature | Timescale | Brightness (mag) | Probe of? |
|---|---|---|---|
| **Core-Collapse SNe** | | | |
| Shock breakout cooling | 3 days | $-17 < M_V < -15$ | Progenitor Radius & Energy |
| Flash spectroscopy | 1 day | $-17 < M_V < -15$ | Ejecta Mass CSM Comp/Extent |
| **SNe Ia** | | | |
| Companion interaction | < 5 days (RG) | $-17 < M_V < -16$ | Nondegenerate Companion Test |
| | 1 day (MS) | $-16 < M_V < -15$ | |
| Early light-curve shape | < 3 days | $M_V < -14$ | Nickel Distribution |
| | | | Progenitor Radius |
| Carbon | Early as possible | N/A | Double Degenerate Test |

## Technical Description

Multiple observational hints in the earliest moments after a SN explosion provide clues to the explosion physics and the progenitor system (Table 4.2), and we describe these further below. Note that each of the "experiments" touched on below will have an efficiency problem (or opportunity, depending on one's perspective). First, many transients discovered by LSST will not be sufficiently young because of the ~ 3-day cadence. Second, those that are young may not be of the type desired. As a result, a relatively small percentage of the observations obtained to identify and characterize young SNe will be useful for this particular program. However, nearly every observation taken as part of this program will be valuable for other efforts such as the "characterization of the transient sky" science case described above.

**Core Collapse SNe:** For several nearby core-collapse (CC) SNe, the progenitor star has been observed directly from pre-explosion, deep *HST* imaging (see, e.g., Smartt 2009). Another route to learning about the progenitor is by observing the shock breakout of the star, which constrains its radius and potentially its circumstellar medium (CSM) but is only visible in the first hours to days after explosion.

*Shock Breakout and Cooling* — After the core of the star collapses, a shock wave travels outwards, reaching the outer envelope of the star over the course of hours. When the shock reaches regions of low optical depth, the supernova is recognizable and begins to shine. This shock breakout and subsequent cooling is most visible in the X-ray and ultraviolet (e.g., Gezari et al. 2008; Modjaz et al. 2008; Soderberg et al. 2008; Schawinski et al. 2008) but can also be visible in blue optical bands (Figure 4.2; e.g., Richmond et al. 1994).

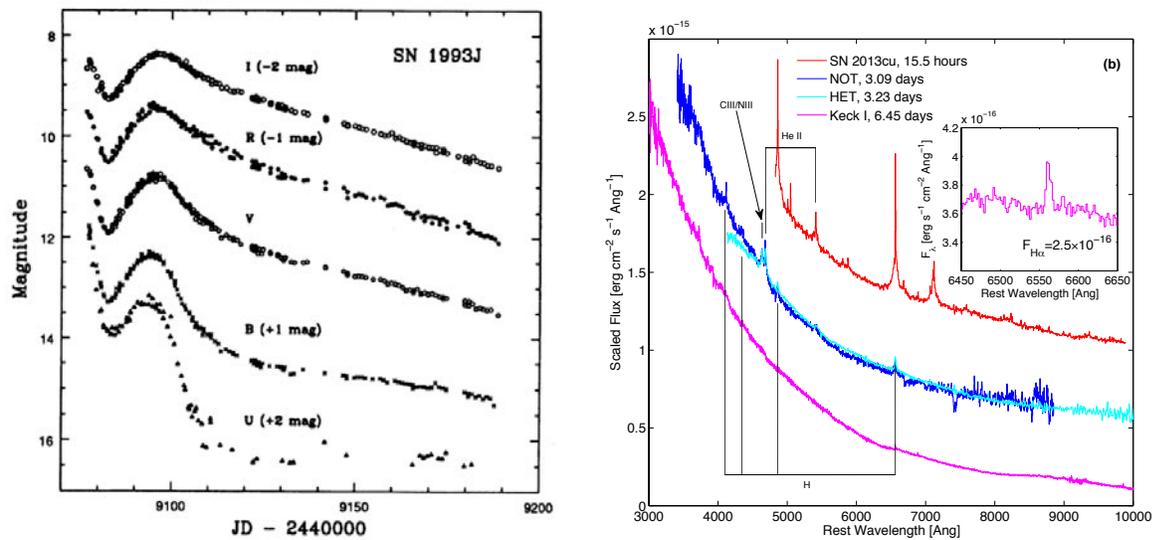

*Figure 4.2. Left panel: Post-shock breakout cooling for SN 1993J, a SN IIb (Richmond et al. 1994), seen as the lightcurves decline in the first days after explosion. The cooling tail is visible for absolute magnitudes of $-17 \gtrsim M \gtrsim -15$ mag. These data can constrain the progenitor star radius and the explosion energy to ejecta mass ratio. Right panel: "Flash spectroscopy" of the Type IIb SN 2013cu, where the earliest observation (~15.5 hours after explosion) shows a hot recombination spectrum, reflecting the flash ionization of the CSM of the Wolf-Rayet progenitor star by the shock breakout from the SN. (Gal-Yam et al. 2014)*

Several analytic cooling models allow us to measure the progenitor star's radius as well as the ratio of explosion energy to ejecta mass (e.g., Nakar & Sari 2010; Rabinak & Waxman 2011). Utilizing these, as well as more detailed hydrodynamic modeling techniques (e.g., Bersten et al. 2012), requires multiwavelength observations as early as possible. Shock cooling measurements have been sparse up until now, and interpretations often push the limits of the data (i.e., rely on early single-band photometry). What is needed is a full suite of optical observations, including multi-band photometry and spectroscopy. This allows for temperature fits over several epochs so that the cooling evolution can be measured to compare with the various cooling models. We note that measuring the progenitor's radius with shock cooling methods is complementary to detecting the progenitor system directly via *HST* imaging, as the imaging method can be complicated by the presence of a binary companion.

*Flash Spectroscopy* — As mentioned above, the shock breakout from the exploding SN produces X-ray/UV light. These high-energy photons can ionize any surrounding CSM, resulting in a recombination spectrum that reflects the CSM composition (Figure 4.2; Gal-Yam et al. 2014). The ability to analyze this fleeting (< 1 day) spectrum, known as "flash spectroscopy," is another benefit of observing the very early evolution of a SN. By measuring the emission line evolution, we can also estimate the physical extent of the CSM. Such nearby CSM probes the final years of the progenitor star prior to explosion. While this technique is still in its infancy, a sample of flash spectroscopy events from various SN types discovered by LSST will provide an invaluable map of the still-uncertain CSM/progenitor connection.

**Type Ia SNe:** The need for a deep understanding of the SN Ia progenitor(s) is particularly acute due to their utility as standardizable candles (Phillips 1993) used to measure the expansion history of the Universe (Riess et al. 1998; Perlmutter et al. 1999). The fact that more luminous SNe Ia occur in late-type galaxies with more star formation while fainter SNe Ia prefer old stellar populations (e.g., Hamuy et al. 1996; Howell et al. 2001) offers a hint that there may be more than one way to make a SN Ia. It is likely that a mix of the single-degenerate (SD) and double-degenerate (DD) scenarios is necessary—SN 2011fe is almost certainly a product of the DD scenario (Nugent et al. 2011), while others that show signs of CSM are most easily explained by the SD model (e.g., Patat et al. 2007; Dilday et al. 2012). The early-time SN Ia data obtained through this program will provide clues both to the progenitor and explosion mechanism.

*Early-Time Lightcurves and Spectroscopy* — One of the least explored aspects of SNe Ia is the behavior of their lightcurves at very early times ($t \lesssim 1$ day), a regime in which few observations have been made with high signal-to-noise ratio (S/N) (Nugent et al. 2011; Foley et al. 2012; Zheng et al. 2013, 2014; Goobar et al. 2015; Firth et al. 2015). Even so, it is clear that SN Ia do not all adhere to the simple "fireball" approximation (in which the lightcurve rises like $t^2$; Arnett 1982) and that the early rise may uncover further SN Ia physical diversity. SNe Ia are expected to have a "shock breakout" similar to CC SNe, but on a shorter timescale (and virtually unobservable), followed by a potential "dark" period before radioactive decay powers the rising lightcurve (Piro & Nakar 2013). The early-time lightcurve of a SN Ia can provide constraints on the radius of the progenitor (Nugent et al. 2011; Bloom et al. 2012), the distribution of $^{56}$Ni (e.g., Piro & Nakar 2013), and the existence of a normal companion star (Figure 4.3; Kasen et al. 2010). Early-time spectra probe the outermost layers of the explosion, providing unique information about the explosion mechanism and the amount of unburnt material (e.g., Parrent et al. 2011; Silverman & Filippenko 2012; Folatelli et al. 2012).

## Needed Capabilities and Estimate of Demand

The goal of this program is to test potential progenitor scenarios and explosion mechanisms of SNe Ia and understand the diversity of shock breakout and flash spectroscopy signatures for CC SNe.

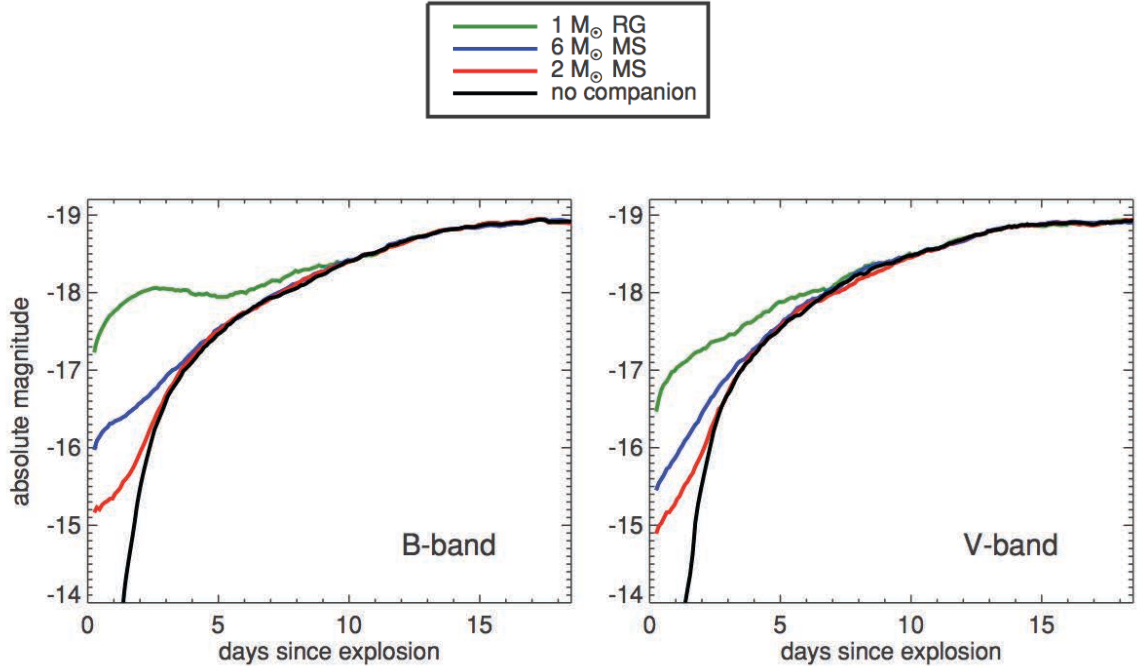

*Figure 4.3. Simulated lightcurves with early shocking due to SN Ia ejecta striking a nondegenerate companion. This is for a viewing angle aligned with the collision region of the degenerate white dwarf and the non-degenerate companion star, which should occur ~ 10% of the time in the SD case. While strong constraints have been placed on red giant companions, the main sequence companion scenario has only been ruled out for individual events. (Reproduced from Kasen 2010)*

To have a sufficiently large sample for each CCSN subclass, we would require a total sample size of ~ 100 very young CCSNe. Each object in this sample would require five non-LSST epochs of photometry in four optical bands within the first week of explosion (e.g., Valenti et al. 2014) in order to characterize the shock cooling temperature decline (estimated total time: 200 hours on a 4m-class telescope). Similarly, five epochs of medium-resolution spectroscopy in the same 1-week time frame would sample the evolution of flash-ionized lines and their temperature evolution (estimated time: 400 hours of spectroscopy on both 4m- and 8m-class telescopes). Later-time, comprehensive photometry to characterize the lightcurves of each CC SN subset would require ~ 10 light-curve points in four bands; if these data were acquired for half of the early CCSN sample, it would require 400 hours of 4m and 200 hours of 8m telescope time. Finally, comprehensive spectroscopy with a medium-resolution workhorse spectrograph for detailed study of half the sample would require ~ 10 addition spectra for 50 SNe, resulting in an additional 250 hours of 4m and 250 hours of 8m time.

For the SN Ia program, we propose to obtain early-time lightcurves for a sample of 1000 SNe Ia; if current evidence holds true, this should result in detections of ~100 red-giant companions and perhaps as many as 10 main-sequence companions. This requires five epochs of *U+B* photometry for each SN, resulting in 1 hour per SN on a 4m telescope. In total, this stage would require 100 nights of 4m time. Each of these SNe would also require a single spectrum on an 8m-class telescope for a total of 50 nights. For those SNe where companion interaction is possibly detected, we propose to fully characterize their lightcurves and obtain additional spectroscopy, requiring an additional 1 hour of photometry and 1 hour of spectroscopy per SN for a total of 10 additional nights of photometry and spectroscopy, respectively. These data would also be useful for constraining the $^{56}$Ni distribution. Additionally, much of these data could be also used for the science goal of characterizing SNe Ia (see section on Type Ia SN Demographics in this chapter).

Our specific recommendations are detailed at the end of this section.

## Science Case: Electromagnetic Counterparts to Gravitational Wave Sources

Since Einstein predicted the existence of gravitational waves (GW) in his theory of General Relativity a century ago, their direct detection has become one of the most highly anticipated events in physics and astronomy. The Advanced Laser Interferometer Gravitational-wave Observatory (aLIGO) began operations in September 2015 and almost immediately detected GW150914, the first known gravitational wave event arising from the merger of two black holes (Abbott et al. 2016a). This landmark discovery confirmed Einstein's predictions and enabled the precise determination of basic parameters for the compact objects, such as their masses and final spin. However, the GW signal had a poor localization of ~ 600 deg$^2$ and a luminosity distance of 410 Mpc with a large uncertainty of ~ 40%, preventing a complete understanding of the merger and its placement within a galaxy (Abbott et al. 2016a).

A coincident electromagnetic (EM) signal would have significantly leveraged the GW detection by providing a more precise redshift and information on the localization, energy scale, and properties of the environment on sub-parsec to kiloparsec scales. Indeed, the tentative detection of an EM signal at gamma-ray wavelengths detected by the *Fermi* satellite significantly reduced the localization uncertainty of GW150914 to ~ 200 deg$^2$ (Connaughton et al. 2016). By the 2020s, the Virgo detectors as well as additional aLIGO detectors will be online and together will form the aLIGO-Virgo (ALV) network, with enhanced localization capabilities compared to aLIGO alone. For instance, ~ 50% of events will have localization uncertainties of ≤ 20 deg$^2$ (Abbott et al. 2016b).

The most promising counterpart of a compact-object merger at optical and near-infrared (NIR) wavelengths is a "kilonova," a relatively long-lived and moderately luminous transient powered by the radioactive decay of heavy elements synthesized in the merger ejecta (Li & Paczynski 1998). Recently, kilonovae were predicted to be faint in the optical, but bright in the NIR, with a transition between these regimes at ~1 μm, with the break corresponding to line blanketing of newly synthesized *r*-process elements. Given their relatively small ejecta masses, kilonovae are expected to fade on 1-week timescales (e.g., Figures 4.4 and 4.5; Metzger et al. 2010; Barnes & Kasen 2013). Their luminosities and temporal evolution map to ejecta masses and velocities; thus, optical and near-IR photometry of kilonovae offers the only way to place robust constraints on these properties. For instance, in 2013, the first kilonova was detected following a short-duration gamma-ray burst, providing an estimate of the ejecta mass and velocity, ~ 0.05 $M_\odot$ and v ≈ 0.1 − 0.3 c (Figure 4.4; Berger et al. 2013; Tanvir et al. 2013). Compared to other transients, kilonovae have a distinct red color with *r*–*H* ≥ 3 mag and a unique rise time of $t_{rise}$ ≤ 4 days; they are also quite faint with $M \approx -15$ mag (Barnes & Kasen 2013). Thus, cuts on color, timescale, and brightness will greatly help to distinguish kilonovae from other types of explosive transients. Finally, the only way to unambiguously confirm that the event is a kilonova is through spectroscopy, which will confirm the composition of the neutron-rich ejecta and explain the origin of heavy elements in the Universe.

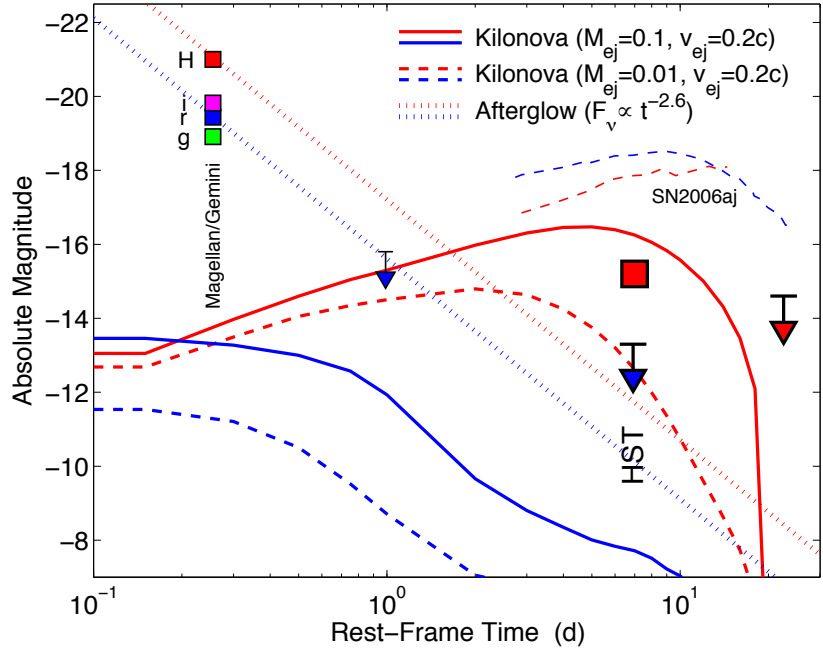

Figure 4.4. Lightcurve of the GRB 130603B afterglow. Dotted curves show the expected r (blue) and H (red) afterglow evolution. HST images taken six days after the GRB (in the rest frame) show a strong excess of light in H relative to the expected afterglow decay. At the same epoch, the object is not detected in deep r images. The late-time flux is consistent with kilonova models (Berger et al. 2013; Tanvir et al. 2013), which are shown as solid and dashed lines (Barnes & Kasen 2013). (From Berger et al. 2013)

## Technical Description

The ALV network will be able to detect kilonovae from compact object mergers to a distance of 200 Mpc. The most robust strategy to detect a kilonova is to image the entire GW localization region multiple times to detect transient emission. The expected event rate is ~ 40 events yr$^{-1}$, and approximately 1/2 of these events are expected to have positional uncertainties of $\lesssim$ 20 deg$^2$. Our observing strategy is twofold: we will obtain (1) multi-band imaging with a wide field of view (~ few deg$^2$) and over several weeks to determine kilonova candidates, and (2) spectroscopy to obtain crucial information about the composition of the ejecta. For imaging, a comparison to kilonova models at 200 Mpc demonstrates that to perform an effective kilonova search, one must obtain at least 3 observations within 20 days after the merger to depths of $r_{AB} \approx$ 25 mag and $y_{AB} \approx$ 24 mag (Figure 4.5). Below, we provide a "straw man" plan of needed capabilities, but we note that this is a relatively young field and our exact observing strategy may change slightly in the coming years.

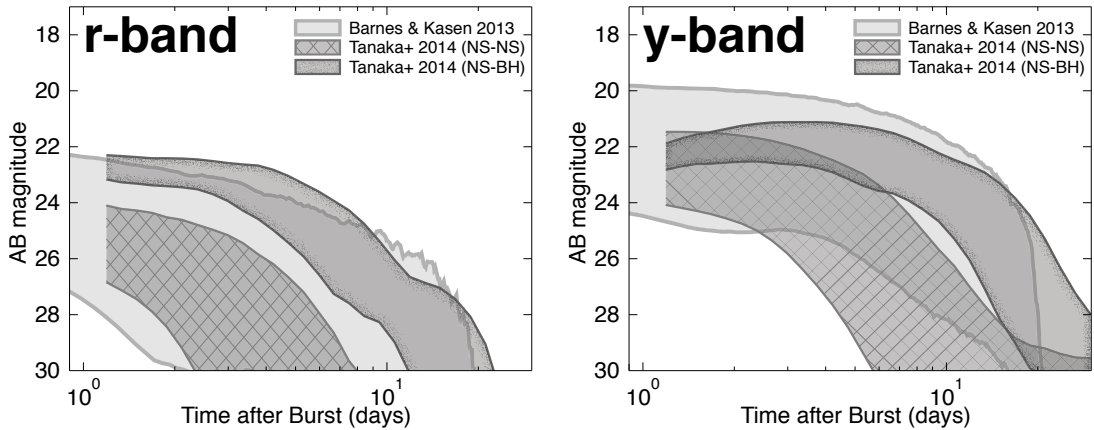

*Figure 4.5. Kilonova lightcurve models for a compact object merger at 200 Mpc. Kilonovae are unique in their distinctive red color, and thus effective kilonova searches require at least two filters to depths of ~ 24 AB mag.*

## Needed Capabilities and Estimate of Demand

The detection of an EM counterpart to a GW source would significantly improve our knowledge of general relativity and open new opportunities related to the physics of the event. While other facilities can possibly detect the EM counterparts, LSST can be the clear leader in this area if planning is done now. In particular, we suggest a coordinated effort where LSST interrupts its nominal observing program to follow GW triggers. When LSST is operating, the typical localizations may be as small as 20 deg$^2$, which would require 3 LSST pointings to cover. Observing each pointing in 2 filters for 75 seconds each will require 7.5 min of integration time, or a total of ~ 10 minutes including overheads. Assuming 3 epochs in 2 filters are needed to distinguish kilonovae from other transients (Cowperthwaite & Berger, 2015), this requires a total of 30 min of LSST time per GW event. Assuming we can trigger LSST on 20 events (half of the yearly expected rate), this amounts to ~ 15 hour per year of LSST time. Given the rapid timescale, this necessitates target-of-opportunity operations with a response time of ≲ 2 days and the ability to determine the filters of observations.

Our knowledge of EM counterparts to GW sources is heavily influenced by the properties of the late-time emission associated with GRB 130603B (Figure 4.4; Berger et al. 2013; Tanvir et al. 2013). Using *HST*, this source was detected in the *H* band but not in the *V* band, with *V*–*H* > 1.9 mag. This is consistent with the kilonova models of Barnes & Kasen (2013). However, recent work performing a more detailed analysis of heavy element opacities predicted a spectral break at > 1 μm (Fontes et al. 2015). If this conclusion holds, LSST will only be able to detect a small fraction of kilonovae in its redder filters and will require contemporaneous imaging in NIR filters. Moreover, if a kilonova is detected in LSST bands (e.g., *y*), NIR photometry would provide essential color information on the source. In particular, searches in the *J* band to a depth of $J_{AB} \approx 23$ mag would significantly improve the likelihood of distinguishing a kilonova from other transients. A new wide-field NIR imager on a > 6m telescope could perform a NIR search in concert with LSST. This facility should be able to observe 20 deg$^2$ to $J_{AB} \approx 23$ mag in a single night. A 3 deg$^2$ imager on a 6m telescope could achieve this in 4.5 hours (1 hour of on-source time and 50% overhead). With 3 epochs, this amounts to 4.5 hours per GW event, or 90 hours per year. We note that if LSST detects candidate kilonovae, in the absence of a wide-field NIR imager, a narrow-field NIR imager on a $\gtrsim$ 6m-class telescope with $J$ and $K$ bands would provide invaluable color information to help vet the candidates. If LSST detects candidate kilonovae for 10 events per year, we would require $\sim$ 30–40 hours per year with a narrow-field imager.

Once candidate EM counterparts are detected, spectroscopy will be necessary both to confirm a source as the counterpart and to determine the composition of the ejecta. For the red kilonova case, we would require a NIR spectrograph on a 30m-class telescope to obtain a high-S/N spectrum of the source. We estimate $\sim$ 45 minutes per candidate including overheads and calibrations. If LSST detects 10 kilonova candidates per year, this amounts to $\sim$ 8 hours on a NIR spectrograph on a 30m telescope. If the counterpart is bright in the optical, an optical spectrograph on a smaller (perhaps 8m-class) telescope would be adequate. Not knowing the true colors of kilonovae, we conservatively estimate 5 candidates per year requiring 1.5 hour each, including overheads and calibrations. This totals $\sim$ 8 hours per year on an optical spectrograph on an 8m telescope. As the kilonova ejecta are expected to have mildly relativistic velocities, we expect any spectral features to be broad. Therefore low-spectral resolution will be adequate. To both reject potential false positives and obtain the most information about the EM counterpart, we need the largest wavelength range possible. Therefore, we recommend a spectrograph that covers all optical and NIR wavelengths from $\sim$ 0.3 to 2.5 $\mu$m on all $\geq$ 8m-class telescopes.

## Overall Recommendations

In summary, the following is an integrated list of the capabilities needed to carry out the science programs described in this chapter.

### 1. Continued robust support for transient broker [Priority 1]

The highest priorities for this science topic are related to identification of transient objects and dissemination of their salient characteristics as quickly as possible to enable follow-up observations with other facilities. The broker is key to this effort, as its determinations will drive the community response. The broker software must rapidly distinguish exotic supernovae, very young SN candidates, SNe Ia across parameter space, and potential GW EM counterparts from the stream of astrophysical alerts. It must then characterize these sources successfully to warrant prompt follow-up observations. Follow-up data should subsequently be fed back to the broker to refine characterization and potentially alter priorities for ongoing observations. The broker should be publicly accessible, with the ability for members of the public to run additional selection algorithms.

LSST GW follow-up observations may be substantially different from normal operations, including different exposure times, different dither patterns, and different available data. As such, the brokers must be tested with these conditions.

## 2. Wide-wavelength coverage, medium-resolution optical/infrared spectroscopy with rapid-response time on 8–30m telescopes [Priority 1]

Prompt spectroscopy of potentially interesting objects (as determined by the broker) is crucial to identifying their nature in a timely way. The most important tool in determining a transient's nature is a broad wavelength range that covers a variety of distinguishing spectral features. Important features exist throughout the entire optical/near-infrared range (0.32–2 μm); a single spectrograph covering this range will guarantee that all wavelengths are covered, that the data are taken simultaneously, and that the telescope efficiency is maximized. The bluest wavelengths are particularly important for measuring shock breakout temperatures and including high-velocity Ca H&K lines. To further increase efficiency and guarantee observations at the most useful times, target-of-opportunity access should be available on as many telescopes as possible.

Some particularly interesting transient classes can only be discovered with moderate-resolution spectra (e.g., by identifying narrow lines produced via CSM interaction and for flash spectroscopy). As the need for very high resolution (i.e., $R > 20,000$) is rare and such spectrographs are typically ill-suited for faint objects, an ideal spectrograph for this purpose would have a resolution of $1000 \lesssim R \lesssim 8000$. Observations of the SN Ia sample would require optical/infrared spectrographs with a resolution of ~ 1000 over the wavelength range 0.32–2 μm deployed on 3–10m-class facilities. We consider it essential to push the wavelength coverage into the blue as far as possible because a) early CC spectra are (roughly) hot black bodies during the shock breakout cooling phase, and wavelengths below ~ 3800 Å will be crucial for measuring temperature and thus constraining the progenitor star radius; and b) high-velocity Ca H&K lines are seen in early SN Ia spectra and may provide hints of the near environment of the SN, but a spectrograph with wavelength coverage to (for instance) only 3800 Å will **not** see these features.

Most existing telescopes have low- to moderate-resolution optical spectrographs, and we expect these will provide a significant amount of data for this science case. However, we specifically recommend ensuring that at least one wide-wavelength coverage, moderate-resolution spectrograph on an 8m-class telescope (e.g., Gen4#3 on Gemini-South) and at least one on a 30m-class telescope are available for LSST science. Such an instrument on a 30m-class telescope is necessary to observe the faintest and/or most interesting objects, and especially kilonovae.

## 3. Observing infrastructure/modes [Priority 1]

This program requires access to a wide range of instruments and facilities. Coordination of observations across the available possibilities will require efficient telescope scheduling software as well as real-time feedback to assess data quality. This will necessitate trustworthy data-reduction pipelines. Automated ingestion of urgent ToOs (where observations need to be taken within a few hours) by 8m (and smaller)-class telescopes with no human in the loop, in conjunction with the transient brokers and observing platforms described above. In addition, observatories should plan for different observing modes, stressing target of opportunity and dedicated follow-up for high-priority targets as identified by the broker and the scheduler. Automated triggering of follow-up observations, depending on these results, would be the icing on the cake.

## 4. Target-of-Opportunity operations for LSST [Priority 1]

Coverage of a typical GW localization area of 20 deg$^2$ would require only 3 LSST pointings. LSST has the opportunity to be the leader in this area, but ToO triggers for LSST will be critical for this effort. The expected rate and the amount of observing time per event is not particularly large given the large scientific benefit.

Because a response of < 2 days is necessary, ToO observations are required. We recommend a system in which LSST automatically breaks its current schedule with a 1-day notice and starts GW follow-up observations using a predetermined algorithm that takes into account several factors such as airmass, moon angle, Galactic reddening, and the GW localization. This algorithm should be determined in advanced with contributions from several groups.

## 5. Rapid optical and NIR photometry from a wide-aperture range [Priority 2]

Multi-band lightcurves are important observational tools in classifying and studying SNe and other transients. Photometric instrumentation is common on telescopes of all sizes (1–10m apertures), especially in the optical region. The intrinsic LSST cadence may be insufficiently long for characterizing exotic transients, especially the faster-evolving varieties. And additional optical filters can provide unique and important information. Some transients will brighten to the point where LSST saturates. Therefore, other telescopes must supplement the LSST photometry to fully understand these objects.

Since many observations will be targeted and the objects will be relatively bright, narrow-field (< 10 arcmin) imagers on 1–6m-class facilities are generally sufficient. However, other studies will want to focus on the transients discovered in the deep drilling fields, where a wide FoV instrument is more efficient. Many such instruments exist, but coordination, cross-calibration, and scheduling issues will have to be resolved.

While NIR photometry is currently less commonly used for transient studies, NIR lightcurves of newly discovered exotic transients would improve color and temperature information for these object classes, probe dust formation in their outer layers and at later times, and help quantify the effects of shock heating on the local ISM.

## 6. Rapid-response ToO spectrographs on a 4m telescope [Priority 2]

Similar wide-wavelength-coverage spectrographs as those described above, situated at a 4m-class facility, would ease the burden on larger-aperture telescopes and be well suited for nearby and early SNe.

## 7. Observing platform/exchanges [Priority 2]

Creation of a software platform that allows research groups to communicate with each other about completed or planned observations to minimize overlap and encourage collaboration. This will improve efficiency and improve temporal coverage of follow-up observations.

## 8. (Wide-field) NIR imagers [Priority 2]

Detection and characterization of EM counterparts to GW sources is significantly aided by NIR photometry. While a small FoV imager on a sufficiently large-aperture telescope could observe potential counterparts discovered by LSST, a wide-FoV imager is necessary to detect particularly red sources and to have the fastest response.

A wide-FoV NIR imager on a > 6m-class telescope could survey the sky in a fashion similar to LSST, providing a NIR transient survey deep enough to significantly enhance the LSST transient datasets. If such a facility existed and if it devoted most of its time to a coordinated survey, it would be sufficient to obtain nearly all of the transient NIR photometry mentioned above in Recommendation 5.

## 9. Optical spectropolarimetry at large-aperture facilities [Priority 3]

Spectropolarimetry has proven to be an invaluable tool for studying the geometry of supernova explosions, ejecta, and circumstellar material. Optical spectropolarimetric capabilities on 10–30m telescopes in the LSST era will allow us to precisely characterize the explosions and immediate environments of exotic transients, as well as constraining their progenitors. Because spectropolarimetry yields three-dimensional geometrical information unobtainable by other means, multi-epoch spectropolarimetry will play an important role in characterizing the transient sky, illuminating potential connections among phenomena whose interrelations are as yet unrecognized.

Spectropolarimeters currently exist at Gemini South, Keck, LBT, and SALT, but these instruments are not always available, especially since some are general-purpose spectrographs that require the explicit inclusion of polarimetric optics. There is currently no planned spectropolarimetric instrument for a 30m-class telescope. Future low- to medium-resolution spectrographs for 10–30m telescopes should be designed with spectropolarimetric capabilities in mind in order to meet the requirements of this science case. Ideally, these instruments would allow for the polarimetric optics to easily move into / out of the beam during the night so that target-of-opportunity observations can be easily performed.

## 10. Galaxy catalog [Priority 3]

As the ALV network will only detect NS-NS mergers to a maximum distance of 200 Mpc, a galaxy catalog with accurate distances can be used as a prior when determining the likelihood of a particular EM source being associated with the GW source. This will be particularly useful when the error boxes are too large to tile and the only feasible search strategy are searches around galaxies within the aLIGO detection volume. We recommend making such a catalog early in the operations of LSST for aiding the detection of EM counterparts to GW events.

# Summary Tables

## Table 4.3. Needed Capabilities

| | Infrastructure | < 3m | 3–5m | 8m | 25m |
|---|---|---|---|---|---|
| **Characterizing Transients** | **Transient Broker**<br><br>**New observing modes, additional ToO opportunities**<br><br>Software to coordinate observations | 0.3–1μm $R \approx$ 5000 single-object spectrograph > 10 x 10 arcmin FOV OIR imager | 0.3–1μm $R \approx$ 5000 single-object spectrograph > 10 x 10 arcmin FOV OIR imager | **0.3–2μm $R \approx$ 5000 single-object spectrograph** > 10 x 10 arcmin FOV OIR imager *0.3–1μm R ≈ 5000 spectropolarimeter* | **0.3–2μm $R \approx$ 5000 single-object spectrograph** > 10 x 10 arcmin FOV OIR imager *0.3–1μm R ≈ 5000 spectropolarimeter* |
| **SNe Ia** | **Transient Broker**<br><br>**New observing modes, additional ToO opportunities**<br><br>Software to coordinate observations | > 10 x 10 arcmin FOV OIR imager | **0.3–2.3μm R ≈ 5000 single-object spectrograph** > 10 x 10 arcmin FOV OIR imager | **0.3–2.3μm R ≈ 5000 single-object spectrograph** *0.3–1.3μm R ≈ 5000 spectropolarimeter* | |
| **Early Sne** | **Transient Broker**<br><br>**New observing modes, additional ToO opportunities**<br><br>**Software to coordinate observations** | > 10 x 10 arcmin FOV OIR imager | 0.3–2μm $R \approx$ 5000 single-object spectrograph | **0.3–2μm $R \approx$ 5000 single-object spectrograph** | |
| **GW EM Counterparts** | **LSST ToO Triggering**<br><br>**Transient Broker**<br><br>*Nearby Galaxy Catalog* | | | **~3 deg² FOV NIR imager**<br><br>0.3–2.3μm R ≈ 5000 single-object spectrograph | 0.3–2.3μm R ≈ 5000 single-object spectrograph |

Entries in boldface type indicate that the capability is **Priority 1 (critical)**.
Roman type indicates Priority 2 (very important).
Italic type indicates *Priority 3 (important)*.

*Table 4.4. Resource Demand*

| | Infrastructure | < 3m | 3–5m | 8m | 25m |
|---|---|---|---|---|---|
| **Characterizing Transients** | **High-performance computing for broker** | 50 hours spectroscopy<br><br>250 hours optical imaging<br><br>250 hours NIR imaging | 100 hours spectroscopy<br><br>125 hours NIR imaging | **300 hours spectroscopy**<br><br>125 hours NIR imaging | **50 hours spectroscopy** |
| **SNe Ia** | **High-performance computing for broker** | 5000 hours optical imaging<br><br>5000 hours NIR imaging | 2250 hours optical imaging<br><br>9750 hours NIR imaging<br><br>**6500 hours OIR spectroscopy** | 2250 hours NIR imaging<br><br>**7750 hours OIR spectroscopy** | |
| **Early SNe** | **High-performance computing for broker** | | 1500 hours optical imaging<br><br>850 hours OIR spectroscopy | 200 hours optical imaging<br><br>**1250 hours OIR spectroscopy** | |
| **GW EM Counterparts** | **High-performance computing for broker**<br><br>**150 hours of LSST time for follow-up** | | | **900 hours NIR imaging**<br><br>80 hours OIR spectroscopy | 80 hours NIR spectroscopy |
| **Total On Sky Time** | | ~ 2.9 years | ~ 5.7 years | ~ 3.5 years | ~ 0.1 yr |

Entries in boldface type indicate that the capability is **Priority 1 (critical)**.
Roman type indicates Priority 2 (very important).
Italic type indicates *Priority 3 (important)*.

# Chapter 5: Rotation and Magnetic Activity of Stars in the Galactic Field Population and in Open Star Clusters

*Suzanne Hawley (University of Washington), Ruth Angus (University of Oxford), Derek Buzasi (Florida Gulf Coast University), James R. A. Davenport (Western Washington University), Mark Giampapa (National Solar Observatory), Vinay Kashyap (Harvard-Smithsonian CfA), Søren Meibom (Harvard-Smithsonian CfA)*

### Executive Summary

LSST will open up new frontiers in the investigation of gyrochronology, magnetic activity and stellar flares, activity cycles and magnetic fields, in stars in both open clusters and the Galactic field. The determination of accurate ages for millions of stars will transform our understanding of the effects of rotation and magnetic fields on stellar evolution and will enable new insight into the formation and evolution of stellar populations, including exoplanet hosts, throughout the Galaxy. To achieve these goals, high-priority ground-based OIR follow-up resources include wide-field (one square degree) imaging with both broadband (*ugriz*) and narrow band (Ca II H and K) filters and highly multiplexed (1000 fiber) spectrographs at moderate (R = 5000) and high (R = 20,000–100,000) resolution. Polarimetry is extremely valuable for magnetic field measurements and requires a high-resolution spectrograph designed for polarimetric observations. Telescopes from < 3m to 25–30m will be needed, with the smaller telescopes mainly used for imaging and the larger ones for spectroscopy. To implement the program outlined here will require significant telescope time over the 10-year period of LSST operations: 700 nights on < 3m telescopes; 1100 nights on 3–5m telescopes; 850 nights on 8–10m telescopes; and 220 nights on 25–30m telescopes. We also require that the individual 15-second LSST images be made available as Level 3 data products (the same as the standard LSST data §5.2.4) and that several LSST special survey ("deep drilling") fields be focused on specific open clusters, including especially the iconic solar age cluster M67, which lies just outside the nominal LSST footprint (§5.3.2). The results of this program will not only advance stellar astrophysics but also synergistically inform other fields such as the nature and evolution of exoplanet system environments and the evolution of the Galaxy.

## Introduction and Background

The Large Synoptic Survey Telescope (LSST) will provide precise ground-based photometric monitoring of billions of stars in the Galactic field and in open star clusters. The lightcurves of these stars will give an unprecedented view of the evolution of rotation and magnetic activity in cool, low-mass main-sequence dwarfs of spectral type GKM, allowing precise calibration of rotation-age and flare rate-age relationships, and opening a new window on the accurate age dating of stars in the Galaxy. Previous surveys have been hampered by small sample size, poor photometric precision and/or short time baselines, so LSST data are essential for obtaining new, robust age calibrations.

The evolution of the rotation rate and magnetic activity in solar-type stars are intimately connected. Stellar rotation drives a magnetic dynamo, producing a surface magnetic field and magnetic activity, which manifests as starspots, chromospheric (Ca II HK, Hα) and coronal (X-ray) emission and flares. The magnetic field also drives a stellar wind, causing angular momentum loss ("magnetic braking"), which slows the rotation rate over time, leading to decreased magnetic activity. More magnetically active stars (larger spots, stronger Ca II HK, Hα and X-ray emission, more flares) therefore tend to be younger and to rotate faster. The rotation-age relationship is known as gyrochronology, and the correlation between rotation, age, and magnetic activity for solar-type stars was first codified by Skumanich (1972). However, the decrease in rotation rate and magnetic field strength over long timescales is poorly understood and, in some cases, hotly contested (e.g., Angus et al. 2015; van Saders et al. 2016). Recent asteroseismic data from the Kepler spacecraft have revealed that magnetic braking for G dwarfs may cease at around the solar rotation rate, implying that gyrochronology relations are not applicable to older stars (van Saders et al. 2016).

In addition, the rotational behavior of lower-mass stars is largely unknown due to the faintness of mid- to late-type M dwarfs. There is reason to believe that M dwarfs cooler than spectral type ~M4 may behave differently from the G, K, and early M stars, since that spectral type marks the boundary where the star becomes fully convective, and a solar-type shell dynamo (which requires an interface region between the convective envelope and radiative core of the star) can no longer operate. Using chromospheric Hα emission as a proxy, West et al. (2008) studied a large sample of M dwarfs from SDSS and showed that magnetic activity in mid–late M dwarfs lasts much longer than in the earlier type stars.

LSST will provide photometric rotation periods for a new region of period-mass-age parameter space. The *Kepler* spacecraft focused on Earth-like planets with Sun-like hosts, thus the majority of its targets were G type, with fewer K and M dwarfs. Unlike *Kepler*, however, any target falling within LSST's field of view will be observed—not just those on a predetermined target list. In addition, due to the large collecting area of LSST, it will be sensitive to a significant population of distant K and M dwarfs. LSST will operate for 10 years, more than double the length of the *Kepler* prime mission. This long time baseline will enable rotation signatures of faint, slowly rotating stars to be detected, populating both low-mass and old regions of the age-rotation parameter space. Thus, LSST will provide an important complementary dataset to *Kepler* (and the upcoming TESS mission).

The LSST data will also allow an unprecedented view of stellar flares and the calibration of flare rate-age relations that may provide an additional method for age-dating M dwarf populations. Kowalski et al. (2009) used sparsely sampled SDSS lightcurves in Stripe 82 to quantify M dwarf flare rates as a function of height above the Galactic plane and showed that flare stars may comprise a younger population than active stars (those showing Hα emission). Long time baselines and monitoring of large numbers of stars are required to obtain good flare statistics, so LSST will be perfectly suited for this study. Such flare statistics are also essential to understand the radiation and particle environment near exoplanet host stars (many of which are M dwarfs), and how it evolves in time. The near-space environment can have significant impact on the habitability of nearby exoplanets.

As coeval, equidistant, and chemically homogeneous collections of stars, open star clusters with different ages are ideal for studying the dependencies of astrophysical phenomena on the most fundamental stellar parameters—age and mass. Indeed, there are few fields in astronomy that do not rely on results from cluster studies, and clusters play a central role in establishing how stellar rotation and magnetic activity can be used to constrain the ages of stars and stellar populations. In particular, clusters provide essential calibration for rotation-age-activity relations, since each cluster gives a snapshot of stellar evolution at a single age, for all masses (e.g., Meibom et al. 2009, 2011a, b, 2015; Giampapa et al. 2006; Gondoin et al. 2012; Gondoin 2013; Wright et al. 2011).

LSST will also enable the use of cluster and field stars as laboratories for investigating magnetic activity cycles (such as the 11-year cycle on the Sun). There is some evidence that younger, more active stars are less likely to show regular cycle behavior, while older stars such as the Sun typically do show regular cycles (Baliunas et al. 1995). Due to the long monitoring times that are required to diagnose activity cycles, it has previously been difficult to carry out a large-scale survey of activity cycle behavior and thus quantify the changes that occur with magnetic dynamo evolution (e.g., as the star spins down). LSST will easily rectify this situation, and indeed the cadence will be well-suited to observing cyclic behavior both in the field and in clusters. The changes that occur in the surface magnetic field (both strength and topology) as a star ages are also not well understood, with fewer than 100 (nearby, bright) stars presently having good measurements. Follow-up observations of stars from a large LSST sample covering a range of ages and masses with well-determined cycle periods will open a new window on the study of magnetic field evolution.

While LSST lightcurves have the potential to answer some of the most fundamental questions regarding the evolution of stellar rotation and magnetism, it is essential that the properties of the target stars be accurately determined. In order to understand the scope of the follow-up observations that will be needed to characterize magnetically active stars that exhibit starspots and flares, we performed a number of simulations. We first describe simulations of field stars at several galactic latitudes, sampled with an LSST cadence, and examine the target densities of stars with detected rotation periods (due to starspot modulation) and detected flares. We then consider open clusters at different ages, predict rotation and flare rates, and discuss the complications of determining cluster membership. Finally we look at the constraints for determining activity cycles and measuring magnetic fields directly. Using the results of these simulations, we outline the follow-up requirements necessary to fully exploit the LSST dataset for stellar rotation and magnetism studies both in the Galactic field and in open clusters.

Although we focused our study here on magnetically active stars, in order to provide well-defined estimates for follow-up capabilities, we note that similar observing strategies and follow-up resources will be valuable for investigations of a wide variety of variable stars, including eclipsing binaries, pulsating stars at a wide range of periods from RR Lyrae to Cepheids to LPVs, cataclysmic variables and novae, and also for other Galactic variability phenomena such as microlensing and planetary transits. Wide-field follow-up imaging and spectroscopic (both moderate- and high-resolution) facilities will be useful for all of these stellar science topics.

## Simulating LSST field star samples

### Cadence Model

We used the *minion_*1016 cadence model, which is the most recent (May 2016) "baseline cadence" being tested for LSST simulations.[1] This model has some deviations from a regular, every few days, cadence, including fewer u and g band observations, and very few observations at high galactic latitude (e.g., b = −80). However, we have adopted it for the purposes of this study and have only considered galactic latitudes between −60 and −10 degrees.

### Field Star Populations

We used the TRILEGAL (Girardi et al. 2005) galaxy simulation code to generate field stellar populations for several representative LSST fields. These fields were centered on the same galactic longitude, *l* = 45 degrees, and four different galactic latitudes: *b* = −10, −20, −40, −60 degrees. Each TRILEGAL field comprises a catalogue of stars with properties including age, effective temperature, gravity, and *ugriz* magnitudes.

Figure 5.1 shows the target distributions for the four fields as a function of apparent magnitude. In total, there are nearly a million stars per square degree in the lowest latitude (*b* = −10 degrees) field, dropping to only a few thousand stars per square degree in the field at the highest latitude studied (*b* = −60 degrees). The right panel shows the distributions of G, K, and M dwarfs for the representative field at *b* = −20 degrees. There are about 10 times as many M dwarfs ($3000 < T < 4000$K) and 6 times as many K dwarfs ($4000 < T < 5000$K) as there are G dwarfs ($5000 < T < 6000$K) in these fields, but the G dwarfs are much brighter (averaging around $g = 20$), while most of the M dwarfs have $g > 24$.

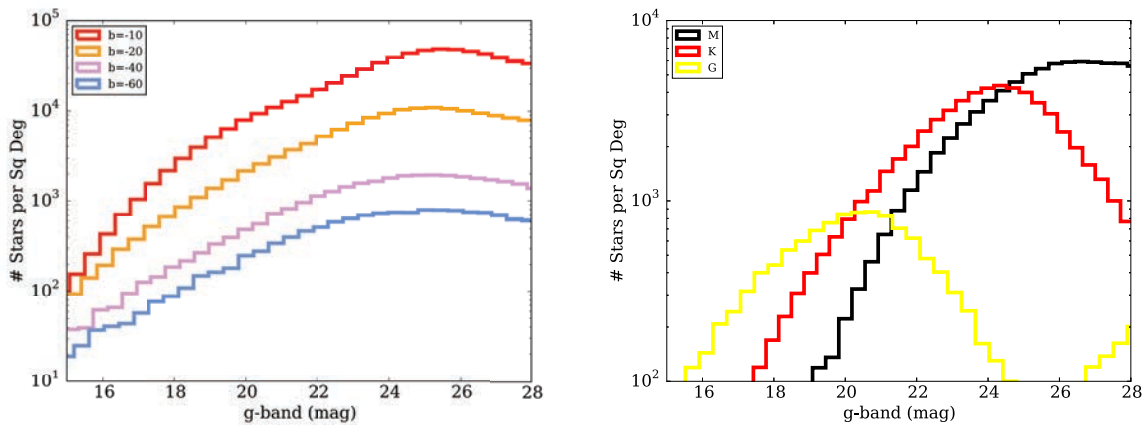

*Figure 5.1. Left: Density of stars along each of the lines of sight as a function of apparent magnitude. Right: Density of G, K, and M dwarfs in the b = −20 TRILEGAL field.*

---

[1] 

## Rotation Simulations

Twenty thousand stars were randomly chosen in each field, and rotation periods were calculated using the Angus et al. (2015) gyrochronology relation, which converts ages and $B - V$ color (calculated from TRILEGAL $g - r$) into rotation periods. Converting $B - V$ color to $g - r$ was done using the transformations from Table 1 of Jester et al. (2005). Code similar to that used in Aigrain et al. (2015) was used to simulate lightcurves by placing dark starspots on a rotating sphere and integrating the total resulting flux over the surface. Stellar flux variations produced by dark active regions on the surface are typically non-sinusoidal, and this starspot model provides a more accurate representation of stellar lightcurves than a simple sinusoid. However, for simplicity we fixed the mean starspot lifetime at 30.5 days for all simulations and did not include differential rotation. Both rapid starspot evolution and differential rotation will result in quasi-periodic lightcurves that will be somewhat more difficult to recover, so our results should be considered "best case."

In order to assign appropriate amplitudes to the simulated lightcurves, we approximated the relation between rotation period, amplitude of variability, and $T_{eff}$, based on the McQuillan et al. (2014) *Kepler* sample. We then assigned amplitudes by drawing values from Gaussians with means corresponding to the mean amplitudes of stars with similar $T_{eff}$ and $P_{rot}$ in McQuillan et al. (2014) and variances given by the variance in each bin. White noise was added to the lightcurves, with amplitudes that depended on the *r*-band magnitude, based on the values provided in Ivezic et al. (2008) and Jacklin et al. (2015).[2] We sampled these lightcurves using the LSST cadence model described above, and attempted to recover their rotation periods using a Lomb-Scargle (LS) periodogram (Lomb 1976; Scargle 1982). LS periodograms were computed for each lightcurve over a grid of 1000 periods ranging from 2 to 100 days. The position of the highest peak in the periodogram was recorded as the measured period.

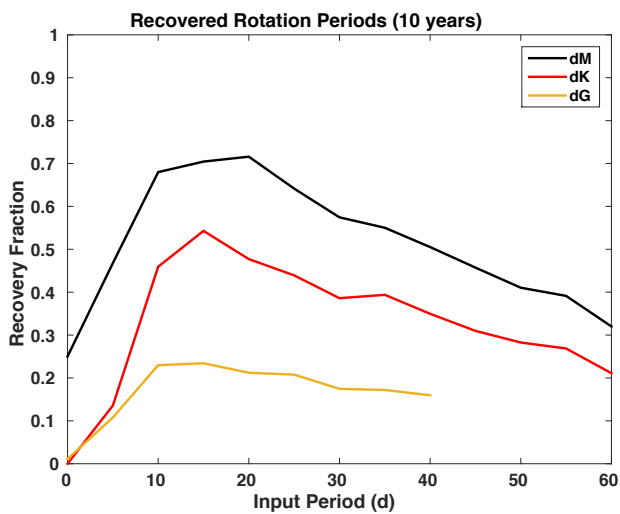

*Figure 5.2. Fraction of recovered rotation periods as a function of input rotation period for G, K, and M dwarfs. Relatively few rotation periods shorter than around 10 days are recovered for all spectral types, likely due to the three-day minimum LSST observing interval. Rotation period sensitivity drops off at longer periods since more slowly rotating stars have smaller variability amplitudes. There are no G stars with rotation periods greater than ~ 40 days, since these stars would be older than the Galaxy. The rotation period sensitivity of LSST peaks at ~ 20 days.*

---

[2] More detailed simulations would include systematic and/or correlated noise sources as determined for real LSST data; however, these are not yet available. Systematic effects will limit robust period recovery, again making our results somewhat optimistic.

We computed the recovery rates of injected periods for 1, 5, and 10 years of LSST monitoring data, and show these rates as a function of period in Figure 5.2 for each spectral type, in the b = −20 field. Successful period recovery was defined by output periods within 10% of the input values. Figure 5.2 illustrates the rotation period sensitivity of LSST, which peaks near 20 days, dropping towards shorter periods due to the relatively large (approximately three-day) interval between observations. The sensitivity also drops towards long periods due to the smaller variability amplitudes (fewer starspots) for slowly rotating stars. Note that the model gyrochronology relations do not predict G stars with periods longer than about 40 days at the age of the Galaxy.

Using the full ten-year dataset, we are able to accurately recover 60–70% of rotation periods for stars with $T_{\text{eff}} > 4500$ K and 70–80% for stars with $T_{\text{eff}} < 4500$ K, brighter than 23rd magnitude. As discussed above, these rates are probably optimistic but are likely within a factor of two of the actual recovery rates once all the mitigating factors are included.

Finally, we scaled the results from the 20,000 star target sample to a one square degree field, and found the density of stars with detected periods (number per square degree) after 1, 5, and 10 years of LSST observing. The density distributions as a function of r magnitude are shown for fields at two different latitudes in Figure 5.3. In each magnitude bin, there are thousands of stars per square degree at low Galactic latitudes and hundreds per square degree at high latitudes.

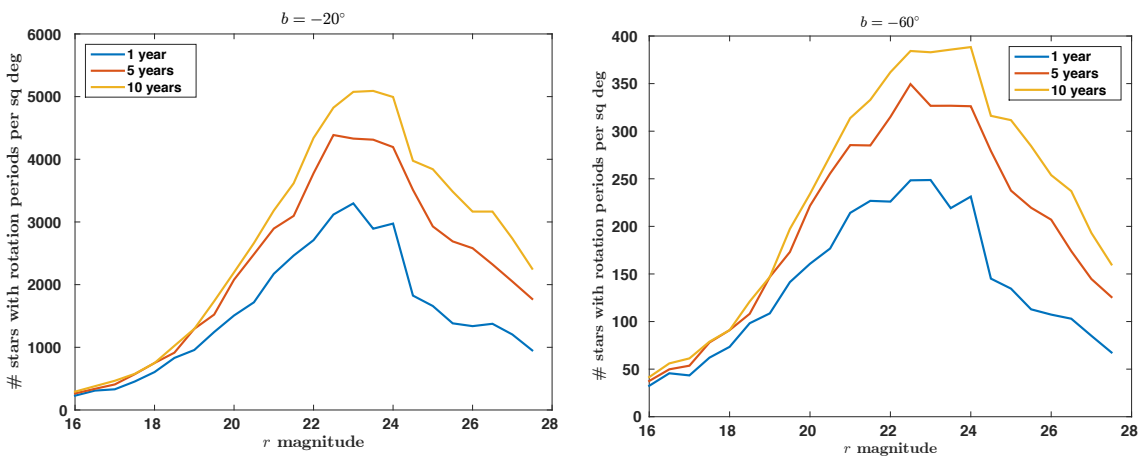

*Figure 5.3. Results of LSST rotation period yield simulations. Left: Density of stars with detected periods at b = −20. Right: Density of stars with detected periods at b = −60.*

### Flare Simulations

Forward modeling of flare star lightcurves allows us to predict the density of detectable flare stars on the sky as a function of limiting magnitude. Estimating this target density is critical for planning systematic follow-up studies of magnetically active stars, as well as for predicting the impact of flares as a contamination source for other variability studies with LSST.

To determine accurate flare yields for stars at the sparse LSST cadence, we first simulated lightcurves with a range of flare rates using a *Kepler*-like cadence of 1 minute. Individual lightcurves contained 1 year of continuous simulated flare data, and were repeated 10 times to create a 10-year lightcurve. The simulations used a typical flare occurrence rate as a function of energy, which is described by a power law:

$$\log \nu = \beta \log \varepsilon + \alpha$$

where $\nu$ is the cumulative number of flares per day, and $\varepsilon$ is the flare energy normalized by the quiescent luminosity of the star. We fixed the slope at $\beta = -2$, a value commonly used for the Sun and nearby flare stars (Hawley et al. 2014). The total number of flares is governed by the power law amplitude parameter $\alpha$ which we varied from $\alpha = 1$ for very active stars to $\alpha = -4$ for inactive stars (Hilton et al. 2011), resulting in the six simulated lightcurves shown in Figure 5.4. The profile of each flare was computed using an empirical model derived from *Kepler* flare observations Davenport et al. (2014). Only classical, single- peaked flares were simulated, though complex flare morphologies resulted for the active star lightcurves due to serendipitous flare overlaps.

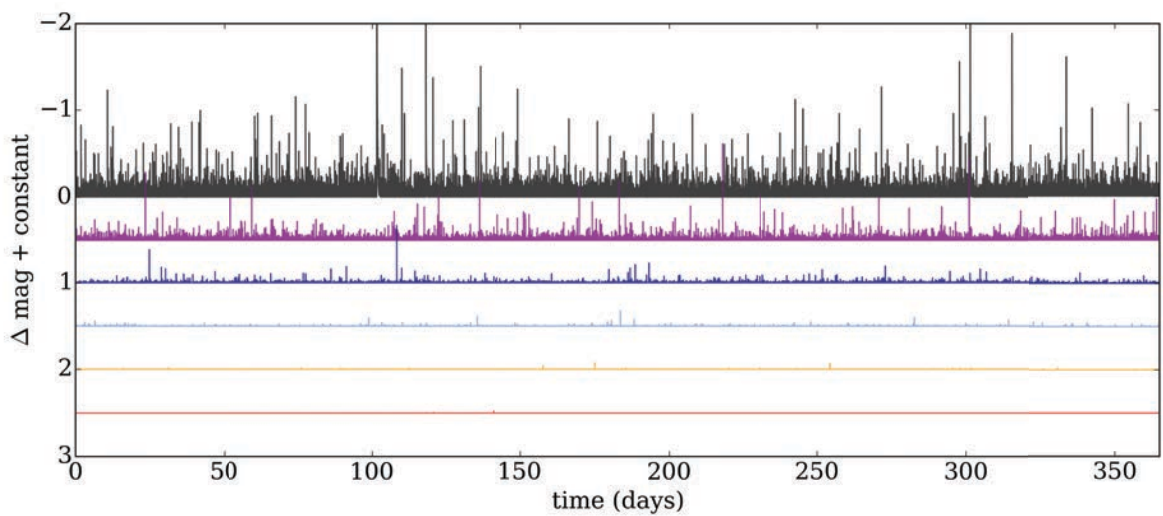

*Figure 5.4. Simulated flare star lightcurves for six levels of flare activity. Each lightcurve is one year in duration, sampled at a one-minute cadence. Lightcurves from top to bottom decrease by an order of magnitude each in their cumulative rate of flares per day, from α = 1 (black line) to α = −4 (red line).*

Within each of the four TRILEGAL fields (see §2.2) we randomly chose 50,000 stars for flare simulation. The quiescent *g*-band magnitude was used to compute the typical LSST photometric uncertainty $\sigma_g$ for each target. Each star was placed into a bin depending on its temperature and age, using $T_{eff}$ bins of 1000 K (roughly G, K and M spectral types) and 3 age bins: < 250 Myr; 250 Myr to 1.5 Gyr; and > 1.5 Gyr. The age and spectral type determine which of the 6 simulated lightcurves is used for the simulation. For each temperature bin, the flare rate parameter ($\alpha$) is decreased with age using a simple ad hoc prescription since the detailed evolution of flare rates with stellar age is still unknown. Indeed, the LSST data will be uniquely able to provide new constraints on the flare rate-age relationship. It will be important to compare ages we obtain from rotation and flare activity to ensemble ages obtained from kinematic population studies.

Once the appropriate lightcurve is chosen for each star (depending on its temperature and age), it is then sampled over 10 years using the cadence model described in §2.1. Only the *u* and *g* band visits result in flare detections, since flares are intrinsically blue events. Thus, each lightcurve is sampled approximately 100 times over the 10-year period, according to the *minion_1016* cadence model (which has many fewer *u* and *g* observations than in the redder filters).

Flares in the sampled lightcurves are defined as single epoch (bright) outliers. We used a conservative limit of a measured flux that is 0.1 magnitudes brighter than the median magnitude of the star (in the *u* and *g* bands) for a confident flare detection. This corresponds to > 10σ and also should reduce false detections from other stellar variability, e.g., from starspots, which is typically < 0.03 magnitudes. We note that if the LSST cadence includes two 15-second exposures as part of the baseline observing plan for each visit, then flares can be verified by comparing the magnitudes in the two exposures. Since flares typically last for several minutes, and large flares last for several hours, the two exposures should both show the flare brightening, which will reduce the number of false positives. Access to the data from the two exposures as part of the Level 3 LSST data products will therefore be essential for identifying real flares and rejecting noise events.

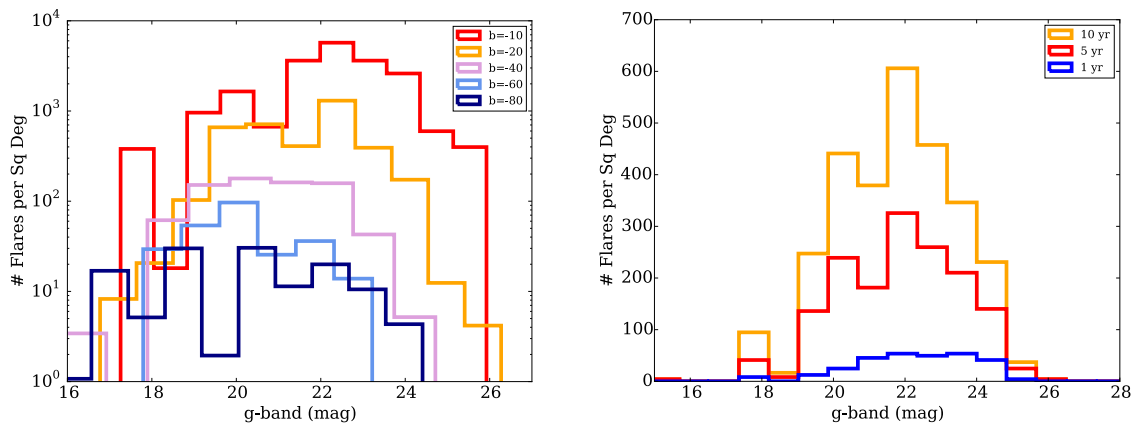

*Figure 5.5. Results of the LSST flare yield simulations. Left: Spatial density of detected flares using 10 years of LSST data versus stellar quiescent apparent magnitude for the four test fields. Right: Flare yield for 1, 5, and 10-year baselines versus apparent magnitude for the b = −20 line of sight.*

For each line of sight we scaled our recovered flare yields from the 50,000 sample stars to a one square degree field. This produced the predicted flare rates for the 10-year LSST survey shown in Figure 5.5. M dwarfs dominate both the total number of stars in each field, and the resulting flare yields, making up more than 97% of the stars with detected flares. There are two reasons: in G and K dwarfs the flare rate drops very rapidly with age, and there are not many young stars even in the lowest latitude field that we modeled. Flare rates for these stars will be much better determined with LSST by studying young open clusters. Also, the flare contrast is reduced in GK stars, so reaching the 0.1 magnitude limit for flare identification requires a very large flare. If the limit for identifying flares can be reduced as the survey progresses, then more GK flares may be found.

The flare rates as a function of galactic latitude are very close (within ∼ 10%) to those predicted by Hilton et al. (2011, see also LSST Science Book), who performed a similar analysis based on flare rates for M dwarfs from ground-based data. The flare rates vary by more than 2 orders of magnitude depending on latitude. Our results show that LSST should detect approximately 1 flare per square degree per exposure at mid to high galactic latitudes and 10–100 at lower latitudes. At lower latitudes, not only are there more flare stars per square degree, and larger rates of flares due to younger stars near the Galactic mid-plane, but we find the quiescent brightness of flare stars reach to fainter magnitudes and further distances. For high-latitude fields, flare stars are only visible at < 2 kpc, while in the lowest latitude fields they may be seen out to ~ 10 kpc in our simulations. However, the largest simulated flares only reach ~ 2 magnitudes in peak amplitude (in *g*), and quiescent stars are only simulated down to 28th magnitude, limiting our ability to find the rare but interesting flare transients that result from very bright flares on very faint stars. It is clear from the simulations that these flare transients (i.e. flares from stars whose quiescent luminosity is below the detection threshold) will be primarily found at lower latitudes.

## Field Stars (Rotation and Flares) Follow-up Observations and Capabilities

Figure 5.6 is a summary plot showing the final results of our field star simulations for rotation and flares. After 10 years of LSST survey observations, the flare stars (nearly all M dwarfs) will number between 100–$10^4$ per square degree, while the stars with measurable rotation periods are almost two orders of magnitude larger, between $10^4$ and $10^6$ per square degree (summed over all magnitudes). The numbers in each category that are found per year increase roughly linearly (see Figures 5.3 and 5.5), so in the first year the samples will be about a factor of 10 less. If stars are restricted to *r* < 24, the rotation sample is reduced by another factor of 2 or so. Thus, in round numbers, there will be 10s–100s of stars per square degree to follow up each year, depending on the latitude being observed.

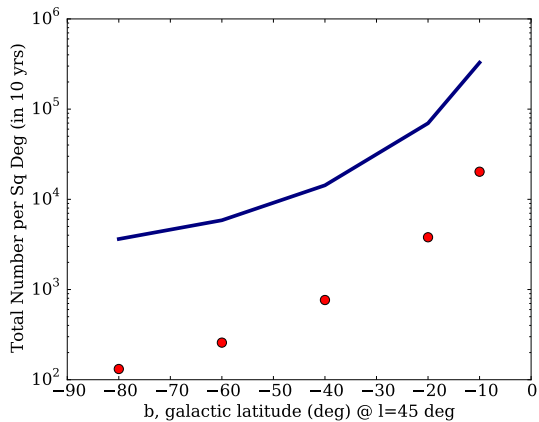

*Figure 5.6. Density of stars with detected flares (red circles) and stars with detected rotation periods (blue line) as a function of galactic latitude, for the full 10-year LSST simulation.*

Follow-up photometry and spectroscopy for the stars with rotation periods will be required in order to confirm their periods, identify them as bona-fide single main sequence dwarfs, obtain stellar parameters (temperature, gravity, metallicity), investigate their magnetic activity properties (e.g., Ca II H and K, Hα emission) and thereby produce a sample that will be useful for gyrochronology and age estimation. We do not expect that it will be necessary to follow up every one of these targets. A thorough characterization of a fraction of them will suffice to gain an understanding of the population of targets and to calibrate relations between their spectroscopic properties and their measured (with LSST) *ugrizy* magnitudes.

Confirming the rotation periods using follow-up photometry will be necessary for two reasons: (1) to measure rotation periods shorter than a few days which may be missed by LSST; and (2) to ensure that the variability appearing in LSST lightcurves is indeed caused by starspot modulations and is representative of a rotational signal. Supplementing LSST data with short-cadence follow-up photometry will be particularly necessary for young populations that are likely to have rapidly rotating stars. By comparing LSST rotation periods with those measured with follow-up photometry, we will determine a typical success rate, e.g., the fraction of LSST rotation periods that show good agreement with the rotation periods measured using follow-up photometry. We anticipate that follow-up photometry of ~ 10,000 targets will be required in order to determine meaningful success rates for LSST periods. This number of follow-up targets was chosen to sufficiently explore the expected range of the field population in temperature, metallicity, and age space. We propose that the targets will be distributed as approximately 100 objects in each of 100 different one-degree square fields. The targets ($V = 16$–24) will span the range of spectral type, period, and age of the full gyrochronology sample and will be distributed in RA, though of course concentrated in the Southern Hemisphere LSST footprint. Each field should be observed for three visits of about 20 minutes each per night for a month, thus 30 hours per field, or 3000 hours total. The brightest targets will be accessible with < 3m telescopes and, for example, would make good use of a facility like LCOGT. However, the majority of the targets will be best observed with a 3–5m facility operated in a survey mode, such as ODI on WIYN or DECam on the Blanco telescope.

Moderate-resolution ($R$ ~ 5000) spectroscopic follow-up will also be needed. Spectra are required in order to establish a sample of well-characterized stars for further study (e.g., to determine activity cycles, magnetic fields, etc.; see below). Multiple spectroscopic observations (~ 3) per star are needed to determine magnetic variability and binarity. The spectra will also be used to train a photometric classification system to characterize a much larger sample of gyrochronology targets. Spectroscopic log $g$, effective temperatures and metallicities will be mapped onto the LSST photometric system using this representative sample of stars and then applied to the full LSST catalog. The number of targets required is driven by our desire to sample each of these physical parameters for stars ranging from late F to M spectral types. We anticipate that LSST photometry will enable determination of $T_{eff}$ to within 200K and metallicity and log $g$ to within ~ 0.2 dex. Given the range of these parameters among our target stars (3000 K, 1.5 dex, and 1 dex, respectively), we would therefore want to populate roughly 1000 bins. With 10 stars per bin, this implies the need for ~ 10,000 stars, or (with 3 spectra per star) 30,000 spectra in total. Ideally there would be significant overlap with the 10,000 stars targeted for photometric follow-up.

The instrument of choice for the spectroscopic follow-up would be a MOS with at least 100 fibers, at R = 5000, covering a one square degree field. Again the brighter targets could use such an instrument on a 3–5m telescope (e.g., WIYN/Hydra, DESI/Mayall) while 6m telescopes such as MMT and Magellan also have existing or planned instrumentation that would be suitable. For the fainter targets, an 8–10m telescope would be required to obtain good S/N (> 20) in reasonable exposure time. Assuming 1–4 hours of exposure time, 3 exposures per field, and 100 fields, this program will require ~1000 hours on each facility (3–5m, 8–10m). Note that our target density is such that we would use only a fraction of a modern MOS with ~1000 fibers in each field. Thus, survey facilities that enable use of fibers for several different programs simultaneously (e.g., such as SDSS) would be well-suited for this program.

Finally, a subset of ~ 100 field stars from the rotation sample will be chosen for more extensive follow-up after the initial characterization. This will include high-resolution spectroscopy at R = 20,000–100,000 to determine radial velocity, rotation velocity ($v \sin i$), and metallicity (including especially Lithium abundance). It is anticipated that these stars will form the activity cycle sample for long term monitoring of Ca II H and K and photometry, and determination of magnetic field properties as described in §4 and §5 below, and the follow-up capabilities and time needed are included under those sections.

For the flare star follow-up, the same type of spectroscopic (R ~ 5000) characterization data are needed as for the rotation sample. Spectra of approximately 10,000 stars will be used to characterize magnetic activity and binarity, and to provide a sample large enough to calibrate flare rate-age relations. Kinematic data for population assignment (thin disk, thick disk) will also provide additional information for mapping the individual stars into age bins. Since these stars are mostly M dwarfs, they will be fainter and require more large-telescope resources, thus we anticipate the 2000 hours will be split as 500 hours on 3–5m telescopes, and 1500 hours on 8–10m telescopes.

A special opportunity for follow-up of rare "superflares" on solar-type stars will also be possible from the LSST alert stream. Such superflares have been identified in *Kepler* imaging data, but are > 100 times more energetic than any flares that have been observed on the Sun. They occur very rarely, perhaps once per 1000 years, but the large sample of G dwarfs being observed with LSST means that there is a chance perhaps a few times per year that such a flare will be observed. In this case, a single-object spectrograph with broad wavelength coverage and modest resolution (R = 100–500) should be immediately deployed to obtain spectra for the rest of the night at high time cadence, in order to follow the flare evolution (typical superflares last less than a day). Multicolor photometry from an imaging camera is also essential. The source of the white light continuum emission for these solar-type superflares is unknown, and such a dataset would provide an unprecedented opportunity to examine the physics of this radiation. Superflares are also important for planetary habitability— including our own! Depending on the brightness of the star, the photometric and spectroscopic follow-up could be obtained with anything from < 3m to 10m telescopes (e.g., LCOGT, SOAR, Gemini, though fainter targets are more prevalent, and therefore more likely). It will be important to verify the flare event by examining both of the 15-second LSST exposures, and to have the LSST transient broker and alert stream operational to direct follow-up resources within a few minutes of the flare event detection. We anticipate perhaps 1–3 events per year, requiring 10–30 nights (100–300 hours) of follow-up observations over the 10-year LSST survey period.

## Open Clusters

Taking advantage of the survey duration, photometric precision, and sky coverage of LSST, we can study variability in open clusters over timescales from days to years. With 0.005–0.01 mag precision between ∼ 16–20 mag (*griz*), studies of stellar rotation, differential rotation, starspot evolution, and magnetic cycle amplitudes and periods will be possible for cool dwarf stars (spectral types F, G, K, and M) with ages up to about 2 Gyr. These stars should also exhibit strong flaring activity, which may extend even to older clusters for the M dwarf population. LSST will have access to several hundred Southern clusters, but only a subset of them are sufficiently rich in stars and at a distance that will allow us to study their cool dwarfs with LSST. There are 227 open clusters in the current LSST footprint that have an estimated membership count of over 100 stars and distances such that M dwarfs are accessible to LSST (i.e., M0 stars have *V* < 20). We have identified three such clusters and use them here as a case study to demonstrate the scope of LSST observations in clusters and the need for follow-up facilities for cluster work.

With an age of 170 Myr and a distance of 1200pc, NGC 5316 represents the early main-sequence phase of cool star evolution. From literature studies we estimate that the cluster currently contains 90 known G dwarf members. We have fit a Kroupa (2001) Initial Mass Function (IMF) to the G dwarf mass distribution and estimate that the cluster was formed with ∼ 240 K dwarfs and ∼1650 M dwarfs. Most of these ∼ 2000 members should be accessible from the deep LSST observations over a wide FOV.

NGC 2477 (820 Myr) and IC 4651 (1.7 Gyr) probe the mid-point and the upper limit of the age-range we can study with LSST. At a distance of 1450 pc, NGC 2477 gives us access to G, K, and M dwarfs IC 4651 at 900 pc allows us to study older K and M dwarfs (the G dwarfs will be too bright). We estimate that NGC 2477 and IC 4651 currently contain 85 and 75 G dwarfs, and IMF fits suggest that they formed with ∼ 230 and ∼ 215 K dwarfs and ∼ 1600 and ∼ 1500 M dwarfs, respectively. That is ∼ 1800–1900 cool dwarfs per cluster.

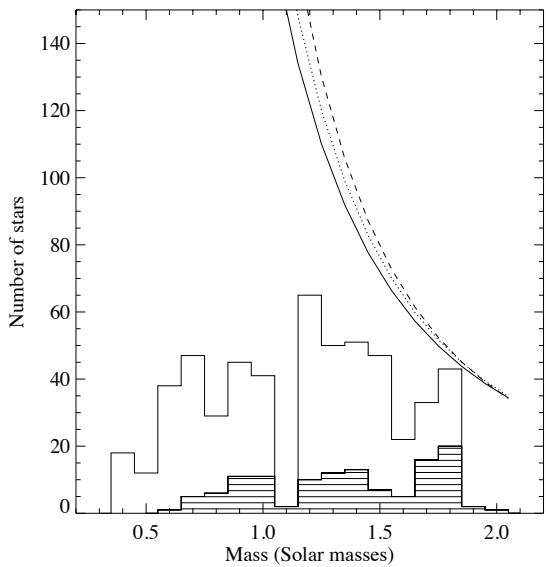

*Figure 5.7. Figure 15 from Meibom et al. (2002), showing the IC 4651 cluster members from radial velocity measurements (shaded histogram), photometry (open histogram), and three possible IMF models (solid, dotted, dashed lines). The Kroupa model used for our membership estimations is the dashed line.*

While dynamical evolution (e.g., mass segregation, dynamical evaporation) is known to remove the majority of low-mass members from the central (core) region of clusters over a timescale of ∼ 1 Gyr (Nordstroem et al. 1997; Meibom et al. 2002), the large FOV of LSST will easily capture the entire cluster as well as the foreground and background field population. Figure 5.7, which is a reproduction of Figure 15 from Meibom et al. (2002), shows the currently known members (histograms) and the predicted fits for three possible IMF models for the relatively old (1.7 Gyr) cluster IC 4651. These data show only members within the central 10 arcmin of the cluster. It is clear that membership determination will require significant follow-up resources such as proper motions, parallaxes, and radial velocities. In particular, Gaia will provide astrometric data for stars brighter than $V \sim 20$, and LSST itself will produce proper motion catalogs, which may be useful for membership determination. However, radial velocities will be needed for at least some fraction of the objects in order to robustly establish cluster membership and binarity status.

## Open Clusters Follow-up Observations and Capabilities

Similar photometric and spectroscopic (moderate-resolution, MOS) follow-up observations as for the field star sample will be needed for the 227 southern clusters in the LSST footprint, to confirm rotation periods and to obtain accurate stellar parameters and magnetic activity levels. Assuming 30 hours of photometric monitoring (spread over 1–3 months, depending on cluster age and therefore sampling cadence) for each cluster gives about 7000 hours total. About 10% of the clusters have very bright members (solar type stars have $V \sim 12$), and the entire cluster population can be captured with < 3m facilities with wide-field imaging capability. The brighter members of all the clusters (G and K stars) can also be observed in broadband filters with < 3m telescopes. Fainter members (M stars to $V \sim 24$) and narrowband filter observations will require follow-up photometry with 3–5m telescopes. We estimate approximately 3500 hours (half the time) on each size of facility.

Selection of cluster candidate members for follow-up spectroscopic observations will be guided by isochrone fits to the color-magnitude diagrams (produced by LSST and follow-up photometry). Stars that fall on the cluster CMD and that have starspot signatures enabling rotation period measurements and/or exhibit flares will be the highest priority for follow-up. Typical contamination by field stars can be as high as 50% for G stars and rise to 80–90% for K and M dwarfs depending on the Galactic latitude of the cluster (see e.g., NGC 5316). The spectroscopy can be carried out on 3–5m telescopes with MOS capability, as for the field stars. Each of the 227 clusters will likely have several thousand candidate members, so the targets can fill a 1000 fiber MOS and each cluster will require about 3 pointings (of 1–4 hours) to get all the candidates. Again most clusters have relatively bright members ($V < 22$) so about 1000 hours on 3–5m telescopes will be sufficient. The fainter candidates ($V = 22$–24) will need about 500 hours on 8–10m telescopes.

From these initial follow-up data, promising candidates can be chosen for membership confirmation with high-resolution radial velocity observations. These will also allow identification of binary stars, which is important since the stellar parameters determined from the photometry are affected. The rotation and activity evolution can also be significantly altered in binary systems. High-resolution (R > 20,000) multi-object spectroscopy on 6–8m telescopes is needed (e.g., an instrument such as Hectoechelle on the MMT), and at least 3 spectra per target. We note that Gaia will be capable of identifying binary stars only out to 250pc, which will exclude nearly all clusters of interest. However, the LSST astrometry may prove very useful for establishing kinematic membership. With 1000 targets per cluster needing the detailed high resolution data, 3 pointings per target to determine binarity, and 1–4-hour exposures, approximately 2000 hours on 8m telescopes with high resolution 1000-fiber MOS will be required.

## Detailed Cluster Evolution Sample

LSST observations of clusters over a range of ages will play an essential role in studying the evolution of stellar astrophysical phenomena (such as activity cycles and flare rates) as a function of stellar mass and in calibrating relationships between stellar age, rotation, and activity, to further develop stellar age-dating methods such as gyrochronology. To this end, we propose that a subsample of 11 open clusters, spanning a range of ages, be observed even more extensively. If possible, these clusters should be targeted with LSST deep drilling (or special survey) fields. If those are not available, then more extensive follow-up observations will be needed.

The detailed cluster evolution sample comprises clusters that were chosen to lie in or near the LSST footprint ($\delta < +5°$), but more than 5 degrees from the galactic equator to mitigate source confusion issues. (However, note that NGC 5316, which is discussed above, is included in the sample although it has $b \sim 0$.) The clusters were also required to have V between 12.5–16 for solar type stars (V between 16.5–20 for the M0 dwarfs) to ensure that the majority of the lower main sequence is accessible to LSST with reasonable SNR and without significant CCD saturation. We further winnowed our list by requiring the clusters to have well-established metallicities. This process results in 10 Southern clusters. We also include the canonical cluster M67, which lies slightly north of the LSST survey limit (declination = +12 degrees), but which would be a very worthwhile target for an LSST special survey. The 11 clusters in the sample are shown from oldest to youngest in Table 5.1. They span ages from ~ 5 Gyr to 100 Myr, enabling the study of a wide range of stellar rotational and activity evolution.

**Table 5.1. LSST Cluster Evolution Sample**

| Cluster | RA | Dec | $l$ | $b$ | $N$ | Dist | $E(B-V)$ | Age | $[Fe/H]$ | $V_{G2V}$ |
|---|---|---|---|---|---|---|---|---|---|---|
| | (hh:mm) | (dd:mm) | (deg) | (deg) | members | (pc) | (mag) | (Gyr) | | (mag) |
| NGC 6253 | 16:59 | -52:42 | 335.5 | -6.2 | 255 | 1511 | 0.2 | 5.012 | 0.43 | 16.26 |
| M 67 | 8:51 | 11:48 | 31.9 | 215.7 | 562 | 908 | 0.059 | 3.90 | 0.00 | 14.67 |
| Ruprecht 147 | 19:16 | -16:15 | 21.0 | -12.7 | 170 | 270 | 0.11 | 2.138 | 0.07 | 12.25 |
| NGC 6208 | 16:49 | -53:43 | 333.8 | -5.7 | 176 | 926 | 0.208 | 1.905 | -0.03 | 15.22 |
| IC 4651 | 17:25 | -49:56 | 340.1 | -7.9 | 325 | 888 | 0.121 | 1.778 | -0.128 | 14.87 |
| NGC 2477 | 7:52 | -38:32 | 253.6 | -5.8 | 1209 | 1450 | 0.291 | 0.822 | -0.192 | 16.44 |
| NGC 1901 | 5:18 | -68:26 | 279.0 | -33.6 | 134 | 406 | 0.021 | 0.708 | -0.331 | 12.87 |
| Collinder 350 | 17:48 | 1:21 | 26.8 | 14.7 | 189 | 302 | 0.302 | 0.513 | 0.00 | 13.07 |
| Ruprecht 110 | 14:05 | -67:28 | 310.0 | -5.6 | 182 | 1241 | 0.312 | 0.501 | -0.234 | 16.16 |
| NGC 5316 | 13:54 | -61:52 | 310.2 | 0.1 | 570 | 1208 | 0.312 | 0.17 | 0.045 | 16.11 |
| NGC 6087 | 16:19 | -57:56 | 327.7 | -5.4 | 379 | 889 | 0.271 | 0.089 | -0.01 | 15.32 |

The required special (deep drilling) observing cadence for the clusters is determined primarily by our desire to measure the full range of rotation periods anticipated in each. Young clusters such as the Pleiades (~ 100 Myr) and M34 (~ 200 Myr) have stars rotating with periods from a few hours to ~ 10–15 days (Hartman et al. 2010; Meibom et al. 2009, 2011a). At the other end of our cluster age distribution (up to 5 Gyr), rotation periods will range from 10–100 days and require the long time-baseline observations provided by the LSST (Gonzalez 2016; Newton et al. 2016). Accordingly, the optimal LSST sampling cadence for clusters would be logarithmic, starting with multiple visits per night for a week, followed by approximately daily visits for a month, and decreasing to the native LSST observing cadence of a few visits per week for the remainder of the project. If these are not carried out as special survey fields with LSST, then these observations would need to be performed on follow-up facilities. This would be an extension of the imaging follow-up described above, adding ~ 1000 hours of additional high cadence imaging on 3–5m telescopes. LSST will still provide observations at its normal cadence of these clusters for the duration of the 10-year survey. Additional observations (see §4) over the ten-year period will enable the detection of activity cycles in cluster members, as well as improved estimation of flare rates.

## Activity Cycles

The cadence of LSST is well adapted to the measurement of long-term variations in the magnetic dynamo, which produce stellar magnetic activity cycles in solar-type stars, similar to the 11-year solar cycle. Using data for field stars, Lockwood et al. (2007) determined a relation between cycle amplitude and chromospheric activity level (log $R_{HK}$) measured from the Ca II HK lines). Mamajek & Hillenbrand (2008) used open cluster data to calibrate this chromospheric activity level with age. From these relations, we find cycle amplitudes > 10 mmag for stars with ages < 250 Myr, and cycle amplitudes of ~ 5 mmag at an age ~1 Gyr. (For reference, the age of the Hyades is approximately ~ 625 Myr while the age of the Pleiades is ~ 100 Myr.) Thus, depending on the actual long-term precision of the LSST photometry, we may be able to measure cycle amplitudes for stars up to about 1 Gyr in age from the survey data alone for relatively bright stars. Of the three open clusters described in detail above, measuring cycles in solar-type stars in NGC 5316 and NGC 2477 should be possible from LSST survey data, but the older cluster IC 4651 is only predicted to have 2.5 mmag cycle amplitude, and thus cycles would likely not be measurable without additional observations.

To obtain cycle measurements for older clusters (ideally to the age of the Sun, and even more ideally in the cluster M67), additional photometry with very high precision will be needed. Clusters targeted for LSST deep drilling fields (see above) could provide these data. It is also of interest to correlate flare activity with HK cycles both in clusters and potentially using active stars that are located in the LSST deep drilling fields chosen for other scientific reasons.

In clusters and the field, programs to monitor Ca II HK (spectroscopically) and photometric variations are needed for studies of activity cycles. Photometric variations in the visible are dominated by cool spots on both short (rotational) and long (cyclical) timescales. Relatively larger photometric cyclic variations are typically seen in the younger active stars. Chromospheric variability on rotational and cycle timescales is also evident in the Ca II H and K resonance lines, which are accessible to ground-based observations in the blue-visible. These features arise from bright magnetic regions on a star. In young, active stars spots tend to be the dominant concentrations of magnetic flux until about an age of 2 Gyr. At ages older than about 2 Gyr photometric amplitude variations become more difficult to detect, declining to only 1 mmag at the age of the Sun. Spectroscopic observations of the Ca II lines are then preferred since cycle variations in the strengths of these features are readily detectable. At all ages and activity levels, observations of photometric variations and spectroscopic observations of Ca II HK variability are complementary since each diagnostic samples distinct and fundamental constituents of stellar magnetism. Therefore, a program of HK monitoring in parallel with photometric monitoring (both with LSST and with follow-up imaging) will provide a more complete picture of the evolution of stellar magnetic fields than either facility alone.

For the long-term cycles program, the original Mt. Wilson HK program initiated and carried out by Wilson (1978) provides a guide to the minimum observational frequency required. For solar-like cycles with periods ~10–11 years, 2–4 observations per month were adequate. For shorter cycle periods of 3–5 years, Wilson and his observers obtained data 5–7 times per month for these stars. In general, it appears that cycle frequency increases with rotation frequency to some power with a steeper dependence for more active stars. Therefore, we can expect that more rapidly rotating stars will generally (though not always) have shorter magnetic cycle periods. Moreover, rotational modulation will add "noise" to the cycle modulation. Hence, the need for more frequent observations during a month for high-activity stars in order to accurately infer long-term cycle properties.

Note that in the critical dynamo regime of the M dwarf stars there is a paucity of photometric data. It is in this region of the H-R diagram where solar-like dynamos, which depend on the presence of an interface region between the outer convection zone and the radiative interior for magnetic field amplification, may undergo a transition when the interior becomes fully convective at approximately spectral type M4. The sparse data that have been obtained so far suggest a range of cycle amplitudes in $V$ of approximately 40 mmag – 100 mmag (see Buccino et al. 2014, and references therein). Given the large number of active M dwarfs that will be identified in the field from the rotation and flare investigations (see above), LSST data will uniquely enable investigation of cycle properties in field stars for this crucial regime of dynamo operation where interiors transition from partial convection to whole interior convection.

### Activity Cycles Follow-up Observations and Capabilities

A sample of ~ 100 field stars will be drawn from the field star sample that has already been characterized with follow-up photometry and moderate-resolution spectroscopy. These stars will be monitored for long-term activity cycle variations in both Ca II H and K and broadband photometry. They will be an important addition to nearby star samples since they will investigate different populations (older, lower mass, and lower metallicity). The 11 clusters in the cluster evolution sample will be similarly monitored for cycle variations and will contain perhaps a few thousand stars in each cluster. The follow-up observations that are required include moderate-resolution spectroscopy (R ~ 5,000) and *ugriz* + narrowband Ca HK photometry with filters approximately 1 angstrom wide. The monitoring observations are needed twice per month in order to characterize stars that have cycle periods of 1–10 years. If the 100 star field sample is chosen from a single 1 degree field, and with an hour per observation, this requires 2 hours per month per field x 12 fields (11 clusters plus one field sample) x 12 months, ~ 300 hours per year or about 3000 hours total of *ugriz*+HK imaging follow-up. To avoid duplicating observations on different telescopes, the sample must be chosen so that the brightest and faintest stars in the observing field are accessible with the same telescope, hence a 3–5m wide field imaging facility such as DECam on the Blanco will be preferred.

The moderate-resolution spectroscopy to determine Ca II HK variations will have the same cadence requirements, but will require 3–5m and 8m facilities depending on the target brightness, as for the initial characterization follow-up (see §2.5). We anticipate 200 hours per year of 3–5m MOS and 100 hours per year of 8m MOS for the field plus cluster samples.

High resolution spectroscopy (R = 20,000–100,000) will also be needed in order to provide additional characterization including radial velocity, rotation velocity, metallicity (including Lithium) and magnetic field properties as described in the next section. The clusters will already have these observations, so they are only needed for the 100 star field sample, hence will only require about 10 hours each on 8m and 25m telescopes.

## Magnetic Fields

Readily accessible chromospheric features such as the Ca II resonance lines are radiative proxies of magnetic field-related activity. A key goal of stellar astrophysics is to obtain direct measurements of magnetic field properties across the H-R diagram. The direct measurement of magnetic fields in late-type stars has been challenging because of tangled field topologies that yield no net circular polarization or only a residual polarization signal in favorable geometrical circumstances. An advance occurred when Robinson et al. (1980) utilized unpolarized, white-light measurements to directly detect the presence of Zeeman broadening in the wings of magnetically sensitive lines as compared to insensitive (or much less sensitive) lines from the same multiplet. This approach, which typically requires spectral resolutions of R ~ 100,000 in the visible, yields a measure of magnetic field strength and fractional area coverage as inferred from a schematic representation of the Zeeman triplet splitting pattern. Extending this method to the infrared offers real advantages in stellar magnetic field measurements given that Zeeman splitting is proportional to wavelength squared, though in practice the gain is proportional to wavelength since natural line (Doppler) widths increase directly with wavelength. The minimum resolution requirements in the infrared would be R ~ 60,000.

The white-light approach described above has been extended with modern spectropolarimetric approaches that are particularly sensitive to the large-scale field and toroidal flux component. An example is the long-term polarimetric monitoring of the solar-type star ξ Boo A (G8 V) by Morgenthaler et al. (2012). These investigators find that the large-scale field on ξ Boo A is characterized by an axisymmetric component that is dominated by its toroidal component. Interestingly, an earlier study of the presumably fully convective, active dwarf M star, V374 Peg, also reveals a magnetic field structure dominated by a strong axisymmetric component in the presence of only weak differential rotation (Morin et al. 2008). Morin et al. note that this finding is in contrast to dynamo theories that require strong anti-solar differential rotation in order to sustain a strong axisymmetric field component while only non-axisymmetric geometries would occur in the presence of weak differential rotation. Clearly, an expansion to a large sample of stars would allow us to explore the broader applicability of this very preliminary finding. In addition to global field topologies, the magnetic field properties of starspots based on molecular features intrinsic to these cool regions along with Stokes V observations are a new area of spectropolarimetric investigation.

Doppler imaging relies on high-resolution spectra to detect the distortions in the cores of absorption line profiles as rotation carries thermal inhomogeneities across the line of sight. Doppler imaging is best applied to stars with high projected rotational velocities so that the signature of a cool spot is resolvable in velocity space as its line-of-sight rotational velocity component traverses the absorption core. These objects also probe dynamo properties in the limit of rapid rotation and, generally, thick convection zones. Through the application of inversion techniques, Doppler imaging yields a mapping with cool spots on the stellar surface as a function of phase. The primary targets of Doppler imaging tend to exhibit large polar spots that appear to be long-lived, relatively stable structures (see reviews by Berdyugina 2005; Donati & Landstreet 2009; Vidotto et al. 2014). Instrumental resolutions of at least 40,000 are needed for Doppler imaging. In addition to Zeeman broadening detection and net circular polarization measurements, Zeeman Doppler Imaging (ZDI) is actively utilized for discerning magnetic field properties in rapidly rotating stars. This technique combines Doppler imaging with Zeeman spectropolarimetry to detect rotationally modulated Zeeman components in a magnetically sensitive line. The resolution requirements are essentially the same as for Zeeman broadening techniques, i.e., R ~ 60,000. Through this approach the poloidal and toroidal components of the large-scale stellar magnetic field can be deduced to unveil the nature of field topologies in active stars. The broad conclusions thus far from ZDI and other magnetic field studies suggest that stars more massive than 0.5 $M_\odot$ and with Rossby numbers near unity are characterized by a mainly non-axisymmetric poloidal component and a significant toroidal component. In contrast, less massive, active stars seem to produce strong large-scale poloidal and axisymmetric fields (see Donati & Landstreet 2009).

Another accessible spectral feature that is uniquely powerful as a radiative proxy for surface magnetic field regions in solar-type stars is the He I triplet line at 1083 nm. This feature (as well as the weaker He I triplet line at 587.6 nm) appears in absorption in active (plage) regions on the Sun and, by implication, in the magnetic regions on Sun-like stars. Since these features are relatively weak, high-resolution spectroscopy is needed to measure their equivalent widths while also disentangling the contamination due to terrestrial water vapor. Typically, resolutions in the range of R ~ 50,000 are desirable. The 1083 nm absorption line is not seen, or appears only very weakly, in the quiet solar (or late-type dwarf stellar) photosphere. The 1083 nm line is spatially correlated with significant concentrations of magnetic flux in the photosphere and sites of X-ray emission in the corona but is otherwise only weakly present in the quiet photosphere. Thus, the absorption equivalent width of the 1083 nm line in the integrated spectrum of the Sun or that of a solar-type star can yield direct information on the filling factor of magnetic regions, when properly calibrated by models or through empirical approaches (Andretta & Giampapa 1995).

The next step in this fundamental area of the solar-stellar connection is to obtain (1) long-term polarimetric studies of the cycle modulation of field topologies for F to M dwarfs and (2) extend magnetic field studies to higher sensitivities in order to obtain information on more nearly solar-like stars with solar rotation periods. The latter will be particularly important in order to understand the extent to which solar global topologies are shared among solar-type stars of different activity levels.

## Magnetic Fields Follow-up Observations and Capabilities

Spectra with resolutions of R ~ 100,000 are required in the visible and R ~ 60,000 in the infrared (where Zeeman splitting is greater for a given field strength). This will require a high-resolution spectrograph on 8m-25m telescopes. We anticipate defining representative subsamples in the 12 fields that comprise the activity cycle sample (11 clusters plus one field star field) for intensive magnetic field follow-up observations. One hour-long observation per month for each field gives ~300 hours per year, split as 200 hours per year on an 8m telescope and 100 hours per year on a 25m telescope. These observations will directly determine how globally averaged field strengths and area coverages change over the course of an activity cycle and will quantitatively establish the correlation between radiative proxies of magnetic activity, such as Ca II H & K, and magnetic field properties.

Polarimetric (Stokes QUV) observations (e.g., with an instrument similar to CFHT ESPaDOns) will be essential to establish the field topology. This will likely require single-object observations of about 4 hours at high resolution so will only be feasible for a small sample, ~ 10 stars in each cluster or about 400 hours total split between 8 and 25m telescopes. The field topologies may also exhibit cycle-dependent properties, however monitoring would be extremely time-intensive, requiring several nights (50 hours) per year or 500 hours for the ten-year survey, for each star. These observations would likely be better carried out on nearby brighter field stars so we have not accounted for them in the time-needed table.

## Summary Tables

### *Table 5.2. Needed Capabilities*

| | **Other Infrastructure** | **< 3m** | **3–5m** | **8m** | **25m** |
|---|---|---|---|---|---|
| **Stellar Rotation (field sample)** | Survey mode operations (data acquisition and reduction pipelines provide reduced data) | Multi-color broadband wide-field imaging (*ugri*) | Multi-color broadband wide-field imaging (*ugri*)<br><br>**Multi-object spectrograph (0.35–1.0 μm, 1000 fiber, R = 5000)** | **Multi-object spectrograph (0.35–1.0 μm, 1000 fiber, R = 5000)**<br><br>*High-resolution spectroscopy (R = 20,000–100,000), 1000 fiber MOS* | |
| **Stellar Flares (field sample)** | Survey mode operations (data acquisition and reduction pipelines provide reduced data)<br><br>Access to Level 3 data products for individual 15- | Single-object rapid follow-up, multi-color imaging (ugri)<br><br>R = 100–500 single-object spectrograph for rapid follow-up spectral | Single-object rapid follow-up, multi-color imaging (ugri)<br><br>R = 100–500 single-object spectrograph for rapid follow-up spectral | **Multi-object spectrograph (0.35–1.0 μm 1000 fiber, R = 5000)**<br><br>*High-resolution spectroscopy (R = 20,000–100,000), 1000* | |

| | Other Infrastructure | < 3m | 3–5m | 8m | 25m |
|---|---|---|---|---|---|
| | second images | monitoring, high time resolution | monitoring, high time resolution<br><br>**Multi-object spectrograph (0.35–1.0 µm 1000 fiber, R = 5000)** | *fiber MOS* | |
| **Activity Cycles and Magnetic Fields (11 open clusters and one 100-star field sample)** | Survey mode Operations (data acquisition and reduction pipelines provide reduced data) | | **Wide-field imaging with narrowband (1Å) Ca II H+K filters**<br><br>Multi-object spectrograph (0.35–1.0 µm, 1000 fiber, R = 5000) | Multi-object spectrograph (0.35–1.0 µm, 1000 fiber, R = 5000)<br><br>High-resolution spectroscopy (R = 20–100,000) 1000 fiber MOS Optical and IR<br><br>Single-slit polarimetry | High-resolution spectroscopy (R = 20,000 – 100,000), 1000 fiber MOS optical and IR<br><br>Single-slit polarimetry |
| **Open Clusters (227 total)** | Survey mode operations (data acquisition and reduction pipelines provide reduced data)<br><br>LSST deep drilling (special survey) fields for selected open clusters with specialized cadence | Multicolor broadband wide-field imaging (*ugri*) | Multicolor broadband wide-field imaging (*ugri*)<br><br>**Multi-object spectrograph (0.35–1.0 µm, 1000 fiber, R = 5000)** | **Multi-object spectrograph (0.35–1.0 µm, 1000 fiber, R = 5000)**<br><br>High-resolution spectroscopy (R = 20,000 –100,000), 1000 fiber MOS Optical and IR | |

Entries in boldface type indicate that the capability is **Priority 1 (critical)**.
Roman type indicates Priority 2 (very important).
Italic type indicates *Priority 3 (important)*.

## Table 5.3. Resource Demand

| | Other Infrastructure | < 3m | 3–5m | 8m | 25m |
|---|---|---|---|---|---|
| **Stellar Rotation (field sample)** | | 1500 hrs *ugriz* imaging confirming rotation periods | 1500 hrs *ugriz* imaging confirming rotation periods<br><br>**1000 hrs R = 5000 MOS** | **1000 hrs R = 5000 MOS characterization** | |

| | Other Infrastructure | < 3m | 3–5m | 8m | 25m |
|---|---|---|---|---|---|
| | | | characterization | | |
| **Stellar flares (field sample)** | Access to Level 3 data products for individual 15-second images | 10–30 nights (100–300 hrs) *ugriz* photometry, R = 100–500 spectroscopy of rare superflares | 10–30 nights (100–300 hrs) *ugriz* photometry and R = 100–500 spectroscopy of rare superflares<br><br>**500 hrs R = 5000 MOS characterization** | 10–30 nights (100–300 hrs) *ugriz* photometry and R = 100–500 spectroscopy of rare superflares<br><br>**1500 hrs R = 5000 MOS characterization** | |
| **Activity Cycles and Magnetic Fields (11 open clusters and one 100-star field sample)** | | | **2000 hrs *ugriz*+HK imaging activity cycle monitoring**<br><br>2000 hrs R = 5000 MOS activity cycle monitoring | 1000 hrs R = 5000 MOS activity cycle monitoring<br><br>10 hrs R = 20,000–100,000 MOS characterization of activity cycle stars<br><br>2000 hrs R = 100,000 MOS mag field monitoring<br><br>200 hrs R = 100,000 polarimetry mag field topology characterization | 10 hrs<br><br>R = 20,000–100,100 MOS characterization of activity cycle stars<br><br>2000 hrs R = 100,000 MOS mag field monitoring<br><br>200 hrs R = 100,000 polarimetry magnetic field topology characterization |
| **Open Clusters (227 total)** | LSST deep drilling (special survey) fields for selected open clusters with high cadence | 5000 hrs *ugriz* imaging rotation period confirmation | 2000 hrs *ugriz* imaging rotation period confirmation<br><br>**1500 hrs R = 5000 MOS characterization** | **500 hrs R = 5000 MOS characterization**<br><br>2000 hrs R = 20,000–100,000 MOS characterization | |
| **Total On Sky Total** | | ~ 2 years | ~ 3 years | ~ 2.5 years | ~ 0.5 year |

Entries in boldface type indicate that the capability is **Priority 1 (critical)**.
Roman type indicates Priority 2 (very important).
Italic type indicates *Priority 3 (important)*.

# Chapter 6: Using Small Solar System Bodies to Understand the Cosmochemical Evolution of the Solar System

*OIR capabilities required for Solar System science enabled by LSST observations*

*David E. Trilling (Northern Arizona University), Cristina A. Thomas (Planetary Science Institute), Lori Feaga (University of Maryland), Henry Hsieh (Planetary Science Institute), Vishnu Reddy (University of Arizona), Scott S. Sheppard (Carnegie Institute for Science/Department of Terrestrial Magnetism)*

**Executive Summary**

LSST will discover millions of small bodies throughout the Solar System. With detailed follow-up observations, these LSST objects can be used to trace the cosmochemical and dynamical evolution of the Solar System. The primary telescope asset needs to maximize Solar System science in light of these LSST discoveries are (1) a wide field optical imager on a 4m telescope that enables short-term orbit refinement so that objects can be observed spectroscopically and (2) a low-resolution ($R \sim 50$–$100$) single-object spectrograph on an 8m telescope with ideal wavelength coverage of 0.4–2.5 microns. Several science cases require additional capabilities such as high angular resolution imaging and optical and infrared spectroscopy with resolutions from $\sim 1000$ up to 25,000. The estimated demand is some 100 nights/year on a 4m, around 50 nights/year on an 8m, and around 15 nights/night on a 25m. Finally, it is critical that these capabilities be available during the main LSST survey, before the objects fade and/or are lost (for Near Earth Objects), or before activity evolves or ceases (for comets and main belt comets); the timescales in both cases are on the order of days to weeks.

## Introduction

There are millions of small Solar System bodies larger than 100 meters—by far the most numerous type of object in our planetary system. These small bodies collectively act as tracers of the dynamical and cosmochemical evolution of the Solar System. Understanding in detail the properties of individual objects and the global properties of the small body populations allows us to understand the formation conditions in the early Solar System; the subsequent mixing of Solar System material; the interactions occurring today within these small body populations; and the dynamical processes by which material presently moves throughout the Solar System. *These measurements provide invaluable clues to the formation and evolution of the Solar System and of planetary systems around other stars.*

The general approach to addressing these topics is to study a relatively small number of small bodies in detail and to carry out broad characterizations of a very large number of objects. For the latter approach, LSST will produce catalogs of unprecedented size, with colors and orbits derived for millions of objects. However, for the former, detailed follow-up observations with a range of non-LSST facilities, with a range of apertures and instrument capabilities, are needed. In this chapter these additional measurements—crucial to understanding the formation and evolution of our Solar System and extrapolating our knowledge to exoplanetary systems—are described.

In order to carry out all of the follow-up studies described in this chapter, the orbits of the Solar System bodies to be studied must be known, but the detailed requirements for the necessary precision vary across science cases, as described below. For bodies with synodic periods around one year—these are bodies whose positions will be measured by LSST essentially every year for 10 years—the orbits in general will be well known, and follow-up studies will be easy to carry out if such observations can be delayed until near the end of the main LSST survey. For these bodies, the orbital uncertainty is generally driven by the frequency and, more importantly, observational arc length spanning the observations, and not by the uncertainty of the individual LSST measurements, where the precision is likely to be an excellent 0.1 arcsec or smaller. However, some follow-up studies described in this chapter must be carried out soon after object discovery. In these cases, since the number of LSST astrometric measurements and LSST-provided observational arc, and therefore orbital knowledge, will be small, some intermediate astrometric observations may be necessary to enable further follow-up observations.

## Science Topic 1: Near Earth Objects

### Science Goal

Near Earth Objects (NEOs) are bodies whose orbits bring them close to the Earth's orbit. The study of NEOs provides scientific constraints on the dynamical evolution of the present-day Solar System as well as on the formation conditions at the time of planetary formation 4.5 billion years ago. Furthermore, the study of NEOs is critical for planetary defense to better understand the impact risk and expected damage from potential impacts.

LSST is expected to detect 90% of NEOs with H = 18.9 (Table 5.1 from LSST Science Book) and 50% of NEOs with H = 22.4. H is the Solar System absolute magnitude (the hypothetical visual magnitude an object would have 1 AU from the Sun, 1 AU from the Earth, and at zero phase angle), and these absolute magnitudes correspond to diameters of 300–1000 meters and 60–200 meters (assuming albedos 0.05–0.50), respectively. From Granvik et al. (2014) this would correspond to 850 new objects discovered with H = 18.9 and 40,000 objects with H = 22.4.

Key questions regarding NEO characterization include the following (after Binzel et al. 2015): (1) To what extent does our current census of the NEO population accurately account for the abundance of low-albedo objects? Low-albedo objects are more difficult to observe, and a significant bias against these objects would affect our understanding of both the compositional distribution and impact risk of NEOs. (2) To what extent are extinct (inactive) comets present in the NEO population? Understanding the volatile (water-bearing) composition of NEOs requires accurate knowledge of the fraction of comet-like objects present in the NEO population. (3) Can discovery surveys and fireball networks augment asteroid-meteorite and asteroid-comet links? Meteorites serve as ground-truth for NEO science, and increases in survey and characterization yield, such as will be provided by LSST and follow-up observations, can lead to increased knowledge of meteorite parent bodies as seen telescopically. Finally, there is the outstanding issue of detecting and refining orbits for potentially hazardous NEOs—objects that could be on impact trajectories, either immediately or in the future. This is not strictly a science issue, and therefore is not addressed in the following sections, but the need for orbit refinement through additional astrometric measurements is critically important for these objects.

## Technical Description

Using the method developed by Cochran (1963) we estimated the number of NEOs with diameter $\geq$ 140 meters that need to be observed to ensure that we are not limited by small number statistics. Using the detection estimates from Granvik et al. (2014), we estimate a sample size of 378 objects to robustly characterize the population in this size range. The 140-meter size range was selected because it is the current goal of the George E. Brown, Jr. Near-Earth Object Survey Act which mandates NASA to detect, track, catalogue, and characterize the physical characteristics of NEOs equal to or larger than 140 meters in diameter with a perihelion distance of less than 1.3 AU from the Sun, achieving 90 percent completion of the survey within 15 years after enactment of the NASA Authorization Act of 2005. Current surveys are discovering about 500 objects/year and will take many years to achieve the mandate.

## Needed Capabilities and Estimate of Demand

Almost all NEOs discovered by LSST will have r magnitudes fainter than 22, and most will be fainter than r ~ 23, but a small fraction will have r < 22. Two main instrument and telescope combinations are needed to answer the above questions.

*1. Global network of 4–8m-class follow-up telescopes.* To enable composition measurements with spectroscopy, newly discovered NEOs to be observed immediately must first be recovered astrometrically so that their orbits are sufficiently well constrained to prevent them from being lost. LSST will be very good at discovering NEOs in the regular survey, but these objects, with observational arcs of just ~ 30 minutes, can have positional uncertainties the next night of 30 arcmin or more. Hence, spectroscopic observations that follow discovery by just a few nights require that the object must first be recovered astrometrically so that the orbit is relatively certain and the positional uncertainty is reduced to the few arcsec needed to put the NEO in the slit of a spectrometer. Although only a small fraction of NEOs could have this kind of large uncertainty the night after discovery, those are interesting targets—generally very small and very close to the Earth, and therefore intrinsically interesting for follow-up characterization. In general, the only way to carry out such spectroscopy is first to recover the objects astrometrically, and a large field of view camera would be needed for many of these objects. Hence, the first need is for a wide-field optical imager with a minimum field of view of 1 degree. Smaller fields of view would mean that the objects with the most uncertain orbits—which are generally the smallest NEOs, and the ones closest to Earth—are preferentially not observed spectroscopically. The current network of asteroid follow-up telescopes is woefully inadequate for this astrometric recovery challenge because the apertures are too small to detect LSST-discovered NEOs. Because exposure times are limited to no more than one minute (because NEOs start to trail in longer exposures), most of the recoveries will be carried out with 4m-class apertures, though some brighter objects can be recovered with telescopes smaller than 3 meters. Telescopes with instruments similar to DECam on the Blanco 4m telescope would be needed to enable meaningful characterization. Each object will require 4–5 observations per night for each of 3–6 nights. Therefore, for ~ 400 objects (the sample size defined above), the total follow-up time is around 167 hours, or around 17 nights. However, for every object that ultimately will be part of the scientific sample, up to 10 objects must be recovered, in order to determine which objects are most interesting. Therefore, the total need is up to around 10 nights/year. These nights are spread across the aperture range, depending on target brightness; perhaps up to 10% of these observations could be carried out with telescopes smaller than 3 meters.

2. *Low-resolution spectroscopy with an 8m-class telescope in the Southern Hemisphere*. A broad wavelength range (either 0.8–2.5 microns or 0.4-2.5 microns), low-resolution (R ~ 50–100) spectrometer similar to the SpeX instrument on the NASA IRTF or X-Shooter (low-resolution mode) on VLT is needed to characterize NEOs. Such an instrument would require accurate ephemeris to point and acquire the recently discovered NEO for characterization. Given that most NEOs discovered by LSST will be fainter than r ~ 22, an 8m telescope is needed for this work, and a telescope in the Southern Hemisphere is strongly preferred. Non-sidereal tracking is essential.

Depending on the viewing geometry and observing conditions, we expect 2–3 targets to be available for characterization each night. For spectroscopic characterization, this would translate to 1–2 hours/target with a total of 4–6 hours/night on an 8m-class telescope with a low-resolution near-IR spectrometer. For 400 objects, the total time need is therefore around 2000 hours, or around 200 nights. Such capability would be able to characterize this population within a year under ideal conditions and two years if one accounts for weather and other possible delays; this time could also be distributed throughout the 10 years of LSST primary survey.

It is critical that these NEO observations (follow-up as well as characterization) be carried out simultaneously with the LSST survey. Objects must be observed within a few days of discovery before they fade to beyond detection limits (due to orbital geometry); these are objects where future LSST observations cannot usefully contribute to orbit refinement since the observing window closes soon after discovery and may not open again for many years. After this few-day window, objects are generally not ever available again for observing. Therefore, the necessary capabilities must be online during the prime LSST survey.

## Science Topic 2: Main Belt Asteroids

### Science Goal

The properties of the main asteroid belt record both the early stages of planetary system formation as well as ongoing physical and dynamical evolutionary processes. Measuring the properties of millions of asteroids with LSST data, and many hundreds of asteroids through follow-up studies, enables us to place constraints on the suggested migration of giant planets early in Solar System history (Minton & Malhotra 2009, 2010, 2011; Roig & Nesvorny 2015). Measuring the properties of individual asteroids provides constraints on the details of the process of asteroid formation, which constrains the microphysics of protoplanet formation and their primordial source regions.

By number, Main Belt Asteroids (MBAs) compose the largest known population in the Solar System today, and therefore, offer unparalleled opportunities to understand the dynamical and cosmochemical evolution of our planetary system. LSST expects to probe MBAs down to ~ 100 meters in diameter with an expected yield of over 5 million MBAs detected by the survey.

The key science themes that can be investigated with LSST-observed MBAs include the following:

1. *The compositions of asteroids in the main belt as a function of location and size trace the global evolution of the main asteroid belt*. There are three components to this project. In the first, MBA families can be identified and defined using photometric colors in combination with their orbital elements (as in Parker et al. 2008; Figure 6.1). The second component is to determine taxonomic type (for hundreds of thousands of asteroids) to infer albedo (Figure 6.2). The third component is to understand space weathering trends on a large scale (e.g., Thomas et al. 2011; Figure 6.3). In general, all of these tasks can be accomplished with LSST multi-band imaging, but increased understanding requires spectral follow-up of objects from a wide range of spectral types, sizes, and locations in the Main Belt; LSST will be the largest provider of these targets. The global evolution of the asteroid belt is the single biggest clue to the process of planetary system formation as it occurred in our Solar System, and can be used by analogy to test planetary system formation theories that are driven by the properties of the exoplanet population.

2. *Individual asteroid properties constrain formation mechanisms*. There are two aspects to this investigation. The first is to derive asteroid mass from gravitational perturbations (single asteroids whose orbits are deflected by other nearby asteroids); this requires high-precision astrometric follow-up of the target in question and other asteroids whose orbits are expected to show or to show signs of the perturbation by the target, and LSST astrometry may in some cases be sufficient, but not in all cases. A preliminary list of the best candidates for mass derivations can be assembled in the ~ 6 months following the LSST survey process. The second aspect of this investigation is to derive asteroid masses from the Yarkovsky effect, which is an important non-gravitational force. This perturbation is the result of anisotropic thermal emission, which causes a secular change in semimajor axis. Since Yarkovsky acceleration depends on many physical properties, we are able to use detectable changes in asteroid orbit to inform our understanding of the asteroid's properties. This approach also requires high-precision astrometric follow-up and is limited to small NEOs and inner MB objects since they are the only populations within the Solar System with the potential to make these changes on currently observable timescales. The former experiment provides constraints on the assembly of the small planetary bodies that accumulate to form planets. The latter experiment constrains how the Solar System is evolving today, with material passing from the main asteroid belt and into near-Earth space, where it can be sampled, or where it can impact planets.

3. *The physical properties of the asteroid population are clues to ongoing evolution*. There are two parts to this theme. In the first, asteroid collision rates are investigated. Initial estimates (LSST Science Book) suggest that there is one catastrophic disruption of a 10m-diameter MBA every day and that LSST would observe one every week (LSST Science Book, section 5.6.1). High-resolution imaging (< 0.2 arcsec/pixel), operated as a target-of-opportunity program, would enable follow-up of these weekly discoveries to investigate the structure of the collisional debris and evolution of the debris cloud (e.g., P/2010 A2). In the second part of this theme, asteroid lightcurves would be measured in order to understand asteroid shapes and consequently the internal structures of a large number of asteroids (through constraints from rotational equilibrium dynamics as well as the identification of binary companions). Some asteroid lightcurves (on the order of $10^4$ or $10^5$) will be known simply from LSST data, but in general these lightcurves will not be dense enough or of high enough precision to carry out this experiment, and dedicated observations may be needed. The ongoing evolution of the asteroid belt is a proxy for understanding how our Solar System is evolving at present. Understanding exoplanetary systems in detail requires that we first understand our own Solar System, and these measurements provide critical constraints on those processes.

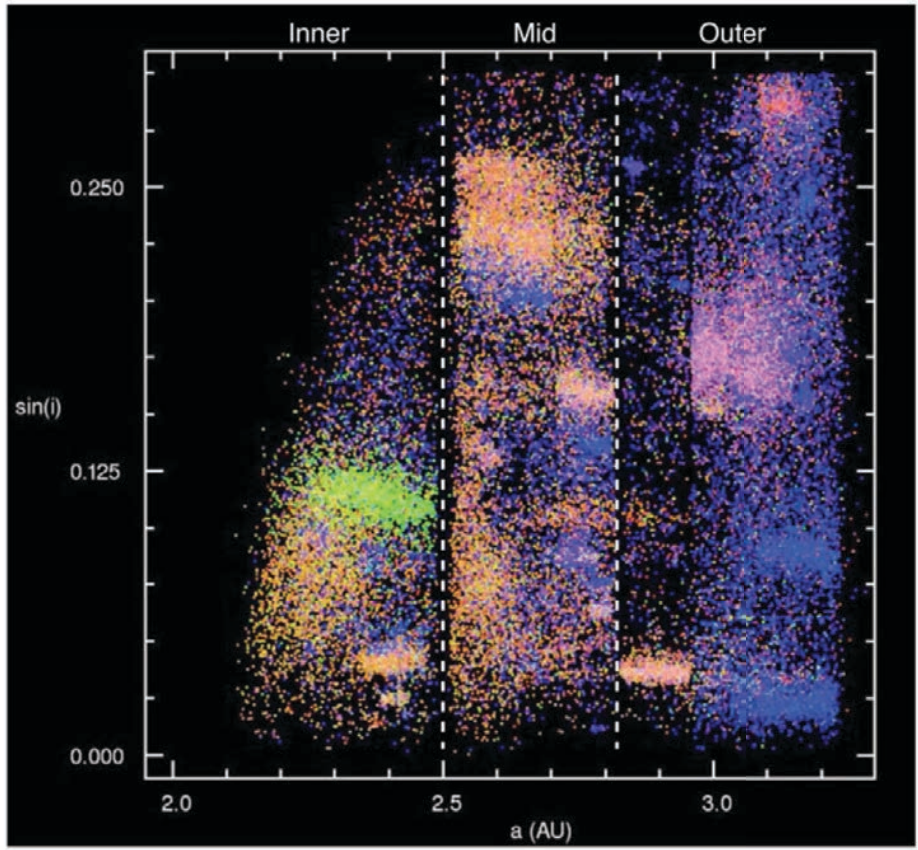

*Figure 6.1. From Parker et al. (2008). Proper a vs. sin(i) for objects observed by the Sloan Digital Sky Survey (SDSS). The figure colors are indicative of the object's spectrophotometric color using the a\* parameter defined in Parker et al.*

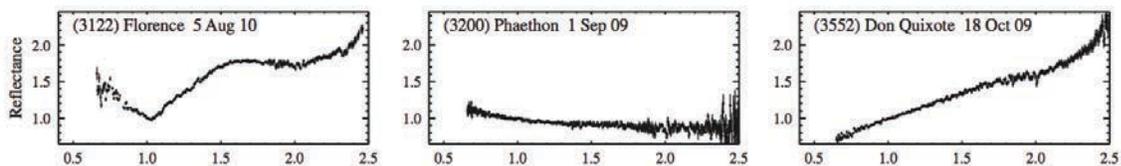

*Figure 6.2. Example near-infrared spectra of asteroids from Thomas et al. (2014). This panel shows normalized reflectance vs. wavelength in microns. The three objects shown here display a range of spectral types, and therefore, surface compositions.*

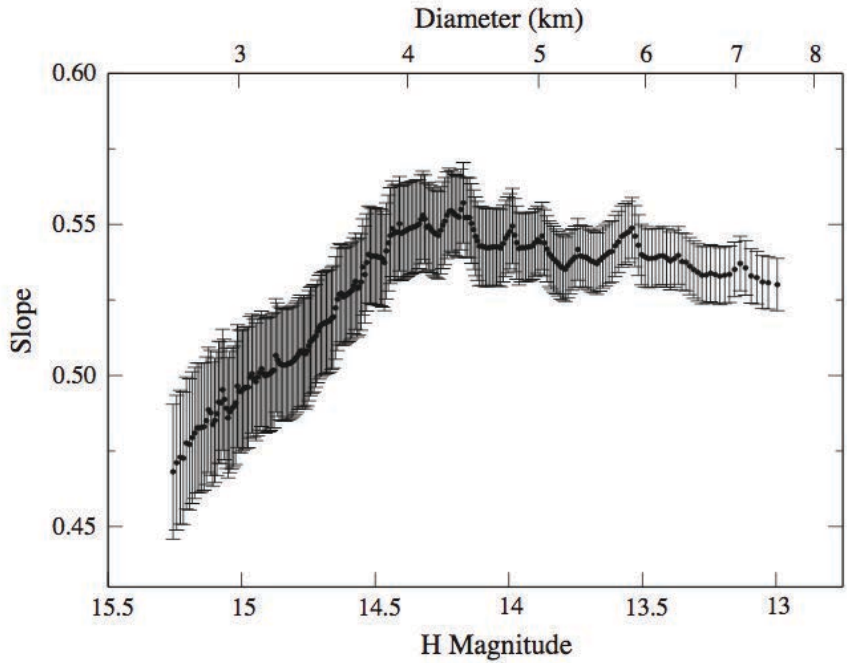

*Figure 6.3. This figure from Thomas et al. (2012) shows an investigation of space weathering in the Koronis Main Belt asteroid family using spectrophotometry from the Sloan Digital Sky Survey Moving Object Catalog. We use diameter as a proxy for surface age. The trend shows an increase in slope with increasing size (older surface age) until saturation that corresponds to an increase in spectral slope due to space weathering.*

## Technical Description

These technical descriptions are numbered in parallel with the science cases described above. Almost all asteroids observed by LSST will have r > 22, and most will have r > 23. A small fraction of asteroids will have r < 22.

1. Compositions throughout the main belt: Low-resolution (R ~ 100) spectral follow-up would require use of 8–10m and, for the faintest targets, 25m-class telescopes. Spectra of ≥ 300 small MBAs over the 10-year LSST survey period are needed to complete the science goals stated above and produce a spectral database similar to existing knowledge for small asteroid sizes. Thus we need to study, on average, a minimum of 30 asteroids each year, requiring ~ 6–9 nights of observing time a year (spread evenly throughout the year).

2. Individual asteroid properties through detailed astrometric measurements: The most important aspect is the long time baseline needed for this study. The minimum requirement is a 10-year observing baseline for each object in the target list (~ 100 objects). To detect any gravitational or Yarkovsky effects on the orbit, we require precise astrometric observations of the targets over a multi-year period. To track these effects over time, we require multiple visits. However, these objects may not always be within the LSST field of regard. Additionally, the observed magnitudes of these objects will change over time, and utilizing additional ground-based resources will enable observers to achieve the required SNR each individual observation. Each object should be observed once every few months (roughly three times per year). This corresponds to ~ 83 nights over 10 years (assuming 20 minutes per observation, 3 times per year for 100 objects), with the nights spread evenly throughout the 10-year period. Astrometric follow-up would require various large (4m, 8–10m) optical telescopes, depending on object brightness, though a small fraction might be accessible to telescopes smaller than 3 meters. Furthermore, radar observations (possible only for a subset of NEOs) would be highly advantageous.

3. Asteroid collisions and lightcurves: These asteroid imaging tasks require two different modes of follow-up. For collision studies, ~ 1 hour per night (to allow for multiple filters) on moderate or large (4–25m) telescopes (preferably with adaptive optics) would be needed. Not all discovered collisions will be followed up. Some 100 collisions could be selected for additional observations with the goal of studying a wide variety of asteroids; a sample of this size allows significant comparisons based on size, type, location, etc. The cadence for each object will have frequent observations (1 per day) for two weeks after discovery of the collision and will taper to once every 3–4 weeks for a total period of a year. Over the 10-year period, this totals ~ 2500–3000 hours of observing, which averages to 1–2 nights per month. For lightcurve studieswhich is relevant to both the Main Belt and Near Earth Object populations—the requirement is around 2 nights per month on various telescopes (4m, 8–10m, with a small fraction available to smaller telescopes) with SDSS filters and the ability to take short exposures (< 1 sec) with fast readout. Over the 10-year survey period, more than 100 targets would be studied in detail.

### Needed Capabilities and Estimate of Demand

Spectral follow-up requires observations in the visible and near-infrared wavelengths. Ideally, these observations would be carried out with an instrument similar to X-shooter on VLT that observes in visible and near-infrared wavelengths at the same time (0.3–2.5 microns). This wide wavelength coverage would enable precise compositional investigations of a range of asteroid spectral types: from primitive objects (e.g., C, D, P) with varied slopes and subtle features in visible wavelengths to the more processed objects (e.g, S, V) with large silicate absorption features in the near-infrared. The minimum acceptable wavelength coverage is ~ 0.7–2.5 microns due to the diagnostic silicate features located around 1 micron. This is a broad feature (0.1–0.2 microns wide), so spectra with resolution of 50–100 are acceptable.

The investigation of collisions requires a rapid response capability (within hour[s] to days; response within seconds is not required). Additionally, studying Main Belt collision events, which can evolve in timescales of ~ days, requires the ability to identify the event in the LSST data quickly in order to carry out follow-up observations.

# Science Topic 3: Main Belt Comets

## Science Goal

Main Belt Comets (MBCs), a subset of "active asteroids," are bodies with asteroid-like orbits in the main asteroid belt that show activity (e.g., tail, coma) as a result of ice sublimation (Jewitt et al. 2015). The discovery of this class of objects (Hsieh & Jewitt 2006) highlights that the division of minor bodies into either asteroid or comet is a false dichotomy and that instead a continuum of volatile contents exists in the population of small bodies of the Solar System. This in turn has revolutionary implications for the cosmochemical evolution of our Solar System, as nearly all bodies in the Solar System are likely to contain some amount of volatile material. This implies that the formation of the Solar System was not a binary process in which objects accreted volatiles or not. Instead, mixing may have been prevalent in the early Solar System, and today's active asteroids are the present-day indicators of those formation processes. The inner Solar System is more volatile rich than was thought just a few years ago, as it is now clear that asteroids throughout the near-Earth and main belt populations contain water, carbon dioxide, carbon monoxide, and other compounds that previously were thought to exist only in the outer Solar System. The origin of life on Earth may be a direct result of these volatile-rich bodies, and the MBC population is likely the largest reservoir of volatile material in the inner Solar system. The primary science goal of this topic is therefore to understand the physical properties of MBCs to understand both the activity mechanisms and the compositions of these bodies.

## Technical Description

LSST will detect some 5 million MBAs (LSST Science Book), roughly one order of magnitude more than the total number of MBAs currently known ($\sim$ 0.7 million). Conservatively, the number of MBCs might also increase by an order of magnitude from the present complement of 10 known objects. In the LSST era, therefore, around 100 MBCs might be known/discovered.

Over the 10–year lifetime of the LSST survey, this implies $\sim$ 10 MBC discoveries per year, each requiring follow-up. In practice, deeper images will reveal a higher ratio of MBCs (i.e., LSST should find objects with faint activity that appear inactive to Pan STARRS1, which is the most sensitive MBC discovery survey to date), but it is unclear how to estimate this number, which will be both a function of number of MBCs that exist, as well as LSST's ability to detect them, which is a non-trivial problem. If LSST is quite good at detecting low levels of activity, then the MBC discovery rate might increase by an order of magnitude, giving up to $\sim$ 100 MBC discoveries per year.

In conclusion, LSST is likely to discover $\sim$ 100–1000 MBCs; this corresponds to $\sim$ 10–100 MBCs discovered per year. To understand the global properties of MBCs, a significant sample size is required such that small number statistics do not limit our understanding. Observing 100 MBCs will increase the number of well-studied MBCs by a factor of $\sim$ 10 from present day and allow us to probe the formation of these objects without being significantly limited by small number statistics, as we are today.

## Needed Capabilities and Estimate of Demand

Several observing modes are required to understand the properties of MBCs, as follows.

1. Deep imaging is needed to ascertain the extent and morphology of activity; regular monitoring is needed to study activity evolution. This work could be carried out with a 4m-class telescope (the default used here). Each object needs 6 hours (1 hour/object/month for a six-month visibility window), for a total of 60 hours for the 10 objects to be observed in year 1. Continued monitoring increases this number to around 150 hours in year 10.

2. Low-resolution (R ~ 850) spectroscopic observations are necessary to characterize gas emission (Jewitt et al. 2015). The most promising 2–3 objects in each year (with magnitudes perhaps around r ~ 22) will be studied with 25m telescopes, since characterization efforts to date with 8–10m telescopes have failed due to very low gas production rates; at a few hours each to reach interesting detection limits, this corresponds to around 1 night/year on the largest telescopes.

3. More than half of LSST MBC discoveries will have nuclei undetectable by LSST itself because the objects are only detected when active, that is, brighter than their dormant state. Therefore, even deeper imaging is required to determine the size and phase function (which is indicative of composition and surface properties) of inactive nuclei. 8m telescopes will be necessary for these observations. Each object may be estimated to require ~ 4 hours (scaling from MacLennan & Hsieh 2012; Hsieh 2014); an additional ~ 1 hour per object is needed for astrometric follow-up, also with a telescope of similar aperture (8m+). This gives a total of 5 hours per object (one visit per object; no monitoring required) with 8m+ telescopes, for 50 hours per year (assuming 10 new objects are followed up per year) on a 8m.

4. Advanced characterization of nucleus colors, rotation rates, and pole orientations will help constrain the physical properties of MBCs and help reveal the activation mechanisms. Dedicated follow-up will be required to measure rotation rates (which will also inform reliability of color determinations) and pole orientations (these are mostly unknown for current MBCs). An estimated one object per year can have its rotation rate measured; this requires ~ 10 hours on a 8m for each object, as rotation rate measurements require sufficient time baseline, for around 100 hours total.

## Science Topic 4: Trans-Neptunian Objects

### Science Goal

Trans-Neptunian Objects (TNOs)—rocky/icy bodies in the outer Solar System, of which Pluto is one of the largest—are some of the most primitive objects in the Solar System. Some of these objects orbit near their formation locations, while others were scattered and captured in the region. Understanding their dynamical properties and compositions constrain models of Solar System formation.

In comparison to the other small body populations described in this chapter, TNOs have generally not been modified since their formation, 4.5 billion years ago. Therefore, these primitive bodies offer a unique window into the frozen state of planetary system formation. The different dynamical subclasses within the TNO population are thought to derive from variations in formation locations and timescales. Understanding in detail the observed differences will place tight constraints on the migration of the giant planets in our Solar System and to what extent proposed large-scale dynamical upheaval occurred when the Solar System was forming (e.g., Tsiganis et al. 2005). Therefore, one of the key science questions that drives observations of TNOs include measuring to what extent the various dynamical classes of TNOs—which have divergent histories and formation locations—have different surfaces and compositions.

Because the TNO population is generally quiescent and TNOs are used as a probe of the early Solar System, understanding evolutionary processes that are presently occurring in the outer Solar System is critical so that the primordial properties can be unearthed. Three indications of post-formation processing are binary TNOs and collisional families (both of which indicate dynamical processing) and color changes as a function of size, which would indicate collisional grinding that is overprinted on the initial formation processes.

## Technical Description

LSST is expected to detect and provide preliminary orbits for some 20,000 trans-Neptunian objects. This will increase the known population by a factor of 10. There are several different classes of TNOs, and to date only the brightest members have been observed in detail. This large number of new objects will allow us to compare the various classes in much more detail dynamically and physically. LSST itself will provide the information for basic population statistics and optical colors of all discoveries. Based on current statistics, some 50–100 LSST TNO discoveries will likely be wide binaries, which will contribute immediately to our understanding of the formation mechanisms of TNO binaries.

## Needed Capabilities and Estimate of Demand

To address the investigations listed above, a number of different capabilities are required. The two primary instrument and telescope combinations are most important, as follows. Most LSST TNOs will have r > 23, and some that are found in stacked images (for example, in deep drilling fields) may be fainter than r ~ 24.

1. Low-resolution (R ~ 300) near-IR spectroscopy on a giant (25m) telescope. 25m telescopes are required to measure the near-infrared features of water ice and other volatiles and organics on faint TNOs; a low-resolution instrument (R ~ 100 in the wavelength range 1–2.5 microns) is sufficient to measure these broad features (Figure 6.4). Each faint object would require around 1 hour on a 25m-class telescope; on an 8m telescope, each object would require more than 1 night of total time. The minimum requirement is at least 10 objects with high-quality spectra per dynamical subclass. There are around 10 subclasses of objects, corresponding to about 10 nights of 25m time (or over 100 nights of 8m time). These observations can be used to understand the compositions of TNOs in each of the dynamical subclasses as well as look for composition changes as a function of size.

2. High-resolution (0.1"/pixel) ground-based imaging (presumably enabled by adaptive optics) on a medium to large telescope will be needed to refine the orbits of newly discovered wide binary TNOs (Figure 5). We expect some 50 to 100 of these wide binaries (LSST Science Book). About 30 minutes of high resolution imaging on a 4–8m telescope twice a month for several months is needed to measure the binary orbital motion. Over 4 months this implies 8 observations, or 4 hours of time. Wide TNO binaries would have orbits of about 2 years and thus would require 4 hours two or three times during that period to see one complete orbit of a binary system. Thus, each object requires about 8 hours of telescope time to refine the orbit. For 50–100 new wide binaries, this corresponds to about 400 to 800 hours or 40 to 80 nights of telescope time on 4–8m telescopes with high angular resolution (i.e., adaptive optics).

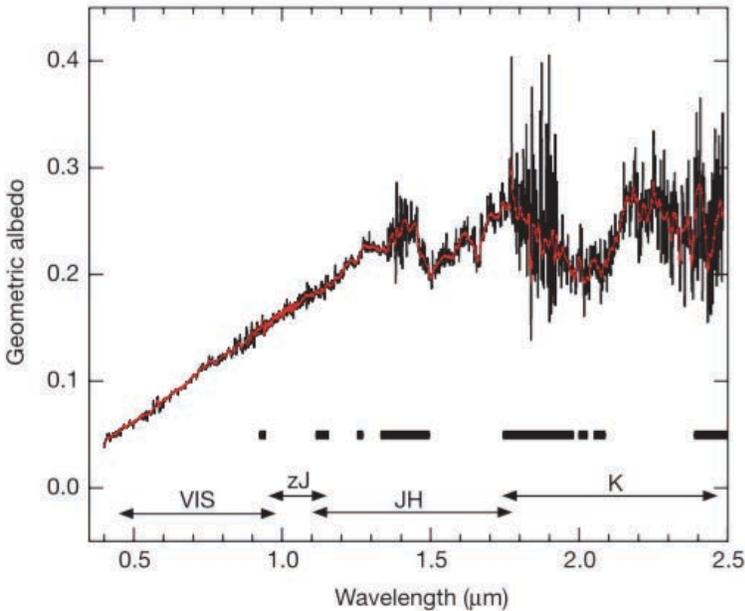

*Figure 6.4. Optical and NIR low-resolution spectra of TNO Quaoar showing water ice absorption at 1.5 and 2.0 microns and crystalline water ice at 1.65 microns (From Jewitt & Luu 2004)*

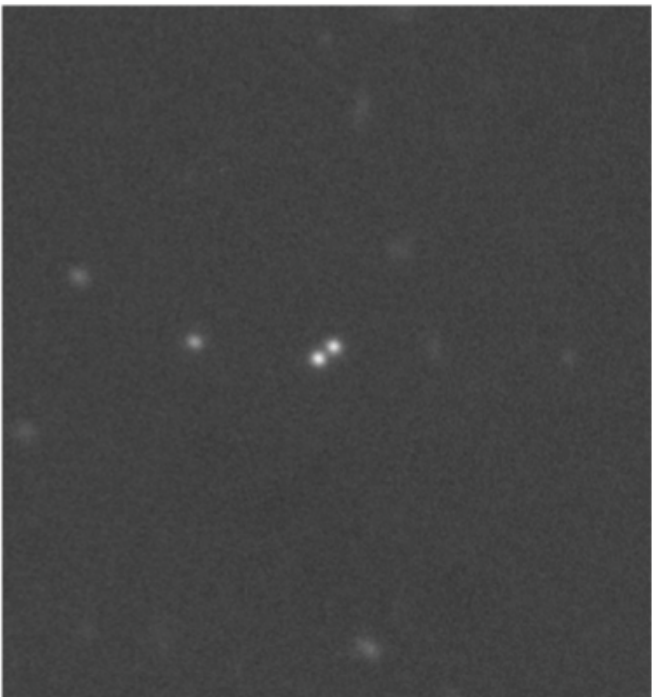

*Figure 6.5. Ultra-wide binary TNO 2007 TY430 from Sheppard et al. (2012). Equal-sized Plutino binary 2007 TY430 is easily seen near the center of this image from the GMOS detector on the Gemini telescope in December 2007. The separation between the components was about 0.7 arcsec.*

# Science Topic 5: Comets

## Science Goal

Comets are primitive objects that originate in the far outer Solar System and undergo substantial evolution as they pass near the Sun. These are bodies with very eccentric orbits that originate in the outer Solar System (in contrast to the main belt comets discussed above, which reside in the main asteroid belt on near-circular orbits). Studying comets is the most effective way to understand the detailed chemical composition of the outer Solar System and indeed the formation conditions of our planetary system as these objects literally reveal their inner secrets through outgassing while being heated by the Sun. Detailed studies of comets (compositions and nuclear properties as a function of orbital distributions) provide high-fidelity information that complements that broad-stroke information gathered from large surveys. For example, molecular composition is thought to be related to the formation distance of the object (e.g., Ehrenfreund et al. 2004). Molecular heterogeneity among comets can reveal evidence for dynamical mixing during the process of planetary system formation or may instead suggest a large range in formation location for these primitive bodies.

## Technical Description

LSST will discover on the order of 10,000 comets with 50 observations or more of each (Solontoi et al. 2010). LSST will also discover comets at large heliocentric distances, with repeated observations over many nights and even years possible. This long baseline allows for inbound and outbound observations of Oort Cloud comets (some of which will be "new"); full sampling of Jupiter Family Comet (JFC) orbits; characterization of the onset of activity; frequent, repeated observations to monitor for outbursts; and potentially high temporal frequency that can be used for rotational period determination. Additionally, follow-up spectroscopy can be useful to search for ices in the brightest comets; to determine the volatile inventory and onset of activity drivers for inbound Oort Cloud comets; and potentially even to estimate bare nucleus sizes if thermal measurements can be made.

The primary comet science goals are the following: determining orbits with reasonable precision such that the comets can be recovered and dynamical class/age be identified; long baseline studies of individual comets across an orbit inbound/perihelion/outbound; nucleus size and color determination prior to onset of activity; characterizing properties and time of onset of activity, compositional and color studies of coma; rotational periods from lightcurves; and monitoring for outbursts.

A typical JFC (like 9P/Tempel 1 or 67P/C-G) will range from 14th magnitude to 23rd magnitude from perihelion to aphelion. A small, active comet surrounded by highly reflective icy dust grains (like 103P/Hartley 2) that comes closer to the Sun (1 AU) is brighter at perihelion (10th mag) but much fainter at aphelion (25th mag).

There is less uniformity among the parameters of Long Period Comets. In this category, recent noteworthy comets include Hale-Bopp and ISON. The former was estimated to be quite large (~ 50 km), was very bright and active at perihelion, and is still active and detectable beyond 25 AU (~ 19th mag). The latter was detected because of its high activity on its first approach to the Sun when it was at 14 AU (~ 22nd mag).

## Needed Capabilities and Estimate of Demand

Some astrometric recovery will be provided through LSST's default observing, but in many cases the need for immediate characterization will first require dedicated astrometric recovery observations with a 4m telescope (though the brightest targets will be accessible to telescopes smaller than 3 meters). Three detections of a comet per night, each of which could require up to 30 minutes, repeated over 3 consecutive nights, gives the best results. This should be repeated over once a month for three months, resulting in a total time of 13.5 hours for astrometric follow-up per comet. Some 100 comets from a variety of dynamical classes will be targeted, implying around 135 nights on a 4m telescope over 10 years. Some 100 comets from a variety of dynamical classes will be targeted, implying around 135 nights on a 4m telescope over 10 years.

Long-term activity monitoring, nucleus size distributions, and composition contribute to our understanding of the origins of comets (and therefore of the Solar System). Frequent custom narrowband observations allow the search for jets, coma development, coma asymmetries, and other morphologies related to comet activity. Ideally, these measurements would be made for around 50 comets, sampling a range of heliocentric distances, orbital classes, and magnitudes. For 50 comets brighter than $23^{rd}$ magnitude at 15 min/filter and 6 filters (a minimum useful experiment) with 3 repeats at different heliocentric distances, the total time needed is 225 hours on a 4m-class telescope.

In order to measure compositions precisely, spectroscopic measurements are essential. A key wavelength range for the ro-vibrational emission bands of the parent molecules (water, carbon monoxide, carbon dioxide, etc.) is 1–5 microns, and resolutions up to R ~ 25,000 are needed to fully characterize the composition of the emission. Brighter comets could be observed with existing 4m and 8m telescopes. Spectroscopy on a sample size of 50 comets on the inbound leg of their orbit would be a significant improvement over our current knowledge. For these objects, 1 hour every 0.5 AU traveled is a sufficient request (~ 6 epochs of observation during the inbound leg), so the total time needed is 300 hours (over 10 years).

Rotation periods of comets typically are 8–48 hours, but sampling intervals of ~ 1 hour are needed if there is no a priori knowledge of the period. The exposure lengths would be 5–30 minutes (depending on brightness and aperture), implying perhaps 5–20 hours total per target. The necessary aperture size depends on the target brightness, but the time estimate, which is driven by the rotation period, is largely independent of target magnitude. For a sample size of 50–100 comet lightcurves, the total time needed is on the order of 1000 hours, over 10 years, and spread across apertures from 2m to 8m.

# Summary Tables

## Table 6.1. Needed Capabilities

| | Infrastructure | < 3m | 3–5m | 8m | 25m |
|---|---|---|---|---|---|
| **NEOs** | **Rapid response (within 24 hours)** | Astrometry (FOV up to 1 deg) | **Astrometry (FOV up to 1 deg)** | **Spectroscopy (R ~ 50–100, vis+NIR)** | |
| **MBAs** | | | **Broadband optical imaging** | Broadband optical imaging; **spectroscopy (R ~ 50–100)** | Broadband optical imaging; spectroscopy (R ~ 50–100) |
| **MBCs** | Short- and long-term monitoring | | **Broadband imaging**; high-resolution imaging | Broadband imaging; high-resolution imaging | High-resolution imaging; R ~ 800 optical spectroscopy |
| **TNOs** | | | High-resolution imaging | High-resolution imaging | High-resolution imaging; NIR spectroscopy (R ~ 50–100) |
| **Comets** | Short- and long-term monitoring | Broadband imaging | **Broadband and narrow band imaging**; *R ~ 25000 IR spectroscopy* | Broadband and narrow band imaging; *R ~ 25000 IR spectroscopy* | |

Entries in boldface type indicate that the capability is **Priority 1 (critical)**.
Roman type indicates Priority 2 (very important).
Italic type indicates *Priority 3 (important)*.

## Table 6.2. Resource Demand

| | Infrastructure | < 3m | 3–5m | 8m | 25m |
|---|---|---|---|---|---|
| **NEAs** | **Rapid response data processing and target identification (within 24 hrs)** | Up to 1 night/yr for astrometry | **Up to 10 nights/yr for astrometry** | Up to 10 nights/yr for astrometry<br><br>**20 nights/yr for R ~ 100 spectroscopy** | |
| **MBAs** | | Up to ~550 hrs total for astrometry and imaging | **450 total hrs of astrometry**<br><br>**500 hrs/yr imaging for monitoring** | **70 hours/year of R ~ 100 VIS/NIR spectroscopy (~ 75% of ≥ 300 object sample)**<br><br>450 hrs of | 30 hrs/yr of R ~ 100 VIS/NIR spectroscopy (~ 25% of ≥ 300 object sample)<br><br>100 hrs/yr imaging |

| | Infrastructure | < 3m | 3–5m | 8m | 25m |
|---|---|---|---|---|---|
| | | | | astrometry total 500 hrs/yr imaging for monitoring | for monitoring |
| **MBCs** | | | **~ 10 nights/year deep imaging for morphology and activity** | 50 hrs/year for nucleus characterization (imaging) 10 hrs/yr for advanced characterization | 1 night/yr (R ~ 800 visible spectroscopy for gas searches and nucleus characterization) |
| **TNOs** | | | | ~ 5 nights/year high angular resolution imaging down to 0.1" with AO | 1 night/yr (NIR spectroscopy of R ~ 100) |
| **Comets** | | Up to 200 hrs total of imaging and lightcurves | **1350 total hrs of broadband imaging** **225 total hrs of narrowband imaging** **~250 total hrs for lightcurves** *150 total hrs of R ~ 25,000 spectroscopy (1–5 microns)* | *150 total hrs of R ~ 25,000 spectroscopy (1–5 microns)* ~ 500 total hrs for lightcurves | ~ 250 total hrs for lightcurves |
| **Total On Sky Time** | | Up to 0.25 yrs | ~ 2.5 yrs | ~ 3.0 yrs | ~ 0.5 yrs |

Entries in boldface type indicate that the capability is **Priority 1 (critical)**.
Roman type indicates Priority 2 (very important).
Italic type indicates *Priority 3 (important)*.

# Chapter 7: The Co-evolution of Baryons, Black Holes, and Cosmic Structure

*Gregory Rudnick (University of Kansas), Adam Bolton (NOAO), Mark Dickinson (NOAO), Dawn Erb (University of Wisconsin-Milwaukee), John O'Meara (Saint Michael's College), Jon Trump (Penn State, Hubble Fellow), Benjamin Weiner (Steward Observatory)*

### Executive Summary

We outline four science projects that will be enabled by LSST and that address the important question of how baryons, black holes, and cosmic structure co-evolve. Our science use cases will lead us to an understanding of (1) the energetics of gas in, around, and beyond galaxies; (2) the spatially resolved rest-frame UV-to-optical properties of bright, lensed high redshift galaxies; (3) the connection between galaxy properties and environment over cosmic time; and (4) the growth of supermassive black holes over cosmic time. Executing these science cases will require several key capabilities, both in instruments and in infrastructure. Our top instrument priorities are as follows, with the number of nights and the priority given in parentheses:

1) a highly multiplexed spectrograph on an 8–10m telescope with ~ 2000–3000 fibers, a large (> 1 deg. diameter) field of view, R ~ 5000, and a wavelength range of  0.36–1.3μm (*critical; 3.2 million fiber hours on source; 330 nights for Subaru/PFS survey with significantly different parameters than the SuMIRe survey, such as a significantly lower stellar mass limit*);

2) A near-infrared (NIR) integral field unit spectrograph (IFU) on a 8–30m telescope with a field of view more than 3 x 3 arcseconds, R ~ 3000–5000, and that is fed by an adaptive optics (AO) system (*critical; 530 nights on an 8m*);

3) a single-slit high-throughput spectrograph on an 8–10m telescope with R ~ 3000–5000 and a wavelength range of  0.36–2.2μm (*critical; 65–130 nights on an 8m*);

4) Multi-object spectrograph on a 30m telescope with a 20 arcmin$^2$ field of view, R ~ 5000, and a wavelength range of 0.3–1μm (*critical; 120 nights on a 30m*);

5) a highly multiplexed optical spectrograph on a 4–6.5m telescope with 100 fibers/deg$^2$, a large (>1 deg. diameter) field of view, and R ~ 1000–5000 (*critical; daily to weekly cadence for ~5 years*);

6) a wide-field optical imager with good throughput in the u-band (*very important*);

7) single-object spectrograph on an 8–30m telescope with a wavelength range of 0.3–1μm and R ~ 50,000 (*important*).

Galaxies are condensations of baryons that grow within an evolving cosmic web of dark matter. As the Universe evolves, this cosmic web grows in density contrast and forms concentrations of dark matter that host galaxies. Baryons flow in and out of these concentrations, and during this process gas is converted to stars, forming the visible markers of cosmic structure, galaxies. The stars within galaxies synthesize heavier elements and expel these "metals" during their deaths. These metals are in turn put back into the gas within galaxies, the interstellar medium (ISM), blown back into the inter-galactic medium (IGM) between galaxies, and populate the gas reservoirs around galaxies known as the circumgalactic medium (CGM). At the heart of every galaxy also resides a supermassive black hole (SMBH), whose properties in the nearby universe are very well correlated with the properties of the host galaxy and that may also influence the evolution of its host galaxy through energy feedback. Understanding galaxy evolution requires understanding this baryon cycle through galaxies, the detailed processes of metal exchange and star formation within galaxies, the interaction of galaxies with their environments, and the role that SMBHs play in galaxy evolution.

LSST will enable large advances in this area by providing the deep and wide-field imaging necessary for the next generation of spectroscopic studies. LSST imaging will allow us to find the rare regions of the sky that are optimal for studying the IGM and CGM, it will allow us to find rare lensed galaxies suitable for detailed spatially resolved spectroscopic studies, and it will provide extremely deep imaging over many square degrees, which will select and characterize galaxies for highly multiplexed spectroscopic studies. Finally, LSST's time domain component will allow us to characterize and identify SMBHs over a large range in cosmic time. Below we outline four science cases that showcase how LSST enables these science goals.

## Science Case: The Energetics and Evolution of Gas in, Around, and Beyond Galaxies

LCDM cosmology has as one of its signatures a filamentary "cosmic web" of HI tracing the dark matter inhomogeneities throughout cosmic time. The bulk of this web appears as the Lyman-$\alpha$ forest, representing overdensities on the order of 0–10. The cosmic web includes information on a number of key parameters, including the temperature, the temperature-density relation, and the photo-ionizing background. As such, when we study the forest, we are studying the thermal history of the Universe. IGM tomography offers the promise of exploring this history in rich detail. As the HI column density of the absorbers increases, we begin to explore the IGM-galaxy interface in the circumgalactic medium (CGM). Studies of the CGM span six orders of magnitude in metallicity, from pristine gas (Fumagalli et al. 2011) to super-solar (Prochaska et al. 2006). The link between HI in absorption and galaxies has been explored with great success using the damped Lyman alpha systems (DLA, Wolfe et al. 2005), along with surveys of bright galaxies near quasar sightlines (Rudie et al. 2013; Steidel et al. 2010).

With all these successes, our complete understanding of the IGM and CGM still eludes us, primarily due to the low density on the sky of bright background sources suitable for absorption-line studies. At moderate to high resolution with 8–10m-class telescopes, the IGM and CGM can only be studied towards quasars or gamma ray bursts at high redshift, limiting our knowledge to primarily 1D skewers through the universe. Telescope aperture and inadequate sky coverage limit our ability to use the considerably more frequent (but fainter) galaxies in limited ways. Even with significant expenditures of telescope time and human capital investment, our best studies of the CGM and IGM (Rudie et al. 2013; Lee et al. 2014) with galaxies to date probe only the very tip of the luminosity and spatial sampling icebergs. Ultimately, one wishes for the ability to study galaxies at high signal-to-noise and moderate spectral resolution so that one can probe the CGM of multiple closely separated galaxies "down the barrel" while simultaneously measuring the IGM in the same line of sight, allowing for a 3D tomographic view of the universe. Such a full 3D sampling of the baryons at z > 2 would provide not only a huge observational leap but also the state-of-the-art dataset to compare to simulations for large scale structure, galaxy/gas accretion and feedback, and various cosmological scenarios.

LSST directly facilitates the next generation of ground-based IGM and CGM studies by providing a transformative increase in the number of moderate to bright galaxies, close separation quasar pairs, and quasar-galaxy and galaxy-galaxy groupings. The gain from LSST for IGM/CGM science is not only an increase in sheer volume probed but also the enabling of the discovery of very rare patches of sky with overdensities of bright objects for follow-up observations in various modes with current and future facilities. As an example, we highlight the key science returned from a multifaceted, multi-facility exploration of a square degree on the sky targeting the 2 < z < 3 universe. This key moment in cosmological time sees the upward march in star formation and galaxy assembly up to its peak and the first hints of the onset of quenching. This exploration directly applies to questions from NWNH including, "How do cosmic structures form and evolve?" and "How do baryons cycle in and out of galaxies, and what do they do while they are there?"

## Technical Description

### *Tomography of the IGM*

At the mean density of the universe, the Jeans length at z ~ 3 is on the order of 1 Mpc, which sets the observational requirements of tomographic reconstruction of the IGM. As shown in Figure 7.1 (taken from Lee et al. 2014), and after accounting for typical LBG colors and the need for a redshift coverage that allows uniform sampling in overlapping lines of sight, we see that we must sample the IGM from galaxies as faint as $r = 26.5$ in order to perform tomographic reconstruction at ~ 1 Mpc scale, the approximate scale at which galaxies individually influence the IGM at z ~ 2.5. This translates into a survey sample of the IGM in over 50,000 galaxy spectra in a single square degree field, and requires a 20+m-class facility. If one decreases the sampling resolution to ~ 4 Mpc, an 8–10m-class facility can observe $r = 24.5$ and brighter galaxies. We must sample at resolutions of a few Mpc or less to probe scales representative of both IGM and CGM inflow/outflow and feedback.

### *A Definitive CGM Survey*

A survey of thousands of galaxies and quasars will redefine our understanding of the CGM, particularly for those galaxies bright enough for higher resolution follow-up. As shown in Figure 7.1, a square degree field on the sky will contain several thousand galaxies (nominally LBGs) and quasars bright enough ($r < 22$) for medium (R ~ 5000) resolution, moderate (~20 or higher) SNR observations on a 30m-class facility, and tens to hundreds on an 8+m. With such data, on a galaxy-by-galaxy basis, we can explore gas kinematics relative to the systemic galaxy velocity with high precision, owing to the resolution and SNR. Our ability to detect weaker absorption features in individual spectra offers a huge advantage over stacks of lower SNR data, in that we can directly link emission line diagnostics of SFR, metallicity, ionization, etc., of the galaxy to the observed absorption over a wide galaxy population. Furthermore, higher resolution data will provide thousands of intervening absorption features as seen in strong HI absorption and metals. As the background source is an extended galaxy, these observations will place new constraints on the extended sizes of absorbers, giving key insights to their true physical nature (e.g., Cooke & O'Meara, 2015). LSST, via deep *u* band coverage, will provide a transformative increase in the number of z ~ 2 galaxies along the line of sight to bright background galaxies, allowing us to directly study the CGM at the peak of galaxy assembly.

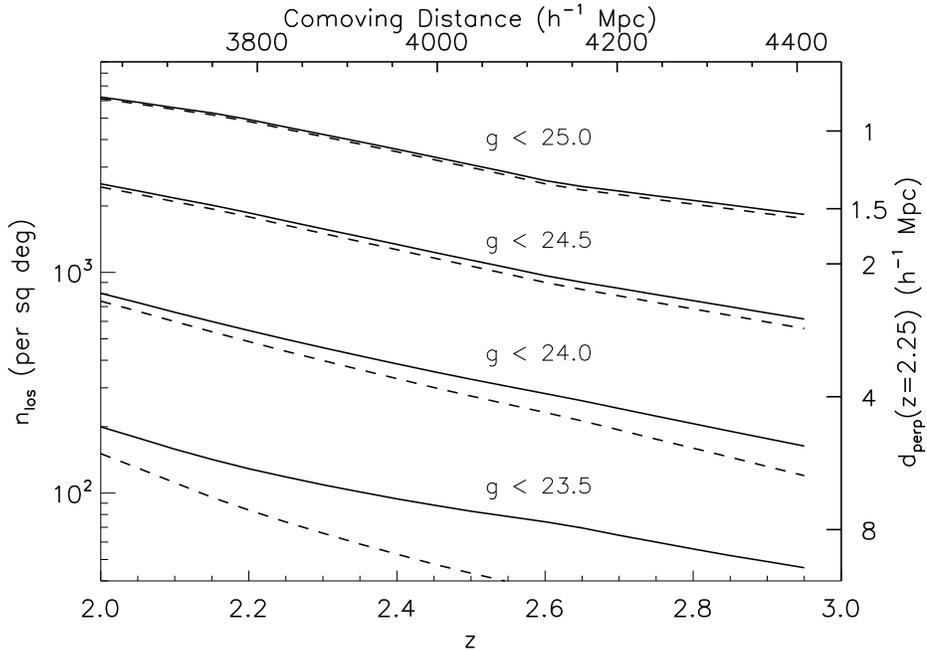

*Figure 7.1. Relationship between the number of galaxies & quasars per square arcminute or inter-galaxy separation, d⊥ and redshift (Lee et al. 2014). The dashed lines are for galaxies alone, while the solid lines are for galaxies and quasars combined. An IGM survey with a 20+m telescope would probe g < 25.5 galaxies and 1 Mpc scales, while a shallower survey with a 8−10 m telescope would reach g < 24 and ~ 4 Mpc scales.*

## Needed Capabilities and Estimate of Demand

The IGM tomography and CGM survey can be carried out on the same square degree of sky. Additional deep *u*-band imaging would need to be performed on the selected field at depths equivalent to the deep drilling field limits. To obtain high (1 Mpc sampling) spatial resolution via spectra of at least 50,000 galaxies, we estimate the time on sky in the following fashion. A TMT+WFOS telescope/instrument combination enables both low (R ~ 1000) and medium (R ~ 5000) resolution observations. We assume a WFOS FOV of 8 × 3 arc minutes, 100 slits per mask, and wavelength coverage down to 3400 Å. Using 4 masks per pointing, with 1 hour exposure time per mask to provide SNR > 5 per spectrum, 144 pointings are required to tile a square degree. This corresponds to 576 hours of on-sky integration. With nominal overheads, this translates to ~ 70 nights. Such a program returns 57600 *r* < 26.5 galaxies, meeting our 1 Mpc sampling goal (the time requirements are significantly reduced if we lower the tomographic resolution). Tomographic maps sampling coarser scales of ~ 4 Mpc over a square degree are possible on 8–10m-class facilities. Lee et al. (2014) estimate a survey of a square degree with a Keck+LRIS configuration to require ~140 hours of observation, i.e., 20 nights of time when overheads are factored in.

In the medium resolution necessary for the CGM program, 72 pointings of 2 masks per pointing and 4.5 hours per mask provides 3600 SNR > 35 galaxies over 1/2 square degree using the above assumptions for TMT+WFOS. This translates to ~ 80 nights. LSST's full sky footprint can be used to select the square degree with the largest density of bright background galaxies and quasars, significantly reducing the 20+m aperture time estimate, as spectroscopy of the $r < 25$ galaxies can be performed in ~ 2−3 hours per mask on 8–10m facilities. Finally, many aspects of the CGM program do not require galaxies to be drawn from the contiguous square degree field, so hundreds of r < 22 galaxies (and galaxy-galaxy pairings) could be observed with 8–10m-aperture facilities upon discovery with LSST, particularly from the deep drilling fields.

Fully exploiting the IGM/CGM science enabled by LSST will require high-resolution (R ~ 50,000) studies of the IGM and CGM as well. The full sky footprint of LSST will dramatically increase the number of bright QSO sightlines with deep imaging for galaxies closely projected on the sky. 8–10m-class facilities can provide SNR ~ 30 spectra for r < 19 quasars in a few hours exposure time, sufficient to obtain ionic column densities and kinematic structures for IGM and CGM gas. 20+-meter facilities open a new parameter space: SNR ~ 1000 spectra of the brightest quasars to trace the metal budget of the IGM over cosmic time and will require exposures of 50–150 hours with a 20+m facility.  Reaching such a high SNR will also require an exquisite control of systematic effects. Finally, LSST will find those very rare quasars with large enough sampling path to obtain EUV lines (e.g., NeVIII $\lambda\lambda770,780$ Å), facilitating for the first time the study of $10^7$ K gas at z > 2.  The exposure time clearly depends strongly on target brightness, but would require on the order of 10 hours per sightline on a 10m-class telescope per target (1 hour per target on TMT)  The number of available targets from LSST is unknown, but a survey of tens of objects would be unprecedented.

## Science case: The evolution of stars, gas, and dynamics in galaxies at high redshift and in high definition

A detailed understanding of the formation of galaxies requires spatially resolved measurements of their stellar populations, star formation activity, feedback, kinematics, chemical abundances and ISM excitation, and SMBH content.  These measurements need to be made for galaxies at z > 1, when the universe was undergoing its most rapid growth in stellar mass density (Dickinson et al. 2003; Rudnick et al. 2003, 2006) and when the Hubble sequence of galaxy morphology was first emerging (Papovich et al. 2005).  Our understanding of high redshift galaxies has been greatly improved by a handful of adaptive optics (AO) assisted observations of large star-forming galaxies at z > 1 (e.g., Föster Schreiber et al. 2011). However, these spatially resolved observations are limited to linear resolutions of hundreds of parsecs, have only moderate signal-to-noise ratios, mainly probe star-forming galaxies, and mainly target only the brightest galaxies at high redshifts.

Gravitationally lensed galaxies with high magnification factors, on the other hand, provide a way to obtain high signal-to-noise measurements of intrinsically faint galaxies, as well as those that are not forming stars. Indeed, studies of strongly lensed galaxies have provided our most detailed information on the kinematics, structure, and evolution of galaxies at high redshift. Careful dissection of these ordinary galaxies, magnified to extraordinary brightness, provides a wealth of information that, we hope, can then be extrapolated to understand the general population of intrinsically similar objects that are too faint to study in such detail. The famous lensed Lyman break galaxy cB-58 (Baker et al. 2004; Pettini et al. 2000, 2002; Riechers et al. 2010; Siana et al. 2008; Teplitz et al. 2000) is a classic example that has been studied in extraordinary detail, and it is joined by a handful of others, but real progress requires intensive study for a larger sample of bright, lensed systems. Lensing also stretches the galaxies, enabling sufficiently deep observations to resolve the spectral properties of galaxies on finer linear scales. (Most seeing-limited measurements for unlensed galaxies only provide information on their integrated light.)

Studying galaxy lenses addresses the NWNH Galaxies Across Cosmic Time panel's science question: "How do baryons cycle in and out of galaxies, and what do they do while they are there?"

By surveying a large fraction of the sky to faint magnitudes in six bands, LSST will become the premier supplier of rare, bright, high redshift galaxies—lensed and unlensed—particularly UV-bright objects that can be selected using LSST data via Lyman break color techniques at $2 < z < 7$. For large samples of truly exceptional objects, or to push down to fainter and more numerous examples, LSST's data will be unmatched. LSST is expected to discover ~120,000 galaxy-galaxy strong lenses, most at $1 < z < 3$; this is an increase of a factor of ~ 50 over the expected yield from DES (Collett 2015), and the very large sample size will enable the creation of a statistical sample of detailed galaxy measurements via the dedicated follow-up of the most highly magnified sources. Bright lenses at still higher redshifts are expected to be much rarer, and LSST's deep imaging in the *i, z,* and *y* bands over a large fraction of the sky will be uniquely important for finding them.

The most exciting and informative data for these distant galaxies will likely come from spatially resolved observations with integral field spectrographs (IFUs). Because bright or lensed objects are rare, single-object IFUs covering relatively small fields of view (FOV; a few arcseconds on a side) will be suitable for most purposes. At near-infrared wavelengths, adaptive optics (AO) can provide higher-order corrections for atmospheric distortions, affording angular resolution at the diffraction limit and the most detailed view of galaxy properties. The high angular resolution provided by AO will also provide essential detail on the host galaxies of gravitationally lensed AGN. At $z < 6$, near-infrared observations can be used to study optical rest-frame nebular emission lines to measure ISM kinematics, chemical abundances, and excitation. Near-infrared absorption line spectroscopy is still more challenging but can be used to measure stellar kinematics and population diagnostics. Substantial progress can be made with a dedicated program to follow up the most promising lensed or intrinsically bright galaxies, using either 8–10m- or 30m-class telescopes.

### *Initial characterization of lensing systems*

At a given apparent magnitude (the key parameter enabling high-SNR observations), gravitationally lensed sources are boosted from a wide range of absolute magnitude (i.e., luminosity). From a sufficiently large pool of bright, lensed galaxies, astronomers will be able to assemble carefully selected samples of galaxies that span a range of intrinsic properties, including broad sampling of the luminosity function down to faint limits that could not be probed any other way. Large samples of lensed galaxies will be discovered with LSST imaging. Some candidates may be identified from standard LSST catalogs plus visual inspection, but lens detectability depends strongly on the search strategy that is employed (e.g., Collett 2015), and optimized selection may require customized algorithms to be run on large collections of LSST cutout images. Additional candidates may be selected spectroscopically from wide-field, massively multiplexed galaxy redshift surveys that target LSST-selected galaxies (see, e.g., Bolton et al. 2004 using SDSS). These samples will require initial screening to confirm and characterize the lens systems before selecting targets for more intensive (and expensive) detailed investigation. This screening can be done primarily with 8–10m telescopes and low- to medium-resolution spectrographs (R = 500 to 3000 at optical wavelengths; R = 3000 to 5000 is optimal for the near-IR). Optical spectroscopy should be sufficient to measure redshifts for most UV-bright lensed galaxies, and many of the foreground lenses as well, although near-IR spectroscopy may sometimes be useful as well, especially for obscured sources. IFUs would be ideal, allowing simultaneous observation of both the foreground and background objects with spatial separation independent of slit angle constraints, but slit spectrographs can suffice. High-throughput and broad-wavelength coverage would be ideal, especially for measuring redshifts rapidly and efficiently. Multiplex is unimportant—these targets will be isolated on the sky. High-resolution imaging to constrain the lensing model will also be important for evaluating magnifications and hence the intrinsic luminosities of the sources. Traditionally this has been done mostly with *HST*, but in the LSST era (post-HST and when JWST is a precious resource) it may be most efficient to use 8–10m telescopes and near-IR AO imaging for this purpose. The AO-fed IFU instrument described below can be used for this, but the throughput of the IFU will likely be lower than a dedicated imager.

### *Spatially resolved galaxy studies with 8–10m telescopes*

At both infrared and optical wavelengths, 8–10m telescopes can yield valuable measurements following up LSST-selected lenses and exceptionally bright unlensed galaxies, and there will be a need for robust IFU and AO capabilities on telescopes in that aperture class in the LSST era. The redshift distribution of the source galaxies in galaxy-galaxy strong lenses discovered by LSST is expected to peak at z ~ 2–2.5 (Collett 2015): the optimal redshift range for ground-based follow-up of both rest-frame UV and optical features.

Rest-frame optical observations of LSST lensed galaxies with 8–10m telescopes will focus on spatially resolved emission line maps, requiring deep observations using a sensitive near-IR IFU with adaptive optics. Spatially resolved line ratios will allow the measurement of metallicity gradients and the detections of spatial variations in density and ionization; these diagnostics will enable the study of galaxy growth through star formation and gas accretion. Because these measurements rely on the detection of weak lines, long integrations with 8–10m telescopes will be required. These observations will also enable kinematic measurements at high resolution, enabling the measurement of rotation, localized velocity dispersion, individual kinematic components, and, if observations are deep enough, the potential detection of emission from outflowing gas. IFU fields of view and AO correction over fields of a few arcseconds are sufficient for most studies of galaxy-galaxy lenses, although the strongest cluster lenses would benefit from wider fields. Single laser guide stars (LGS) can be used, but will lead to radial PSF degradation that would complicate scientific analysis. Multi-conjugate adaptive optics (MCAO) and other multi-laser AO systems would provide more uniform correction over the area of interest. Near-IR spectral resolutions of 3000–5000 are important for resolving the OH night sky forest and for kinematic measurements in the galaxies themselves.

### *Spatially resolved galaxy studies with giant telescopes*

Giant Segmented Mirror Telescopes (GSMTs) will offer a major advance in capability from 8–10m telescopes, thanks to an order of magnitude increase in collecting area, and 3–5× better diffraction-limited angular resolution (and better still compared to JWST, although probably with a significantly worse Strehl ratio). At 2 μm, the diffraction-limited performance of a 30m telescope corresponds to linear scales of 100–150 pc at $0.5 < z < 6$ without accounting for the stretching effects of gravitational lensing. For strongly magnified high redshift galaxies, GSMTs will measure structure and spectral properties on scales of 10s of parsecs. This will be particularly powerful to study star clusters and clumpy structure in distant galaxies. For smooth structure, a GSMT will provide much higher SNR per fixed linear scale than would smaller telescopes,[3] enabling measurement of weaker spectral features beyond the reach of 8–10m telescopes. Examples include spatially resolved measurements of nebular abundances using weak line indices, the mapping of stellar abundances and other stellar population diagnostics through absorption lines at both near-IR and UV wavelengths, and spatially resolved stellar kinematics. GSMTs will also open larger samples of fainter lensed (and unlensed) galaxies for intensive study.[4] These are the sorts of objects that LSST's six-band imaging over very wide sky areas will find in abundance that precursor surveys from DECam or HSC will not.

The instrumental requirements for extending studies of lensed galaxies to GSMTs are similar to those for IFU spectroscopy and AO imaging on 8–10m telescopes, although MCAO and other laser constellation methods will become essential for image correction over the necessary field of view on the giant telescopes. IFUs that can be configured to multiple spaxel scales would enable measurements at the highest diffraction-limited angular resolutions, as well as "binned" measurements on coarser scales but with higher SNR for measuring intrinsically weak/faint features. Higher spectral resolutions (R > 5000) may also become interesting to take advantage of the greater GSMT collecting area to measure more detailed velocity structure in absorption and emission lines, particularly for dwarf galaxies with very high magnifications.

## Technical Description

Near-IR IFU observations will focus on the brightest lensed galaxies. We use predictions from Collett (2015) to estimate the number of galaxy-galaxy lens systems found by LSST (Figure 7.2). The table below provides the estimated number of lensed galaxies brighter than $r_{obs} = 23$ and $r_{obs} = 21$ as a function of redshift; number counts for the rare, brightest galaxies should be considered uncertain as these numbers were computed for a single realization of the catalog. Multiple realizations must be computed for a more reliable estimate. The code to do this, however, is well tested and does not require new significant computational resources.

---

[3] Gravitational lensing conserves surface brightness, so the SNR per diffraction-limited beam for smooth galaxy structure is not better for a GSMT than for an 8–10m telescope, although the angular resolution itself is higher.

[4] Seeing-limited measurements (or AO measurements at fixed angular scale) on a 30m telescope will reach 1.4 mag fainter than the same exposure time on an 8m. This can dramatically increase the number of LSST-selected lenses that can be studied (see Technical Description, below). Diffraction-limited observations of internal substructures (e.g., star clusters within high redshift galaxies) on a 30m telescope will achieve 3.8x better angular resolution and 2.9 mag fainter flux limits compared to an 8m telescope.

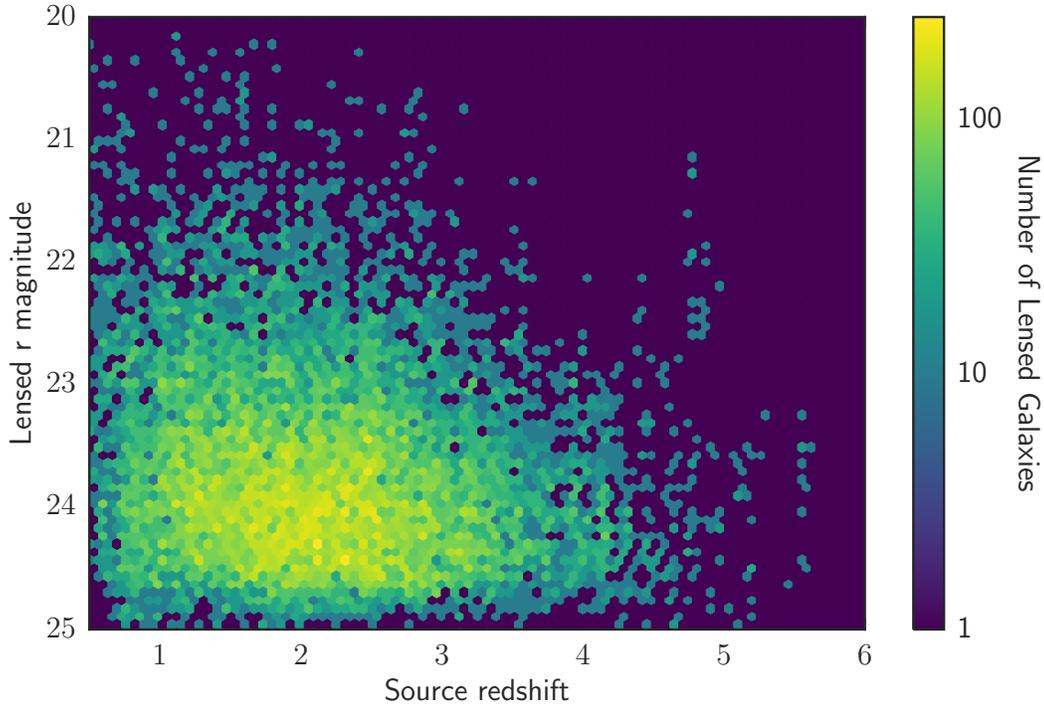

*Figure 7.2. Predicted number of galaxy-galaxy strong lenses as a function of r magnitude and source redshift in a realization of the LSST wide survey. Follow-up observations will focus on the rare, brightest objects, necessitating the wide area of LSST. Numbers estimated from the catalog of Collett (2015).*

| Redshift Range | N($r < 23$) | N($r < 21$) |
|---|---|---|
| 1–1.5 | 4200 | 160 |
| 1.5–2 | 4300 | 170 |
| 2–2.5 | 3300 | 90 |
| 2.5–3 | 2400 | 90 |
| 3–3.5 | 800 | 10 |

Galaxies with $r < 21$ will provide the highest-quality data (MS1512-cB18 has $r = 20.4$). To evaluate the evolution of galaxy properties with mass and redshift, an ideal survey will be divided into at least three bins in redshift, stellar mass, and metallicity. With at least 10 galaxies in each bin, ~ 270 galaxies are required, to be drawn from the brightest lensed galaxies found by LSST.

### Needed Capabilities and Estimate of Demand

The initial screening can be done primarily with 8–10m telescopes and low- to medium-resolution spectrographs (R = 500 to 3000 at optical wavelengths; R = 3000 to 5000 is optimal for the near-IR) with a broad spectral range sufficient to ensure that lines are detected regardless of the source redshift. Assuming 270 lenses and a 50% contamination, we will have 540 targets, which clearly depends on the quality of the lens selection technique. Assuming 1 hour per target, 50% overheads, and 30% weather losses, we require 130 nights for the initial screening. Clearly this time can be reduced by increasing the purity of the candidate sample. Therefore, significant time savings can be realized if new techniques are developed that allow for more robust selection of lenses from LSST data.

These observations require a near-IR IFU with AO on an 8–10 m telescope. Estimated exposure times assume observations with Gemini NIFS, with a 3 x 3 arcsec field of view. Typical predicted Einstein radii are 0.9–1.9 arcsec, so this FOV will cover the entire galaxy for a significant fraction of the LSST sample. However, unless the FOV can be increased, additional time will be required for background sky observations. Additionally, the sky coverage at high Galactic latitude for the Gemini facility AO system, ALTAIR, is rather small, due to the need for relatively bright natural guide stars.

Observations in at least two bands (usually $H$ and $K$) are required to measure line ratios. We estimate 2 hours on target per object, for a total of ∼ 1080 hours on source, or ∼ 460 nights accounting for nodding to sky, 30% additional overheads, and 30% weather loss. This number is halved if a larger IFU can be accessed that allows the source to be dithered within the IFU and hence removes the need for nodding to sky.

The NIFS ITC suggests a ∼5 sigma detection of an emission line flux of ∼ 3 x $10^{-16}$ erg cm$^{-2}$ s$^{-1}$ in 2 hours in the $K$-band, and of ∼ 9 x $10^{-16}$ erg cm$^{-2}$ s$^{-1}$ in 2 hours $H$. Both calculations assume the lines are well placed between the night sky lines. These are reasonable numbers for the strong lines in the brightest lensed galaxies; for reference, emission line fluxes in cB58 range from ∼ 1 − 15 x $10^{-16}$ erg cm$^{-2}$ s$^{-1}$ for the lines of interest.

The continued availability of near-IR IFUs with LGS-AO is critical to enable this important LSST-enabled science program. Gemini's NIFS and ALTAIR AO system can fill this role, but their effectiveness could be substantially improved by increasing the NIFS field of view (roughly halving the required observing time) and by improving AO throughput to enable the use of fainter natural guide stars, which in turn can increase the sky coverage and thus the sample of lenses that are suitable targets for detailed follow-up studies.

## Science case: Understanding the connection between galaxy properties and environment over cosmic time

An enduring problem in galaxy evolution is understanding how environment influences galaxy properties. Indeed, even the most sophisticated theoretical models today do a poor job at simultaneously modeling the environmental dependence of the quenching of star formation, the buildup of metals, and the assembly of stellar mass (Hirschmann et al. 2014). LSST nominally provides (1) a large enough wavelength range when coupled with future NIR space missions (e.g., WFIRST) to characterize the spectral energy distributions (SEDs) of these galaxies; (2) large enough areas to build sample sizes sufficient to probe highly dimensional parameter space. Furthermore, the LSST Deep Drilling Fields (DDFs) will enable significant progress toward advancing this field by providing (3) multi-band imaging of the sky to depths sufficient to detect passive galaxies at z < 2. Key to capitalizing on the opportunity afforded by LSST DDFs is a large spectroscopic survey on a highly multiplexed 8–10m telescope. Spectroscopy is required both to obtain astrophysical information about the galaxies and to measure environments, as moderate density environments like groups are poorly recovered when using broadband photometric surveys (Cooper et al. 2005.) A large survey conducted with a wide-field, multi-object spectrograph on an 8m telescope is required to fully exploit this science in the era of LSST.

This project address the NWNH Galaxies Across Cosmic Time Panel's science questions "How do baryons cycle in and out of galaxies, and what do they do while they are there?" and "How do cosmic structures form and evolve?

## Technical Description

Our goal is to track the evolution of a large fraction of the cosmic stellar mass all the way out to z = 2. This can be accomplished by (1) performing a spectroscopic survey targeting a sample of galaxies at z < 2 that is mass-complete to $\log(M_{star}/M_\odot) = 10.0$, which is both the mass of the main progenitor of a present-day M$^\star$ galaxies, and the mass below which intrinsic quenching can be neglected (Peng et al. 2010); (2) using clustering measurements at moderate halo mass to link descendants and progenitors via their host dark matter halos; (3) measuring masses for massive halos ($\log(M_{dyn})$>13) directly via velocity dispersions; (4) inferring the joint distribution and evolution of the star formation rate, stellar mass, dynamical mass, gas-phase metallicity, redshift, environment, and star formation history.

### Large N, low S/N

Our survey strategy combines traditional methods—in which astrophysics are possible with moderate signal-to-noise individual spectra—with Bayesian statistical modeling that allow one to determine the distribution function of physical parameters from large numbers of spectra whose signal-to-noise is sufficient only for obtaining a redshift. This technique has been impressively demonstrated with data from Shu et al. (2012), who used noisy spectra from the Baryon Acoustic Oscillation Survey (BOSS) to derive the galaxy velocity dispersion distribution function from spectra for which the velocity dispersion could not be individually measured.

### Limiting magnitude

The characteristic mass for galaxies at z ∼ 0 is $\log(M^\star/M_\odot) = 11.0$ (Bell et al. 2003), and Torrey et al. (2015) used the Illustris simulation to determine that 80% of the main progenitors of such galaxies at z = 2.0 have $\log(M_{star}/M_\odot) > 10.0$. This sets the mass limit of our survey.

We estimate the magnitude limit corresponding to this stellar mass limit in two ways. First, we use the 3D-HST catalog in COSMOS (Skelton et al. 2014), which includes extremely deep NIR imaging from UltraVISTA (McCracken et al. 2012). We select galaxies with z ∼ 2 and

log($M_{star}/M_\odot$) ~ 10.0. 95% of these galaxies have J(AB) < 24.4. We compare this estimate to that obtained by using the Bruzual & Charlot (2003) stellar population models of this stellar mass and find that this magnitude limit is similar to a $z_{form}$ = 2.5 Simple Stellar Population (SSP) observed at z = 2 (Figure 7.3; black dotted line), where $z_{form}$ is the redshift at which stars star forming in this model. The 4000 Å break is in the *z*-band at z = 1 and *z*-band magnitude for our fiducial model at z = 1 is 24.25. Even though these same objects in the *J*-band will have a magnitude of 23.25, we will observe z ~ 1 objects in the *z*-band, as the much brighter NIR sky makes the observations there more expensive. On the other hand, at z ~ 2 the $z_{form}$=2.5 SSP model predict galaxies with *z*-band a magnitude of 26, which makes observations in the *J*-band more feasible, as the 4000 Å break lies in the *J*-band at z = 2. Therefore the optimal strategy involves using different wavelength regimes to select and estimate spectroscopic exposure times for galaxies at different redshifts.

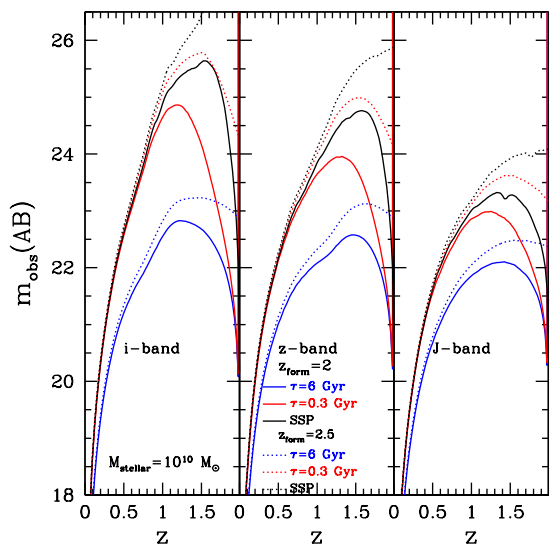

*Figure 7.3. The observed magnitudes for a $M_{star}$=$10^{10}M_\odot$ galaxy with one of three star formation histories. Black: Simple stellar population (SSP); blue: an exponentially declining SFH with τ = 6 Gyr; red: an exponential with τ = 0.3 Gyr. The solid lines correspond to $z_{form}$ = 2 and the dotted lines to $z_{form}$ = 2.5, where $z_{form}$ is the redshift at which stars start forming. The fiducial limits for our survey are z-band = 24.25 at z = 1 and J-band = 24.4 at z = 2.*

### Galaxy selection

The LSST Science Book[5] outlines a plan for a set of deep drilling fields (DDFs) with straw man magnitude limits assuming 1% of the total survey time is spent on a single DDF pointing. The resultant 5-σ magnitude limits in *z* and *y* will likely be 28.0 and 27.0, respectively.[6] The wide LSST survey will have a limiting magnitude in z-band of 26.0, which will be sufficient for target selection given that our faintest targets in the z-band will be ~ 25.0. However, performing our survey in the DDFs will allow us high-precision (S/N > 10) characterization of the SEDs and photometric redshifts of our targets and will allow us to identify and characterize the SEDs of galaxies well below our spectroscopic limit and thus to understand the environments of our targets. Given the large area (~ 100 $deg^2$) of the DDFs, there will ample sky area to choose from. We will therefore perform this survey in the DDF fields. We will supplement the LSST imaging with data from the Spitzer SERVS survey of some of the DDFs (Mauduit et al. 2012) and NIR imaging from Euclid[7] (YJH 5-σ depths of 24.0) and from the WFIRST High Latitude Survey[8]

(zYJH 5-σ depths of 26.5–26.9). Euclid imaging will be useful for SED characterization of brighter objects. WFIRST will probe significantly fainter and hence will allow precision SED characterization for galaxies down to the spectroscopic limit of our survey. While these other missions will be useful for SED characterization, they will not be necessary for target selection thanks to the very deep LSST imaging in the DDFs.

## Needed Capabilities and Estimate of Demand

### Required facilities

A continuum redshift of an object at $z = 24.25$ or $J = 24.4$ clearly requires an 8–10m-class telescope. For example, the Gemini Observations of Galaxies in Rich Early Environments (GOGREEN; PI Balogh) Large Program uses the GMOS spectrograph with newly updated red-sensitive chips on Gemini-S to obtain redshifts of passive galaxies with $z = 24.25$ at $z = 1-1.5$ and requires 15-hour exposures to do so. To measure passive galaxy redshifts via the 4000 Ang break at $z = 2$, we require a spectrograph with a wavelength coverage out to 1.3μm at the long side. As we will have low redshift interlopers for which we would like to get spectra, we will need a wavelength coverage at the blue end down to approximately 0.5μm, which is sufficient to get the 4000 Å break at $z > 0.25$ and Hα at all redshifts.

The resolution of the spectrograph is set primarily by the need to mitigate the effects of the night sky lines, which are dominant at $\lambda > 0.8$μm. This is critical for obtaining passive galaxy redshifts at our faintest levels. This requires a resolution of ~ 3500 in the NIR and ~ 2500 in the optical, which is similar to that of the Subaru Prime Focus Spectrograph (PFS; Takada et al. 2012). This corresponds to a velocity dispersion of 85 km/s. For the smallest galaxies, this may make it difficult to measure velocity dispersions. If R = 5000, as for a future spectrograph, then we would have a velocity resolution of 60 km/s, which should be adequate to measure velocity dispersions of our lowest-mass galaxies, although the low S/N in each spectrum for these galaxies (see below) would require an ensemble analysis of the velocity dispersions.

To study the effect of galaxy environment on galaxy properties, it is necessary to probe large volumes (areas) to contain many of the most massive halos, while also having a high density of spectra, to well sample these regions and obtain velocity dispersions. This naturally lends itself to a highly multiplexed fiber system feeding multiple spectrographs, similar to PFS.

### Exposure times

Our base exposure time is driven by the need to obtain a redshift for each object. No ETC exists for a fiber spectrograph with these parameters on an 8–10m telescope and so we base our exposure times on the estimated times required for the PFS SuMIRe project (Takada et al. 2012). They plan to obtain continuum redshifts to J = 23.4 in 3 hours. We have shown that our magnitude limit: approaching J = 24.4 for our survey, results in a factor of 6.25 longer exposure times, or 18.75h per pointing. This is consistent with the strategy of the ongoing Gemini Observations of Galaxies in Rich Early Environments (GOGREEN; PI Balogh) Large Program using the GMOS spectrograph on Gemini-S to obtain redshifts of passive galaxies with $z = 24.25$

---

at z = 1 − 1.5, requiring 1-hour exposures to do so. This will result in S/N > 5 for all objects at our magnitude limit, sufficient for a 4000 Å break redshift.

Another exposure time constraint comes from the necessity of measuring emission line fluxes, absorption line indices, and velocity dispersions from the spectra. This requires accurate modeling of the continuum as well as decomposing absorption and emission lines (e.g., Rudnick et al. 2000; Moustakas et al. 2011). This will not be possible for individual galaxies at our magnitude limit, but the parameters for an ensemble of galaxies can be determined by jointly modeling their spectra (Shu et al. 2012). We can evaluate the total information in the spectra of a sample of galaxies via the S/N in the composite spectrum (or "stack"). If we stack 16 galaxies at our magnitude limit, we can expect to achieve a stack with S/N = 20, which is enough to measure line strengths and fit line profiles. This sets the minimum number of galaxies in any bin of parameter space.

### *The survey size*

We require spectra of enough galaxies to populate the full range of parameter space. A baseline estimate results from considering the following properties and bin sizes: stellar mass (5 bins), gas-phase metallicity (3 bins), halo mass from galaxy-galaxy velocity dispersion for groups and clusters (3 bins), internal galaxy velocity dispersion (3 bins), SFR (4 bins), and luminosity weighted stellar age for passive galaxies (3 bins), and distinguishing centrals and satellites (2 bins).

While optimizing the number of galaxies per bin requires a full simulation, we can make an estimate of the survey size by requiring 16 galaxies per bin, which gives us an effective S/N = 20 in the stack for our faintest sources. This results in 25,920 objects per redshift bin. If we assume 5 redshift bins at $0.5 < z < 2$ of $\Delta z = 0.3$ each, then the total survey will contain $1.3 \times 10^5$ galaxies with spectroscopic observations. Our survey at $z < 0.5$ will be anchored by existing SDSS spectroscopy and the eventual DESI Bright Galaxy Survey (BGS).

To determine the area of the survey, we integrated the redshift-dependent stellar mass functions from Muzzin et al. (2013) at $M_{star} > 10^{10} M_\odot$ to determine the area necessary to obtain 26,000 galaxies in each redshift bin. This are plotted in Figure 7.4. Our survey will need to be 4 deg$^2$ to obtain the number of galaxies that we require. This is similar in size to the DEEP2 survey (Newman et al. 2013), but that survey only extends to z = 1.4 and has a stellar mass limit of $\log(M_{star}/M_\odot) > 11.1$ at z = 1 (Cooper et al. 2012.) We plot the volume probed vs. the mass limit compared to other surveys in Figure 7.4. Photometric samples can offer larger numbers or area, but photometric redshifts, even from excellent data like the UltraVISTA (Muzzin et al. 2013), are almost entirely untested by spectroscopy down to our stellar mass and magnitude limits.

### *The time request*

For a spectrograph of fixed sensitivity and spectral resolution, the total survey time necessary depends on the telescope aperture, field of view, and multiplexing, which are all related. Here we scope out the survey time required for a PFS-like spectrograph discussed above ($N_{fiber}$ = 2400, field diameter = 1.3 deg$^2$), on an 8m telescope. In addition to estimates of nights, we will also give an estimate of fiber hours, which is scalable for different degrees of multiplexing.

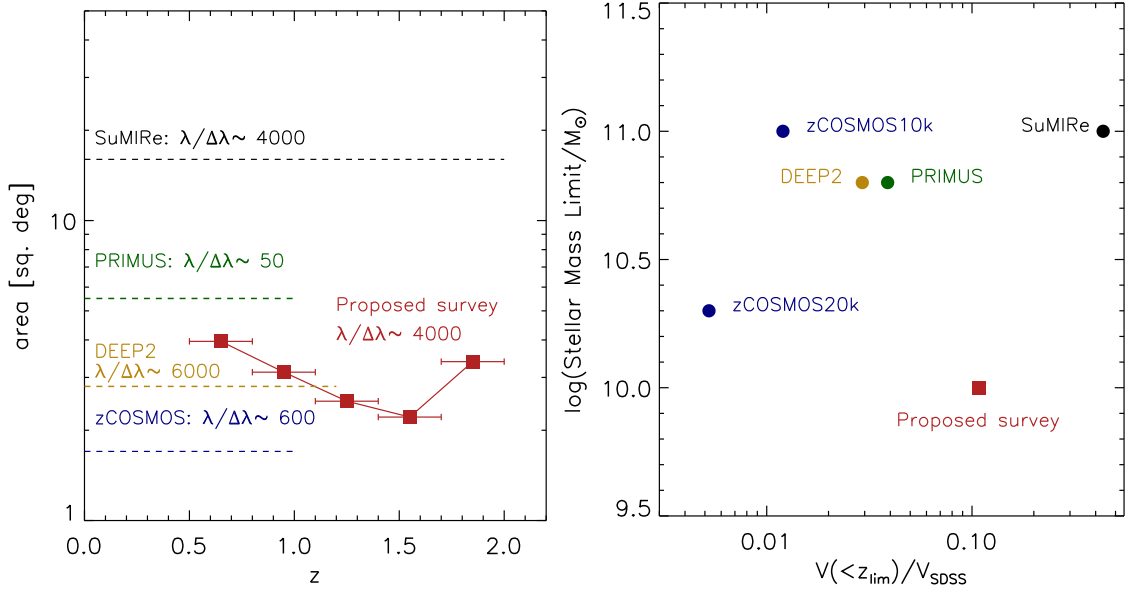

*Figure 7.4. Left panel: The area needed for our planned survey, compared to the area of other surveys with their redshift extents indicated. This area is derived directly from the required volumes needed to obtain enough galaxies to sample the multidimensional parameter space. Right panel: The enclosed volume of various spectroscopic surveys vs. the stellar mass completeness limit (for passive galaxies) of those surveys. Our proposed survey probes 10% of the total SDSS volume but with a stellar mass limit that is an order of magnitude below other surveys of comparable area.*

The surface density of $\log(M_*/M_\odot) > 10.0$ targets at $0.5 < z < 2.0$ is 44,600 galaxies per deg$^2$ and 59,100 galaxies per single PFS pointing. Assuming that we can only place 80% of the fibers on unique sources, we will need 31 fiber configurations to target every galaxy above our mass limit within a single 1.3 deg$^2$ field of view. To cover the 4 deg$^2$ required by our proposed survey (Figure 7.4) will require 3 PFS fields, assuming no field overlap. For 3 fields, with 31 setups each, and 18h per setup, we have a total integration time of 1670h. Assuming 20% overheads and 8 hours per night, this results in 326 nights. Equivalently, this corresponds to 3.2 million fiber hours of integration time. Our brighter targets will not require the full 18h for a redshift, and exposing to a fixed S/N limit would save some time, although most of the sources will be faint due to the slope of the number counts. However, we keep this design for a constant exposure time, as the high S/N spectroscopy of the brighter sources will enable us to test our Bayesian modeling of the low S/N spectra. This is an ambitious request but will be the definitive galaxy evolution spectroscopic survey in the age of LSST, not to be surpassed for decades.

Our time request clearly depends on the sampling of parameter space, but the numbers given here can be directly scaled by changing the number of galaxies per parameter bin, or the number of bins. We argue that a key requirement is the $z = 2$ limit and the stellar mass limit.

It is important to note that the photo-z training set survey (Chapter 8) and this one will overlap to some extent in their selection. Indeed, 84% of our targets would be brighter than the i = 25.3 limit of the photo-z calibration sample. Also, 12% of i < 25.3 sources satisfy our redshift and stellar mass cuts. The i < 25.3 sample has a significantly higher surface density and therefore could be used to fill fibers not occupied by our sparser targets. This implies that the two projects could benefit from overlap and time savings.

## Statistics and Analysis Infrastructure

### Hierarchical Bayesian spectrum analysis forgalaxies

For the faintest galaxies in the spectroscopic survey, the S/N will be insufficient for reliable estimates of anything beyond redshift and crude spectral type on an object-by-object basis. The aggregate spectroscopic dataset for this large population will nevertheless have extensive information content. Classical approaches to extracting this information typically focus on "stacking" (i.e., averaging) of large numbers of spectra to obtain a single high-S/N output. However, stacking analysis does not yield any information about the population variance or correlations within the stacked sample. Furthermore, measurements of non-linear parameters (such as velocity dispersions or absorption-line diagnostics) from stacked spectra bear a non-trivial relationship to the average value of those parameters within the population (Shu et al. 2012).

An alternative approach that can recover population means, variances, and correlations from many low-S/N spectra (or any other data) is the framework of "hierarchical Bayesian" analysis. At the most basic level, this approach is simply an application of the "law of total probability," which states in differential form that

$$p(a) = \int db\, p(a|b)\, p(b) \ .$$

Consider a spectrum represented by the data vector d. To measure a physical parameter vector *a* from this data, the classical approach in the high-S/N regime is to express the likelihood of the parameters given the data, which is by definition equivalent to the probability of the data given the parameters:

$$\mathcal{L}(\mathbf{a}|\mathbf{d}) \equiv p(\mathbf{d}|\mathbf{a}) \ .$$

The form of p(d|a) can in principle be written down given an understanding of the observational process, the associated noise, and the mapping by which the physical parameters of interest project into the frame of the data. The "maximum likelihood" (ML) estimator for the parameter *a* can then be obtained through appropriate means, and these estimators can be binned for many objects into histograms for the study of population demographics.

In the low-S/N regime, ML estimators of physical parameter *a* will be unreliable on an object-by-object basis. The hierarchical Bayesian approach overcomes this by modeling the probability density of the population in physical parameter space (which in any event is more interesting scientifically than the particular parameters a of a randomly selected galaxy!). This probability is expressed as

$$p(\mathbf{a}|\mathbf{t}) \ ,$$

where *t* is a vector of hyperparameters that characterize the population pdf in physical parameter space (for example, the mean and variance of a Gaussian).

With this expression for the population pdf, a simple application of the law of total probability gives an expression for the likelihood of the hyperparameters $t$ given the data:

$$\mathcal{L}(\mathbf{t}|\mathbf{d}) = p(\mathbf{d}|\mathbf{t}) = \int d\mathbf{a} \, p(\mathbf{d}|\mathbf{a}) \, p(\mathbf{a}|\mathbf{t}) \ .$$

Thus, by working not with a point-estimate of the physical parameters $a$ for a single object, but rather by integrating over all possible values of $a$, observational constraints on a single galaxy are propagated directly to constraints on the population as a whole.

This approach is of course of limited utility with a single object alone, but with many objects $i$ drawn from the same population, likelihoods can be multiplied (or log-likelihoods added) to give constraints on the population that are informed by the entire observational sample:

$$\ln \mathcal{L}(\mathbf{t}| \{\mathbf{d}_i\}) = \sum_i \ln \mathcal{L}_i(\mathbf{t}|\mathbf{d}_i) \ .$$

This population-wide likelihood can then either be (a) used to find ML estimators for the hyperparameters $t$ or (b) converted into a posterior pdf for $t$ through the use of suitable priors, depending on the preferences of the investigator (with the latter approach being the truly Bayesian). Examples of the application of hierarchical Bayesian analysis to the luminosity and velocity-dispersion demographics of luminous red galaxies observed in large numbers at low S/N within the SDSS-III Baryon Oscillation Spectroscopic Survey (BOSS: Dawson et al. 2013) can be found in Shu et al. (2012) and Montero-Dorta et al. (2016).

While the basic formulation of the hierarchical Bayesian approach is simple, the full application of this technique to the analysis of large spectroscopic galaxy survey populations will require significant development and/or adoption of statistical formalism and analysis software for (1) modeling complex distributions in high-dimensional physical parameter spaces; (2) enabling discovery of significant phenomenology that will not be visibly apparent in the way it would be with high-S/N spectroscopy; and (3) propagating sufficient physical-parameter likelihood samples for large catalogs of objects.

### *Spectro-photo cross-correlation*

Our current strategy uses spectroscopy to determine halo mass of objects, a necessary component of environment studies. There are two ways to modify this by cross-correlating the spectroscopy with sources having only photometric redshifts. For galaxies that would nominally be in the spectroscopic sample, such a technique could help us to recover the halo masses and potentially environments, in a statistical way, for the full mass-limited sample. This has the potential to permit more sparsely sampled, and therefore less time-consuming, spectroscopic surveys. In the survey mentioned above, this implies that fewer fiber setups will be needed per pointing, translating directly to time gains. This technique will not replace spectroscopy for the determination of inherently spectroscopic parameters such as velocity dispersion and metallicity, but it gives the possibility of reducing the number of galaxies with spectroscopic observations. Another way this can be used is to cross-correlate our spectroscopic sample with galaxies below the formal spectroscopic limit, thus correlating their photometric properties with the spectroscopic properties and halo masses of the brighter sample.

There has been some work in the area of cross-correlating spectroscopic and photometric surveys (e.g., Matthews & Newman 2012; Mubdi et al. 2015), but these works have focused on reconstructing the redshift distribution and not the halo mass and environments of galaxies. By investing in the development of such techniques, we can save on some of the cost of our spectroscopic survey and push the reach of such a survey to significantly fainter galaxies and lower stellar masses.

## Science case: Mapping the growth of black holes over cosmic time

Essentially every massive galaxy hosts a supermassive black hole (SMBH) in its center. The mass of the SMBH correlates to the mass of the galaxy bulge, indicating that SMBH growth in active galactic nucleus (AGN) phases must have corresponding episodes of galaxy star formation. But the details of this coeval SMBH–galaxy growth remain poorly understood. Quantifying SMBH mass growth over time is extremely difficult, as the vast majority of $z > 0.3$ black hole masses are uncertain by a factor of $> 3$. The physics of SMBH accretion is also poorly understood: the Shakura & Sunyaev (1973; hereafter SS73) thin-disk model remains the standard accretion-disk theory (and is the fifth-most cited paper in astrophysics), but there is growing evidence that SS73 fails to match observations, especially by failing to capture the effects of mass and accretion rate on disk structure (e.g., Abramowicz et al. 1998; Narayan & McClintock 2008). Our understanding of black hole growth, and accretion theory in general, is fundamentally limited by a lack of SMBH mass and accretion-disk measurements.

The *synoptic* observations of LSST are especially valuable for measuring the growth of SMBHs—especially if coupled to *wide-field, massively multiplexed spectroscopic monitoring*. Accretion-disk reverberation mapping (RM) directly probes the physics and accretion structure near the black hole, a critical part of the NWNH Physics of the Universe objective. Pushing accretion-disk RM to its limit with LSST would probe SMBH accretion shortly after seed formation at Cosmic Dawn, another key NWNH objective. This science case addresses the NWNH Galaxies Across Cosmic Time questions: "How do black holes grow, radiate, and influence their surroundings?" and "What were the first objects to light up the universe, and when did they do it?"

### Spectroscopic monitoring for black hole mass demographics

Reverberation mapping (RM) probes physically connected regions around the SMBH: see Figure 7.5. The emission of outer regions is observed to "lag" behind inner-region emission, with a light travel "lag time" $\tau$ related to size as $R = c\tau$. The most common RM application is to measure broad-line emission time delays in response to continuum variations, using the broad-line size and velocity to measure black hole mass with the virial theorem. Broad-line RM is the only direct method for SMBH masses beyond the local universe. But to date, the vast majority of $z > 0.01$ SMBH masses to date come from the indirect "single-epoch" estimator and are uncertain by at least a factor of $\sim 3$ (Krolik 2001; Shen & Kelly 2012). The current state of unreliable SMBH masses severely limits our understanding of SMBH demographics (e.g., Kelly & Shen 2013) and makes it impossible to distinguish coincidental BH/galaxy coevolution from causal AGN fueling and feedback (Jahnke & Maccio 2011; Sun et al. 2015). Key for understanding BH/galaxy coevolution is measuring evolution in its scatter: models of causal coevolution predict decreasing scatter as a function of redshift, while coincidental shared fueling of AGN/SF predict constant scatter. Only RM masses have uncertainties small enough to robustly measure evolution of the BH/galaxy relation scatter: for example, a modest "piggyback" RM campaign of ~ 40 AGNs would enable an observational test of causal versus coincidental coevolution models (see **Technical Description** for more details), similar to that being done with SDSS (Shen et al. 2015).

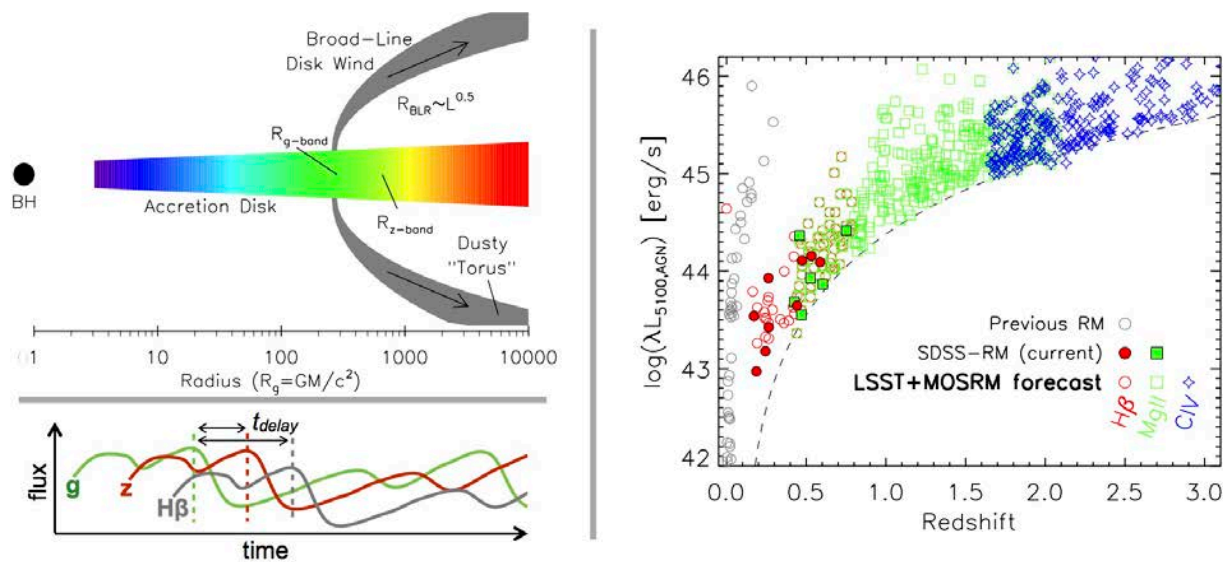

Figure 7.5. *Left: A cartoon illustrating accretion-disk and broad-line reverberation mapping (adapted from Trump et al. 2011). The bluer/inner and redder/outer emissions reverberate with a time lag related to size as $R = c\tau$. Continuum RM measures the accretion-disk size and temperature profile, while broad-line RM directly estimates black hole mass. Right: Forecasted BH masses measured by combining LSST with simultaneous multi-object spectroscopic reverberation mapping (MOSRM), using the DESI spectrograph and 4m telescope with a weekly cadence: see Technical Description for more details. Combining LSST with simultaneous spectroscopic monitoring in this design would result in ~ 400 reliable BH masses spanning 0 < z < 3: entirely uncharted space in BH demographics and AGN/galaxy coevolution.*

## *Light echoes of accretion-disk physics*

RM can directly probe accretion-disk structure by measuring continuum time lags between the inner/bluer and more distant/redder emission to map accretion-disk size and temperature profile. Following SS73 as a basic (but observationally uncertain) framework, the measured time lag between continuum emission at rest-frame wavelengths $\lambda_1$ and $\lambda_2$ (for example, observed in the $g$ and $z$ bands) relates to accretion-disk size and temperature profile as

$$\tau = \frac{\Delta R}{c} = \frac{R_0}{c}\left(\lambda_2^\beta - \lambda_1^\beta\right)M_{BH}^\gamma \dot{M}_{BH}^\delta$$

In the SS73 model, $\beta = 4/3$, $\gamma = \delta = 1/3$, and $R_0 = [(45G)/(16\pi^6 hc^2)]^{1/3}$. Only a handful of Seyfert 1 AGNs currently have accretion-disk RM measurements: these and microlensing observations find a temperature scaling $\beta = 4/3$ consistent with SS73 but also measure accretion-disk sizes that are too large by a factor of 3–4 (e.g., Collier et al. 1998; Morgan et al. 2010; Fausnaugh et al. 2016). Because these previous studies were limited to a few quasars in a narrow range of mass and luminosity, the causal physics behind the discrepant disk sizes is unclear: i.e., it is unknown if they are caused by a different $R_0$, $\gamma$, or $\delta$ compared to the SS73 expectation. Understanding accretion-disk physics—particularly how they scale with mass and accretion rate—critically requires continuum RM on a large AGN sample spanning a wide range of mass and luminosity. On its own, LSST intra-band time lags measure the accretion-disk temperature profile across *ugrizy*. But the greatest gain for understanding accretion physics comes from coupling LSST to simultaneous spectroscopy and direct RM masses, for the first observational tests of how accretion disk size depends on black hole mass and accretion rate.

### *A new frontier in AGN reverberation mapping*

BH masses measured by simultaneous spectroscopic monitoring can also be used to calibrate LSST for purely *photometric* RM of $> 10^6$ AGNs. For example, photometric broad-line RM could be executed by modeling the broad-line contribution in each LSST filter (Chelouche et al. 2014). If accretion-disk RM measurements calibrate a tight $M_{BH}$ scaling (following the SS73 $M_{BH}^{1/3}$ expectation or otherwise), then photometric accretion-disk RM could be used as a direct mass estimator for $> 10^6$ AGNs across the LSST sky coverage.[9] This would include low-luminosity Seyfert AGNs to luminous quasars, and the deep drilling single-night depth ($i < 25$) would reach $\log(M_{BH}/M_\odot) > 6$ at $z \sim 4$: direct BH masses near the expected massive end of direct-collapse SMBH seeds. Similarly, a $L^\delta$-$R$ relation between AGN luminosity and photometric accretion-disk size could be a standard candle for redshift-independent luminosity distances to $z \sim 7$. Photometric RM would be a transformative new method, with potentially millions of direct BH mass measurements. But it is only calibrated and enabled by coupling LSST to simultaneous multi-object spectroscopic monitoring.

### *Transient AGN events*

Black hole growth is fitful and turbulent, characterized by many kinds of transient flares related to changes in accretion rate and/or disk geometry. Tidal disruption events of infalling stars occur with a rate of $\sim 10^{-5}$/yr/galaxy (Stone & Metzger 2016). Other transients include "changing-look" AGNs, which dramatically brighten or dim due to changes in accretion rate (Denney et al. 2014; LaMassa et al. 2015). LSST will also discover an unknown (but almost certainly non-zero) number of gravitational wave powered inspiraling SMBH binaries, detectable by their semi-periodic variability (e.g., Graham et al. 2015).

### Technical Description

To forecast the number of BH masses achieved by LSST with simultaneous multi-object spectroscopic reverberation mapping (MOSRM), we scale from observed lightcurves in the SDSS-RM project (Shen et al. 2015), currently operating with the BOSS spectrograph and 2.5m SDSS telescope. Figure 7.5 (right panel) presents the forecast return from a DESI-like spectrograph and telescope (Mayall 4m), empirically calculated by extending the SDSS-RM lightcurves to the S/N achieved by a larger aperture. Starting with LSST imaging in a deep drilling field with a ~daily cadence, adding a dedicated monitoring program with a DESI-like spectrograph on a weekly cadence would return an unprecedented ~ 400 AGNs with accurate black hole masses and accretion-disk sizes: a factor of ~ 10 (for SMBH masses) to ~100 (for accretion-disk sizes) larger than any current or planned efforts before LSST. A more modest campaign using only ~ 1% of the DESI fibers, "piggybacking" within a larger program, would still provide accurate black hole masses for ~ 40 quasars spanning $0.5 < z < 2.5$: sufficient to enable the first robust ensemble tests of whether BH-galaxy coevolution is *causal* with active feedback or *coincidental* with stochastic fueling from a shared gas supply.

---

[9] There will be ~$10^7$ AGNs of $i < 24.5$ in the LSST footprint: see Section 10.1.3 of the LSST Science Book, which derives this number from the Hopkins et al. (2007) luminosity function. About half of these AGNs will have detectable variability in ~ 6 months of LSST observations, of which ~ 30% may have observed-frame accretion-disk lags detectable by a few-day cadence. While the number of objects suitable for such an analysis depends critical on the ability to detect such lags with the LSST cadence, such a program would result in $> 10^6$ AGNs with measured accretion-disk lags.

The AGNs targeted by an LSST + DESI-like survey are expected to have broad-line lags of 10s to 100s of days, and *ugrizy* accretion-disk lags a factor of a few shorter (on the order of a few days). This means that photometric monitoring on a daily cadence would be most effective, along with spectroscopic monitoring on a ~weekly cadence. Deep drilling field photometry on a slower three-day cadence would roughly halve the number of detected photometric accretion-disk lags. Meanwhile, the five-year duration is required to detect the longest broad-line lags, although this is more important for the spectroscopy than the photometry. A shorter one-year duration for the photometry would retain ~ 80% of the accretion-disk and broad-line lags, so long as the spectroscopic monitoring still continued for 5 years.

### Needed capabilities and estimate of demand

Fully exploiting LSST for black hole growth and accretion-disk physics critically requires *simultaneous monitoring* with a *wide-field* multi-object spectrograph. There are ~ 100 AGNs/deg$^2$ of $i < 21.5$, at least half of which would have detectable variability for reverberation mapping with a weekly cadence spanning 5 years. This represents ~ 400 RM masses in a ~ 7 deg$^2$ DESI field that fits within a single LSST deep drilling field. A single DDF would only be viewable for half of each year, implying that there would be 125 (5 years x 25 weeks/year) separate epochs. Such a campaign in a single DDF would yield an order of magnitude increase in the number of precise black hole masses. Here the $i < 21.5$ depth is matched to one-hour single-epoch exposures with a spectrograph on a 4m telescope. QSOs in this magnitude range will have (observed-frame) broad-line lags of 10s to 100s of days, such that a long monitoring duration is more important than a rapid cadence. (That said, a rapid cadence produces smaller uncertainties in the measured lags.) This results in a total time commitment of 87,500 fiber hours spread over five years of monitoring in a single field. A shorter one-year total duration would result in only ~ 80 QSOs with RM masses, preferentially removing high-redshift and massive/luminous QSOs. Moderate spectral resolution (R ~ 2000) is sufficient for resolving the broad emission lines and for enabling coarse velocity-resolved RM of inflow/outflow. Since the density of relevant AGNs is ~ 100/deg$^2$, a wide field is more important than high fiber density. The DESI spectrograph is the ideal existing instrument for supporting LSST with MOSRM (presumably available only after the dark energy survey concludes): the 4m Mayall could reach an equatorial LSST deep drilling field (e.g., COSMOS), while moving DESI to the 4m Blanco would access the full range of LSST sky coverage.

Even in the dedicated monitoring program forecast in Figure 7.5, a MOSRM program would use only ~ 20% of the total DESI fibers. Thus AGN RM is best suited to being part of a larger "fiber-sharing" survey that accommodates repeated one-hour observations in the same field. A more modest "piggyback" survey using only ~ 1% of the total DESI fibers would still achieve the ~ 40 RM masses spanning $0 < z < 2$, enabling unprecedented precision of ~ 0.1 dex for BH/galaxy scatter per unit redshift: a robust test of causal versus coincidental coevolution.

For detecting transient AGN events, the most critical capability is an Event Broker to identify, characterize, and prioritize variability events unlike the typical stochastic variability of AGNs. Understanding the physics driving flares, TDEs, and binary SMBH inspirals requires rapid (within ~ 24 hours) ToO follow-up with a moderate-resolution spectrograph, ideally with a very broad wavelength (like VLT/XShooter or the planned Gemini "Gen 4#3" instrument). Fiber-sharing programs within larger multi-object spectrograph surveys would also facilitate rapid ToO follow-up, especially for bright transient events.

# Summary Tables

## Table 7.1. Needed Capabilities

| | Infrastructure | < 3m | 4–6.5m | 8m | 25m |
|---|---|---|---|---|---|
| **IGM/CGM** | | | Wide-field imager with u-band sensitivity (if field not in DDF) | **0.3–1.0μm R ~ 5000 multi-object spectrograph** <br><br> *0.3–1μm R ~ 50,000 single-object spectrograph* | **0.3–1.0μm R ~ 5000 multi-object spectrograph** <br><br> *0.3–1μm R ~ 50,000 single-object spectrograph* |
| **Lensed Galaxies** | Techniques to find lenses. Perhaps production of stacks containing best seeing portion of data. Or fit models to frames (Tractor) and examine residuals. | | | **optical-NIR (0.36–2.2μm) single-slit spectrograph with R ~ 3000–5000. AO optional** <br><br> **NIR MCAO IFU, small FOV (3–10 arsec), R = 4000–5000** | **NIR MCAO IFU, small FOV (3–10 arsec), R = 4000–5000** |
| **Galaxy Environments** | Techniques for hierarchical Bayesian analysis of low S/N spectra to constrain highly dimensional parameter space <br><br> Techniques for cross-correlation of spectroscopy and photometry | | | **0.36–1.3μm highly multiplexed spectrograph with R ~ 3000–5000** | |
| **SMBH Demographics** | Event brokers to identify AGN—both from standard variability and rare events. Including support for developing them and running them. | | **R ~ 1000–5000 optical spectrograph. FOV of a > 1 degree diameter is best. High multiplexing.[10]** | | |

Entries in boldface type indicate that the capability is **Priority 1 (critical)**.
Roman type indicates Priority 2 (very important).
Italic type indicates *Priority 3 (important)*.

---

[10] This science can also be done with an 8–10m telescope.

## Table 7.2. Resource Demand

All conversions from hours on target to number of nights assume 30% weather losses and 8 hours per night.

| | Infrastructure | < 3m | 4–6.5m | 8m | 25m |
|---|---|---|---|---|---|
| **IGM/CGM** | | | | **IGM: 20 nights Keck/LRIS program**[11]<br><br>*High-resolution spectroscopy: ~ 3 hours per object.*[12] | **IGM: 576 hours on target. 120-night TMT/WFOS program**[13]<br><br>**CGM: 648 hours on target. 140-night TMT/WFOS program**[14]<br><br>*High-resolution spectroscopy: ~ 50–150 hours per object*[15] |
| **Lensed Galaxies** | | | | **Screening: 540 hours on target. 130-night program.**[16]<br><br>**Characterization: 1080 hours on target (not counting sky). 460-night Gemini/NIFS program.**[17] | **Characterization: 1080 hours on target (not counting sky). 460-night TMT/IRIS.**[18] |
| **Galaxy Environments** | | | | **3.2 million fiber hours. 330 nights Subaru/PFS program.**[19] **Survey area is 4 deg$^2$ and can be covered in multiple fields.** | |
| **SMBH Demographics** | | | Daily cadence for ~1 year | | |

---

[11] Results in a reduced spatial resolution of the tomography map (4Mpc vs. 1Mpc with TMT/WFOS).

[12] This assumes R < 19 sources and is based on Keck/HIRES sensitivities.

[13] The conversion from hours on target to number of nights assumes 30% overheads.

[14] The conversion from hours on target to number of nights assumes 30% overheads.

[15] For S/N = 1000 spectra. Exposure time depends on source brightness.

[16] Assumes 50% overheads. This number is directly proportional to the contamination of the lens candidate samples, so improvement in techniques for finding lenses will lower this request.

[17] Assumes NIFS field-of-view, which requires sky offset exposures and 30% additional overheads. Increasing NIF field-of-view to allow dithering within IFU would cut total time by a factor of 2.

[18] This program can either achieve much higher S/N at same magnitude or same S/N at fainter magnitude. Baseline estimate is for 1-hour exposures.

[19] Assumes field of view and multiplexing of PFS/Subaru with 80% fiber placement efficiency and 20% observational overheads.

| | Infrastructure | < 3m | 4–6.5m | 8m | 25m |
|---|---|---|---|---|---|
| | | | with 1-hour exposures per epoch. 100 fibers/deg$^2$ needed per visit. | | |
| **Total** | | | 87,500 fiber-hours | ~ 2.7 years | ~ 3.1 years |

Entries in boldface type indicate that the capability is **Priority 1 (critical)**.
Roman type indicates Priority 2 (very important).
Italic type indicates *Priority 3 (important)*.

# Chapter 8: Facilitating Cosmology Measurements from LSST

*Jeffrey Newman (study lead, University of Pittsburgh), Adam Bolton (NOAO), Will Dawson ((Lawrence Livermore National Laboratory), Mark Dickinson (NOAO), Ryan Foley (UCSC), Eric Gawiser (The State University of New Jersey, Rutgers), Elise Jennings (Fermilab), Eric Linder (Lawrence Berkeley National Laboratory), Rachel Mandelbaum (Carnegie Mellon University), Phil Marshall (SLAC National Accelerator Laboratory), Tom Matheson (NOAO), Chad Schafer (Carnegie Mellon University), Sam Schmidt (UC Davis), Anja von der Linden (SUNY), Ben Weiner (Steward Observatory)*

### Executive Summary

Almost all LSST studies of cosmology can be enhanced by the addition of spectroscopic information. The most critical needs, as they affect essentially all cosmological probes (but particularly weak lensing, large-scale structure, and galaxy cluster studies), are training of photometric redshifts (i.e., obtaining spectroscopic samples to refine photo-$z$ algorithms) and photo-$z$ calibration (i.e., characterizing the actual biases and errors of photo-$z$ algorithms). The former will require a wide-field (>20 arcmin, with > 1 degree preferred), highly multiplexed (>2000×), medium-resolution ($R > 4000$ in the red), broad-wavelength coverage (0.4–1.0$\mu$m minimum, 0.3–1.5$\mu$m preferred) multi-object spectrograph on an 8m-class telescope. A minimum of 1.1 years of observing time would be necessary for this campaign (assuming a Subaru/PFS-like instrument; other planned spectrographs for 4–8m telescopes would take longer). Calibration of photometric redshifts via cross-correlation techniques will require overlap of a DESI-like galaxy and quasar survey with the LSST footprint, which is already planned to occur. A number of cosmological probes can utilize such a spectrograph (or, in some cases, instruments with smaller fields of view but otherwise similar capabilities) to mitigate potential systematics in cosmological measurements or to provide new constraints on modified gravity theories; these activities will provide very important contributions to cosmological studies taken as a whole.

For efforts to constrain cosmology using strong gravitational lensing, adaptive optics imaging and IFU spectroscopy on 8–30m telescopes will be critical (and hence very important for cosmology taken as a whole). Instrumental requirements include an effective resolution of ~0.1 arcsec FWHM, field of view of 4 arcsec diameter, and (for spectroscopy) $R \sim 4000-5000$ over a wavelength range of 1.0–2.2$\mu$m. Total time requirements are roughly 30 hours for imaging and 100 hours for spectroscopy (split between 8–10m telescopes and GSMTs according to brightness) to cover a sample of 100 particularly high-quality strong lens systems. Supernova cosmology will be critically dependent upon spectroscopy of thousands of near-peak supernovae to investigate supernova physics and constrain the properties of supernovae that are not associated with a visible host galaxy. Additionally, spectroscopic redshift measurements for hosts of supernovae whose spectra were not obtained before they faded can greatly enhance the size of the samples used for constraining cosmology, providing very important contributions to supernova cosmology. Direct supernova spectroscopy will require broad wavelength coverage (0.3–1.0$\mu$m minimum, ~0.3–2.5 $\mu$m preferred), modest-resolution ($R \gtrsim 100$), high-efficiency single-object spectrographs on 4m, 8–10m, and GSMT telescopes, with total time requirements over the life of LSST of 300–900 telescope-nights (split roughly 20% / 60% / 20% between 4m, 8–10m, and GSMT telescopes). Supernova host spectroscopy is most efficiently conducted with very large field, highly multiplexed spectrographs similar to those required for photometric redshift training and calibration. Redshifts can be obtained for the majority of supernovae in the LSST "deep drilling" high-cadence fields with ~ 1.5 nights of observations per year per field on a DESI-like spectrograph on a 4m telescope.

## Introduction

A key driver for LSST is to explain the accelerating expansion of the Universe. The primary possibilities are that there is a new, unknown component that dominates the energy density in the Universe today (commonly labeled "dark energy") or else that Einstein's theory of General Relativity (GR) fails to provide an accurate description of the action of gravity on large scales (a class of models generally referred to as "modified gravity" theories). Many of the planned cosmological measurements from LSST can also be used to constrain neutrino masses and dark matter properties. **By strengthening LSST cosmological constraints and mitigating potential systematics, the work described in this chapter will help LSST to address a key objective identified in the** *New Worlds, New Horizons* **report, studying the Physics of the Universe.**

The constraining power of almost all LSST probes of cosmology (as well as other extragalactic work with LSST) will be greatly strengthened by the addition of spectroscopic redshift measurements for training photometric redshift algorithms. Obtaining such samples requires spectrographs that maximize multiplexing, areal coverage, wavelength range, resolution, and telescope aperture, as described below. Studies of cosmology using strong gravitational lensing also requires adaptive optics IFU observations of the highest-priority lens systems in order to obtain precision source positions and tighten constraints; instruments currently available on 8–10m telescopes or planned for ELTs are well suited for this work.

## Science Case 1: Multi-object Spectroscopy for Training and Calibrating Photometric Redshifts

*Jeffrey Newman, Will Dawson, Elise Jennings, Anja von der Linden, Rachel Mandelbaum, and Samuel Schmidt*

LSST dark energy constraints, as well as almost all LSST extragalactic science, will be critically dependent upon photometric redshifts (a.k.a. photo-$z$'s), i.e., estimates of the redshifts of objects based only on flux information obtained through broad filters. In this section, we in prin describe the utilization of spectroscopy for photometric redshifts for two separate purposes:

- *Training, that is, making algorithms more effective at predicting the actual redshifts of galaxies, reducing random errors. Essentially,* **the goal of training is to minimize all moments of the distribution of differences between photometric redshift and true redshift,** *rendering the correspondence as tight as possible, and* *hence maximizing the three-dimensional information in a sample; and*

- *Calibration, the determination of the actual bias and scatter in redshifts produced by a given algorithm (for most purposes, this reduces to the problem of determining the actual redshift distribution for samples selected based on some set of properties). Essentially, **the goal of calibration is to determine with high accuracy the moments of the distribution of true redshifts of the samples used for a given study**.*

Different datasets will be needed for each of these purposes. However, as will be described below, the same instrumental capabilities—and often the same datasets—needed for photometric redshift training or calibration can also contribute to LSST cosmology in a variety of other ways; we will describe some of these applications at the end of this section.

## Science Goals

**Training:** Photo-$z$ methods generally use samples of objects with known $z$ to develop or refine algorithms, and hence to reduce the random errors on individual photometric redshift estimates. This will result in smaller errors on cosmological parameters of interest and will enable analyses in narrower redshift bins. Larger and more complete training sets result in smaller RMS errors in photo-$z$ estimates, increasing LSST's constraining power. With a perfect training set of galaxy redshifts down to the magnitude limits of science samples (necessary as the range of galaxy SEDs has been found to vary with luminosity at both $z = 0$ and $z = 1$), we could achieve system-limited performance; this would improve the Dark Energy Task Force figure of merit from LSST lensing + BAO studies by ~25%, with greater impact in other areas (e.g., studies of galaxy clusters). **Better photometric redshift training will improve almost all LSST extragalactic science, and hence addresses a wide variety of decadal science goals.** To enable this, we need secure spectroscopic redshifts for as wide a range of galaxies as possible down to the $i = 25.3$ magnitude limit of the LSST weak lensing "gold sample" (LSST Science Collaborations 2009), at minimum, or deeper (reaching the limits of LSS samples) if possible.

**Calibration:** Similarly, secure spectroscopic redshifts are needed for *calibration*, i.e., the empirical determination of photo-$z$ bias and scatter. **Inadequate calibration will lead to systematic errors in almost all extragalactic science cases with LSST and hence many decadal science priorities**. Extremely high completeness (> 99.9%) in the spectroscopic samples used for training would enable LSST calibration requirements to be met directly. However, existing deep redshift samples lack secure redshifts for a systematic 20%–60% of their targets; it is therefore quite likely that future deep redshift samples will not solve the calibration problem.

Instead, we can use cross-correlation methods to calibrate photo-$z$'s. These techniques correlate the positions on the sky of objects with known $z$ with the positions of the galaxies whose redshift distribution we aim to characterize. We can exploit the fact that bright galaxies trace the same underlying dark matter distribution as fainter objects, enabling the determination of the $z$ distribution for purely photometric samples with high accuracy using spectroscopy of only the most luminous objects at a given redshift. Fundamentally, photo-$z$ calibration via cross-correlations requires redshifts for large numbers (> 100,000) of objects over a wide area (> 100 sq. deg.) of spatially overlapping sky, spanning the full redshift range of LSST targets of interest.

## Technical Description

**Training:** A previous white paper on *Spectroscopic Needs for Imaging Dark Energy Experiments* (Newman et al. 2015) has explored in detail the minimum characteristics a photometric redshift training set should have. We summarize those conclusions here. In short, we require:

**Spectroscopy of at least 30,000 objects down to the survey magnitude limits**, in order to characterize both the core of the photo-*z*/spectroscopic-z relationship and outliers (cf. Ma et al. 2006; Bernstein & Huterer 2009; and Hearin et al. 2010); this will require large exposure times on large telescopes.

**High multiplexing**, as obtaining large numbers of spectra down to faint magnitudes will be infeasible otherwise.

**Coverage of as broad a wavelength range as possible**, in order to cover multiple spectral features, which is necessary for getting the required **highly secure** redshifts. At minimum, spectra should cover from ~4000 to ~10,000 Angstroms, but coverage from 0.3 to 1.5μm would be advantageous.

**Moderately high resolution ($R \gtrsim 4000$) at the red end**, critical as it enables secure redshifts to be obtained from the [OII] 3727 Angstrom doublet alone. This resolution also enables sensitive spectroscopy over the ~ 90% of the spectrum that lies between the skylines (cf. Newman et al. 2013).

**Field diameters $\gtrsim 20$ arcmin**, needed to span multiple correlation lengths to enable accurate clustering measurements. Larger (> 1 deg.) fields would be even better, particularly at low redshifts.

And finally, **coverage of as many fields as possible (~ 15 minimum)**, in order to minimize the impact of sample/cosmic variance.

**Calibration:** As described in Newman et al. (2015), cross-correlation calibration of photometric redshifts for LSST should require spectroscopy of a minimum of ~ $5 \times 10^4$ objects total over multiple independent > 100 square degree fields, with coverage of the full redshift range of those objects whose photometric redshifts will be calibrated (for LSST dark energy science, this is essentially $0 < z < 3$).

## Needed Capabilities and Estimate of Demand

**Training:** In principle, a number of spectrographs currently in existence or being planned have sufficient wavelength coverage and spectral resolution to obtain secure redshifts over a wide range of galaxy properties for photo-z training. However, the time required will depend on the instrumental and telescope characteristics. If sky coverage is low, the limiting factor will be how many tilings of the sky will be needed to cover > 15 fields that are 20 arcmin in diameter. If multiplexing is low, the limiting factor will be how many tilings are needed to reach 30,000 spectra. Finally, if both of these factors are high enough that each field need only be observed once, the limiting factor will be how much exposure time it takes the telescope to achieve the desired depth. Formulae for exposure timescales in these scenarios are given in Newman et al. (2015). Roughly one hundred hours of Keck/DEIMOS exposure time would be sufficient to achieve the same signal-to-noise for i = 25.3 objects that DEEP2 obtains at i ~ 22.5; at that magnitude, DEEP2 obtained secure redshifts for ~75% of targets. A spectrograph with broader wavelength range or higher spectral resolution would be expected to do at least as well as this at equivalent signal-to-noise.

Newman et al. (2015) tabulates the properties of a variety of current and upcoming spectrographs and estimates the total time they would require to complete the proposed survey. It would take more than 10 years with Keck/DEIMOS, 5 years with Mayall/DESI, ~ 1.8 years with TMT/WFOS, just over 1 year with Subaru/PFS, or as little as 4–5 months on GMT or E-ELT (depending on the final characteristics of instruments, whose design is still in flux). If less telescope time is available, it will be necessary to either allow spectroscopic redshift failure rates to increase or to reduce the depth of the sample; it is likely that the former would have smaller impact on photometric redshift training. By design, this training sample would also be sufficient to meet LSST calibration requirements **if** spectroscopic redshift failure rates of order ~ 0.1% could be achieved. However, based on past experience (with 20–60% failure rates), we expect to need less direct methods for calibrating photometric redshifts.

We note that the results of this survey—a set of galaxies spanning the full range of galaxy properties down to the LSST magnitude limit with a maximal amount of spectroscopic information—will enable a wide variety of galaxy evolution science going well beyond just the training of photometric redshifts. Planned surveys (e.g., PFS/Sumire) will also overlap with the desired samples, but not approach the required depths for high redshift success at the faint end of the LSST weak lensing sample. A number of applications for such a sample are discussed in Chapter 7, the sample described in that chapter has considerable overlap, but (if IR spectral coverage is available) somewhat shorter estimated exposure times. Ideally, this spectroscopy would occur early in the lifetime of LSST, but training redshifts will be useful whenever they are obtained. Photometric redshift training spectroscopy will be of critical importance for cosmology studies with LSST.

**Calibration:** LSST photometric redshift calibration requirements should be met by the overlap between LSST and planned baryon acoustic oscillation experiments. For instance, DESI should obtain redshifts of $\gtrsim 30$ million galaxies and QSOs over the redshift range $0 < z < 4$ over more than 14,000 square degrees of sky (Levi et al. 2013). It is expected to have > 4000 square degrees of overlap with LSST.

However, the DESI survey will cover only the northern portion of the LSST sky. Photometric redshift performance may not be the same there as elsewhere in the LSST footprint (e.g., due to airmass differences), which could yield a miscalibration when applied to LSST as a whole. This risk can be mitigated by DESI-like spectroscopy in the south. There are plans to conduct a DESI-like survey with the 4MOST instrument, which would fulfill this need well. If this does not happen, it would be extremely valuable to have access to a DESI-like spectrograph (with wide field of view, ~ 5000x multiplexing, full optical coverage, sufficient resolution to split [OII], and a ~ 4m-diameter telescope aperture) in the Southern Hemisphere. With such an instrument, covering the non-DESI LSST footprint would take comparable observing time to the original DESI survey. Access to DESI-like data in the south will be of critical importance for cosmology studies with LSST.

## Other Applications of the Required Instrumentation

The same instrumentation needed for a photometric redshift training survey would be well suited to address a number of other important issues affecting cosmological measurements with LSST. We describe a key example here.

**Intrinsic Alignment Studies:** Some of the strongest LSST constraints on dark energy are expected to come from measurements of the apparent shearing of galaxy images by weak gravitational lensing. Characterizing and mitigating systematic uncertainties will be critical to ensure that LSST weak lensing measurements are not systematics dominated. The photometric redshift training and calibration datasets will address one important systematic for weak lensing, but the multi-object spectroscopy can also help characterize and constrain another systematic: intrinsic alignments (IA) of galaxy shapes with the cosmic web, which are a contaminant to weak lensing measurements (Joachimi et al. 2015).

Existing methods of mitigating this systematic all have limitations, making it important also to explore intrinsic alignment effects directly, using redshifts to determine which galaxies are in physical proximity to each other. A direct measurement of IA would require a spectroscopic or spectro-photometric dataset that covers substantial contiguous areas ($\lesssim 1$ deg$^2$) while sampling a *representative* galaxy sample of tens of thousands of galaxies at minimum. Larger-field spectrographs such as Subaru/PFS could provide the necessary areal coverage and sampling as part of a photo-$z$ training survey. If the photo-$z$ training survey covers smaller fields, supplemental spectroscopy with the same sort of instrument required for training work could be necessary.

Because high redshift precision is not needed, an alternative approach would be to supplement LSST photo-$z$'s with many-band narrowband imaging or low-resolution spectroscopy; however, high signal-to-noise at faint magnitudes would be necessary, requiring new instrumentation on large telescopes. Another alternative would be to constrain IA using cross-correlations with spectroscopic galaxies instead of employing autocorrelation measurements (cf. Blazek et al. 2012 and Chisari et al. 2014), which could utilize the same data needed for the photo-$z$ cross-correlation calibration method. Having both photo-$z$ training and calibration sets available, and making sure they are well-optimized for IA purposes, will provide us with multiple potential routes to developing and testing IA models, maximizing the chances of success. Constraining IA models will be of critical importance for weak lensing studies with LSST; given lensing's strong contributions to the LSST figure of merit, that makes them critical for cosmology considered as a whole as well.

## Applications Requiring Smaller Fields of View

Although the most urgent cosmological needs for multi-object spectroscopy require wide fields of view (of order 0.5–1 degree) to span the typical scale lengths of large-scale structure ($\sim 5-10 h^{-1}$ Mpc comoving) multiple times, certain science cases require a dense placement of spectra in smaller $\sim 5-10$ arcmin fields of view (though other requirements are similar to those for wide-field multi-object spectroscopy). Because the relevant galaxies are relatively tightly packed on the sky and multiplex requirements are not high, existing or planned slit spectrographs such as Keck/LRIS, Keck/DEIMOS, Gemini/GMOS, and TMT/WFOS may be suitable. If fibers can be packed tightly enough, a wide-field multi-object spectrograph could also contribute to these activities, but may be less optimal. We discuss a few example applications of such spectroscopy below. These activities will generally be very important for LSST cosmology considered as a whole.

**Galaxy Cluster Studies:** Cosmological measurements based upon the cluster mass function may be one of the most powerful probes of cosmology in the LSST era, if cluster masses can be measured with sufficient accuracy and precision (Dodelson et al. 2016; Krause & Eifler, 2016). Galaxy clusters also provide avenues for probing gravity as well as the properties of dark matter. Multi-object spectroscopy plays a major role in all of these:

Photo-$z$ systematics specific to cluster fields could affect cluster weak lensing analyses, which anchor the absolute cluster mass calibration (Applegate et al. 2014). Galaxies in a cluster field are more likely to be at the cluster redshift than would be expected based upon posterior probability distributions calculated without any reference to position. Spectroscopic surveys of $\sim 500$ galaxies per cluster selected by their photo-$z$'s to be behind clusters at a range of redshifts can be used to characterize the accuracy of $p(z)$ distributions in cluster fields. FOVs of $\gtrsim 20$ arcmin are required for clusters at $z_{Cl} \sim 0.2$, but smaller fields are sufficient at higher redshifts, and optical wavelength coverage is sufficient for clusters at $z \lesssim 1$, so instrumental requirements are not as challenging as for photo-$z$ training. Because the primary goal is validation of methods of incorporating cluster presence in redshift distributions, this spectroscopy does not require highly complete redshift measurements over the entire sample if the data suffice to rule out the specific cluster redshift, allowing shorter exposure times than for photo-$z$ training.

The combination of weak lensing with dynamical mass probes (e.g., measurements of galaxy velocities within clusters) can be used to test non-GR theories of gravity, which provide an alternative explanation for the cosmic acceleration commonly attributed to dark energy. The same spectroscopic observations used to characterize photo-$z$ around clusters can be used to measure infall velocities around clusters and hence provide constraints on the origin of cosmic acceleration, so long as the FOVs extend to $\gtrsim 2h^{-1}$Mpc from the cluster centers (corresponding to $\sim 7.5$ arcmin at $z = 0.3$).

The structure and evolution of clusters can be sensitive to the properties of dark matter (though see Peter et al. 2013 for challenges). For example, the effects of self-interacting dark matter can leave signatures in post-merger clusters. LSST will identify hundreds to thousands of cluster mergers, but kinematic information is critical for reconstructing merger histories. Instruments with FOVs of $\sim 5-15$ arcmin and dense multiplexing capabilities are well-matched to this application. Observational efficiency will be maximized if objects within $\sim 5$ arcseconds can be simultaneously targeted.

**Strong Lens Environment Characterization:** Achieving stage IV accuracy in constraining cosmology with time delay lenses (see below) requires a model for the mass in the immediate environment of the lens and along the line of sight. LSST will provide photometric redshifts and stellar mass estimates for all galaxies in the fields of each cosmographic lens. However, for some lens fields, photometric information may not be enough; multi-object spectroscopic data can significantly improve the accuracy of the models.

Predicting the external convergence for an individual lens to few-percent uncertainty requires a full model for the ~ 30 galaxies that will contribute the most to this signal, making them prime targets for spectroscopy (McCully et al. 2016). Velocity dispersion measurements for the lens galaxies can help break the lens radial profile degeneracy, a key source of systematic uncertainty; such measurements of other foreground galaxies enables more effective forward modeling, reducing the residual bias in external convergence estimates by a factor of two (Collett et al. 2013).

The technical requirements for this observing program are less stringent than those for photometric redshift training. Hence, the same instruments could be used for this work (if fibers can be packed densely enough):

- *Spectral coverage: as broad as possible, preferably from the u band through the near-infrared*
- *Spectral resolution: can be low, $R$ of a few hundred to a few thousand (with $R >$ 2000 necessary for velocity dispersion measurements)*
- *Field of view: at least 5 arcmin, preferably more*
- *Number of fields: at least 100, with ~ 10−100 targets per field*

Observational costs may be reduced by embedding these observations in photo-$z$ training and/or galaxy evolution surveys. However, because of the association of strong lensing with massive galaxies in a limited redshift range, requiring the presence of strong lens systems would bias redshift distributions and cause surveys not to be a fair sampling of the Universe.

Additional spectroscopy can be useful for characterizing the general effects of weak lensing on strong lensing measurements; this has similar requirements as modeling intrinsic alignments and can find strong synergy with photo-$z$ training surveys.

**Ambiguous blends:** Roughly 14% of the objects detected in the LSST survey will be ambiguous blends of two or more galaxies (Dawson et al. 2016). These blends are a potential source of systematics for photometric redshift algorithms, which assume that all objects are individual galaxies. The impact of these blends can be predicted given knowledge of galaxy redshift distributions, colors, sizes, and clustering; this should be provided by the photometric redshift training survey. One could use overlapping space-based imaging, ideally in both field and cluster environments, to identify a sample of ambiguous blends and measure their redshifts (either in parallel to or after the photo-$z$ training survey). Because the objects used to test algorithms need not be a representative galaxy sample, this work could be done with smaller field-of-view instruments than those needed for photometric redshift training; a number of current (e.g., Keck/DEIMOS) or planned (e.g., TMT/WFMOS) spectrographs could fulfill this need.

# Science Case 2: AO Imaging and/or IFU Spectroscopy for Strong Lensing Cosmography

*Phil Marshall, Tommaso Treu, Curtis McCully, and Eric Linder*

Our ability to do cosmography with either time delay lenses or multiple source plane "compound" lenses is dependent on follow-up observations to constrain the mass models: without these data, strong lenses will not be able to be used for precision cosmology. In this section we outline what will be required in order for us to be able to exploit the LSST strong lens sample.

## Science Goals

The primary route to cosmology from strong lensing is time delays in galaxy-scale lensed quasars and supernovae. Galaxy scale compound lenses (i.e., systems with two sources at different redshifts) have also been suggested. We expect to be able to compile samples of several hundred lensed AGN and lensed SN systems with accurately measured LSST time delays (Liao et al. 2015) and dozens of compound lens systems (Collett 2015).[20]

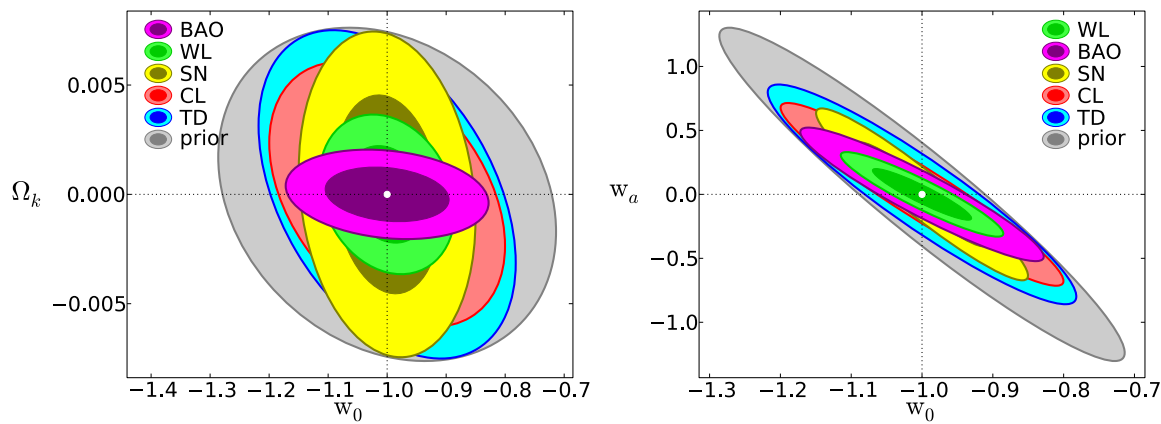

Figure 8.1. LSST time delay lens cosmological parameters forecasts ("TD," blue ellipses) compared to forecasts from the Dark Energy Task Force (Albrecht et al. 2006) for four other Stave IV probes of cosmology: Baryon Acoustic Oscillations ("BAO"); Weak Lensing ("WL"); Type 1a Supernovae ("SN"); and Cluster counts ("CL"). The parameters considered are the Dark Energy equation of state parameter $\omega\alpha$ evaluated $\omega = P/\rho$ today ($\omega o$); its evolution with the scale factor of the Universe in a linear model, i.e., the parameter for the model $\omega(a) = +\omega o\ x\ (1 - \alpha)$; and the curvature of the Universe expressed as a fraction of its critical density, $\Omega k$. The prior probability distribution assumed for cosmological parameters is shown in grey. A sample of 100 lenses is assumed, each one providing a 5% accurate time delay distance. (Figure reproduced from Coe & Moustakas 2009)

---

[20] The LSST strong lens sample will be compiled semi-automatically, using algorithms developed and implemented by the LSST DESC that are run on the DM Level 1 and 2 data products, and whose products are then assessed—including visually—by the DESC analysis team. Some of the technology involved in this process could be developed and operated in collaboration with the Galaxies and Strong Lenses collaborations.

Figure 8.1, reproduced from Coe & Moustakas (2009) shows approximate forecast cosmological parameter constraints from a sample of 100 lenses, where each provides a time delay distance accurate to 5%. Two lenses to date have been shown to provide 6% distance precision (Suyu et al. 2013): this was only achievable with a combination of deep, high-resolution imaging from *HST* (to enable the Einstein rings to be modeled), and high-fidelity spectra from 10m-class telescopes yielding lens velocity dispersions that anchor the mass model. LSST data on their own will provide the opportunity for this analysis by providing a large sample of accurately measured time delays. However, without these follow-up data, the mass models can only be constrained to 20–30% accuracy; to enable this cosmological probe to reach useful limits, the additional observations described here are essential.

## Technical Description

To be useful as probes of cosmological distances, galaxy scale lenses need very well constrained mass models. These constraints will come from two types of targeted follow-up observations:[1]

1. **High-Resolution Einstein Rings Imaging** *due to the source AGN or SN host galaxy. Image quality of ~ 0.1" or better provides the Einstein ring constraints on the lens mass model density profile slope (via the arc thickness) that are needed to turn each of these systems into a 5% precision distance (Meng et al. 2015).*
2. **Spatially Resolved Spectroscopy of the Lens Galaxy,** *enabling measurement of the stellar velocity dispersion field to break the degeneracy between the predicted time delays and the lens mass density profile, calibrating each system to enable it to provide a 4% accurate distance.*

We now assess the technical requirements of each of these observations.
1. *High-Resolution Einstein Ring Imaging:*
   - *Spectral coverage and resolution: imaging in two or three bands is recommended, to enable clean lens galaxy subtraction.*
   - *Angular resolution: the higher the resolution the better, but at least 0.1 arcsec FWHM.*
   - *Depth: the host galaxies of the lensed sources are faint (i ~ 23–25). Brighter sources will be prioritized when compiling the follow-up sample, based on analysis of the survey images, and systems requiring exposure times of up to 1 hour would be considered, with fainter systems being discarded (cf. Meng et al. 2015).*
   - *Field of view: at least 4 arcsec, to capture the Einstein ring without dithering.*

2. *Spatially Resolved Spectroscopy of the Lens Galaxy:*
   - Spectral coverage and resolution: $R \approx 4000-5000$ over a wavelength range of 1.0–2.2 microns (to cover the Calcium triplet at rest frame at $\sim 8500-8700$ Angstroms, or CO at $1.5-1.6\mu m$).
   - Angular resolution: 0.1–0.2 arcsec, to resolve the lens galaxy well.
   - Depth: the lens galaxies will have brightness in the range $i \sim 19-22$. Again, systems requiring exposure times of an hour or less will be considered, with fainter systems being discarded.

- Field of view: at least 4 arcsec, to capture the lens galaxy within the Einstein ring without dithering.

The above requirements are derived from end-to-end simulations of the kind carried out by Meng et al. 2015, which will be refined before proposals to next-generation facilities are submitted.

## Needed Capabilities and Estimate of Demand

AO-assisted imaging and integral field spectroscopy on GSMTs would be best for providing the lens mass model constraints detailed in the previous section. We will need capabilities such as those that are currently available on Keck and that will be available on all three of GMT, TMT, and E-ELT (albeit with somewhat different technical specifications). OSIRIS on Keck is the current best option, even though the field of view is a bit too small and the current AO system at Keck has relatively low strehl at $1\mu$m. The Keck system will be upgraded, but TMT should be much better: IRIS on TMT would provide the capability we need. Similar data would be needed for the compound lenses. *JWST* could also contribute to this work, but is likely to be highly oversubscribed; utilizing ground-based resources will therefore be key.

The lightcurves will be accumulated by LSST over the lifetime of the survey, but we expect accurate cosmography to be possible after the first five years. Prior to this, the same facilities could be used to good effect improving the models of lenses found in shallower precursor surveys such as DES, KIDS, and HSC. We expect to have good time delays for about 30–40 systems before the LSST survey begins: these would be the targets for high-resolution follow-up before 2024, with an additional 60–70 targets coming from the LSST survey being pursued between 2024 and 2028 and beyond. The demand is therefore likely to be around 10–20 systems per year; we assume the higher value below, to help with planning.

While the targeted observations outlined below would be narrow field, they would enable a considerable amount of ancillary science, notably in the areas of dark matter substructure (from perturbations to the imaged rings) and AGN host galaxy structure. To summarize, the needs for LSST strong lensing studies are:

1. *High-Resolution Einstein Ring Imaging:*
   - *Proposed facilities: GSMTs for fainter rings, 10m-class telescopes for brighter ones*
   - *Observations needed: Targeted snapshot (e.g., 200–2000-second exposure time with TMT) imaging in the optical and near infrared*
   - *Total time required: ~ 30 hours, depending on balance between facilities (assuming 20 mins per system, on average), or ~ 6 hours per year.*

2. *Spatially-Resolved Spectroscopy of the Lens Galaxy:*
   - *Proposed facilities: GSMTs*
   - *Observations needed: IFU spectra in the near infrared.*
   - *Total time required: ~ 100 hours, depending on balance between facilities (assuming 60 mins per system, on average), or ~ 20 hours per year.*

Both the imaging and spectroscopy will be critical for LSST strong lensing cosmology and hence very important for LSST cosmology considered as a whole.

# Science Case 3: Wide Field and Single-Object Spectroscopy for Supernova Cosmology

*Ryan Foley, Tom Matheson, and Jeffrey Newman*

## Science Goals

Observations of Type Ia supernovae (SNe Ia) led to the discovery that the Universe's expansion is currently accelerating (Riess et al. 1998; Perlmutter et al. 1999). SNe Ia continue to be a mature and important cosmological tool (e.g., Suzuki et al. 2012; Betoule et al. 2014; Rest et al. 2014). Further observations of SNe Ia will be critical to improved understanding of the nature of dark energy, perhaps the most puzzling open problem in all of physics. In this section, we consider ways in which additional data can enable or enhance LSST supernova cosmology measurements; broader studies of the properties and physics of supernovae are discussed in Chapter 4.

LSST will detect and observe $\sim 10^6$ SNe Ia out to $z \approx 1$ (LSST Science Book). With these data, we will be able to measure precise distances and constrain cosmological parameters. However, dark energy constraints are not currently limited by statistics, and even $10^3$ SNeIa are more than necessary to reach the current systematic floor (Betoule et al. 2014; Scolnic et al. 2014). While LSST will certainly reduce some systematic uncertainties (such as those related to calibration), those related to astrophysics (differences in the SN observables, the unknown nature of dust, etc.) can also be addressed with the proper auxiliary information.

There are two approaches to using large samples of SNe Ia for cosmology. The first, which has been the standard for more than two decades, is to use a sample of spectroscopically confirmed SNe Ia. The second, which has only just begun to be used for cosmology (Campbell et al. 2013), is to use photometrically classified SNe Ia. The former guarantees that the sample is "pure," consisting of only genuine SNe Ia, while the latter is more "complete," but at almost certainly an equal or lower purity level. SN cosmology with LSST is expected to primarily use photometric samples.

It has been shown that knowing the redshift of a photometric SN significantly improves its classification and will also reduce distance scatter somewhat. Nonetheless, we can make a first pass at classification with only lightcurve information (with perhaps including a prior using photometric redshifts). The expected path to cosmology will likely require host-galaxy redshifts. While a subset of $\sim 10\%$ of the SNe will be "hostless" (having a host galaxy fainter than the detection limit of the reference image), most host galaxies could be targeted to obtain spectroscopic redshifts after the SN has faded, which can be done efficiently using multi-object spectrographs.

For the hostless SNe, on the other hand, we must generally obtain a redshift from the SN itself. Hence, if one desires an unbiased sample of all SNe Ia, obtaining spectra of hostless SNe is a necessity. Furthermore, spectroscopy of SNe themselves provides tests of photometric classification, improved distance precision (Bailey et al. 2009; Blondin et al. 2011; Foley & Kasen, 2011), and additional information about the explosion physics. Spectra of a subset of near-peak SNe Ia will therefore be necessary to perform the most precise cosmology analysis.

## Technical Description

We consider three different applications of spectroscopy for supernova cosmology separately: targeted spectroscopy of SN Ia hosts after supernovae have faded; targeting near-peak supernovae during the course of other wide-field spectroscopy; and single-object spectroscopy of individual supernovae, both to measure redshifts for hostless supernovae and to improve our understanding of supernova physics. All of this work will be dependent upon continued support for a transient broker to prioritize SNe Ia for follow-up observations (see Chapter 4); it will be important to be able to obtain samples of near-peak or historical SNe Ia on demand whenever spectroscopic resources are available.

**Wide-field Multi-object Spectroscopy:** Spectra of host galaxies are useful for all SNe Ia. For supernovae without spectroscopy, they can determine the redshift, but even for those of known redshift, spectra can provide information about star-formation rates or other host properties that can be used to reduce distance errors (e.g., Pan et al. 2014).

Most of the host galaxies of the supernovae found by LSST will be faint enough to require significant exposure times on large telescopes; observing a non-negligible fraction thus requires multi-object spectrographs. Extrapolating from SN Ia detection rates in the DES deep fields, we expect to discover roughly 100 SNe Ia $deg^{-2}$ $year^{-1}$ in each LSST deep drilling field. The SNe in the deep drilling fields will have much better lightcurves than those detected in the main LSST survey, and hence will be the most useful for constraining cosmology. In principle one could wait until the end of the survey to obtain SN host redshifts (when the density in deep drilling regions would be roughly 1000 $deg^{-2}$, or equivalently ~$10^4$ per field), but it would be impractical to wait that long.

Since the aim of SN host spectroscopy is to measure galaxy redshifts, the instrumental requirements are very similar to those for photometric redshift training spectroscopy, as described in the Technical Description Section of Science Case 1. Because the supernova host samples are dilute within fields that are ~ 10 square degrees each, a wide field of view (preferably >1 sq. deg.) is essential. Hence, as some LSST deep drilling fields are at too low declination for effective observations from the North, maximizing host galaxy samples will require a new wide-field multi-object spectrograph with capabilities like those described in the Technical Description Section of Science Case 1 in the South.

**High-Throughput, Wide-Wavelength Optical(/NIR) Spectroscopy on Large Telescopes:** Spectroscopy of SNe themselves will be critical to the success of the LSST SN Ia cosmology program. Most LSST SNe will be found near the magnitude limit of the search images, requiring a $\gtrsim$ 8m telescope to obtain high-S/N (~ 5–10 per resolution element) observations. The source density is low enough that high-throughput, "single-object" spectrographs are best suited for this work, unless other high-priority sources could fill the vast majority of available fibers (q.v. below).

Since SN features are relatively broad, a low-resolution (R > 300) spectrograph is adequate; it is most important to have broad wavelength coverage. This both aids in supernova type identification and makes comparisons between samples at different redshifts easier. Ideally, the spectrograph would cover all optical and NIR wavelengths from roughly 0.3 to 2.5 μm.

However, a spectrograph covering the full optical range (~ 0.3–1 μm) would be adequate. It will also be important to minimize contamination by second-order light, which can significantly distort SN spectra. The more telescopes have such spectrographs available, the greater will be the fraction of the LSST supernova sample that can be studied spectroscopically.

**Real-Time Fiber Allocation:** DES has been successful at allocating fibers to active transients during AAT observing runs (Yuan et al. 2015). This has yielded spectra of roughly as many SNe as all other DES spectroscopic programs combined. It would be valuable if LSST supernova spectroscopy could similarly piggyback on other spectroscopic campaigns on robotically positioned wide-field multi-object spectrographs.

## Needed Capabilities and Estimate of Demand

**Wide-field Multi-object Spectroscopy:** A reasonable strategy for supernova host spectroscopy would be to observe all available host galaxies in the LSST deep drilling fields roughly once a year. The density would then be roughly 100–300 $\deg^{-2}$ (some galaxies will not yield redshifts after a single pass, while other hosts will have multiple supernovae over the course of the survey but need observing only once). The maximum density would fill the available fibers on WHT/WEAVE or 4MOST, but would occupy only ~ 50% of fibers on DESI or ~ 20% on PFS. As a result, it would be most effective to combine multiple science programs, with a large fraction of fibers being reserved for SN host galaxies; this is currently being done successfully for bright supernovae found by DES using AAT/AAOmega (Yuan et al. 2015).

The majority of LSST supernova hosts will be brighter than $r = 24$, and almost all will be at $z < 1.6$. DESI or an equivalent spectrograph should be able to measure redshifts for the great majority of such objects in ~8 hours of observation time, hence requiring 1.5 nights per year to cover each LSST deep drilling field (the DESI field of view is modestly smaller than LSST's, but shifting field centers for each year's campaign should enable the vast majority of hosts to be targeted over time). Smaller field-of-view spectrographs are inefficient for this work as they would need to tile the field rather than covering all hosts simultaneously; it would take Subaru/PFS roughly the same amount of time to achieve an identical signal-to-noise on the same set of supernova hosts that DESI would cover, despite being on an 8m telescope instead of a 4m.

**High-Throughput, Wide-Wavelength Optical(/NIR) Spectroscopy on Large Telescopes:** Characterizing hostless supernovae and obtaining sufficient spectroscopy for detailed exploration of supernova physics should require ~ $3 \times 10^3 - 10^4$ objects in total. The typical exposure time with an efficient spectrograph on a telescope of suitable aperture for a given object should be ~ 30 minutes. Hence, total exposure times will be ~ 1500−5000 hours, or ~ 300−900 nights (including weather losses). Crudely, we expect ~ 20% of LSST supernovae of interest to be observable with a 4m telescope, ~ 60% to require an 8m, and ~ 20% to be at high-enough redshift that observations with a GSMT are strongly preferred. Hence, the total instrumental need corresponds to 60–180 4m nights, 180–540 8m nights, and 60–180 GSMT nights over the course of the 10-year LSST survey. To enable this, it will be important that every large-aperture telescope possible in the Southern Hemisphere should be outfitted with a multi-purpose instrument that can enable this work; existing examples of such spectrographs have been in extremely high demand, and that should only increase in the LSST era.

**Real-Time Fiber Allocation:** We can expect roughly 10–20 near-peak SNe Ia per square degree. Hence, it would be beneficial if whenever wide-field, robotically positioned, fiber-fed spectrographs are aimed at LSST fields, they could allocate ∼ 15 fibers deg$^{-2}$ to observations of active SNe, with targets to be identified shortly before the time of observation. In this case, the overall exposure time, etc., are set by the primary survey on which that SN spectroscopy is piggy-backing and cannot be estimated separately.

## Summary Tables

### Table 8.1. Needed Capabilities

| | Infrastructure | < 3m | 3–5m | 8m | 25m |
|---|---|---|---|---|---|
| **Photometric Redshift Training** | **Support for photo-z development personnel** | | **0.4–1μm minimum, 0.37–1.5 μm preferred; R > 4000–5000 at red end, 5000–20000x multiplexing, > 1 deg$^2$ FOV** , extreme exposure times OR | **0.4–1μm minimum, 0.37–1.5 μm preferred; R > 4000–5000 at red end, 25005000x multiplexing, ∼ 1 deg$^2$ FOV** OR | **0.4–1μm minimum, 0.37–1.5 μm preferred; R > 4000–5000 at red end, 500–1000x multiplexing, > ∼0.1 deg$^2$ FOV** |
| **Photometric Redshift Calibration** | Co-location of LSST data and DESI-like datasets | | **>500 sq. deg. of overlap with DESI or DESI-like surveys spanning full LSST footprint** | | |
| **Weak Lensing (inc. intrinsic alignment studies)** | | | **As for photometric-redshift training and/or calibration** | **As for photometric-redshift training and/or calibration** | As for photometric-redshift training and/or calibration |
| **Cluster studies: photo-z training and cross-checks, modified gravity and dark matter tests** | | | As for photometric-redshift training, but ∼ 500x multiplexing and ∼ 0.1 deg$^2$ FOV acceptable; dense packing of slits/fibers necessary | As for photometric-redshift training, but ∼ 500x multiplexing and ∼ 0.1 deg$^2$ FOV acceptable; dense packing of slits/fibers necessary | As for photometric-redshift training |
| **Strong lensing cosmography** | | Optical imaging to monitor time variation | Optical imaging to monitor time variation | **0.1" or better resolution imaging over 4" FOV; R ≈ 4000–5000** | **0.1" or better resolution imaging over 4" FOV; R ≈ 4000–5000** |

| | Infrastructure | < 3m | 3–5m | 8m | 25m |
|---|---|---|---|---|---|
| | | | | **spectroscopy over a wavelength range of 1.0–2.2 with 0.2" or better resolution and 4" FOV** | **spectroscopy over a wavelength range of 1.0–2.2 with 0.2" or better resolution and 4" FOV** |
| **Supernova studies via single-object spectroscopy** | **Transient brokers** | **High-throughput, broad-wavelength (~0.35–1µm minimum, 0.3–2.5µm goal) spectroscopy with R > 100** | **High-throughput, broad-wavelength (~0.35–1µm minimum, 0.3–2.5µm goal) spectroscopy with R > 100** | **High-throughput, broad-wavelength (~0.35–1µm minimum, 0.3–2.5µm goal) spectroscopy with R > 100** | **High-throughput, broad-wavelength (~0.35–1µm minimum, 0.3–2.5µm goal) spectroscopy with R > 100** |
| **Supernova studies via multi-object spectroscopy** | Ability to add SN targets to spectroscopy focused on other science in near real-time | | Multi-object spectrograph with broad wavelength coverage, wide field, and rapid redesign of observations | Multi-object spectrograph with broad wavelength coverage, wide field, and rapid redesign of observations | Multi-object spectrograph with broad wavelength coverage, wide field, and rapid redesign of observations |
| **Supernova host redshifts** | | | <0.4–1+µm, R > 4000–5000 at red end, ~5000x multiplexing, >1 deg$^2$ FOV | <0.4–1+µm, R > 4000–5000 at red end, 2500–5000x multiplexing, ~1 deg$^2$ FOV | |

Entries in boldface type indicate that the capability is **Priority 1 (critical).**
Roman type indicates Priority 2 (very important).
Italic type indicates *Priority 3 (important).*

## *Table 8.2. Resource Demand*

| | Infrastructure | < 3m | 3–5m | 8m | 25m |
|---|---|---|---|---|---|
| **Photometric Redshift Training** | | | **~5 years (inc. weather loss) with DESI-like spectrograph** OR | **~1 year (inc. weather loss) with PFS-like spectrograph** OR | **~5 months (inc. weather loss) with best-case instrumentation** |
| **Photometric Redshift Calibration** | | | **Currently planned DESI & 4MOST BAO surveys** | | |

| | Infrastructure | < 3m | 3–5m | 8m | 25m |
|---|---|---|---|---|---|
| **Weak Lensing (inc. intrinsic alignment studies)** | | | **See photo-z training & calibration (may require additional time)** | **See photo-z training & calibration (may require additional time)** | **See photo-z training & calibration (may require additional time)** |
| **Clusters: photo-z training, modified gravity and dark matter tests** | | | | ~100–1000 hours | ~100 hours |
| **Strong lensing cosmography** | | 0–4000 hours over 5 years | 0–4000 hours over 5 years | **~30 hours of imaging and ~100 hours of spectroscopy over 5 years (split between 8+ and 25+m telescopes)** | **~30 hours of imaging and ~100 hours of spectroscopy over 5 years (split between 8+ and 25+m telescopes)** |
| **Supernova studies via single-object spectroscopy** | | | **60–180 nights over 10 years (inc. weather losses)** | **180–540 nights over 10 years (inc. weather losses)** | **60–180 nights over 10 years (inc. weather losses)** |
| **Supernova studies via multi-object spectroscopy** | | | Set by observations being piggy-backed on | Set by observations being piggy-backed on | Set by observations being piggy-backed on |
| **Supernova host redshifts** | | | 15–30 nights per year per deep drilling field <span style="color:red">OR</span> | 15–30 nights per year per deep drilling field | |
| **Total** | | ~ 1 year | ~ 5 DESI-yrs + ~ 2 add'l years | ~ 1 PFS-yr + ~ 1.4 add'l years | ~ 5 months with ideal MOS + ~0.3 add'l year |

Entries in boldface type indicate that the capability is **Priority 1 (critical).**
Roman type indicates Priority 2 (very important).
Italic type indicates *Priority 3 (important)*.

# Chapter 9: Infrastructure for a Time Domain Follow-up System and the Evolution of Observing Paradigms

*Rachel A. Street (LCOGT), Steve Ridgway (NOAO), David Ciardi (IPAC, Caltech), Adam Bolton (NOAO), Tom Matheson (NOAO), Jay Elias (NOAO), Chad Schafer (Carnegie Mellon University), Erik Tollerud (STScI), Bryan Miller (Gemini Observatory)*

### Executive Summary

The volume of LSST transient alerts and the need to obtain follow-up observations rapidly (minutes–hours) will make both general-purpose and specialized alert brokers, capable of federating LSST's alert stream with additional data and classifying alerts where possible, indispensible. Building an alert broker that operates at the LSST scale and rate of alerts is a multi-disciplinary problem that encompasses astronomy and computer science. The development of target and observation manager software will add significant value to such brokers by enabling independent science teams to select targets from the alerts, conduct observations, and collate all available data, as well as (optionally) sharing this information with other teams. Rapid dissemination of such follow-up data would be facilitated by public data reduction pipelines for time domain follow-up observations. The full suite of software functionality required to maximize time domain science will evolve with time, and it will be necessary to continue conversations with the community to understand their needs and the division of tasks between alert broker and target/observation managers. We recommend stimulating that evolution by learning from trial systems developed for current surveys and associated follow-up programs. Efficient use of OIR facilities will further benefit from project teams sharing data and results, and from open-access triage programs to help classify targets. To enable LSST follow-up observations to be made on a range of timescales, we encourage observatories to optimize their infrastructure and time allocation processes in favor of time domain programs, including queue-scheduled modes, and to provide rapid, quick-look reductions of data products. Many of these findings will also improve static science output in the LSST era, as follow-up programs of large statistical samples of objects will become increasingly commonplace.

A number of practical implementation ideas have been raised, but as these will impact the policies and resources of a range of institutions, they will require significant further discussion, ideally at a dedicated workshop. We therefore make the following recommendations.

*Critical*
- Development of both general-purpose and specialized public alert broker(s) that will interact with LSST's Alert Service and complement LSST's limited alert filtering service
- Development of target and observation management software
- Increasing the availability of follow-up telescopes in queue-scheduling modes, spanning a range of apertures, instrumentation, and geographical locations
- Support for national-level data archives that include follow-up observations in addition to LSST data products

- A workshop aimed at developing a plan for making a suite of follow-up facilities accessible for real-time, large-volume, time domain observations
- Development and testing of software and hardware infrastructure that facilitates follow-up programs for LSST, using existing surveys as a proxy

*Very important*
- Development of general-purpose time domain telescope scheduling software
- Development of data reduction pipelines for key follow-up facilities. We note the premature state of analysis software for spectropolarimetry will be problematic for stellar science in particular.
- Funding for triage observing programs on suitable facilities
- Study of the viability of a follow-up survey capable of filling in gaps in LSST lightcurves

*Important*
- Coordination between time domain follow-up teams and the sharing of data obtained in response to LSST alerts, incorporating the capability within LSST software resources

## Science Goals

LSST will deliver transient alerts with initial rates expected to be 10,000 per 30-second visit on the sky. Although the number of these alerts that represent truly unknown phenomenon will substantially decrease once the variable sky is thoroughly mapped, LSST will continue delivering true transient discoveries at unprecedented rates and to r ~ 24.7.

These discoveries will be delivered as alerts, each with a package of LSST-recorded history—if any. Many discoveries will require additional observations, not expected from LSST, for the purposes of identification and prioritization. Some targets will require time-critical follow-up—only rapid processing of alerts and assignment of observing assets can provide timely responses.

This situation is analogous to smaller-scale projects that already exist to respond to alerts from current surveys, with targets ranging from microlensing to supernovae to Near Earth Objects (NEOs). We can therefore draw on their experience and evaluate the lessons learned, the requirements, and the tools available to determine their applicability in the LSST era.

Whereas other chapters in the report detail the instrumentation required for a range of science, the mere existence of suitable facilities is not enough—it is also necessary to ensure they can be brought to bear on well-chosen targets in a timely manner. The scale and diversity of the LSST alert feed makes this a non-trivial task. In this chapter, we identify areas where investment in the observing infrastructure is needed to mount effective and efficient follow-up programs.

A notional model for alert handling is described in Figure 9.1. Alerts from LSST (and other time domain surveys) will be handled by a function widely described in the community as a "broker."[21] A broker provides to subscribers the subset of alerts of most relevance to their

---

[21] In this chapter, we refer to the alerts issued by LSST as the LSST Alert Service. The Service should interface with an ecosystem of both general-purpose and specialized community brokers, which could serve either the public or a private community. LSST itself will only provide a limited service to filter its

science program, including relevant ancillary data. Brokers do this by associating alerts with other data, and by rapid and fully automated filtering according to numerous (possibly user-provided) algorithms and criteria. Brokers populate a comprehensive database (needed for external use as well as for future broker operation) with their classifications, as well as issuing secondary alerts to subscribers, according to their preferences. These subscribers may be individual scientists, teams, or additional brokers working at a level of finer detail. They may also be Target/Observation Managers, which are prioritizing targets, allocating follow-up to available resources, and tracking assignments and progress.

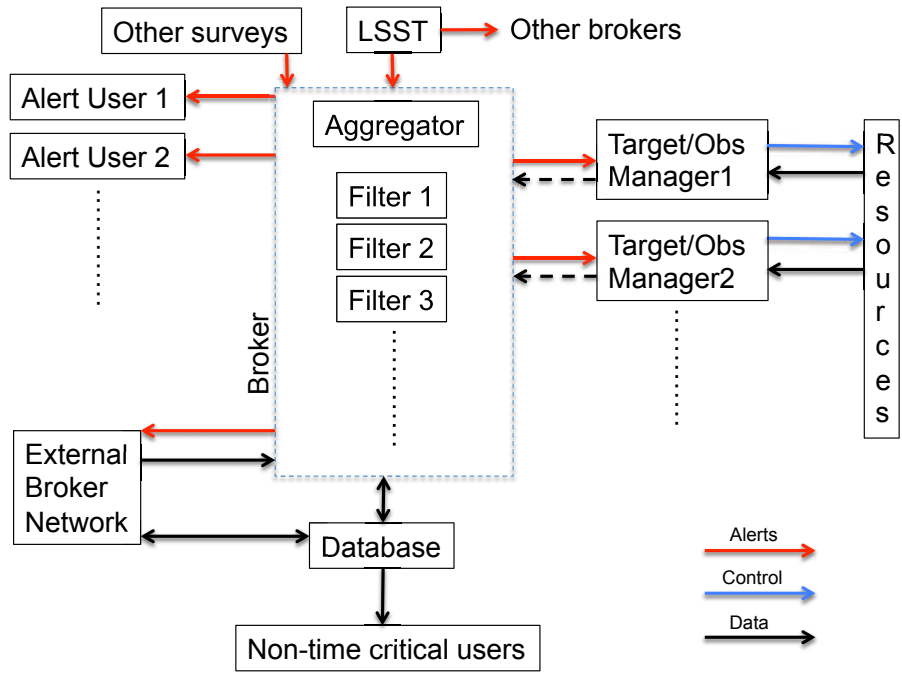

*Figure 9.1. Schematic diagram of a notional model for alert handling and data flow, recognizing multiple sources of alerts, a network of brokers (both general-purpose and specialized, public and private), PI/team users, and multiple options for coordinated follow-up, rapid or otherwise. The dashed box represents the core, essential community broker services—rapid response to incoming alerts, aggregation with other information, filtering (which implicitly depends on classifying, though perhaps at a high level), and retransmission of filtered alerts to subscribers. Note that alerts also contain associated data. One or more of these community brokers should exist to receive and process alerts from LSST's Alert Service. Dashed arrows indicate data may be returned from follow-up programs at the user's discretion.*

In the remainder of this chapter, we will discuss in detail each element of the alert response process. Necessary time domain infrastructure often receives less attention than more traditional means of support for astrophysics, such as telescopes and instrumentation. We stress that this rapid-response functionality is equally important. We also identify some aspects of common own alert stream. The "transient broker" discussed in Chapter 4, would be one such community broker, and more than one such service may exist. Community brokers could also issue alerts relevant to variable stars (Chapter 5), AGN (Chapter 7), and other phenomena.

collaborative and data handling practices that will need to evolve to maximize the science return from LSST.  Investment in these areas will also produce a high return in static astronomy.

## Technical Description

### Alert Broker(s)

LSST itself will provide a limited alert filtering service that notionally may enable up to 500 users to select up to 20 alerts of interest per visit by implementing basic selection criteria. Although this service will be sufficient to enable a broad range of LSST time domain science, a fuller suite of broker services will be required to maximize this science.

LSST will provide its full Alert Service (the full stream of alerts identified in each visit on the sky) to only a small number of clients, thus timely wide dissemination of alerts will depend on the existence of one or more central dedicated database(s) and dispatch service(s), which we refer to as a "broker."  However, we note that the term "broker" is overloaded and that its expected functionality has still to be fully defined.  In this context, we refer to a broker as *a software platform that serves and (possibly) archives the LSST alert stream, federates the alerts with additional astronomical data (the "aggregator" function), characterizes and classifies targets (to the extent possible), and rebroadcasts filtered alerts for further consumption by the astronomical community.*

A number of systems with broker functionality are under development or have already been developed, notably ANTARES (Saha et al. 2014), the PTF Marshall, SNEx,[22] NEOCP,[23] NEO Exchange[24] (Lister et al. 2016), ExoFOP,[25] and the Spitzer Microlensing Portal.  Though many of these systems are specific in focus—specializing in a certain class of object or phenomena—it is striking that they share much common functionality, including a searchable database to track the evolving status of a large number of targets and cross-matching of those targets against existing catalogs and subscriptions to multiple alert feeds. The volume and diversity of LSST alerts, and the worldwide user base, argues for the development of an automated, general-purpose broker, such as ANTARES. That said, it is likely that multiple brokers will coexist in the LSST era, as some science communities already depend on the existing brokers. A good example is the NEO Confirmation Page (NEOCP) system run by the Minor Planet Center (MPC1[26]) as a clearinghouse for survey-alerted asteroids currently in need of follow-up observations. The MPC develops, supports, and runs the specialized analysis software required to link follow-up observations with known objects and re-compute their orbits. Similarly specialized modeling functions are performed by the microlensing event service ARTEMiS2[27] (Dominik et al. 2008). While LSST will compute asteroidal orbits independently (see the LSST Data Products Definition Document), the full range of specialized modeling functions are beyond LSST's scope.  However, other brokers could provide added-value data products to LSST's products.

---

[22] http://supernova.exchange
[23] http://www.minorplanetcenter.net/iau/NEO/toconfirm_tabular.html
[24] http://lcogt.net/neoexchange/
[25] https://exofop.ipac.caltech.edu/
[26] http://minorplanetcenter.com/
[27] http://www.artemis-uk.org/

The LSST Alert System may therefore operate in an "ecosystem" of brokers where data sharing would facilitate the efficient classification of LSST alerts as well as target selection and the coordination of follow-up observations. More work is needed to establish common protocols of communication (for example VOevents[28]) between brokers and other data services such as SIMBAD[29] and NED.[30]

Most of the science study groups supporting this study emphasized the need for an alert broker, though in some cases the exact functionality required has yet to be fully defined. Broker developers should engage directly with the science community as soon as possible to quantify their needs and expectations for community brokers. A few use-case examples illustrate the potential range in broker capabilities that the community may require:

- SNe science requires candidate SNe targets to be identified so that they can be observed at early times (see Chapter 6, Early Evolution of Supernovae). Many SNe science cases will require spectroscopic follow-up observations, and LSST's photometric cadence will be insufficient to full characterize SNe candidates without additional photometric data. Real-time, multi-color lightcurves will be necessary for accurate SNe classification. Access to deep imaging thumbnails around the targets will be important to evaluate the host galaxy characteristics. These requirements drive the services and cross-matched catalogs that a broker will need to fully enable SNe science. (See Chapters 4 and 8 for further details.)

- Solar System science needs to link objects found at different sky positions in images taken at short intervals, requiring a specialized analysis pipeline. This needs to run on rapid timescales in order for follow-up to be feasible, as the uncertainty of the target location increases quickly as a function of time ($\sim$ hours–days), and the object is lost. (See Chapter 6 for further details.) LSST will update orbits for Solar System bodies every 24 hours. Even if the LSST Alert System filters out previously known objects, it will still produce a substantial alert feed requiring follow-up observations to fully constrain the orbital parameters. The analysis of these data is not within LSST's purview.

- Some aspects of stellar astrophysics will depend on the ability to identify periodicities within the lightcurves, in contrast to most transient targets, which will be identified from other metrics. A broker sufficiently general to support science with periodic variables will therefore need to compute a wide range of statistics from the lightcurves and catalog data in order to classify alerts. However, LSST's relatively sparsely sampled lightcurves will have aliases for periodic targets, at least at the beginning, and transient targets can be difficult to distinguish at early phases because, by definition, the signature has only just started to manifest == low signal-to-noise. There is good existing work on lightcurve classification (e.g., Brink et al. 2013), but much of it has focused on specific variable types, often at the expense of all others. More work will be needed to build a robust classifier capable of identifying all types of variability in sparsely sampled lightcurves. This could be done in the context of either specialized broker-like services for specific types of variables, truly general-purpose brokers, or individual groups performing their own classification through interacting with the LSST dataset directly.

With an alert broker(s) a critical component to the success of LSST science, support for it entails resources for development, maintenance, and operations as well as user support.

### Auxiliary Datasets

A number of datasets exist that would provide added value to LSST alerts, in both the selection and study of targets of interest. Examples include photometry, spectra, and imaging from existing catalogs covering a range of passbands. Co-location of auxiliary datasets with the LSST's data products (including its Alert Service) is beyond LSST's current scope. However, to increase the science output of LSST's Alert Service, community brokers or other facilities could investigate approaches to efficient cross-matching of alerts.

Follow-up observations or other auxiliary data could also be contributed back to community brokers by users making observations in response to broker alerts. Such data sharing would make the alert classifications more reliable for all.

### Target Selection Filters and Classification

Enabling users to reliably filter the targets they are interested in from the flood of LSST alerts will add significant scientific value to LSST's data stream. It is likely that filters and classification tools of increasing sophistication will continue to be developed by the community both in the run-up to and during the survey and that the science return will be maximized by facilitating the application of these filters. Given the data volume, enabling community-developed software to be run at the location of community broker (rather than obliging users to download data) will be important for efficient broker operations. A model in which the general community will contribute development of broker functionality will be most effective if support can be provided both for broker-based classification and filtering, as well as for the development and testing of community software. For example, the Kepler mission provides a successful model for stimulating community software development through the award of Guest Observer grants, while the Astropy project has demonstrated that astronomers can take an active role in open-source development programs.

## Coordinating Follow-up Observations

A major aspect of time domain projects is the coordination of follow-up observations across a range of manually operated, remotely controllable, and fully automated facilities. Existing programs, such as the Spitzer Microlensing Program, have demonstrated that a large follow-up program can be coordinated this way through a Target/Observation Manager (TOM, Figure 9.2). Depending on the project, this Manager service may be a human and/or software designed to interact with and complement the functionality of the alert broker. The interface accepts targets selected from the alert stream by pre-defined filters. Observation requests for those targets can be submitted via an Application Programming Interface (API) to robotic facilities, while observers on remotely operated or manual telescopes can choose targets from online tables. Alternatively, targets can also be submitted to non-robotic telescopes that are equipped to accept them. Coordination between facilities can be improved by enabling both robot and human operators to indicate when a target is selected for observation and whether the observation has succeeded or failed.

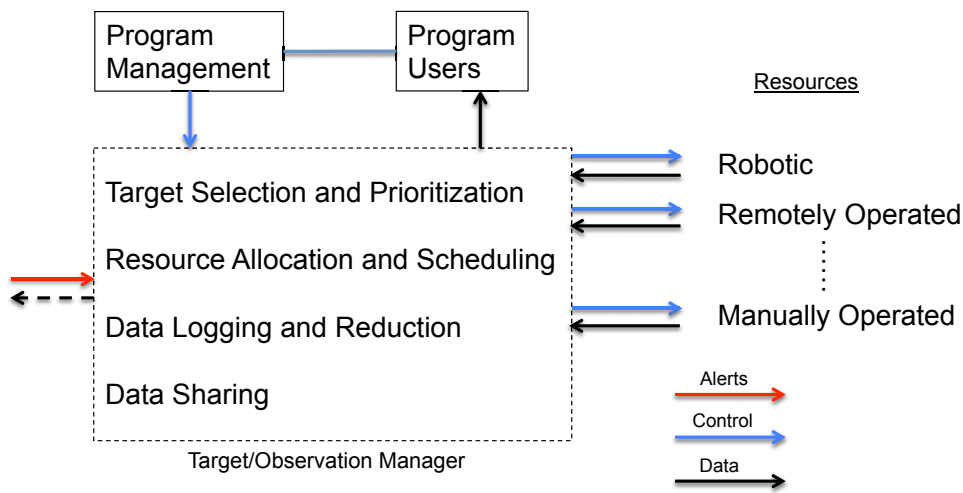

*Figure 9.2. Likely and possible components of a Target/Observation Manager. Localization of data reduction, data sharing policies, and other details will probably vary with facility and user agreements.*

This basic workflow and general-purpose TOM software could fit well with a wide range of projects, regardless of science goals, if general user-configuration to customize the target selection filters, the online data display, and other project-specific aspects are developed. Current projects have already created multiple interfaces fulfilling this role, including RoboNet (Tsapras et al. 2009), SNEx, NEOexchange, and the PTF Marshall. It could be cost-effective to develop a single Web-based platform that enables users to build and configure their own TOM interface, analogous to the way systems such as Wordpress enable users to build their own blog. Such a system could also provide widget-based tools for interacting with widely available robotic and remotely operated telescopes.

The division of roles between the broker and TOM interface does vary between current projects, and the optimum design for LSST follow-up is not yet known. The needs of some science communities have been established by current time domain surveys, whereas for others this mode of operation is relatively new. For instance, supernova and microlensing science has in recent years depended upon their communities responding to alerts from surveys such as the Palomar Transient Factory and OGLE (see Chapter 6). In contrast, this responsive mode of observing is relatively new to areas such as stellar magnetic activity (Chapter 4). Cross-disciplinary discussions between current follow-up programs have highlighted lessons learned only as a result of actually running full end-to-end survey response programs, leading to "v2.0" re-designs. For this reason we recommend supporting programs to follow up current survey alerts to stimulate the development and robust testing of all aspects of a system designed to scale to LSST.

## Observation Scheduling

Time domain astrophysics benefits from follow-up observations on a wide range of timescales, from target-of-opportunity overrides with immediate effects to long-term monitoring observations over the course of months or years. Often, the targets (and their ephemerides, if any) are not known in advance and in some cases may be unknown just hours before observations are needed. For these reasons, increasing the queue-observing modes available at LSST follow-up facilities will increase the science output, e.g.:

- Target-of-opportunity override (ToO): data taking must start immediately (e.g., early time SNe, microlensing caustic crossing)
- Rapid-response: data required within 1–2 hrs (not necessarily interrupting ongoing data taking)
- Intermediate-response: observations to be conducted within a specified time window, often to be repeated at a given interval
- Long-term monitoring: observations to be repeated at intervals over a period exceeding the normal six-month allocation semester

Some telescope-operating institutions have also expressed a willingness to make their facilities available for LSST follow-up, and a range of ideas have been raised for a Time Exchange Program among such facilities.  This would be valuable, since it would increase the range of facilities available and may also serve to "level the playing field" for smaller institutions with less access to resources. However, a workable mechanism has yet to be established and raises a range of issues that can only be resolved through discussions between potential participating institutions. A community workshop to discuss the relevant issues, and possible implementation methods, would be a low-cost way to mature this general vision.

### Access to Geographically Distributed Facilities

Some transient targets demand immediate observations or 24-hour monitoring (e.g., microlensing caustic crossing, early-time supernova). The only way this can be guaranteed from Earth is to ensure suitable follow-up resources are accessible in a range of longitudes, both in terms of the availability of appropriate instruments and adequately flexible time allocation as well as real-time observation-request mechanisms. Telescopes are required in both hemispheres to follow up equatorial and NEO targets, owing to their rapid motion.

The scientific community should collectively consider these benefits when developing possible partnerships and time-exchanges. Many institutions worldwide have access to undersubscribed telescopes, most often in the < 2m-aperture class. Their relative ubiquity across all longitudes and in both hemispheres could provide a robust global network ideal for time domain follow-up, if they were operated in coordination (see below for comments on encouraging institutions to participate in the development of such a network).  These facilities could be employed to conduct a significant fraction of the imaging follow-up required for many LSST projects and thereby relieve some of the burden on larger-aperture facilities.

## Implementation of Queue-Mode Observations—Operation and Robotization of OIR Facilities

To maximize time domain science, telescopes should be able to carry out multiple programs on a given night with adaptable scheduling rather than having fixed nights for individual projects. For example, long-term monitoring projects may only need one hour of time per night but need the same observations made every night for several months. Fast-response transient programs may only need a few hours of data a few times per year but be unable to predict exactly when. Queue-mode operations, dynamic scheduling of observations depending on the environmental requirements, instead of classical operations, assigning specific nights to specific projects, are a particularly efficient way to enable time domain science. Efficient ways to obtain queue observations can include any or all of service observing (observatory employee observes on behalf of the scientists), remote observing (scientists can observe at sites remote from the observatory), and robotic observing (no human interaction is needed to obtain the observations), and may require some evolution of traditional time allocation processes.

Large facilities can be operated safely and cost-effectively via a remote-observing mode (e.g., Keck, Gemini base-facilities operations), where the telescope itself is run by a full-time Operator and the astronomer controls the instruments via a remote-desktop or online portal. Remote or robotic observing enables time domain programs on telescopes without service observing by dramatically reducing the amount of travel to telescope sites. Sending personnel physically to the telescope for rapid-response or monitoring projects can be prohibitively expensive or impossible. Even on a telescope with a service observing mode (e.g., Gemini), remote observing capabilities can increase science output by allowing investigators to eavesdrop or monitor their service mode observations.

Increasing the accessibility of existing remote-observing platforms to more institutions, particularly for those instruments identified as "critical" by the science cases within this document, enabling remote observing at a wider range of facilities (which might require enhanced Telescope Operator roles and additional software infrastructure), or roboticizing operations altogether would increase the efficiency of time domain observations. An additional benefit is that the reduction in travel needed by these observing modes helps reduce carbon emissions that exacerbate climate change

Most queue-mode observing currently requires telescope operators to consider the requirements of a number of different programs and to find an efficient way to dovetail their observation requests into a practical sequence each night. Given the range of time domain observation types, this is necessarily a highly dynamic, real-time process, with potential interruptions from ToOs and rapid-response programs. Furthermore, the schedule must take local environmental and logistical concerns into account, which are themselves time-variable: weather, instrument availability, technical issues. This is a challenging problem that can create a potentially prohibitive workload for observatory staff. However, current software tools have demonstrated that this problem is not intractable (e.g., Lampoudi & Saunders, 2013; Lampoudi, Saunders & Eastman 2015; Miller & Norris 2008; Saunders et al. 2014), and we recommend the development of freely available software tools for this purpose.

The costs and technical challenges of converting an existing telescope to remote or full robotic-operation can be mitigated if an online knowledge base is developed with contributions from institutes who have already roboticized existing telescopes (e.g., LCOGT, Caltech, Tel Aviv University). This substantial knowledge base is particularly applicable to the relatively smaller-aperture telescopes that many US institutions have some access to.

It is possible to develop software interfaces to enable even customized Telescope Control Systems (TCS) to communicate with external systems (via remote control or robotic operation) by establishing a common communication protocol. LCOGT is presently implementing such a system to link the pre-existing Wise 1.0 m Telescope in Israel to its wider robotic network. Similar protocols could be developed to interface with widely used commercial TCS such as ASCOM and ACP and then implemented at a range of facilities.

Roboticization of some facilities would require additional hardware, for example, upgrades to dome control and safety systems, installation of weather systems and webcams, pointing encoders, remotely operable switches, etc. One way to encourage institutions to participate in the follow-up network would be to provide a grant opportunity to which institutions could apply if they agree to allocate a minimum amount of time to LSST follow-up.

It is important to distinguish between the automation of telescope operation to increase flexibility and speed of observations and automation aimed primarily at dispensing with on-site support or operations staff. For LSST follow-up, the former aspect is what is most important.

## Data Processing

Here we consider two distinct groups of data products: those produced in the course of follow-up observations and those produced by LSST itself.

### Observatory Data Reduction Pipelines and Archives

Many LSST alerts will trigger requests for follow-up observations from a wide range of facilities and instruments, and the resulting data will be vital to evaluate the classification of that alert and to determine the course of future observations and analysis. In the case of transient phenomena, any delay in the delivery of data can result in the loss of critical information. It is therefore ideal if the facilities performing the follow-up observations deliver at least "quick-look" reduced data products as rapidly as possible. At minimum, support should be given to the development and operation of archives to serve the raw data products, but this will require that every user develop their own pipeline to reduce those data. A more efficient solution would be to support the development and operation of data reduction pipelines by key follow-up facilities, which those institutions could then run centrally for all data.

### Data Sharing

As alerts stream in from LSST, inevitably filters from different projects will select the same targets. In some cases, subsequent data (either from LSST or a follow-up facility) may contradict the initial classification or indicate the target is of low priority to the project's science goals. A competing project with similar science goals may also select the target, only to request the same follow-up, leading to duplication of effort and wasted telescope time. Alternatively, the target may prove to be of great interest to a project with different goals, which may not prioritize it based on the initial data alone. Either way, it would be much more efficient for the project to share the additional information and hence assist the coordination of follow-up observations community-wide.

This may prove to be especially important if the rate of alerts from LSST places high demand on follow-up facilities. Data sharing is much more likely if user-friendly tools are available to enable the upload of additional information to the alert broker.

### Simulated and Commissioning LSST Data Products

Simulated real-time LSST data, along with documentation and software tools to handle it, would help the community develop software and test follow-up strategies. Such simulated data would have the most value in advance of the survey, to minimize the possible targets of interest that will be missed if the time domain community is not ready *before* the survey starts. Of course, testing with real data is always more valuable, and some science results are possible from the commissioning data also, so support for the LSST's distribution of some early commissioning data would be a particularly valuable resource for LSST's time domain community.

## Sociological Change

### Handling the Alert Stream—Triage Programs

Accurately selecting desired targets from the LSST alert stream may be complicated by false positives caused by both data processing and astrophysical phenomena. Many projects may need to conduct preliminary "triage" observations following an initial LSST alert in order to prioritize targets for substantial follow-up and eliminate false positives. While the exact nature of the triage depends on the science case, these observations could burden follow-up resources owing to the sheer number of alerts. In the traditional model of competing science projects, teams have tended to keep triage targets private for fear of being scooped, leading to multiple teams repeatedly observing the same few targets while neglecting others. This could be averted by the teams making public data on confirmed false positives, although there is little incentive to do so and preparing catalogs for release can be substantial work. Furthermore, observatory Time Allocation Committees (TACs) (and external reviewers) tend to downweight triage proposals that simply help to select targets for characterization elsewhere. Nevertheless, triage observations will increase the efficacy of target selection for a range of LSST science and therefore could contribute significantly to the efficient use of follow-up resources. We recommend exploring measures to change these paradigms. Options include the following:

- Funding triage observing programs on suitable facilities (e.g., multi-filter imaging on < 4m telescope, single-shot R ~ 1000 spectra) to which anyone in the US community could submit requests for (limited) observations. The resulting data could go public immediately and, if not ingested into LSST's own database, then instead (in addition) be uploaded to one or more community brokers;
- Support for national-level archive facilities to assist researchers in preparing data on false alarms for release on rapid timescales.

## Incentives for Data Sharing and Coordinated Follow-up Programs

Once targets are selected, they often require substantial telescope time to obtain the follow-up data necessary to characterize the phenomenon and sometimes observations repeated many times over the course of months. This means follow-up programs need to handle both new alerts and monitor existing targets, and the time required in total ultimately limits the number of targets that can be effectively followed up. For example, a single target requires spectra (exposure time ~ 1 hour) taken every 2–3 days for ~ 3 months plus imaging in ugriz (=10 min exposures × 5 filters ~ 1 hour including overheads). A single telescope with an 8-hour night can therefore follow up to ~ 4 targets/night. Ongoing follow-up programs to current surveys already receive more alerts than can be followed up, even after false alarms are eliminated. Table 9.1 compares the current alert rate with that expected from LSST for three example science cases.

*Table 9.1.*

*Summaries of alert rates received from current time domain surveys in different science areas, compared with the number that can be practically selected for follow-up by programs on the LCOGT network and the expected alert rate from LSST (LSST Science Book). Note that the alert rate of candidate targets can be substantially higher than the confirmed discovery rate for each class of object.*

|  | Supernovae | Solar System Objects | Microlensing |
|---|---|---|---|
| **Total alerts/year (currently)** | ~ 700 | ~ 5400 | ~ 2100 |
| **New alerts/day (currently)** | ~ 2 | ~ 100 | ~ 8 |
| **Alerts followed up/day** | 40–60 | ~ 10 | ~ 10 |
| **Total alerts/year (LSST)** | 10,000 | 557,000 | ~ 10,000 |

This issue may be exacerbated when competing science teams request follow-up for the same subset of alerts, potentially leading to duplicated observations while simultaneously neglecting other targets. Though we note that the discovery rate for some classes of objects decreases substantially as a factor of time (e.g., variable stars; Ridgway et al. 2014), this is not true for transient events. This underlines the need to reliably select and prioritize targets, and to efficiently follow them up.

There are therefore benefits to coordinating the follow-up of alerts both between science teams and across aperture classes of telescopes, so that larger-aperture facilities perform the observations only they can make.

However, we recognize the reluctance of some teams to share data and target lists, and the importance of restricting access to them in some fields, as receiving credit for publishing the work is often a key component to continued funding and careers. In these cases, alternative options such as the following could be explored:

- Developing tools to enable observatories to identify similar observations requested for the same target by multiple teams. The observations would be conducted only once but the data made available to both teams.
- Encouraging the rapid public sharing of data on false positives.
- Encouraging teams to share which targets they have requested follow-up observations for.

It will be important to encourage the timely sharing of data obtained on a target wherever possible. This will involve some sociological change in some fields where the norm has been to restrict access to target lists and/or data, but by opening a dialog with the community, LSST could spur a shift towards a more open data policy.

### "Common Resource" Follow-up Surveys

Several science working groups have recognized that a small number of specific facilities could perform a substantial fraction of the follow-up they require, for example a wide-field (> 1 sq. deg.) optical imager on a 4m-class telescope. Since most science goals will require time series observations and queue-mode scheduling, this raises the possibility of dovetailing their observation requirements into a full-time follow-up survey strategy for a dedicated facility. This may work best for non-transient targets and could begin once they have been identified from the first 1–2 years of LSST alerts. Alternatively, a survey could be conducted from a site longitudinally separated from LSST that would fill in gaps in the lightcurve coverage: this would also benefit transient science. It may be that this approach is more efficient than several teams independently organizing follow-up of a subset of targets, and we recommend that the option be explored. If it proves to be viable, then the data should be made publicly available as a common resource.

## Evolution in the Resource Allocation Process

Most resources in astronomy, in particular telescope time, are allocated via a committee review process that is determined by the controlling institution(s). This review process is an important step in allocating scarce resources to the most scientifically valuable projects, but the traditional model of allocating time has limitations for time domain programs, as discussed above. A number of alternative models may be envisaged, for example:

- Introducing queue-scheduled programs at facilities currently scheduled in blocks
- Target-of-opportunity programs for specific resources open to everyone with immediate review, possibly with certain restrictions (similar to the current Swift ToO program)
- Expanded allocations of time open to the whole US community
- "Resource exchange" between teams or facilities, either in advance or dynamically
- Some percentage of a range of resources dedicated in advance to an Open Access program to which anyone can apply at any time, with immediate review (an expanded version of the open target-of-opportunity)

Note that while open-access programs are particularly important to enable research at a wide range of institutions, as opposed to comparatively few major centers, there are potential advantages to larger institutions as well. For example, it could provide direct access to a wider range of resources and provide a mechanism to encourage the involvement of foreign partners with resources located in complementary longitudes.

All of these possible models have their own challenges, and since resource allocation remains the prerogative of the controlling institution(s) and/or their sources of funding, this topic deserves further discussion. It should therefore be one of the topics raised at the workshops on the accessibility of follow-up resources.

## Capabilities Required to Maximize Time Domain Science

The following list of recommendations was drawn up through discussion with the six science working groups convened in support of this study:

### 1. One or more community alert brokers with configurable target filtering that can accept real-time feedback of information from users

It is critically important for at least one general-purpose broker to be available in time for testing on LSST commissioning data and remain available throughout the operation of the project. It is likely that it will need to continually adapt to the needs of the community during its operation, as we learn to accurately classify targets and identify new phenomena. The design of other resources, such as the resource coordinators and software developed by science projects, will depend on the functionality, performance, and data products from the alert broker. However, the exact requirements from some science groups are unclear at this time. The software "ecosystem" will take substantial time to develop and will benefit from having access to simulated and commissioning data for testing purposes.

### 2. Queue-mode time allocation on facilities of all aperture classes, including open-access target-of-opportunity, rapid-response, and long-term monitoring programs

Facilitate telescope access (including possible service mode, remote-operation, or robotization strategies) and time exchange programs between observatories, and data sharing facilities and policies. This requirement brings with it the need for dynamic, robust scheduling software to enable observatories to dovetail observations specified by multiple projects in real time. The software must operate locally for manual, remotely, and robotically operated facilities of all apertures and instruments. A workshop is proposed to develop requirements and incentives for participation.

### 3. "Common-resource follow-up" surveys
Explore the viability of dedicated follow-up surveys that provide data servicing a number of science goals, and where the data are made publicly available.

### 4. Development of instrument data pipelines
Support for observatories to develop data reduction pipelines for their instruments together with online archive facilities to serve at minimum "quick look" data products on rapid (~ 1 hr if not faster) timescales, including spectroscopic and spectro-polarimetric reductions (discussed in detail in Chapter 4, Magnetic Fields Follow-up Observations and Capabilities).

# Recommendations

In the interests of a national system that will maximize time domain science in the LSST era, we make the following recommendations:

## Critical

***Development of a public alert broker system***
***Development of target and observation management software***
Both of these components will be essential to running any efficient, large-scale follow-up program for LSST alerts. We note that the community represents a software-development resource that is highly skilled and motivated and should therefore be encouraged to contribute to this necessary software infrastructure. This can be achieved by funding software development and testing programs as well as by providing "toolkit" platforms that enable users to apply their software to LSST data. At the same time, the development of an alert broker that can process the LSST alert stream has challenges beyond the field of astronomy alone. There are key questions that can best be addressed by computer scientists working with astronomers to resolve this multi-disciplinary problem, so support should be available across the relevant fields.

***Increasing the availability of follow-up telescopes in queue-scheduling modes, spanning a range of apertures, instrumentation, and geographical locations***
It will be important, particularly for transient science, to have rapid and flexible access to appropriate instruments at several longitudes to characterize fast-evolving phenomena. Moving objects will require observations from both hemispheres. This could be achieved in several ways, e.g., time exchange programs or building new instruments for existing telescopes. It will also incorporate efforts to make facilities more widely accessible in service-, remote- and partly or fully roboticized modes.

***Support for national-level archive facilities***
The LSST Data Access Center, and/or other data archives, should make both simulated and commissioning datasets available to the community, ahead of main science operations.

***Support a workshop aimed at to develop a plan for making a system of follow-up facilities, suite for real-time, large-volume, time domain observations***
***Support the development and testing of software and hardware infrastructure that facilitates follow-up programs for LSST, using existing surveys as a proxy***
Robust testing of such systems can be conducted now using current survey and follow-up facilities and would offer valuable training opportunities that will influence future development.

## Very important

***Development of general-purpose time domain telescope scheduling software***

***Development of data reduction pipelines for key follow-up facilities***

***Funding for triage observing programs on suitable facilities***

***Explore the viability of a follow-up survey capable of filling in gaps in LSST lightcurves***

<span style="color:#4472C4">**Important**</span>

***Facilitating and encouraging coordination between follow-up teams and the sharing of data obtained in response to LSST alerts, incorporating the capability within LSST software resources***

Additional data may also be contributed by users making observations in response to broker alerts. We believe this should be encouraged and facilitated, as the data is likely to make the alert classifications more reliable for all. For some areas of astronomy this may involve a considerable change in accepted practices, which raises legitimate concerns. However, we feel that these can be mitigated and that LSST offers an opportunity to start this discussion.

# Chapter 10: Computing and Data Resources for Maximizing LSST Science

*Adam S. Bolton (NOAO), Rachel A. Street (LCOGT), Erik Tollerud (STScI), David Ciardi (IPAC, Caltech), Steve Ridgway (NOAO), Tom Matheson (NOAO), Jay Elias (NOAO), Chad Schafer (Carnegie Mellon University), Bryan Miller (Gemini Observatory)*

*Executive Summary*

LSST is the most data-intensive project in the history of optical astronomy. The quantitative scale of the LSST dataset will bring a qualitative shift in astronomical research methods, with more astronomers executing much or all of their analysis on remote systems. The scientific return from both "pure-LSST" and "LSST++" research will depend critically upon the success of data systems and computing practices of unprecedented scale and complexity in the field. Much, but not all, of the essential capability is being delivered by the Data Management systems of the LSST project. While this study process has focused primarily on prioritizing observing capabilities necessary to maximize LSST science, we can also identify major areas of need for the development of data, computing, and community resources beyond the LSST project boundary.

We make the following recommendations:

*Critical*
- Conduct a systematic study to prioritize the computing, software, and data resources required to maximize the science return of LSST. This study should account for the capabilities being delivered by the LSST project and other efforts, the demands of forefront LSST-enabled research, and the opportunities presented by new technology. Additional studies should be conducted at periodic intervals in the future to account for ongoing developments in science and technology.
- Related to the previous recommendation, support the development of high-priority tools and systems that utilize astronomical data and computing resources beyond the scope of LSST operations to conduct compelling LSST-enabled science.

*Very Important*
- Support coordination, standardization, and broad community adoption of data-analysis and data-exploration tools and services to work with multiple datasets at the scale of LSST. Include robust mechanisms for community feedback to ensure that development is responsive to science-user needs and interests.
- Support the training of scientists at all career stages in the analysis techniques and computing technologies that will be necessary in the LSST era.
- Increase the viability of career paths in astronomical software, data handling, and pipeline reduction, and encourage and recognize publications of software.
- Support cross-disciplinary workshops to facilitate the cross-pollination of ideas and tools between astronomy and other fields.

# Background and Context

The LSST construction project is currently building the baseline software infrastructure necessary for the scientific exploration and analysis of the data products that the LSST telescope and pipelines will deliver. This infrastructure includes a Web portal and workspace environment for flexibly specified data queries and visualization (the Science User Interface Toolkit, or SUIT), capabilities for running iPython notebooks in colocation with the LSST data archive, and distributed parallel databases for querying entire LSST catalogs efficiently (the Qserv system). Capabilities to access LSST data and services directly through application programming interfaces (APIs) will also be provided, as will capabilities to directly import the pipeline processing modules of the LSST software stack. This infrastructure will be run at LSST data centers and will be backed by access to computing resources, filesystem storage, and personal database space at a level of 10% of the LSST operational capacity.

These LSST-provided capabilities are being developed to be responsive to a broad range of science use cases. However, much LSST-enabled science will require additional computational, data, and software resources beyond the scope of the LSST project itself. The aim of this chapter is to lay out the broad categories of these additional resources, highlight some of the particular considerations that apply to each category, and derive a set of recommendations (given above) for further support and study in the context of LSST data systems, the LSST survey, and the broader landscape of software and computing. We stress that the prioritization among these resources must ultimately be driven by the scientific priorities and associated observing capabilities laid out in earlier chapters of this report. At the same time, we emphasize that the evolution of science goals and technological possibilities over the operational lifetime of LSST will require a degree of flexibility in this prioritization.

# Technical Considerations

## Present and Future of Large Astronomical Data Archives

Science analysis in the LSST era will be shaped by an ongoing trend towards flexibility in the design and methods of access to large astronomical data archives, with an increasing emphasis on virtualization and cloud-based solutions (either commercial or privately hosted) and on bringing the user's analysis into colocation with the data archive. The LSST SUIT and its supporting infrastructure represents one among a number of new archive-interface technologies currently being developed within this paradigm. At NOAO, the Data Lab project is developing similar capabilities to provide data discovery, interactive exploration, and automated programmatic analysis of the large survey holdings of the NOAO science data archive (such as the Dark Energy Camera imaging surveys DES and DECaLS). Building on the legacy of the Catalog Archive Server system of the Sloan Digital Sky Survey, the SciServer initiative at Johns Hopkins University is developing a collaborative platform for large-scale data-driven research across multiple scientific domains. The Gaia collaboration is developing the Gaia Added Value Interface Platform (GAVIP) to provide many of these same flexible data analysis capabilities to the data from Gaia as it becomes public.

These various modern archive development initiatives are being pursued within a range of different contexts with regard to institutions, timescale, datasets, and user community missions. Nevertheless, given the common technological challenges, the astronomical community would be well served by maximizing collaboration and coordination among these efforts. By borrowing specific implementations where feasible and appropriate, teams can leverage each other's developments to reach service milestones sooner and more cheaply. By sharing lessons learned about best practices in these new modes of archive development and operation, all teams involved can improve the quality of service that they deliver to their user bases. By standardizing back-end protocols, different centers can maximize interoperability between their archival holdings and services. And by developing uniform front-end look-and-feel conventions, the interface skills that users develop through one archive can be generalized to other data centers.

The code-to-data paradigm of these new data-archive technologies raises multiple operational questions for the LSST era: How to "sanitize" the uploaded software and prevent malicious attack? How to structure the data for effective and efficient searching? How to train users in the most efficient computing techniques and avoid software introducing excessive overheads? What computing languages and software packages (e.g., Astropy) should be supported? These questions must be addressed at both the institutional level and at the community-wide level, with the associated resources for study and implementation to be driven by scientific community priorities.

Within the context of LSST, these new archive-interface technologies also raise questions about the scalability of baseline infrastructure to projects beyond the scope of LSST support. The collective science opportunities provided by LSST will likely entail computational requirements that exceed the 10% community allocation of LSST resources. Therefore, the scientific maximization of LSST data will likely require the capability to connect the user-interface infrastructure to computing and storage capacity that lies beyond the operational scope of LSST. This additional capacity may come through competitively allocated resources available in colocation with the LSST data centers, through national-grid computing allocations, through commercial cloud providers, or through the users' own institutions.

## Combining LSST with Other Survey Datasets

For a large number of science use cases, cross-matching LSST catalogs with other astronomical data will be necessary for classification, modeling, and false-positive rejection. Other major survey programs before and during the LSST era will deliver large and homogeneous datasets that are scientifically complementary to LSST in this regard. Some datasets, such as Gaia, will be integral to the production of calibrated LSST data products and will necessarily be co-located with the main LSST production systems. Other datasets, while external to the generation of LSST data products, will be of great scientific interest for co-analysis with the LSST database. Some examples are given in Table 10.1. Particular scientific analyses of these datasets may require large-scale spatial cross-matching, joins across massive databases, and joint analysis of global photometric and astrometric systems. Capabilities to enable this work—whether by co-locating auxiliary datasets with LSST, or by enabling geographically distributed cross-matching—are beyond the scope of LSST project construction and operations. However, the astronomical community has a strong interest in leveraging and coordinating with the development activities of LSST and other projects to enable the broadest range of "LSST++" analyses with the greatest efficiency.

*Table 10.1.*

*Some of the major survey programs expected to produce data products before and during the LSST, which will be a critical element in LSST data reduction, assessment of alerts, and large-scale analyses.*

| Catalog | Size | Data Type(s) |
|---------|------|--------------|
| Gaia | > 1PB | Catalog, Time series, Spectra |
| Pan-STARRS | > 2PB | Catalog, Images, Time Series |
| WISE/NEOWISE | 428TB | Catalog, Images, Time Series |
| 2MASS | 27TB | Catalog, Images |
| SDSS | 61TB | Catalog, Images, Spectra |
| DECam | > 400TB | Catalog, Images |
| DESI | 100–200TB | Spectra, Catalog |

## Large-Scale Computing

A major scientific theme for LSST is large-scale population analysis: both the discovery of rare phenomena and the compilation and characterization of statistically significant samples of a wide range of objects. Some such studies may require reprocessing across the entire LSST imaging footprint to generate new catalogs, quantify recovery of simulated sources, or provide customized estimators of noise and data quality. Given the size of the LSST database, the most ambitious scientific analyses of the LSST era can be expected to push the envelope of accessible supercomputing capacity. This capacity will increase between now and the conclusion of the LSST operations phase, with analyses that would be prohibitive today becoming feasible with the supercomputers of the future. The following questions arise with regard to supercomputing in the LSST era:

- What are the technological implications of a requirement to make LSST data accessible for community-driven supercomputing applications?
- What additional, non-standard demands will LSST data and science users place upon high-performance computing support staff?
- What is the model for development and scale-up of computationally intensive projects making use of LSST data?

## Simulated Data and Commissioning Data

In addition to the main survey data, both simulated and commissioning data products should be made available through the LSST archive. Access to simulated LSST data, along with documentation and software tools to handle it, will make it possible for the community to develop software and test follow-up strategies. Doing this in advance of the survey is extremely important, particularly for transient science. Important events could be missed if teams are not ready to respond to alerts *before* the survey starts. Testing with real data from commissioning will be even more valuable, and may provide early science results. Above-baseline resources would likely be required to make both of these data products available to the community in a full-featured and timely way.

### Technological Wildcards

Technology is subject to rapid evolution. In the era of LSST, the astronomical community will need to stay abreast of cutting-edge developments, and to evaluate new technology opportunities as they arise, in areas such as:

- Containerization (Docker,[31] Shifter, etc.)
- New hardware technologies such as solid-state-drives (SSDs) and the increased use of Graphical Processor Units (GPUs)
- New commercial data service offerings (e.g., commercial cloud computing)
- New commercial software capabilities
- Developments in network transfer protocols and bandwidth
- Evolving models for data release and publication
- Developments in other academic fields
- Emergence of new science-appropriate programming languages (e.g., Julia)

## Community Considerations

### Software Development

The importance of software in astronomy is set to increase as we enter the LSST era, along with the imperative to recognize the importance of the associated development work. The LSST project encompasses a major software effort, which coexists alongside other major project- and community-based software efforts such as Astropy, DESDM, and DESI. This raises several broad questions:

- How does maintenance and support of the LSST software stack coordinate with other major software efforts?
- What are the roles of universities and research centers in seeding and sustaining major astronomical software efforts?
- How can the development of common standards be balanced with the imperative for rapid development of running code?

### User Support and Training in the LSST Era

Just as not all astronomers have been required to be experts in instrumentation in order to conduct observational programs in the past, so in the future it must be possible for astronomers who are non-experts in data systems to carry out diverse scientific research. At the same time, expert users must be empowered to perform non-standard analysis of LSST data. The field must develop new "support astronomer" roles for providing expert assistance to community users across the spectrum of data-intensive astronomical research, along with metrics for measuring the success of this support and providing effective feedback. Likewise, training opportunities (schools, workshops, online materials) for astronomers at all career stages will be essential for developing broad-based capabilities for data-intensive science in the LSST era.

---

[31] http://www.docker.com

## Cross-Disciplinary Collaborations

Scientific analysis and computing based on LSST data will not be confined within the astronomical community; rather, the effort will necessarily encompass multiple interdisciplinary connections to statistics, computer science, and other fields with Big Data experience complementary to astronomy. It is essential for the astronomical community to foster and sustain this interdisciplinary collaboration, to capture and capitalize on the innovation that it delivers, and to normalize and recognize bona fide interdisciplinary work (especially within academia.)

## Career Paths for Data and Software Astronomers

Specific expertise is needed for development in all of the areas described above. Therefore, there is a clear need to establish viable career paths for the specialists who will be developing all of these capabilities. This may be through traditional university faculty tracks like today's instrumentalists, through project-specific contract funding, or through long-term service appointments at national labs and centers. These alternatives have different implications for how effectively these data-intensive astronomers can contribute to the broader community and how well they can be retained within astronomy when their skillsets are in high demand in private industry. Agencies such as the NSF, NASA, and private foundations could lead by placing special focus on providing *long-term* funding support for astronomers pursuing these new roles. Additionally, journals and professional societies can contribute by establishing frameworks within which data- and software-related contributions can be credited on a par with refereed scientific publications as metrics for success and advancement.

# Chapter 11: Findings and Recommendations

As a discovery machine and exploration portal for the astronomy and physics communities, LSST will enable many discoveries based on LSST data alone. At the same time, the scientific legacy of LSST will be richer and more diverse when supporting ground-based OIR resources are available to complement LSST data and follow up LSST discoveries.

The previous science-based chapters illustrate the diverse ways in which LSST data will be used to pursue important astrophysical problems that have strong synergy with the science described in *New Worlds, New Horizons*. Table 11.1 summarizes the primary capabilities identified in the science-based chapters and their rough 3-tier priority: Priority 1 (critical), Priority 2 (very important), and Priority 3 (important).[32] The quantitative flowdown from detailed science cases to needed capabilities leads us to the following findings and recommendations regarding telescopes, instruments, observing infrastructure, and computing requirements. Additional needs identified in the study that fall outside our formal charge (Appendix B) are described in Appendix C.

## Findings

### Diverse facilities needed to maximize LSST science

Table 11.1 illustrates the diversity of telescope apertures and instruments that are required to maximize LSST science. As is evident from the table, there are important roles for both small- and large-aperture facilities in the LSST era. While much of the science described in the Solar System and Stars chapters (Chapters 5 and 6) requires current aperture facilities (from less than 3m to 10m in diameter) for imaging and spectroscopic follow-up, Galaxy Evolution (Chapter 7) science would benefit greatly from GSMT aperture facilities. Transient science (Chapter 4) values facilities of all apertures.

### Need for "workhorse instruments" that enable a broad range of LSST science

Several "workhorse" capabilities are critically important to multiple science areas. These high-priority, high-demand capabilities, which would enable a broad array of LSST science (Table 11.1), include the following. The instrument characteristics described below were identified in workshop breakout panels that included representation from the study groups that prioritized that capability. The indicated demand, which is based only on the example science cases considered here, almost certainly underestimates the true demand for these resources from the entire US community.

---

[32] The summary tables at the end of each science chapter provide a more complete list of the capabilities identified in this study.

| Capability | Telescope Aperture | | | |
|---|---|---|---|---|
| | < 3m | 3–5m | 8–10m | ≥ 25m |
| **Optical Imager (Wide-field)** | Solar System<br>**Stars**<br>Transients<br>*Dark Energy* | **Solar System**<br>**Stars**<br>Milky Way<br>Transients<br>*Dark Energy* | **Solar System**<br>**Stars**<br>Transients<br>Galaxy Evolution | Transients<br>*Solar System* |
| **NIR Imager** | | Transients | Transients<br>*Milky Way* | Transients |
| **AO IFU R ~ 5000** | | | **Galaxy Evolution**<br>Dark Energy | **Galaxy Evolution**<br>Dark Energy |
| **OIR MOS R = 5000 0.35–1.3 micron** | | **Stars**<br>**Galaxy Evolution**<br>**Dark Energy** | **Stars**<br>**Milky Way**<br>**Galaxy Evolution**<br>**Dark Energy** | **Galaxy Evolution**<br>Dark Energy<br>*Milky Way* |
| **Optical SOS R = 1k–5k 0.35–2.5 micron** | Stars | **Solar System**<br>Stars<br>Transients | **Solar System**<br>**Transients**<br>**Galaxy Evolution**<br>Stars<br>*Milky Way*<br>*Dark Energy* | **Transients**<br>Solar System |
| **Optical SOS R > 20,000** | | | Stars<br>*Transients*<br>*Galaxy Evolution* | Stars<br>*Transients*<br>*Galaxy Evolution* |
| **OIR MOS R > 20,000** | | | **Milky Way**<br>**Stars** | **Stars**<br>*Milky Way* |

Entries in boldface type indicate that the capability is **Priority 1 (critical)** for that science topic.
Roman type indicates Priority 2 (very important).
Italic type indicates *Priority 3 (important)*.

**Wide-field optical imaging on 3–5m telescopes** was called out as a high priority for the study of small bodies in the Solar System (Chapter 6), stellar rotation and activity (Chapter 5), Milky Way science (Chapter 3), and transients (Chapter 4). An imager such as DECam on the Blanco 4m telescope at CTIO, which has a 2.2 degree diameter field-of-view, would meet the needs of many of these (broad- and medium-band imaging) science cases. As an indication of the high demand for this capability, carrying out the science cases on Solar System small bodies and stellar rotation and activity alone would require ~ 5.5 years of observing time (Table 11.2).

**Wide-field optical multi-object spectroscopy on 3–5m and 8–10m telescopes** was called out as a requirement for photometric redshift training and investigations of a number of potential systematics in cosmological measurements (Chapter 8), studies of galaxy evolution and environments and circumgalactic medium (CGM) tomography (Chapter 7), Milky Way and local dwarf galaxy stellar spectroscopy (Chapter 3), reverberation mapping of active galactic nuclei (Chapter 7), and studies of stellar rotation and activity (Chapter 5).

These studies could be carried out with a wide-field multi-object spectrometer (MOS) on an 8m-class telescope. The required resolution is R ~ 5000 in the red and R ~ 2500 in the blue. The wavelength coverage must extend to at least 0.37 micron in the blue and at least 1 micron in the red. (Extragalactic and cosmological science would benefit from wavelength coverage extending as blue as 0.35 microns and as red as 1.3–1.5 microns.) The field-of-view must be at least 20 arcmin, and ideally larger than 1 degree, in diameter. High multiplexing is critical (at least 2500) for the cosmology science case. Some of the needed observations could, in principle, be carried out on a smaller-aperture facility (3–5m) with a similar instrument but much longer exposure times. A key open question regarding this option is whether sky subtraction with a fiber spectrograph can be carried out with high-enough fidelity to achieve the required sensitivity.

Among existing capabilities, PFS on Subaru and DESI on the Mayall could fulfill most of the above requirements (assuming sufficiently effective sky subtraction and that extremely large time allocations could be obtained for DESI). However, both facilities are located in the Northern Hemisphere (as is the Maunakea Spectroscopic Explorer at CFHT, currently in the design phase), and access to these capabilities is limited to a very restricted portion of the US community. Another potential opportunity is the Southern Spectroscopic Survey Instrument (SSSI), a project recommended for consideration by DOE's Cosmic Visions panel (see arxiv.org/abs/1604.07626 and arxiv.org/abs/1604.07821 for the reports). SSSI is envisioned as a DESI-like instrument, providing highly multiplexed spectroscopy (thousands of fibers) over a wide field-of-view and covering the full optical wavelength window, on a 4–6 m or larger telescope in the Southern Hemisphere. DOE could potentially fund such an instrument in time for first light around 2023–2024.

The demand for this facility is extremely high (Table 11.2). Three of the above science cases alone would require years of time on a PFS-like facility: ~ 1 year for photometric redshift training, ~ 1 year for studies of galaxy evolution and environments, and ~ 6 years for studies of the Milky Way halo. The PFS field of view is a significant limitation for the Milky Way halo survey. The surface density of stars down to the desired magnitude limit is 125 per square degree, so only ~ 150 targets can be observed in a single PFS pointing. To study a million halo stars with PFS then requires 6000 pointings at 3 hours each, for a total time of 6 years. A wide-field MOS on an 8m telescope with a much larger field-of-view of 7–8 square degrees would be able to complete the study in less than a year.

As a result, it is unclear whether such studies could be carried out in a timely way on a facility such as PFS that is already engaged in meeting the needs of other communities.

**Broad-wavelength, optical-infrared spectroscopy at moderate resolution (R = 2000 or larger) on an 8–10m telescope.** This single-object spectroscopic capability was called out as a high priority for the study of transient phenomena (Chapter 4), the study of flaring stars (Chapter 5), the characterization of lensed/lensing galaxies for studies of galaxy evolution (Chapter 7), and the study of rotating Solar System objects (Chapter 6). Broad wavelength coverage is

valuable for characterizing not only time-variable systems that have features of interest in both the optical and near-infrared (NIR) but also new phenomena, where it is unclear at what wavelengths the features of interest will lie.

*Table 11.2. Illustrative Demand for Selected Capabilities in Example Science Cases*

| Capability | Telescope Aperture | | |
|---|---|---|---|
| | 3–5 m | 8–10 m | > 25 m |
| **Optical Imager (Wide-field)** | **~ 2.5 yrs with Blanco/DECam** for Solar System science case<br><br>**~ 3 yrs** for Stars science case | | |
| **AO IFU R ~ 5000** | | **~ 1.3 yrs with Gemini/NIFS** for Galaxy Evolution science case | **~ 1.3 yrs with TMT/IRIS** for Galaxy Evolution science case |
| **OIR MOS R = 5000 0.35–1.3 micron** | | **~ 8 yrs with Subaru/PFS** equivalent for Milky Way, Galaxy Evolution, Dark Energy science cases | **~ 0.7 yrs with TMT/WFOS** for Galaxy Evolution science case |
| **Optical SOS R = 1k–5k 0.35–2.5 micron** | | **~ 3.5 yrs** for Transients science case<br><br>**~ 0.4 yrs** for Galaxy Evolution science case | |
| **OIR MOS R > 20,000** | | **~ 10 fiber-yrs** for Milky Way science case<br><br>**~ 550 fiber-yrs** for Stars science case | |

Estimated on sky time needed to carry out the specified science case. Poor weather, maintenance, and other overheads would increase the number of calendar nights needed to carry out each program.

These studies require a wavelength range of 0.36–2.5 microns. The short wavelength limit provides access to the Balmer jump for studies of stellar flares on G-type stars. It also enables studies of young and/or interacting supernovae (SNe). A blue limit of 0.34 microns is highly desired, whereas a blue limit of 0.4 microns would significantly reduce the scientific productivity of the spectrograph. The long wavelength requirement is set by the need to measure thermal emission from low albedo Near Earth objects (NEOs) to directly measure their diameters. The minimum required resolution is R ∼ 2000, although R = 5000 is desired. A minimum resolution of R = 2000 is needed for accurate SN redshifts. Higher resolution (R = 5000) spectra will resolve narrow interstellar medium (ISM) or circumstellar medium (CSM) lines in SN spectra and are more optimal for stellar spectroscopy in general. A prism mode that enables very low resolution (R ∼ 100) spectroscopy would be valuable for Solar System science. Because most features in the spectra of Solar System small bodies are broad, a low-resolution mode allows for fast, efficient compositional studies. Rapid acquisition (< 2 minutes) is important for any time-variable object.

As an indication of the high demand for this capability (Table 11.2), the transient science case alone would require a total of ∼1200 nights over 10 years (approximately 30 nights/year to characterize the transient sky, ∼ 80 nights/year to investigate Type Ia supernova demographics, and ∼ 10 nights/year to characterize the early evolution of supernovae).

**High-resolution optical spectroscopy on an 8–10m telescope** is required for studies of stellar rotation and activity (Chapter 5), the IGM and CGM (Chapter 7; tens of nights), SNe with CSM interactions (Chapter 4), and chemical abundance studies of Milky Way halo stars (Chapter 3).

These studies require a resolution of R = 20,000–100,000 and a wavelength range of ∼ 0.3–1 micron. Given the target densities (e.g., bright halo stars; 20 per square degree), a multi-object capability of 10–20 targets per square degree would be valuable. To enable polarimetry, which is highly desired for the study of transients and stellar activity, slits are preferred over fibers.

As a rough indication of the high demand for this capability, the study of Milky Way halo stars requires spectroscopy of ∼ 10,000 stars with ∼ 3 hours of integration time per star or ∼ 30,000 slit- or fiber-hours. The proposed study of stellar activity cycles in open clusters requires 2 million fiber-hours.

### Critical need for existing, planned, and future capabilities

Some of the needed capabilities identified in Table 11.1 are **currently available**. It is important to continue to support these into the LSST era. Relatively rare, high-demand capabilities include (Table 11.3) the following:

- *Wide-field imaging on 3–5m telescopes, which is available in the Southern Hemisphere, e.g., through DECam on the Blanco 4m telescope at CTIO*
- *AO-fed diffraction-limited imaging and integral field spectroscopy, which is available through Gemini/NIFS*

Several science cases also make use of instrumentation that is currently standard on many facilities. Specific examples called out in the previous chapters include the following:

- *Transient sky (Chapter 4): low- to moderate-resolution, single-object optical spectrographs; single-object optical and NIR photometry*
- *Stellar rotation and activity (Chapter 5): single-object, multi-color imaging (< 5m); single-object R = 100–5000 optical spectroscopy (3–5m)*

Support costs for these capabilities include those associated with routine operations as well as timely repair and refurbishment.

Other high-priority capabilities are **currently in the planning stages**. The need for broad wavelength coverage, optical-infrared spectroscopy on an 8–10m telescope could be met by the Gen 4#3 instrument currently under development for Gemini (Table 11.3). Similarly, some of the need for optical high-resolution spectroscopy would be met by GHOST on Gemini (as well as other existing instruments such as HIRES on Keck). It is critical that development plans for these capabilities proceed in a timely way so that the capabilities are available when LSST operations begin.

*Table 11.3. Possible Implementation Pathways for High-Demand Capabilities*

| Capability | Telescope Aperture | | | Infrastructure |
|---|---|---|---|---|
| | 3–5 m | 8–10 m | > 25 m | |
| Optical Imager (Wide-field) | Blanco/DECam | | | |
| AO IFU R ~ 5000 | | **Gemini/NIFS** (larger FOV highly desired) | **Significant US participation in GSMT** | |
| OIR MOS R = 5000 0.35–1.3 micron | | **SSSI** in South, **Subaru/PFS, European MOS telescope** (South), **or MSE** (North) | **Significant US participation in GSMT** | |
| Optical SOS R = 1k–5k 0.35–2.5 micron | | **Gemini/Gen4#3** | | |
| Public alert broker system | | | | **Develop existing prototypes** |
| Target and observation manager software | | | | **Build on the experience from existing systems** |

Some high-priority capabilities are **not currently available** to the broad community, e.g., wide-field optical multi-object spectroscopy on 3–5m and 8–10m telescopes. Similar capabilities are available only to a restricted portion of the US community (e.g., PFS on Subaru). Given the long lead time to develop any new capability, there is an urgent need to investigate possible development pathways now so that the needed capabilities can be available *in* the LSST era (but not necessarily when LSST operations begin).

Possible pathways to this capability (Table 11.3) include the following:
- *Implementing a new wide-field, massively multiplexed optical spectrograph on a*

Southern Hemisphere 6–8m telescope (e.g., Magellan). Because such a capability (a "Southern Spectroscopic Survey Instrument," or SSSI) is a top priority in the US Department of Energy's Cosmic Visions reports (arxiv.org/abs/1604.07821, arxiv.org/abs/1604.07626), funding from the Department of Energy could be available for the construction of an instrument, or conceivably even an instrument/telescope combination. Community access to such a capability would enable the surveys described above to begin early in the LSST era. Implementation on a Northern Hemisphere telescope (e.g., Keck, Gemini, or a Magellan-like telescope on San Pedro Martir) would be less ideal due to the limited overlap with LSST but would be much better than lacking this capability entirely.

- *Obtaining US community access to the PFS instrument on the Subaru telescope in order to propose and execute new large surveys such as those described in this report. The advantage of utilizing PFS is that its specifications meet most community requirements, it is already in development, and it should be deployed before LSST imaging begins. However, there are also substantial disadvantages. Firstly, adequate time to carry out the extensive surveys described here may not be available on PFS, due to other demands on the Subaru telescope. Additionally, only part of the LSST footprint is accessible from Maunakea, limiting the capability of PFS to complete some of the surveys described in previous chapters. Community access to the telescope, not just to data from currently planned PFS projects, is vital, because the planned projects only address a minority of the scientific goals described in this report.*

- *Joining other highly multiplexed spectrograph projects that are currently under discussion or development. ESO has established a working group on the future of multi-object spectroscopy in the era of LSST and other missions (R. Ellis, chair). They are reported to be converging on a 10–12m telescope with a 5-degree$^2$ field-of-view and medium- and high-resolution spectroscopic modes; joining such a project could fulfill the community's needs for wide-field multi-object spectroscopy, if sufficient time could be obtained and the telescope built in a timely fashion. Similarly, the proposed 11m Maunakea Spectroscopic Explorer (MSE) telescope would be highly efficient at conducting the surveys described here, with survey speeds generally twice those provided by PFS. However, like PFS, MSE would be limited to the Northern half of the LSST footprint.*

## Parallel need for time domain–related infrastructure investments

As described in Chapter 9 (time domain infrastructure), in addition to having instruments *available* for follow-up programs, infrastructure developments are also critical to enable the community to *use* the instruments effectively, for both time domain science and the large static domain, follow-up programs that LSST data will enable. Important infrastructure includes flexible ways to apply for telescope time, queue-mode scheduling for critical facilities on a range of timescales, and modes of telescope control and data analysis.

Critical elements of the follow-up system include
- *a public alert broker system and data archive*
- *target and observation manager software capable of tracking large numbers of candidate targets, coordinating their follow-up observations, and compiling*

> *additional data in real time*
> - *an accessible network of facilities equipped and scheduled in modes that support time domain follow-up observations spanning a range of timescales, apertures, instrumentation, and geographical locations*
> - *general-purpose telescope scheduling software for time domain observations*
> - *software to produce quick-look (or better, fully reduced) data and support for archives to disseminate it*

Different science goals are likely to require a range of functionality from the above elements of the follow-up system, but currently the requirements on each are not well defined in many science areas. It is important to improve our understanding of these requirements in the near future to develop an effective follow-up system for LSST. To this end, we can learn from follow-up programs that respond to *current* survey discoveries. In the process, these efforts will develop the tools and techniques that will be essential to maximize the science return of LSST.

One lesson learned from ongoing efforts is that building a general-purpose alert broker that operates at the LSST scale and rate of alerts has challenges beyond the field of astronomy alone. It is a multi-disciplinary problem that encompasses astronomy and computer science.

Time domain science demands highly flexible and time-critical modes of observation. For some facilities, meeting these demands will require changes to procedures for time allocation, execution of observations, and data delivery. Further discussion with the operators of observing facilities is needed to develop workable, cost-efficient procedures. Options for remote- or robotic-observation modes can significantly increase the efficiency of time domain programs and should be explored. Some science goals require rapid reduction and dissemination of the resulting data products and, therefore, the development of data reduction pipelines and online archives.

To identify genuine candidates and filter out false positives, it will be critical to triage targets selected from LSST alerts (i.e., perform follow-up observations and analyses to confirm target classification and determine the next appropriate steps). Although triage programs increase the science yield for many programs, they are rarely supported by time allocation committees. It is important to identify ways to support triage activities. In addition, the total demand for follow-up observations may be reduced by consolidating the needs of multiple programs into a dedicated follow-up survey. Further work is required to explore the feasibility of this option.

Follow-up programs should avoid duplication of effort and the waste of telescope time on repeated observations wherever possible. The best way to achieve this is to facilitate and encourage coordination and data sharing among follow-up teams.

## Further work needed to understand and support computing requirements

Substantial computing resources will be required to enable the community to fully exploit LSST data products and to support the infrastructure necessary for LSST follow-up programs. Technologies evolve rapidly and data system capabilities will improve significantly during the development and operation of the survey. The community's science-driven requirements will also likely evolve as they prepare for LSST and adopt new technologies over time. It is therefore extremely important to regularly review data system capabilities.

The Big Data era will bring analysis techniques and technologies that are unfamiliar to many in astronomy. Community training programs are needed to bridge the gap. The scale of future datasets also presents an increasing workload for scientists in software, data handling, and pipeline reduction roles. These tasks are crucial to the overall science output of the community but are often undertaken by people in soft-money positions, impacting their research output and career prospects. Long-term career tracks for scientists specializing in these roles will be very valuable.

## Our high-priority capabilities overlap strongly with OIR System Report findings

This study, which investigated community needs from a purely science-based approach, comes to many of the same conclusions as the OIR System Report and endorses those findings. Our report also extends the discussion begun in the OIR System Report by providing example science cases that are closely connected to the science priorities of *New Worlds, New Horizons* and that are worked out in detail to illustrate the specific kinds of capabilities required and the level of demand for these.

**Recommendation #3** from the OIR System Report states that "the National Science Foundation should support the development of a wide-field, highly multiplexed spectroscopic capability on a medium- or large-aperture telescope in the Southern Hemisphere to enable a wide variety of science, including follow-up spectroscopy of Large Synoptic Survey Telescope targets. Examples of enabled science are studies of cosmology, galaxy evolution, quasars, and the Milky."

We find that wide-field highly multiplexed optical spectroscopy on 3–5m and 8–10m telescopes is a high-priority capability for LSST science. Among the science cases considered here, it is critical to LSST-based studies of dark matter, the Milky Way and the Local Group (Chapter 3), stellar rotation and activity (Chapter 5), galaxy evolution (Chapter 7), and cosmology (Chapter 8). Because the specific science cases considered here would alone require many years of a capability like PFS/Subaru to accomplish their science goals, they would greatly benefit from the development of a new spectroscopic capability.

**Recommendation #4a** from the OIR System Report states that "the National Science Foundation should help to support the development of event brokers, which should use standard formats and protocols, to maximize Large Synoptic Survey Telescope transient survey follow-up work."

Our study confirms that brokers are important to maximize LSST science.

**Recommendation #4b** from the OIR System Report states that "the National Science Foundation should work with its partners in Gemini to ensure that Gemini South is well positioned for faint-object spectroscopy early in the era of Large Synoptic Survey Telescope operations, for example, by supporting the construction of a rapidly configurable, high-throughput, moderate-resolution spectrograph with broad wavelength coverage."

We find that a high-throughput spectrograph with broad wavelength coverage (0.36–2.5 micron) and moderate resolution (R = 2000 or larger) is a high priority for LSST science. It is a critical capability for understanding the transient universe (Chapter 4), characterizing small bodies in the Solar System (6), and galaxy evolution (Chapter 7). The Gem4#3 instrument currently under development for Gemini is a valuable opportunity to provide this capability.

**Recommendation #4c** from the OIR System Report states that "the National Science Foundation should ensure via a robustly organized U.S. Optical and Infrared (OIR) System that a fraction of the U.S. OIR System observing time be allocated for rapid, faint transient observations prioritized by a Large Synoptic Survey Telescope event broker system so that high-priority events can be efficiently and rapidly targeted."

While we did not delve deeply into the question of *how* observing time should be allocated to support observations of transient phenomena, our study strongly supports the need for a robustly organized US OIR System (of telescopes, instruments, and other observing infrastructure) to maximize the science from LSST discoveries of transient phenomena.

**Recommendation #4d** from the OIR System Report states that "the National Science Foundation should direct its managing organizations to enhance coordination among the federal components of medium- to large-aperture telescopes in the Southern Hemisphere, including Gemini South, Blanco, the Southern Astrophysical Research (SOAR) telescope, and the Large Synoptic Survey Telescope (LSST), to optimize LSST follow-up for a range of studies."

Our study identifies potential roles for Gemini South, the CTIO Blanco, and SOAR in providing capabilities that will be critical for LSST science (Table 11.4).

As described above, the future Gemini instruments Gem 4#3 and GHOST, both of which are single-object spectrometers, could fulfill the needs for workhorse broad-wavelength, high-throughput spectroscopy and high-resolution optical spectroscopy. Because the former would be used to follow up time-critical phenomena, deploying the instrument early in the LSST mission is necessary to maximize LSST science. A basic, workhorse instrument, deployed early in the LSST mission, is greatly preferred to a multi-mode instrument that arrives later in the mission.

Continued wide-field imaging on the Blanco (DECam) would provide a high-priority capability. Because of its wide field-of-view, the Blanco is also a potential platform for highly multiplexed optical spectroscopy.

With its ability to support more than one instrument at a time, the SOAR telescope could play multiple roles in the LSST era:

- *Time domain follow-up, i.e., time-critical observations made in immediate response to LSST observations*
- *Complementary cadence observations, e.g., to supplement LSST observations of Solar System objects*
- *Targeted follow-up of individual objects, i.e., non-time-critical observations made at complementary wavelengths, resolutions, etc.*

*Table 11.4. Potential Roles for Federal Components of Southern Hemisphere Facilities*

| Telescope | Attributes | Potential Roles |
|---|---|---|
| Gemini South | **Large aperture** (8m)<br>**High angular resolution**<br>Small FOV | Broad-wavelength, modest resolution spectroscopy (Gen 4#3);<br>High-resolution optical spectroscopy (GHOST);<br>AO-fed IFU (NIFS) |
| SOAR | **Rapid slewing**<br>**Suite of OIR instruments always mounted**<br>Modest aperture (4.1m)<br>Small FOV (7 arcminutes) | Time domain follow-up;<br>Complementary cadenced observations;<br>Non-time critical follow-up of individual sources |
| CTIO Blanco | **Large FOV**<br>Modest aperture (4m) | Wide-field imaging |

**Recommendation #5** from the OIR System Report states that "the National Science Foundation should plan for an investment in one or both Giant Segmented Mirror Telescopes in order to capitalize on these observatories' exceptional scientific capabilities for the broader astronomical community in the Large Synoptic Survey Telescope era, for example, through shared operations costs, instrument development, or limited term partnerships in telescope or data access or science projects."

Our study identifies multiple important roles for GSMTs in maximizing LSST science. The science cases considered here place high priority on AO-fed integral field unit (IFU) imaging and spectroscopy, NIR imaging, NIR spectroscopy, and broad wavelength high-throughput spectroscopy with GSMTs.

**Recommendation #2** from the OIR System Report states that "NSF should direct NOAO to administer an ongoing community-wide planning process to identify the critical Optical and Infrared System capabilities needed in the near term to realize the decadal science priorities. NOAO could facilitate the meeting of a system organizing committee, chosen to represent all segments of the community, which would produce the prioritized plan. NSF would then solicit, review, and select proposals to meet those capabilities, within available funding."

Our study takes some initial steps in this direction, having carried out a community-based study of the OIR capabilities that are needed to realize decadal science in connection with the use of LSST.

# Recommendations

Our study's key findings lead to the following recommendations, grouped into 3 bins by importance and development status of the resource.[33] The 3 bins are listed in priority order below. The recommendations relate to the capabilities that were found to have particularly high priority and high demand, typically from multiple communities (Tables 11.1 and 11.2). A more complete census of capabilities motivated by this study and their demand can be found in the Findings section of this chapter and in the individual science chapters.

Taking full advantage of LSST data also entails OIR system infrastructure developments as well as computing and analysis resources. We investigated these topics less fully than the telescope- and instrument-related resources, and further work is needed to specify these needs. We make several recommendations along these lines in a fourth bin below.

## Critical resources in need of a prompt development path

**Develop or obtain access to a highly multiplexed, wide-field optical multi-object spectroscopic capability on an 8m-class telescope, preferably in the Southern Hemisphere.** This high-priority, high-demand capability is not currently available to the broad US community. Given the long lead time to develop any new capability, there is an urgent need to investigate possible development pathways now, so that the needed capabilities can be available in the LSST era. Possibilities include implementing a new wide-field, massively multiplexed optical spectrograph on a Southern Hemisphere 6–8m telescope, e.g., as in the Southern Spectroscopic Survey Instrument, a project recommended for consideration by the DOE's Cosmic Visions panel (arxiv.org/abs/1604.07626 and arxiv.org/abs/1604.07821); open access to the PFS instrument on the Subaru telescope in order to propose and execute new large surveys; alternatively, joining an international effort to implement a wide-field spectroscopic survey telescope (e.g., the Maunakea Spectroscopic Explorer at CFHT or a future ESO wide-field spectroscopic facility) if the facility will deliver data well before the end of the LSST survey.

## Critical resources that have a development path

**Deploy a broad wavelength coverage, moderate-resolution (R = 2000 or larger) OIR spectrograph on Gemini South.** The Gen 4#3 instrument is an ideal opportunity. It is critical that development plans for these capabilities proceed in a timely way so that the capabilities are available when LSST operations begin. A basic, workhorse instrument, deployed early in the LSST mission, is greatly preferred to a multi-mode instrument that arrives later in the mission. A wavelength range of at least 0.36–2.5 microns would provide the highest scientific impact.

**Ensure the development and early deployment of an alert broker, scalable to LSST.** Public broker(s), and supporting community data and filtering resources, are essential to select priority targets for follow-up. The development of an alert broker that can process the LSST alert stream has challenges beyond the field of astronomy alone. These challenges can be effectively tackled by computer scientists working with astronomers on this multi-disciplinary problem, and support is needed to enable effective collaboration across the relevant fields.

---

[33] Although this study found strong synergy between LSST and GSMT, no GSMT-related recommendations are included here because they are beyond the scope of our formal charge (see Appendices A and B). The Findings section of this chapter illustrates the need for GSMT capabilities, and further details are given in Chapters 3–8.

## Critical resources that exist today

**Support into the LSST era high-priority capabilities that are currently available.** Wide-field optical imaging (e.g., DECam on the Blanco 4m at CTIO) is one valuable, but relatively uncommon, capability, as is AO-fed diffraction limited imaging (e.g., NIFS on the 8m Gemini telescope). Other important capabilities are standard on many facilities. Those called out in this report include

- *single-object, multi-color imaging on < 5m facilities*
- *single-object R = 100–5000 spectroscopy on 3–5m facilities*

Support costs for these capabilities include those associated with routine operations as well as timely repair and refurbishment.

## Infrastructure resources and processes in need of timely development

**Support OIR system infrastructure developments that enable efficient follow-up programs.** Two of LSST's strengths are the large statistical samples it will produce and LSST's ability to provide rapid alerts for a wide variety of time domain phenomena. An efficient OIR system can capitalize on these strengths by (i) developing target and observation management software and increasing the availability of (ii) follow-up telescopes accessible in queue-scheduled modes, as well as (iii) data reduction pipelines that provide rapid access to data products. Following up large samples will be time and cost prohibitive if on-site observing is required and/or large programs and triage observations are not part of the time allocation infrastructure. To develop and prioritize community needs along these lines, we recommend a study aimed at developing a follow-up system for real-time, large-volume, time domain observations. As part of this study, discussions with the operators of observing facilities (e.g., through targeted workshops) are important in developing workable, cost-efficient procedures.

**Study and prioritize needs for computing, software, and data resources.** LSST is the most data-intensive project in the history of optical astronomy. To maximize the science from LSST, support is needed for (i) the development and deployment of data analysis and exploration tools that work at the scale of LSST; (ii) training for scientists at all career stages in LSST-related analysis techniques and computing technologies; (iii) cross-disciplinary workshops that facilitate the cross-pollination of ideas and tools between astronomy and other fields. We recommend a follow-on systematic study to prioritize community needs for computing, software, and data resources. The study should account for the capabilities that will be delivered by the LSST project and other efforts, the demands of forefront LSST-enabled research, and the opportunities presented by new technology.

**Continue community planning and development.** It is critical to continue the community-wide planning process, begun here, to motivate and review the development of the ground-based OIR System capabilities that will be needed to maximize LSST science. The current study focused primarily on instrumentation. Further work is needed to define the needs for observing infrastructure and computing, as described above. Regular review of progress (and lack thereof) in all of these areas is important to ensure the development of an OIR System that does maximize LSST science. Studies like these form the basis for a development roadmap and take a step in the direction envisioned by the Elmegreen committee that "a system organizing committee, chosen to represent all segments of the community ... would produce the prioritized plan. NSF would then solicit, review, and select proposals to meet those capabilities, within available funding."

# Appendix A: Relevant OIR System Study Recommendations

In April 2015, a National Research Council committee led by Debbie Elmegreen (Vassar) delivered the report, *Optimizing the U.S. Ground-Based Optical and Infrared System in the Era of LSST*. The Kavli Foundation–funded study and workshop described in the present document were motivated by the need to follow up on Recommendations 2, 3, and 4a–d of this OIR System Report. Our findings also connect with Recommendation 5 of this report.

### Recommendation 2

NSF should direct NOAO to administer an ongoing community-wide planning process to identify the critical Optical and Infrared System capabilities needed in the near term to realize the decadal science priorities. NOAO could facilitate the meeting of a system organizing committee, chosen to represent all segments of the community, which would produce the prioritized plan. NSF would then solicit, review, and select proposals to meet those capabilities, within available funding.

### Recommendation 3

The National Science Foundation should support the development of a wide-field, highly multiplexed spectroscopic capability on a medium- or large-aperture telescope in the Southern Hemisphere to enable a wide variety of science, including follow-up spectroscopy of Large Synoptic Survey Telescope targets. Examples of enabled science are studies of cosmology, galaxy evolution, quasars, and the Milky Way.

### Recommendation 4a

The National Science Foundation should help to support the development of event brokers, which should use standard formats and protocols, to maximize Large Synoptic Survey Telescope transient survey follow-up work.

### Recommendation 4b

The National Science Foundation should work with its partners in Gemini to ensure that Gemini South is well positioned for faint-object spectroscopy early in the era of Large Synoptic Survey Telescope operations, for example, by supporting the construction of a rapidly configurable, high-throughput, moderate-resolution spectrograph with broad wavelength coverage.

### Recommendation 4c

The National Science Foundation should ensure via a robustly organized U.S. Optical and Infrared (OIR) System that a fraction of the U.S. OIR System observing time be allocated for rapid, faint transient observations prioritized by a Large Synoptic Survey Telescope event broker system so that high-priority events can be efficiently and rapidly targeted.

### Recommendation 4d

The National Science Foundation should direct its managing organizations to enhance coordination among the federal components of medium- to large-aperture telescopes in the Southern Hemisphere, including Gemini South, Blanco, the Southern Astrophysical Research (SOAR) telescope, and the Large Synoptic Survey Telescope (LSST), to optimize LSST follow-up for a range of studies.

### Recommendation 5

The National Science Foundation should plan for an investment in one or both Giant Segmented Mirror Telescopes in order to capitalize on these observatories' exceptional scientific capabilities for the broader astronomical community in the Large Synoptic Survey Telescope era, for example, through shared operations costs, instrument development, or limited term partnerships in telescope or data access or science projects.

# Appendix B: Letter of Request from NSF/AST

August 26, 2015

Dr. C. M. Mountain, President, Association of Universities for Research in Astronomy
Dr. D. Silva, Director, National Optical Astronomy Observatory
Dr. S. Kahn, Director, Large Synoptic Survey Telescope Project

Dear Matt, Dave, and Steve:

Pursuant to the recognition of a compelling need to develop a strategic plan for OIR astronomy in the era of LSST, the NSF Division of Astronomical Sciences (AST) commissioned a National Research Council report, entitled "Optimizing the U.S. Ground-Based Optical and Infrared System," which was delivered in April 2015 (hereafter referred to as the OIR System Report).

The OIR System Report includes recommendations for a number of significant activities to be carried out by the National Optical Astronomy Observatory (NOAO). All of the Report recommendations and conclusions directed at NOAO, and several others as well, are within the scope of NOAO's purpose and mission. In principle, AST agrees with all instances in the Report which state that NSF should direct NOAO to carry out a specific task.

In this letter, AST asks NOAO in collaboration with the LSST Project Office (LSSTPO), under the management of Association of Universities for Research in Astronomy (AURA), to pursue an initial implementation of OIR System Report recommendations 2, 3, and 4a-d through a joint NOAO-LSST workshop on fully realizing LSST-enabled science. In accordance with a draft description for the workshop that was provided by NOAO and LSSTPO, we expect the primary goal of the workshop would be to develop a report, organized by six to eight representative science cases for LSST, that quantifies the resources needed to accomplish each case in the era of LSST. The report would assess and prioritize potential OIR System resources such as telescope apertures, wavelength ranges, instrument capabilities, number of observing nights, software and computing power, and data management and data product service to the community. The report would then highlight ways that existing and planned resources could be positioned to accomplish the science goals, and it will identify high priority future investments for OIR infrastructure (including IT infrastructure). For example, this may include suggestions for efficient cross-disciplinary implementation such as specific multiple programs that could efficiently be conducted simultaneously on massively multiplexed spectrographs, partnerships between facilities, and/or data sharing.

AST fully endorses the workshop that has been described. NSF understands that a proposal will be submitted to the Kavli Foundation to support the workshop.

Nigel Sharp, [nsharp@nsf.gov](mailto:nsharp@nsf.gov), and Vernon Pankonin, [vpankoni@nsf.gov](mailto:vpankoni@nsf.gov), will be the NSF/AST points of contact for this activity.

Sincerely,

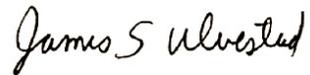

James S. Ulvestad
Division Director

cc: V. Pankonin, N. Sharp

# Appendix C: Additional Resource Needs

## Requirements on LSST Deep Drilling Fields

The galaxy evolution study report (Chapter 7) identified the following needs related to the LSST deep drilling fields (DDFs):

1. The CGM/IGM science case requires deep *u*-band imaging to the same limits as the DDF. This is necessary to identify faint *u*-band dropout galaxies that could be associated with absorption systems. If the best field for IGM studies—as determined by the number of bright background sources—is in the wide-field LSST survey, the addition *u*-band imaging to a 5-σ depth of 28.0 mag would be necessary.

2. The galaxy evolution survey requires imaging to ~ 2.5 mag deeper than the spectroscopic selection, i.e., to 27.5 magnitude in the *zY* bands and to 28 magnitude in *ugri*. The survey will make spectroscopic observations in at least five widely separated locations across the sky to control for cosmic variance. Therefore, we need at least 5 extragalactic DDFs with 5-σ magnitude limits of 28 in *ugri* and 27.5 in *zY*.

3. The SMBH demographics program requires daily cadence for one year on a single DDF. Not every band needs to be completed each day as long as there are 2–3 bands each night. Significant losses in the number of accretion disks that can be analyzed happen only if there are a significant number of days where no data is taken: a slower three-day cadence would roughly halve the number of detected accretion-disk lags. Expanding the daily cadence for five years would also expand the number of accretion-disk lags by ~25%. But if there is a trade-off between cadence and duration, daily cadence for one year is definitely preferred over a five-day cadence for five years.

The stellar rotation and magnetic activity study described in Chapter 5 requires that several DDFs focus on specific open clusters, including the iconic solar age cluster M67, which lies just outside the nominal LSST footprint.

## Requirements on LSST Level 3 Data Products

The stellar rotation and magnetic activity study described in Chapter 5 requires that the individual 15-second LSST images be made available as Level 3 data products.